\documentclass[twocolumn]{aastex62}

\usepackage[binary-units]{siunitx}
\usepackage{enumitem}
\usepackage{graphicx}
\usepackage{bm}
\usepackage{mathtools}
\usepackage{siunitx}
\usepackage{threeparttable}
\usepackage{rotating}

\graphicspath{{./}}

\shorttitle{AMISS I}
\shortauthors{Keenan et al.}

\begin{document}

\title{The Arizona Molecular ISM Survey with the SMT: Survey Overview and Public Data Release}

\correspondingauthor{R. P. Keenan}
\author[0000-0003-1859-9640]{Ryan P. Keenan}
\altaffiliation{NSF Graduate Research Fellow}
\affiliation{Steward Observatory, University of Arizona, 933 North Cherry Avenue, Tucson, AZ 85721, USA}
\email{rpkeenan@arizona.edu}

\author[0000-0002-2367-1080]{Daniel P. Marrone}
\affiliation{Steward Observatory, University of Arizona, 933 North Cherry Avenue, Tucson, AZ 85721, USA}

\author[0000-0002-3490-146X]{Garrett K. Keating}
\affiliation{Center for Astrophysics, Harvard \& Smithsonian, 60 Garden Street, Cambridge, MA 02138, USA}

\author[0000-0001-6439-8140]{Evan C. Mayer}
\affiliation{Steward Observatory, University of Arizona, 933 North Cherry Avenue, Tucson, AZ 85721, USA}

\author{Kevin Bays}
\affiliation{Arizona Radio Observatory, 933 North Cherry Avenue, Tucson, AZ 85721, USA}

\author{John Downey}
\affiliation{Arizona Radio Observatory, 933 North Cherry Avenue, Tucson, AZ 85721, USA}

\author{Lochlann C. Dunn}
\affiliation{Arizona Radio Observatory, 933 North Cherry Avenue, Tucson, AZ 85721, USA}

\author{Joanne C. Flores}
\affiliation{Arizona Radio Observatory, 933 North Cherry Avenue, Tucson, AZ 85721, USA}

\author{Thomas W. Folkers}
\affiliation{Arizona Radio Observatory, 933 North Cherry Avenue, Tucson, AZ 85721, USA}

\author{David C. Forbes}
\affiliation{Arizona Radio Observatory, 933 North Cherry Avenue, Tucson, AZ 85721, USA}

\author{Blythe C. Guvenen}
\affiliation{Arizona Radio Observatory, 933 North Cherry Avenue, Tucson, AZ 85721, USA}

\author{Christian Holmstedt}
\affiliation{Arizona Radio Observatory, 933 North Cherry Avenue, Tucson, AZ 85721, USA}

\author{Robert M. Moulton}
\affiliation{Arizona Radio Observatory, 933 North Cherry Avenue, Tucson, AZ 85721, USA}

\author{Patrick Sullivan}
\affiliation{Arizona Radio Observatory, 933 North Cherry Avenue, Tucson, AZ 85721, USA}

\begin{abstract}

The CO(1--0) line has been carefully calibrated as a tracer of molecular gas mass. However, recent studies often favor higher $J$ transitions of the CO molecule which are brighter and accessible for redshift ranges where CO(1--0) is not. These lines are not perfect analogues for CO(1--0), owing to their more stringent excitation conditions, and must be calibrated for use as molecular gas tracers. Here we introduce the Arizona Molecular ISM Survey with the SMT (AMISS), a multi-CO line survey of $z\sim0$ galaxies conducted to calibrate the CO(2--1) and CO(3--2) lines. The final survey includes CO(2--1) spectra of 176 galaxies and CO(3--2) spectra for a subset of 45. We supplement these with archival CO(1--0) spectra from xCOLD~GASS for all sources and additional CO(1--0) observations with the Kitt Peak 12m Telescope. Targets were selected to be representative of the $10^9\,{\rm M}_\odot\leq M_* \leq 10^{11.5}\,{\rm M}_\odot$ galaxy population. Our project emphasized careful characterization of statistical and systematic uncertainties to enable studies of trends in CO line ratios. We show that optical and CO disk sizes are on average equal, for both the CO(1--0) and CO(2--1) line. We measure the distribution of CO line luminosity ratios, finding medians (16th--84th percentile) of 0.71 (0.51--0.96) for the CO(2--1)-to-CO(1--0) ratio, 0.39 (0.24--0.53) for the CO(3--2)-to-CO(1--0) ratio, and 0.53 (0.41--0.74) for the CO(3--2)-to-CO(2--1) ratio. A companion paper presents our study of CO(2--1)'s applicability as a molecular gas mass tracer and search for trends in the CO(2--1)-to-CO(1--0) ratio. Our catalog of CO line luminosities is publicly available.

\end{abstract}



\section{Introduction} \label{sec:intro}

Millimeter-wavelength spectral lines provide a trove of information about cold interstellar gas. In particular, millimeter-wavelength rotational lines of the CO molecule allow us to study the cold, dense molecular phase of the interstellar medium (ISM) that fuels star formation \citep{kennicutt+12,bolatto+13,saintonge+22}. Surveys of the 2.6~mm CO $J=1\rightarrow0$ rotational transition (hereafter CO(1--0)) have helped uncover the molecular gas properties of the Milky Way \citep{dame+85,dame+01,dame+22}, local galaxies \citep{kennicutt98,saintonge+11a,saintonge+17,bolatto+17,sorai+19,wylezalek+22}, and the high redshift universe \citep{ivison+11,daddi+15,pavesi+18,riechers+20}. The CO(1--0) line has a low excitation temperature and critical density, and therefore arises in any environment where the CO molecule is present. Extensive work has gone into calibrating the CO(1--0) luminosity to molecular gas mass conversion factor \citep{bolatto+13}, and how it varies in different environments \citep{downes+98,genzel+12,sandstrom+13,carleton+17,accurso+17,gong+20,dunne+22}. 

In practice, modern facilities can often detect the 1.3~mm CO $J=2\rightarrow1$ (CO(2--1)) line with greater efficiency than CO(1--0) leading to a growing number of studies focused exclusively on this line \citep[e.g.,][]{leroy+09,bothwell+14,cairns+19,colombo+20,leroy+22}. CO(2--1) and higher transitions are even more prevalent in high redshift studies, where CO(1--0) is very faint and cannot be observed from the ground in certain redshift intervals \citep{daddi+10,bauermeister+13b,carilli+13,tacconi+13,freundlich+19,boogaard+19,valentino+20,liu+21}.

Lines higher in the ladder of CO rotational transitions require steeply increasing energy and density to excite. The ratios of flux in these lines depend on density, temperature, and optical depth of the molecular clouds where the emission originates \citep{gong+20,kamenetzky+17,leroy+17}. Nonetheless, many extragalactic studies use CO(2--1) and CO(1--0) emission as interchangeable tracers of molecular gas mass. Resolved studies of CO(1--0) to CO(2--1) ratios in the Milky Way \citep{sakamoto+97,sawada+01} and nearby galaxies \citep{leroy+09,koda+12,vlahakis+13,koda+20,denbrok+21,yajima+21,leroy+22} are increasingly suggesting this assumption is not valid. 

As CO(2--1) becomes a fixture of millimeter astronomy, it is pertinent to better understand the connection between the well-calibrated CO(1--0) line and the more observationally convenient lines at higher $J$. To this end, we present the Arizona Molecular ISM Survey with the SMT (AMISS). AMISS is a survey of CO(2--1) and CO(3--2) emission from galaxies with existing CO(1--0) measurements from the extended CO Legacy Data base for the GASS \citep[xCOLD~GASS;][]{saintonge+11a,saintonge+17}. AMISS was conducted with two primary aims:
\begin{enumerate}[noitemsep,nolistsep]
    \item To determine the relationships between galaxy-integrated flux of the CO(1--0), CO(2--1) and CO(3--2) emission lines and measure how the ratios between these lines depend on other galaxy properties;
    \item To understand how CO(2--1) traces the total molecular gas abundance of galaxies and utilize this line to characterize the molecular gas properties of a large sample of local galaxies for comparison to high redshift studies.
\end{enumerate}
These aims necessitate a large, homogeneous set of observations covering multiple CO transitions and a wide range of galaxy properties. \citet{denbrok+21} and \citet{leroy+22} find that compilations of archival data from multiple sources are fundamentally limited by differences in calibration, which can introduce systematic uncertainties comparable in magnitude to trends between CO line ratios and galaxy properties. AMISS is designed to utilize one telescope per line -- the IRAM 30m for CO(1--0) and the Arizona Radio Observatory's Submillimeter Telescope (SMT) for CO(2--1) and CO(3--2) -- with all observations for a given line conducted under a standardized set of procedures and reduced in a consistent manner. This eliminates most systematic effects in the data and allows for straightforward computation of line ratios. 

This large, homogeneous set of multi-line data makes AMISS unique among the growing number of large extragalactic CO surveys. These data will serve as a guide for translating between surveys conducted in CO(1--0) and CO(2--1) or CO(3--2), and a low redshift anchor for high redshift molecular gas studies which use different CO lines for different redshift ranges.

In this paper we present the AMISS sample and describe the construction of our multi-$J$ CO line catalog. In a companion paper we use this data to study variations in CO line excitation across the star forming galaxy population \citep{keenan+24b}. A future paper will use the completed survey to determine the redshift zero molecular gas abundance using CO(2--1). 

The remainder of this paper is structured as follows: in Section~\ref{sec:survey} we describe our survey design and target selection. In Section~\ref{sec:obs} we describe our observations and in Section~\ref{sec:data} we describe the processing and analysis of the data. We describe our synthesis of AMISS data with the xCOLD~GASS CO(1--0) data in Section~\ref{sec:xcg}. Because careful accounting of the uncertainty in our line fluxes is necessary for many of the core science goals of AMISS, in Section~\ref{sec:errors} we  characterize the sources of statistical and systematic error in our CO line fluxes. Section~\ref{sec:results} presents our spectra, a catalog of CO line luminosities, and the public release of our data. Section~\ref{sec:results} also uses the AMISS data to determine the average CO emission disk profile, assess the prevalence of extended molecular gas disks falling outside IRAM 30m beam, and determine typical ratios amongst the three lowest energy CO lines. We conclude in Section~\ref{sec:conclusion}. Throughout this work we assume a flat $\Lambda$CDM cosmology with $H_0=70$ km s$^{-1}$ Mpc$^{-1}$ and $\Omega_m=0.3$.


\section{Survey Description}\label{sec:survey}
\begin{figure*}
    \centering
    \includegraphics[width=\textwidth]{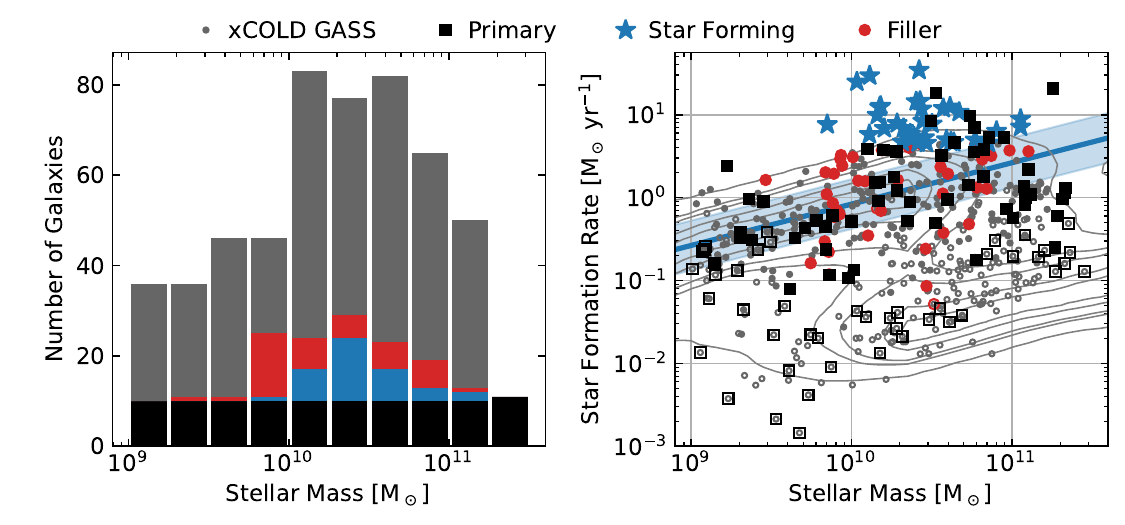}
    \caption{Left: the stellar mass distributions of the various components of our CO(2--1) sample are shown in black (primary sample), blue (star forming sample), and red (filler), along with the distribution of the xCOLD~GASS parent sample in gray. Right: the samples from the left panel are shown distributed across the stellar mass-SFR plane. Filled markers indicate that a source is detected in CO(2--1) (for the AMISS targets) or CO(1--0) (for the xCOLD~GASS targets) with a signal to noise ratio of at least 3. Open markers indicate non-detections. Contours show the distribution of SFR at a given mass for all SDSS galaxies at $z<0.05$. The blue line and filled region show the $z\sim0$ main sequence, as defined by \citet{speagle+14}.}
    \label{fig:samp21}
\end{figure*}

\begin{figure}
    \centering
    \includegraphics[width=.45\textwidth]{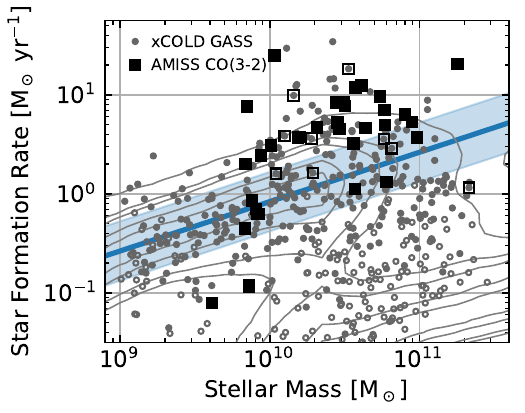}
    \caption{Black squares show the distribution of galaxies targeted by AMISS for CO(3--2) observations in stellar mass and SFR. Filled markers indicate that a source is detected in CO(3--2), while open markers indicate non-detections. The remaining plot elements are described in Figure~\ref{fig:samp21}}
    \label{fig:samp32}
\end{figure}

\begin{figure}
    \centering
    \includegraphics[width=.45\textwidth]{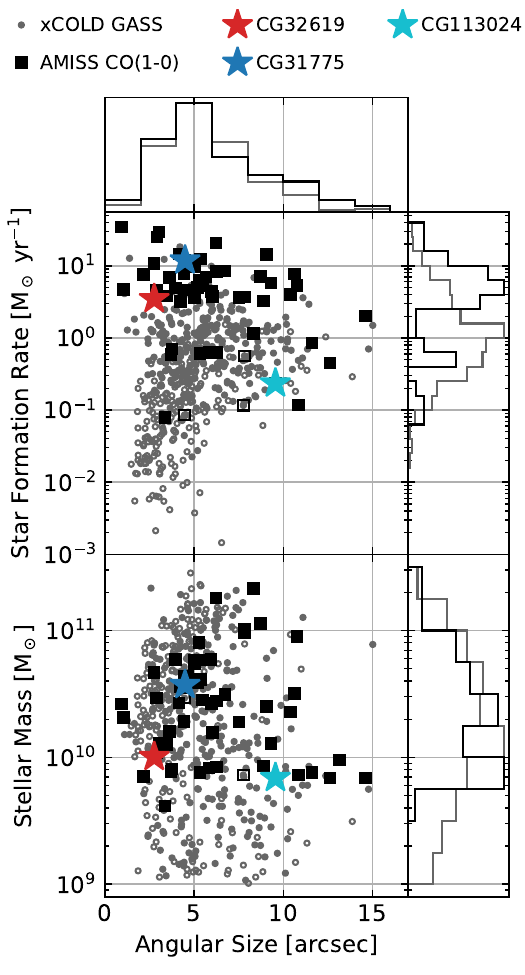}
    \caption{Larger panels show the joint distribution of optical half light radius with star formation rate (center left) and stellar mass (lower left) for the xCOLD~GASS (gray circles) and AMISS (black squares) CO(1--0) targets. Filled points correspond to detections, while open points correspond to galaxies undetected by xCOLD~GASS. We highlight the locations of xCOLD~GASS galaxies 31775 and 32619, which were independently observed with the ARO 12m by \citet{wylezalek+22}, with blue and red stars. We also highlight xCOLD~GASS 113024, which showed a significant discrepancy between the xCOLD~GASS CO(1--0) and AMISS CO(1--0) data. The histograms at the edge of each axis show the normalized distribution of the CO(1--0) detected galaxies from xCOLD~GASS (gray) and AMISS (black).
    \label{fig:samp10}}
\end{figure}

Targets for the AMISS survey were selected from a parent sample of 532 galaxies with CO(1--0) observations from xCOLD~GASS \citep{saintonge+11a,saintonge+17}. xCOLD~GASS was designed to produce a representative sample of CO(1--0) measurements for galaxies with stellar masses above $10^9$~M$_\odot$. These data have been used to study the scaling of CO(1--0) emission with galaxy properties \citep{saintonge+11a,saintonge+11b,saintonge+16,saintonge+17} and determine the cosmic molecular gas abundance in the local universe \citep{fletcher+21}. All xCOLD~GASS observations were conducted with a single telescope -- the IRAM 30m -- and a uniform observing protocol, providing a homogeneous set of CO(1--0) measurements to build upon. For 68\% of xCOLD~GASS targets simultaneous CO(2--1) observations were obtained using the multi-band functionality of the 30m receiver. Because the 30m beam is four times smaller at 230~GHZ than 115~GHz, these measurements cover only the centers of many galaxies. A wide range of ancillary data is also available for these sources, including galaxy properties from the Sloan Digital Sky Survey \citep{kauffmann+03,brinchmann+04}, star formation rates from \citet{janowiecki+17}, and HI data from the extended GALEX Arecibo SDSS Survey \citep[xGASS][]{catinella+18}. 

We proceeded to observe CO(2--1) and CO(3--2) in subsets of the xCOLD~GASS galaxies using the Arizona Radio Observatory's (ARO) Submillimeter Telescope\footnote{\href{https://aro.as.arizona.edu/?q=facilities/submillimeter-telescope}{https://aro.as.arizona.edu/?q=facilities/submillimeter-telescope}} (SMT) located on Mt. Graham, Arizona. As the 115 GHz CO(1--0) data comes from the IRAM 30m telescope, the 10m SMT provides a matched beam size for CO(3--2) at 345 GHz. This eliminates aperture corrections as a source of uncertainty when calculating line ratios. For the 230 GHz CO(2--1) line, the SMT beam is larger than either the CO(1--0) or CO(3--2) beams, however for most targets the expected gas disk is small relative to the size of the beam, and the needed aperture corrections small. We describe our aperture correction procedure in detail in Sections~\ref{ss:errors-beam} and \ref{ss:results-disk}. Sixty-one percent of our targets also have IRAM 30m CO(2--1) spectra from xCOLD~GASS covering the inner part of the galaxy.

In addition, we obtained new CO(1--0) observations for a subset of targets using the ARO's 12m ALMA Prototype Antenna\footnote{\href{https://aro.as.arizona.edu/?q=facilities/uarizona-aro-12-meter-telescope}{https://aro.as.arizona.edu/?q=facilities/uarizona-aro-12-meter-telescope}} located on Kitt Peak, Arizona. The 12m telescope provides a much larger beam than that of the IRAM 30m, which allows us to check for flux missed by the original xCOLD~GASS observations. These observations were motivated by the recent finding of \citet{wylezalek+22} that xCOLD~GASS source 31775 has significant flux missed by the 30m beam, and their suggestion that this may be a common phenomenon. These data allow us to validate our aperture correction procedure, and ensure our measured CO line ratios are not biased by missing CO(1--0) flux in the xCOLD~GASS dataset.

\subsection{CO(2--1) Sample}\label{ss:samp21}

Our primary survey consists of CO(2--1) observations of 101 galaxies selected by dividing the xCOLD~GASS sample into bins of stellar mass, spaced by $0.25$~dex from $10^9$ to $10^{10.5}$~M$_\odot$, and randomly drawing 10 objects from each bin for CO(2--1) follow-up\footnote{Because the highest mass bin contained only 11 total galaxies, all 11 of these sources were included, resulting in a final sample of 101 galaxies instead of 100.}. As the parent sample was also constructed on the basis of mass, this procedure results in a simple selection function dependent only on the stellar mass of each galaxy. Therefore, applying the appropriate weights to objects from each mass bin makes it possible to re-construct distribution functions for other galaxy properties, including the CO luminosity function \citep{saintonge+17,fletcher+21}. We refer to these 101 objects as our ``primary sample''.

Early in our survey it became evident that the CO(2--1)/CO(1--0) line ratio was correlated with star formation rate (SFR). To improve the sampling of the high-SFR end of this trend, we also observed a ``star forming sample'', consisting of all xCOLD~GASS galaxies with SFRs greater than $4.5$~M$_\odot$~yr$^{-1}$ not already included in our primary survey. This sample consisted of 34 additional objects. A further 8 objects met the SFR criterion but had already been included in the primary sample.

Finally, our survey also includes CO(2--1) observations for an assortment of 39 filler targets not selected for either sub-sample. These observations were taken when no higher priority targets could be observed. Typically filler targets were selected because they had bright CO(1--0) emission, leading to short estimated observing times for a CO(2--1) detection. Later in the survey, a few fainter targets with low SFRs were also observed to further fill out the SFR parameter space.

Figure~\ref{fig:samp21} shows the distribution of our sample along with the parent xCOLD~GASS sample. The sample spans a stellar mass range from $10^9$ to $10^{11.5}$  M$_\odot$ and an SFR range from $10^{-3}$ to $10^{1.5}$ M$_\odot$~yr$^{-1}$. We detect CO(2--1) in most targets on and above the main sequence, with detections extending into the green valley. However few quiescent galaxies are detected in either AMISS or the xCOLD~GASS sample.

\subsection{CO(3--2) Sample}\label{ss:samp32}

We initially planned to observe our primary CO(2--1) sample in the CO(3--2) transition as well. However, there was insufficient time with submillimeter observing conditions during our survey to achieve this goal. Instead we targeted a subset of our CO(2--1) targets, selected primarily to be detectable in modest integration times (typically less than 10 hours). Our final sample includes a total of 45 galaxies with CO(3--2) observations. Figure~\ref{fig:samp32} shows the distribution of our CO(3--2) targets in the stellar mass-SFR plane.

\subsection{CO(1--0) Sample}\label{ss:samp10}

For our CO(1--0) targets, we estimated the integration time required to detect the CO flux reported by xCOLD~GASS at 5 sigma, and proceeded to observe any object that 1) had been observed in our CO(2--1) survey as of April 2022 and 2) could be detected in less than 1 hour. This resulted in a sample of 38 galaxies. We also re-observed the two xCOLD~GASS sources previously observed with the 12m by \citet{wylezalek+22} -- xCOLD~GASS IDs 32619 and 31775 -- to allow direct comparison with their results, along with two fainter targets from our CO(2--1) sample which showed evidence of discrepancies between CO(1--0) and CO(2--1) in either line profile or integrated flux. Finally we observed a handful of additional sources -- mostly bright sources observed in CO(2--1) after April 2022 -- bringing our total 12m-CO(1--0) sample to 56 galaxies.

Figure~\ref{fig:samp10} shows the distribution of r-band half light radii of the xCOLD~GASS sample and the AMISS CO(1--0) targets. Both samples span a wide range of optical sizes, which correlate with CO disk diameter \citep{leroy+09}. The selection based on CO(1--0) flux biases the AMISS sample towards objects with high SFRs, but otherwise the distribution is comparable in angular size and stellar mass to xCOLD~GASS.


\section{Observations}\label{sec:obs}
We observed CO(2--1) and CO(3--2) using the SMT over the course of six semesters: March 2020--June 2020, October 2020--February 2021, March 2021--June 2021, March 2022--June 2022, October 2022--February 2023 and March 2023--June 2023. We observed CO(1--0) with the ARO 12m telescope from February to April 2022, with a short follow-up in May 2023.

\subsection{CO(2--1) Observations}\label{ss:obs21}

CO(2--1) observations were carried out with the SMT's facility 1.2~mm receiver, with central frequencies ranging from 219.6 to 228.3 GHz. We used the Forbes Filterbank (FFB) backend consisting of four 500 MHz filterbanks. We used two filterbanks per polarization, arranged in series to cover a 1~GHz bandpass for all observations. Sources were observed in the lower sideband with the receiver typically tuned to place the line near the center of the bandpass. Some narrower lines were observed offset from the center to facilitate observations of multiple sources with a single tuning. The 1.2~mm receiver is sideband-separating, with typical rejections better than 15~dB, and no bright lines fall in the image sideband, therefore image contamination should not be a concern for the CO(2--1) spectra.

Targets were observed in ``beam switching'' mode, moving the nutating secondary mirror between $+2$ and $-2$~arcminutes at a rate of 2.5 or 2.6~Hz\footnote{In observations during December 2022 a resonance between a mechanical element in the cryostat and the 2.5~Hz switch rate resulted in standing waves in the data, and alternative switch rates of either 1.0~Hz, 2.4~Hz, or 2.6~Hz were explored. Data from January 2023 onward were collected with a 2.6~Hz switch rate.} and moving the source between the $+2$ and $-2$ arcminute beam every 90 seconds. Telescope pointing was checked every two hours using bright line or continuum sources, and focus was checked approximately every 12 hours when sufficiently bright sources were available. When no focus measurement was possible a standard value was used. The focus of the SMT is sufficiently stable that these standard values will result in less than 2\% reductions in flux \citep{keenan23thesis}. Data were not corrected for optical losses, which are instead accounted for in the flux calibration step (Section~\ref{ss:data21-fluxcal}) by point source sensitivities derived from observations taken in the same observing mode.

Integration times were six minutes per scan, and total observation times ranged from approximately one hour for bright sources to tens of hours for the faintest sources. Each source was observed either until the CO(2--1) line was detected with a signal to noise ratio greater than 3 or a $3\sigma$ upper limit of $L_{\rm CO(2--1)}<10^{8}$ K km s$^{-1}$ was reached.\footnote{We targeted detection SNRs greater than 5, however accounting for baseline uncertainties reduces our SNR below this target. Final SNRs of 5 (4) or better were achieved for 67\% (91\%) of detected detected galaxies.} Targets where a CO(1--0) line was detected by xCOLD~GASS, or where some evidence ($<3\sigma$) of the CO(2--1) line was seen at the $10^8$ K km s$^{-1}$ limit were typically observed to deeper limits. Due to time constraints at the end of our survey, four faint targets were observed to less constraining upper limits, between $10^{8.3}$ and $10^{8.5}$ K km s$^{-1}$.

\subsection{CO(3--2) Observations} \label{ss:obs32}

We observed CO(3--2) with the SMT 0.8~mm receiver, tuned to central frequencies from 329.6 to 342.4 GHz. The SMT 0.8~mm receiver is a double-sideband system. Sources were typically observed in the lower sideband leading to possible contamination by the $J=4\rightarrow3$ transitions of HCN and HCO+ in the image sideband for some receiver configurations. However, these lines are expected to be 10-100 times fainter than CO \citep{gao+04}, and thus should have little effect on our measured CO(3--2) luminosities given the modest signal to noise ratio of our detections.

CO(3--2) observations were otherwise conducted in a similar manner to CO(2--1) observations. Whenever possible targets were observed until a line was detected. Thirty-four of our 45 CO(3--2) targets were detected with ${\rm SNR}>3$. While the remaining targets typically showed evidence of a line at lower SNR, but could not be re-observed to achieve a significant detection. Integration times ranged from 1 to 12 hours per source.

\subsection{CO(1--0) Observations} \label{ss:obs10}

Our CO(1--0) observations were conducted with the 3~mm band of the 12m's four-band receiver. The backend was the ARO Wideband Spectrometer (AROWS) in its widest configuration, providing 4~GHz of bandwidth per tuning. The large bandwidth made it possible to observe most sources with only a handful of tunings. Sources were observed in the upper sideband, and central frequencies for all observations were selected from one of 110.0, 110.5, 111.0, 113.0, and 113.5 GHz. The 12m receiver is sideband-separating, with typical image rejections better than 20~dB.

Targets were observed in beam switching mode, alternating the secondary mirror position between $\pm2$~arcminutes at a switch rate of 1.25~Hz. Integration times were three minutes per scan, and total observation times were typically 45 to 75 minutes, with longer observations for a handful of faint targets. Pointing and focus were checked every ninety minutes using continuum observations of a planet or quasar.


\section{Data Processing}\label{sec:data}

\begin{figure}
    \centering
    \includegraphics[width=0.45\textwidth]{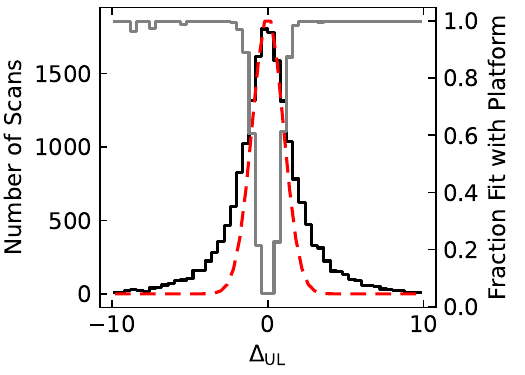}
    \caption{The black histogram shows the distribution of $\Delta_{\rm UL}$, the normalized difference between the upper and lower frequency half of individual spectra, for each scan that passed our final set of quality checks. A normal distribution is shown for comparison (red line, normalized so the area within $|\Delta_{\rm UL}|<1$ matches that of the observed distribution). The gray line and left axis show the fraction of scans for which our baseline fitting method favored the platforming model over a constant baseline, as a function of $\Delta_{\rm UL}$.
    \label{fig:platforming}}
\end{figure}

\begin{figure*}
    \centering
    \includegraphics[width=\textwidth]{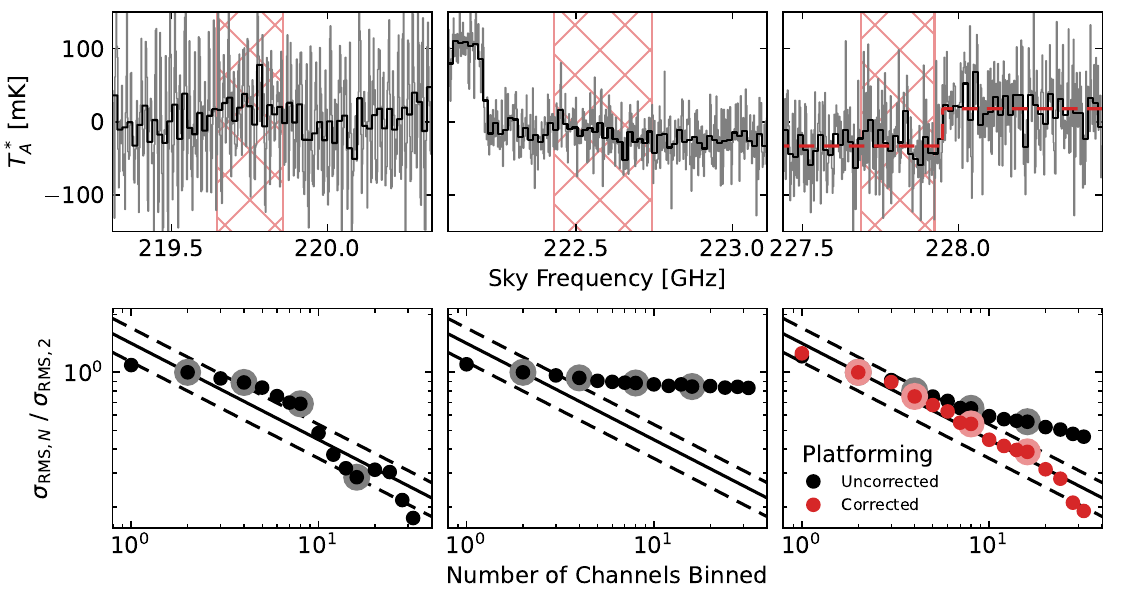}
    \caption{Upper row: three examples of scans with irregular behaviors. The left panel shows a scan with high frequency oscillations across the bandpass, the center panel shows a scan with a sharp discontinuity towards the lower frequency edge of the band, and the right panel shows a scan with platforming at the center of the band. Data is shown at both 1~MHz (gray) and 16~MHz (black) resolutions. The light red hatched regions show where the data has been masked to exclude potential line emission. In the right panel, we also show the best fitting baseline (red dashed line), which can be used to remove the platforming.
    Lower row: scaling of the channel RMS in each spectrum after binning to a specified number of channels. Data have been normalized to the RMS at $N=2$ binning. The solid line shows the expected drop in noise level relative to $N=2$, and the dashed lines show a $\pm20\%$ deviation. The four larger points at $N=2$, 4, 8, and 16 indicate the bins used in our automated testing. Other values of $N$ are shown for illustrative purposes only. In the right panel, we show the results both before (black) and after (red) removing the platform from the data, showing that baseline correction restores the expected drop in RMS with $N$.
    \label{fig:badscans}}
\end{figure*}

\begin{deluxetable}{c|c}
\tablecaption{\label{tab:tests} Scan-level quality checks of SMT CO(2--1) data}
\tablehead{
    \colhead{Test} & \colhead{\hspace{1cm}N Failures (\%)}\hspace{1cm}}
\startdata
    $|\frac{\sigma_{\rm RMS}}{\sigma_{\rm rad}}-1|>0.1$ & 425 (2\%) \\
    $|\sqrt{\frac{2}{N}} \frac{\sigma_{{\rm RMS},N}}{\sigma_{{\rm RMS},2}}-1|>0.2$ & 3587 (16\%) \\
    $|\Delta S|>1\sigma$ & 4 (0\%) \\
    Inspection & 336 (1\%) \\
    \hline
    Total Rejected & 4138 (18\%) \\
    \enddata

    \tablecomments{The first four rows give the number and percentage of scans that fail the following checks: Row 1 -- $\sigma_{\rm rad}$ and $\sigma_{\rm RMS}$ differ by more than 10\%; Row 2 -- the scaling of the RMS noise when binning $N$ channels differs from the expected $\sqrt{N}$ by more than 20\%; Row 3 -- the inclusion of the scan alters the final integrated line flux by more than $1\sigma$; Row 4 -- our visual inspection of the data revealed other problems with the scan. Data can fail these tests for a variety of reasons, and are removed from further analysis. Row 5 gives the total number of scans removed from our analysis.}
\end{deluxetable}

Data reduction was performed using the purpose built Computations for Heterodyne Analysis, Observations, and Science (\textsc{chaos}) pipeline. CO(1--0), CO(2--1), and CO(3--2) data were all processed in a similar manner, with variations to account for the particularities of the different telescopes and receivers. We describe the CO(2--1) data reduction in detail in Section~\ref{ss:data21}, as the majority of our data are for this line. Sections~\ref{ss:data32} and~\ref{ss:data10} summarize modifications made for the CO(3--2) and CO(1--0) data.

\subsection{CO(2--1) Data Reduction Pipeline} \label{ss:data21}

ARO observations are recorded in \textsc{class} files in units of antenna temperature $T_A^*$. We exported each scan to a FITS format which was then analyzed using our \textsc{chaos} pipeline.

\subsubsection{Scan Validation}\label{ss:data21-validation}

While the majority of recorded scans appear to be of good quality, a number of issues can affect individual scans or sets of scans. Here we describe the quality checks performed to ensure the reliability of our data. For all of these checks, we began by masking the region expected to contain signal. For objects with a CO(1--0) detection from xCOLD~GASS, we used a mask of $1.5\times {\rm FWHM}$ where FWHM is the line width reported in \citet{saintonge+17}. For objects with no xCOLD~GASS detection we applied a mask of 300 km/s centered on the optical redshift of the galaxy.

The most common of issue found in our data was platforming at the center of the bandpass, caused by slight gain variations during the course of a scan. To assess the prevalence of this effect, we computed the average and standard deviation of unmasked channels in the higher- and lower-frequency halves of our spectrum, and then determined the ``normalized difference'' between the two halves:
\begin{equation*}
    \Delta_{\rm UL} = \frac{\mu_{\rm U}-\mu_{\rm L}}{\sqrt{\sigma_{\mu_{\rm U}}^2+\sigma_{\mu_{\rm L}}^2}}
\end{equation*}
where $\mu$ and $\sigma_\mu$ are the mean and its standard error and the subscripts denote the upper and lower frequency sections. Figure~\ref{fig:platforming} shows the distribution of $\Delta_{\rm UL}$ for the 18,925 scans that passed all other quality checks. If no scans exhibit platforming, the distribution of $\Delta_{\rm UL}$ should be Gaussian with a standard deviation of one. The actual distribution deviates significantly from this expectation, even for small values of $|\Delta_{\rm UL}|$, indicating that platforming affected a significant fraction of the data, and may be present at a low level even for scans where it was not readily apparent. Approximately 57\% of all scans show $|\Delta_{\rm UL}|>1$, a significant departure from the expected 32\% for the platform-free case.

The high incidence of platforming drove our method for fitting and removing baselines. We fitted each scan with two baseline models, the first was a constant offset from zero, and the second was a platforming model with a different offsets for the high-frequency and low-frequency halves of the scan. We used a reduced $\chi^2$ test to select between the two fits and then removed the best fitting baseline from each scan. Figure~\ref{fig:platforming} shows the fraction of scans for which the platforming model was favored as a function of $\Delta_{\rm UL}$. For $|\Delta_{\rm UL}|>1$ the platforming baseline was nearly always favored.

The 1~GHz bandpass of the SMT provides a relatively narrow region over which to fit baselines for the upper and lower half of each spectrum. Therefore we only included constant terms in our two baseline models. We found that, aside from platforming, the baselines from our data tended to be flat, and the need for higher order baseline corrections was minimal. Any scans with poor baseline behavior were flagged and removed in subsequent data cuts.

After baseline correction, we computed the RMS noise in the unmasked regions of the spectrum, $\sigma_{\rm RMS}$, and compare this to the expected noise in the spectrum based on the radiometer equation:
\begin{equation}\label{eq:sigrad}
    \sigma_{\rm rad} = \frac{\sqrt{2} T_{\rm sys}^*}{\sqrt{\Delta \nu \Delta \tau}}
\end{equation}
where $\sigma_{\rm rad}$ is the expected noise level, $T_{\rm sys}^*$ is the system temperature, accounting for atmospheric losses, $\Delta \nu$ is the channel width of the spectrum, and $\Delta \tau$ is the on-source integration time. We flagged scans where $\sigma_{\rm RMS}$ and $\sigma_{\rm rad}$ differed by more than 10\%.\footnote{Channels in the SMT filterbanks are partially overlapping, resulting in correlation between adjacent channels. To account for this we scale $\sigma_{\rm RMS}$ up by a factor of 1.2 before performing the comparison.} This resulted in the flagging of 2\% of scans. Visual inspection of flagged scans suggests that large discrepancies between $\sigma_{\rm RMS}$ to $\sigma_{\rm rad}$ are typically caused by short (frequency) period baseline oscillations, which occasionally appear in the data. Rapid changes in the atmosphere due to, e.g., clouds passing through the line of sight can also cause discrepancies, as can poor baselines.

Next, we tested that the RMS in the unmasked regions of each spectrum decreased as $\sqrt{N}$ when $N$ channels were binned together. This test can identify spectra with poor baselines, where channel-to-channel variations are partially caused by baseline oscillations rather than thermal noise. Because adjacent channels in the SMT filterbanks overlap partially in their frequency response and are therefore correlated, we used the spectrum binned by $N=2$ as our reference, and compared the noise level in spectra binned by $N=2$, 4, 8, and 16. For sources where the $N=16$ spectra contained fewer than 20 unmasked channels, we limited the test to $N=2$, 4, and 8 channels. Spectra where the noise at any of these binning levels deviated from expectations by more than 20\% were flagged. This resulted in flagging of 16\% of the data. Visual review of scans flagged by this test suggests that large baseline oscillations, uncorrected platforming, discontinuities in locations other than the bandpass center, and short period standing waves can all cause this test to fail. Figure~\ref{fig:badscans} shows examples of scans with various defects, along with the behavior of RMS with binning for each scan. 

We also compared the final integrated line flux for each source in our survey (described in Section~\ref{ss:data21-intfluxes}) to the integrated flux recomputed excluding one scan at a time. This procedure allowed us to search for anomalous features in individual scans that resulted in outsized contribution to the final flux measurement. If the inclusion/exclusion of any given scan would have altered the final line flux by more than $1\sigma$ we flagged that scan. This resulted in the removal of only 4 scans.

Finally, each spectrum was inspected by eye. Scans with obvious defects that had been missed by the automated tests were flagged. Occasionally flags set by the automated checks were deemed to be erroneous, and the affected scans were unflagged. This was most commonly the case when a scan that showed no obvious baseline problems had failed the re-binning test only at the $N=16$ binning level.

Table~\ref{tab:tests} summarizes the number of scans that failed each of the data cuts. In total, we flagged 4126 scans, amounting to 18\% of our data, for exclusion from subsequent analysis. We believe that our data cuts were generally conservative -- many scans are flagged for marginally exceeding our flagging thresholds, but those that are truly contaminated tend to be well beyond the thresholds.

\subsubsection{Flux Calibration}\label{ss:data21-fluxcal}

\begin{figure}
    \centering
    \includegraphics[width=.45\textwidth]{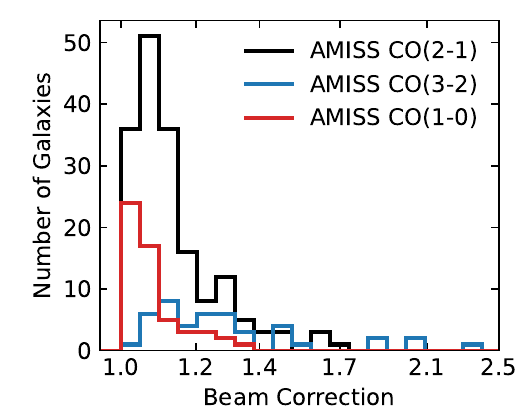}
    \caption{The distribution of aperture corrections ($1/K$) for sources observed by AMISS in CO(2--1) (black), CO(3--2) (blue), and CO(1--0) (red).}
    \label{fig:bc}
\end{figure}


\begin{deluxetable*}{l|ccc|ccc}
\tablecaption{Instrumental parameters of the ARO receivers used in this study: forward main beam efficiencies, half-power beam widths, and point source sensitivities \label{tab:pss}}
\tablehead{
    & \multicolumn{3}{c|}{H-polarization} & \multicolumn{3}{c}{V-polarization} \\
    Date Range & $\eta_{\rm fmb}$ & FWHM & $\chi_{\rm ps}$ & $\eta_{\rm fmb}$ & FWHM & $\chi_{\rm ps}$}
    \startdata
        \multicolumn{7}{c}{SMT 1.2~mm Receiver}\\
        \hline
        2020 Mar--2021 June & 70\% & 31'' & 61 Jy K$^{-1}$ &66\% & 31'' & 65 Jy K$^{-1}$ \\
        2022 Feb--2022 June & 68\% & 31'' & 63 Jy K$^{-1}$ &68\% & 31'' & 63 Jy K$^{-1}$ \\
        2022 Oct--2023 June & 65\% & 31'' & 66 Jy K$^{-1}$ &64\% & 31'' & 67 Jy K$^{-1}$ \\
        \hline
        \multicolumn{7}{c}{SMT 0.8~mm Receiver}\\
        \hline
        2020 Mar--2021 June & 62\% & 22'' & 78 Jy K$^{-1}$ &62\% & 22'' & 78 Jy K$^{-1}$ \\
        2022 Dec--2023 Jan & 55\% & 25'' & 107 Jy K$^{-1}$ &55\% & 22'' & 87 Jy K$^{-1}$ \\
        2023 Jan--2023 June & 55\% & 22'' & 87 Jy K$^{-1}$ &55\% & 22'' & 87 Jy K$^{-1}$ \\
        \hline
        \multicolumn{7}{c}{ARO 12m 3~mm Receiver}\\
        \hline
        2023 Jan--2023 June & 74\% & 47'' & 48 Jy K$^{-1}$ &74\% & 47'' & 48 Jy K$^{-1}$
    \enddata
    \tablecomments{Beam sizes are given for representative frequencies of 230.5, 345.8 and 115.3~GHz}
\end{deluxetable*}

Accepted scans were converted from $T_A^*$ to flux units following
\begin{equation}\label{eq:Snu}
    S_{\nu} = \frac{\chi_{\rm ps}}{K} T_A^*
\end{equation}
where $\chi_{\rm ps}$ is the point source sensitivity \citep{kramer+08} and $K$ is the fraction of the source flux which couples with the beam of the telescope ($K=1$ for a point source). Table~\ref{tab:pss} gives the values of $\chi_{\rm ps}$ for each receiver used in this study.

We computed the coupling term following the procedure outlined in \citet{saintonge+12} and \citet{lisenfeld+11}. For each source, we assumed that the CO emission is distributed in an exponential disk with a half light radius and inclination equal to those measured from the SDSS r-band images of the galaxy. We multiplied the resulting flux distribution by the spatial response pattern of the telescope beam -- which we modeled as a two dimensional Gaussian -- and integrated over the full extent of the source. The ratio between the spatially integrated fluxes with and without the response of the telescope applied was then our coupling term $K$. The coupling depends on both the source properties and on the telescope beam size, thus we computed separate values of $K$ for each source and and each telescope beam considered in the survey. Figure~\ref{fig:bc} shows the distribution of $1/K$ computed for each of our three target lines. The expected emitting regions for our CO(1--0) and CO(2--1) targets were typically small relative to the 12m and SMT beams at 115 and 230 GHz, and only minimal corrections were required. The SMT beam at 345 GHz missed some flux for the largest sources in our survey. We discuss the uncertainty in the coupling term in Section~\ref{ss:errors-beam}, and validate the use of an exponential disk model in Section~\ref{ss:results-disk}.

We did not apply the aperture correction ($1/K$) to spectra themselves, but did account for it in determining final integrated fluxes.

\subsubsection{Final Line Fluxes, Spectra, and CO Luminosities}\label{ss:data21-intfluxes}

For each scan we measured an integrated flux 
\begin{equation}\label{eq:sdv}
    S\Delta v = \Delta v_{\rm ch} \sum_{\rm win} S_{\nu,{\rm ch}}
\end{equation}
where $\Delta v_{\rm ch}$ is the channel width, $S_{\nu,{\rm ch}}$ is the flux density in a channel, and the sum is conducted over a window containing all channels with line emission. The integration windows were determined by jointly inspecting the xCOLD~GASS CO(1--0) spectra, preliminary versions of our AMISS CO(2--1) spectra, and, when available, HI spectra from GASS. Integration limits  were drawn manually to encompass all visible flux in either CO line. HI and CO spectral profiles can differ considerably, and so the HI spectra were only used to help differentiate between noise and faint emission when the extent of one or both CO lines was ambiguous. When no CO detection was apparent a fixed region of 300 km/s around the optical redshift was used. 

To verify that defining integration limits by visual inspection does not bias our results, we defined an alternative set of integration limits derived from fits to the line profiles. For each source we fitted either a single or double Gaussian function to the CO(1--0) and CO(2--1) spectra, and took the points where the fitted model reaches 5\% of the peak as integration limits. The fluxes derived using these limits were consistent with our manually selected windows, with an average (median) difference of 0\% (1\%) and an RMS deviation of 8\%. However, the edges of spectral lines were typically sharper than was captured by Gaussians, which caused the fits to return unnecessarily wide integration regions with little or no additional signal. 

For scans fit with a constant baseline the statistical uncertainty in the flux is given by 
\begin{equation}
    \sigma_{S\Delta v} 
    = \sigma_{\rm ch} \Delta v_{\rm ch} \sqrt{N_{\rm win} + \frac{N_{\rm win}^2}{N_{\rm bl}}}
\end{equation}
where, $\sigma_{\rm ch}$ is the uncertainty in the flux density of a single channel, taken to be $\sigma_{\rm rad}$ from Equation~\ref{eq:sigrad}, $N_{\rm win}$ is the number of channels in our summing window, $N_{\rm bl}$ is the number of channels used to fit the baseline. The $N_{\rm win}^2/N_{\rm bl}$ term accounts for the fact that the zero level of the spectrum is not perfectly known, and is particularly necessary for correctly characterizing the uncertainty when a spectral line takes up a significant fraction of the spectrometer bandwidth. For scans fit with a platform baseline the uncertainty is given by (see Appendix~\ref{ap:blerr})
\begin{equation}
    \sigma_{S\Delta v}
    = \sigma_{\rm ch} \Delta v_{\rm ch} \sqrt{N_{\rm win} + \frac{N_{\rm win,U}^2}{N_{\rm bl,U}} + \frac{N_{\rm win,L}^2}{N_{\rm bl,L}}}
\end{equation}
where the U and L subscripts indicate the numbers of channels in the upper or lower halves of the spectrum.

We determined the integrated line flux for each source as the inverse-variance weighted average of the scan-by-scan integrated flux. We constructed final spectra by averaging together all scans using the same inverse-variance weights.\footnote{Measuring the final integrated flux by averaging together fluxes extracted from each scan gives identical values to extracting a single flux from the final averaged spectrum.} Finally we converted each flux into a line luminosity using the redshift listed by xCOLD~GASS and
\begin{equation}
    L_{\rm CO}^\prime = 3.25\times10^7 S\Delta v \frac{D_L^2}{(1+z)\nu_{\rm CO}}\,,
\end{equation}
where $L_{\rm CO}^\prime$ has units of K~km~s$^{-1}$~pc$^2$ when $S\Delta v$ is given in Jy~km~s$^{-1}$, $D_L$ is the luminosity distance in Mpc, $z$ is the redshift, and $\nu_{\rm CO}$ is the rest frequency of the CO line in GHz.

\subsection{Modifications for CO(3--2) Data Reduction}\label{ss:data32}

\begin{deluxetable}{c|c}
\tablecaption{Scan-level quality checks of SMT CO(3--2) data\label{tab:tests32}}
\tablehead{
    \colhead{Test} & \colhead{\hspace{1cm}N Failures (\%)}\hspace{1cm}}
\startdata
    $|\frac{\sigma_{\rm RMS}}{\sigma_{\rm rad}}-1|>0.15$ & 490 (10\%) \\
    $|\sqrt{\frac{2}{N}} \frac{\sigma_{{\rm RMS},N}}{\sigma_{{\rm RMS},2}}-1|>0.2$ & 651 (13\%) \\
    $|\Delta S|>1\sigma$ & 2 (0\%) \\
    Inspection & 156 (3\%) \\
    \hline
    Total Rejected & 1059 (21\%) \\
    \enddata

    \tablecomments{The number and percentage of CO(3--2) scans failing our data quality checks. Rows are the same as Table~\ref{tab:tests} and tests are described in Section~\ref{ss:data21-validation}. Data failing these tests are excluded from subsequent analysis.}
\end{deluxetable}

Here we describe the relevant modifications to our pipeline and the results of our data quality checks for the AMISS CO(3--2) data. 

The CO(3--2) data were reduced in largely the same manner as the CO(2--1) data. However, the beams of the 0.8~mm receiver -- particularly in horizontal polarization -- are partially truncated in the relay path through the elevation axis of the SMT. This manifests as a standing wave in horizontal polarization spectra which raises the RMS noise relative to the expected value and caused a large fraction of the horizontal polarization data to fail our comparison of $\sigma_{\rm RMS}$ and $\sigma_{\rm rad}$. The standing wave is only a few channels wide, and over the full width of our spectral lines the effect averages away. We therefore overrode this flag in our horizontal polarization data and, unless other obvious defects were present, included these these data in our analysis.

For observations in December 2022 and January 2023, the horizontal polarization feed of the 0.8~mm receiver was further misaligned, resulting in a beam pattern $\sim15$\% broader than in optimal alignment and a different point source sensitivity (Table~\ref{tab:pss}). We accounted for the different coupling between our sources and this larger beam by re-scaling the affected scans by the ratio of the coupling corrections for the nominal and misaligned beams. We show in Section~\ref{ss:errors-jackknife} that this correction results in consistent fluxes between the two polarizations.

Windows for extracting the CO(3--2) flux were derived using the AMISS CO(2--1) and xCOLD~GASS CO(1--0) spectra, rather than from the CO(3--2) directly, as the lower $J$ lines have higher signal to noise.

Table~\ref{tab:tests32} summarizes the results of our various quality checks on the CO(3--2) data. In total, we flagged 1059 scans, amounting to 21\% of our data.

\subsection{Modifications for CO(1--0) Data Reduction}\label{ss:data10}

\begin{figure}
    \centering
    \includegraphics[width=0.45\textwidth]{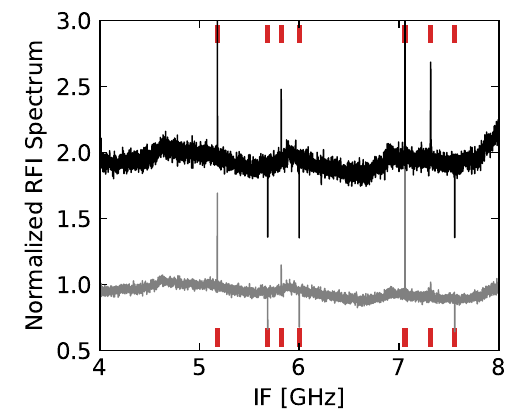}
    \caption{Channel-by-channel 68th (gray) and 95th (black) percentiles of all 12m-AROWS CO(1--0) scans after normalizing by the scan RMS and masking known astrophysical signals. Channels where the 95th percentile deviates significantly from the expected value of 2 are flagged as bad, and indicated by red hashes at the top and bottom of the plot. Most flagged regions are a single channel wide, however the feature at 7.32 GHz IF is approximately 5 channels wide and appears to be caused by a local broadcasting station.
    \label{fig:rfi}}
\end{figure}

\begin{deluxetable}{c|c}
\tablecaption{Scan-level quality checks of 12m CO(1--0) data \label{tab:tests10}}
\tablehead{
    \colhead{Test} & \colhead{\hspace{1cm}N Failures (\%)}\hspace{1cm}}
\startdata
    $|\sqrt{\frac{1}{N}} \frac{\sigma_{{\rm RMS},N}}{\sigma_{{\rm RMS},1}}-1|>0.2$ & 120 (4\%) \\
    $|\Delta S|>1\sigma$ & 0 (0\%) \\
    Inspection & 27 (1\%) \\
    \hline
    Total Rejected & 139 (5\%) \\
    \enddata

    \tablecomments{Rows are the same as Table~\ref{tab:tests} and tests are described in Section~\ref{ss:data21-validation}. Data failing these tests are excluded from subsequent analysis.}
\end{deluxetable}

Differences between the ARO 12m telescope and the SMT necessitated a number of revisions to our analysis procedure for the AMISS CO(1--0) data. Here we describe the modifications to our pipeline and the results of our data quality checks for the 12m data. 

First, the 12m telescope site is subject to radio frequency interference (RFI) at the intermediate frequency of the receiver, which manifests as strong but spectrally narrow lines many individual scans. To identify channels most strongly affected by this RFI, we converted each scan from its sky frequency to the intermediate frequency in the 3~mm receiver. We took the absolute value of each channel and normalized by the expected scan RMS. We then measured the 68th and 95th percentile of values in each channel of the spectrometer, excluding scans where a given channel was expected to contain CO(1--0) flux from a source. Figure~\ref{fig:rfi} shows the results of this procedure. Channels free from RFI should show 68th and 95th percentile values of approximately one and two respectively. A handful of channels deviate from this expectation by a significant amount and can be easily flagged. We identified seven regions contaminated by RFI features. Six of these regions were a single channel (625 kHz) wide, while the seventh showed significant contamination in about 5 adjacent channels. We clean the 11 RFI contaminated channels by replacing them with values from adjacent channels on a scan-by-scan basis. In most cases, our tunings placed the RFI away from the location of the target spectral line, and this correction was purely cosmetic. For the few instances where the RFI features fell near our signal, we note that single channels corresponds to 1.6 km/s, while the spectral lines we observed are typically hundreds of km/s wide, the RFI cleaning procedure therefore had negligible impact on our measured fluxes or uncertainties.

Second, the digital spectrometer at the 12m is not subject to the platforming effects seen at the SMT and simultaneously provides a much wider bandpass. Because of this, we fitted all scans with a linear baseline instead of the constant or platform baselines used for the SMT data. Additionally, adjacent channels in the spectra are non-overlapping, and therefore statistically independent, so we used the RMS of the unbinned ($N=1$) spectrum as our reference when checking that the RMS in each scan fell as $\sqrt{N}$ when $N$ adjacent channels were binned together.

Finally, windows for extracting the CO(1--0) flux were derived using the AMISS CO(2--1) and xCOLD~GASS CO(1--0) spectra, rather than from the new AMISS CO(1--0) spectra directly, as these other datasets generally have higher signal to noise.

Table~\ref{tab:tests32} summarizes the quality checks performed on the CO(1--0) data. In total, we flagged 139 scans, amounting to 5\% of our data, for exclusion from subsequent analysis. The wider bandwidth and digital signal processing in the 12m spectrometer result in a much lower fraction of scans being rejected compared to the SMT data.


\section{Reprocessing of xCOLD~GASS Spectra}\label{sec:xcg}

In addition to new CO observations taken for AMISS, we processed the public xCOLD~GASS spectra to ensure that all luminosities used in our analysis are derived from methods that are as uniform as possible. For example, we ensured that line fluxes were integrated over matched spectral windows and confirmed detections/non-detections of faint lines by comparing of the multiple available spectra.

We began by gathering the reduced xCOLD~GASS CO(1--0) and, when available, CO(2--1) spectra for each AMISS target from the xCOLD~GASS website.\footnote{http://www.star.ucl.ac.uk/xCOLDGASS/data.html} The xCOLD~GASS spectra are presented in units of Janskys, obtained by multiplying the native $T_A^*$ values by point source sensitivities tabulated by IRAM. 

We computed source-beam coupling factors for each source in the same manner as Section~\ref{ss:data21-fluxcal} and used them to correct for the flux falling outside the beam. We also applied a correction for beam efficiency degradation in wobbler switched observations caused by pointing the secondary mirror off axis \citep{zarghamee+85,greve+96}.\footnote{the beam switched AMISS observations are also subject to this effect, however, ARO beam efficiency measurements are made in the same observing mode and account for this loss implicitly, without further need for correction.} According to IRAM documentation, this effect results in a 20\% loss in efficiency for a beam throw of $\pm$120" and an observing frequency of 230~GHz.\footnote{https://publicwiki.iram.es/Iram30mEfficiencies, version dated 2016-11-03} Following \citet{lamb99}, we scaled this value to the frequencies and beam throws of the xCOLD~GASS observations as
\begin{equation}\label{eq:xcg_corr}
    \frac{\eta}{\eta_0} = 1 - 0.20 \Big(\frac{\lambda}{1.3~{\rm mm}}\Big)^2 \Big(\frac{\Delta \alpha}{120"}\Big)^{-2}
\end{equation}
where $\eta/\eta_0$ is the ratio of the on-axis beam efficiency and the beam efficiency with the secondary mirror offset by angle $\Delta \alpha$ at observing wavelength $\lambda$. We then divided each xCOLD~GASS spectrum by this  factor. This correction was generally small, 1-3\% (depending on beam throw) for CO(1--0) data and 5-11\% for CO(2--1) data.

We found a discrepancy between the flux scale of the xCOLD~GASS 30m data and that of our SMT and 12m data. In short, comparison of beam-corrected luminosities measured with the 30m and the ARO telescopes suggests the 30m flux scale is lower than that of the ARO telescopes by 10\% at $\sim$115~GHz and 24\% at $\sim$230~GHz. 
To correct this discrepancy, we scaled the IRAM spectra by factors of 1.12 for CO(1--0) and 1.32 for the CO(2--1), which forces the median CO luminosities derived from IRAM and ARO observations to match. We give a more extensive discussion of the evidence for this discrepancy and our procedure for computing the correction in Appendix~\ref{ap:xcg_flux}.

After re-scaling the xCOLD~GASS data, we extracted a portion of each spectrum centered on the CO line and covering five times its width. To account for differences in signal masks between the xCOLD~GASS and AMISS data processing, we fitted and removed a constant baseline from each spectrum, masking the same velocity range as in Section~\ref{ss:data21}. We calculated integrated fluxes and corresponding CO luminosities over spectral windows identical to those used for the AMISS observations. Our revised xCOLD~GASS luminosities are presented along with AMISS measurements in Section~\ref{sec:results}.


\section{Estimation of Systematics and Errors}\label{sec:errors}

Telescope calibration can contribute significant uncertainty to measurements of CO line ratios. Multiple studies have found that systematic differences in the calibration of line flux measurements between different surveys to be a limiting factor in understanding trends in the ratios of the low-$J$ CO lines \citet{denbrok+21,leroy+22}. 

In this section we carefully examine the uncertainties and systematics in our data. Section~\ref{ss:errors-jackknife} validates our statistical uncertainties by comparing between subsets of our own data. Sections~\ref{ss:errors-fluxcal}, \ref{ss:errors-be}, and \ref{ss:errors-beam} describe and quantify the uncertainties introduced by telescope calibration. Section~\ref{ss:errors-othertels} compares our results with measurements from other facilities. Section~\ref{ss:errors-summarry} summarizes the total statistical and calibration uncertainty for our sample.

\subsection{Validation of Line Fluxes and Statistical Uncertainties}\label{ss:errors-jackknife}

\begin{figure*}
    \centering
    \includegraphics[width=\textwidth]{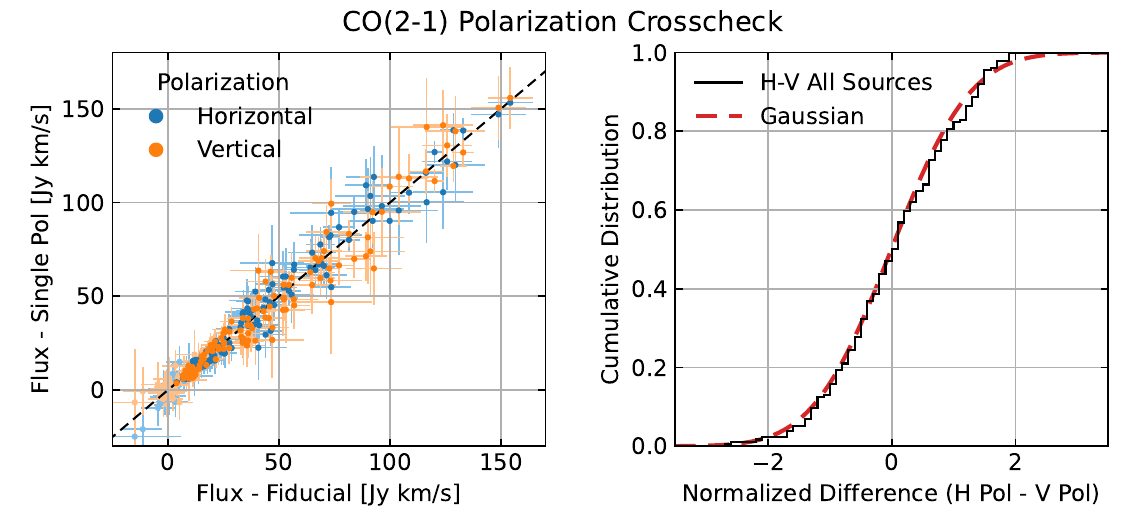}
    \caption{Left: The integrated line flux for each of our CO(2--1) targets using only horizontal (blue) and vertical (orange) polarization data plotted as a function of our fiducial flux measurements using both polarizations. Lighter points correspond to non-detections, while darker points are detected with SNR greater than 4.
    Right: The cumulative distribution of variance-normalized differences between horizontal and vertical polarization ($\Delta_{HV}$, Equation~\ref{eq:hv}), for all sources is shown by the black histogram. The red curve shows a standard normal distribution -- the expected distribution for $\Delta_{HV}$ if our errors are estimated correctly.
    \label{fig:pol}}
\end{figure*}

\begin{figure*}
    \centering
    \includegraphics[width=\textwidth]{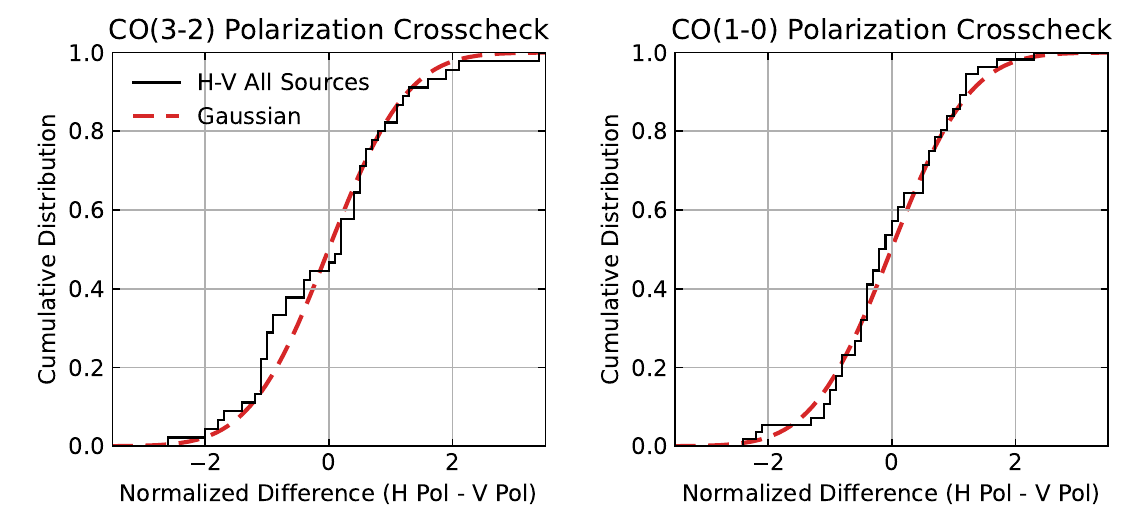}
    \caption{Cumulative distributions of $\Delta_{HV}$ for sources in our CO(3--2) survey (left panel) and our CO(1--0) survey (right panel) are shown by the black histograms. Red curves show a standard normal distribution -- the expected distribution of $\Delta_{HV}$ if our errors are estimated correctly.
    \label{fig:pol32and10}}
\end{figure*}

To verify that our CO line fluxes are reliable and that we have correctly accounted for statistical errors in our measurements, we perform a cross-validation analysis by splitting our dataset into two independent halves and comparing the fluxes measured in each.

All SMT and 12m observations are dual polarization, with the horizontal (H) and vertical (V) polarization data recorded and saved separately. We therefore compute the integrated CO line flux for each target of our survey using only H and only V polarization data, and compare these two measurements. We have verified that there is no instrumental correlation between the spectrometer modules which could introduce correlated noise between the polarizations.

To compare the two sets of measurements, we define the quantity
\begin{equation}\label{eq:hv}
    \Delta_{HV} = \frac{(S\Delta v)_H-(S\Delta v)_V}{\sqrt{\sigma_{(S\Delta v)_H}^2 + \sigma_{(S\Delta v)_V}^2}}.
\end{equation}
If our data processing is free of systematic differences between the polarizations, and our uncertainties $\sigma_{(S\Delta v)_H}$ and $\sigma_{(S\Delta v)_V}$ have been estimated correctly, then $\Delta_{HV}$ will be normally distributed with a mean of zero and a variance of one. On the other hand, systematic offsets in the calibration of the two polarizations can cause the mean of the distribution to differ from zero, while under- or over- estimations of our statistical uncertainties will respectively widen or narrow the width of the distribution.

Figure~\ref{fig:pol} shows the measured distribution of $\Delta_{HV}$ for each galaxy targeted in our CO(2--1) survey. The observed distribution closely follows the expected normal distribution. We perform a Kolmogorov-Smirnov test to determine whether the observed distribution of $\Delta_{HV}$ differs from the expected Gaussian. We find a $p$-value of 0.50, providing little evidence for a difference between the two distributions. This confirms that our statistical uncertainties are calculated correctly and no unknown measurement errors are present. 
We find that the mean $\Delta_{HV}$ value is slightly offset ($0.08$) from zero. Removing this offset would improve the $p$-value of the Kolmogorov-Smirnov to 0.94, however, the offset is comparable to the standard error of the mean for our sample size of 176 and therefore not significant. Small offsets between the two polarizations, if real, can most likely be attributed to uncertainty in the values of $\chi_{\rm ps}$ for each polarization. Our final flux measurements average over any such difference, and we take no further steps to correct it.

Distributions of $\Delta_{HV}$ for our CO(3--2) and CO(1--0) data are shown in Figure~\ref{fig:pol32and10}. Note that the CO(2--1) sample is about 4 times larger than either of these other datasets and therefore provides the best statistics for crosschecking. 

For the CO(3--2) observations a Kolmogorov-Smirnov test returns a $p$-value of 0.46, providing little evidence that the distribution of $\Delta_{HV}$ differs from the expected standard normal distribution. However, repeating the test without correcting the for larger beam size of the H polarization data from December 2022 to January 2023 gives a $p$-value 0.025, indicating a discrepancy between the H and V polarization calibration at $>95\%$ confidence when this effect is not accounted for.  

For our CO(1--0) observations the distribution of $\Delta_{HV}$, again shows excellent agreement with the expected normal distribution. The $p$-value for the Kolmogorov-Smirnov test is 0.92, providing no suggestion that the two distributions differ.

Overall, the results of this crosscheck show that our measurements are internally consistent and that our statistical uncertainties are correctly characterized. However, these results do not constrain systematic uncertainties or temporal variations in the flux scale used to calibrate our data. We explore these uncertainties in the next section.

\subsection{Spectral Line Flux Monitoring and Calibration Uncertainties}\label{ss:errors-fluxcal}

\begin{figure*}
    \centering
    \includegraphics[width=.49\textwidth]{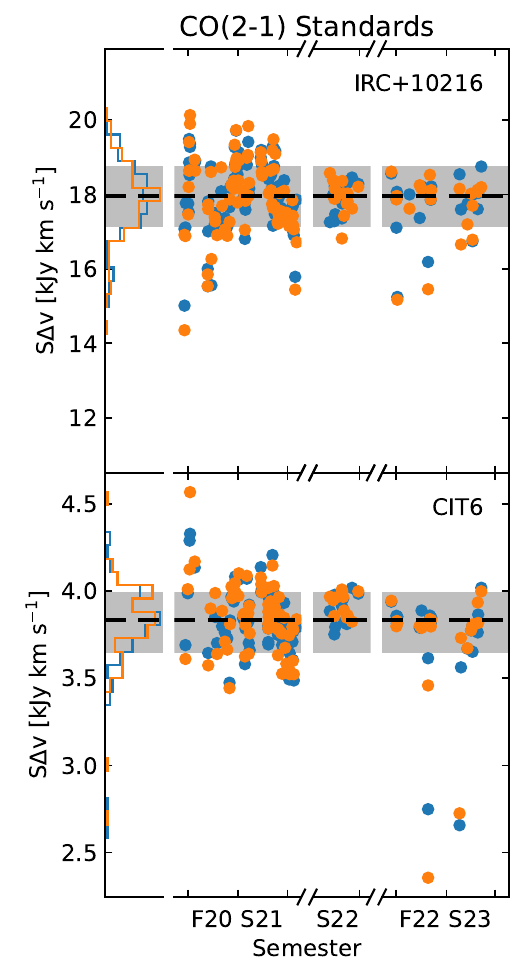}
    \includegraphics[width=.49\textwidth]{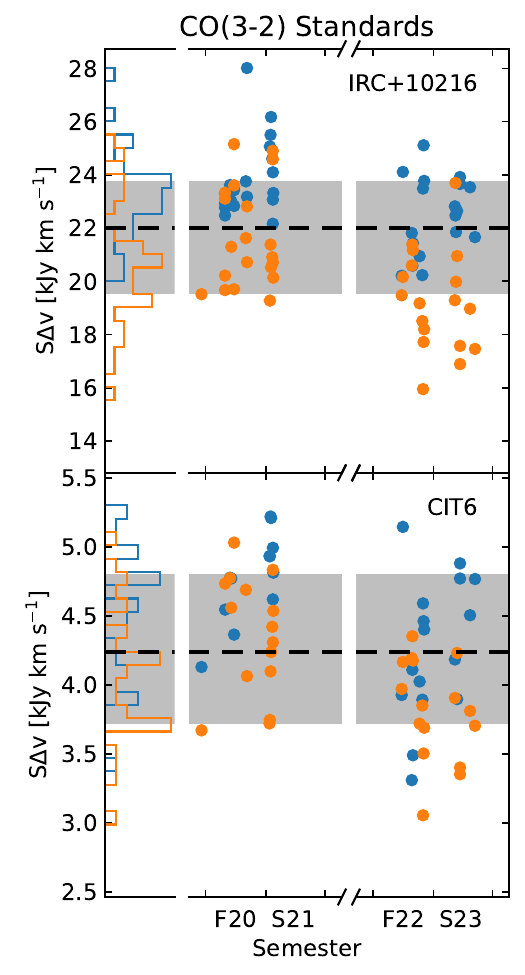}
    \caption{Left: Integrated flux of the CO(2--1) lines of IRC+10216 (top) and CIT6 (bottom) as measured with the SMT 1.2~mm receiver over the course of our program. Results for the horizontal and vertical polarizations are shown in blue and orange respectively. Dashed lines indicate the median flux over the full time frame and the filled regions show the 16th-84th percentile range.
    Right: CO(3--2) fluxes for the same sources, monitored with the SMT 0.8~mm receiver. A smaller time range is covered due to the limited windows in which the 0.8~mm receiver was available at the SMT.}
    \label{fig:fluxcal}
\end{figure*}

To quantify systematic uncertainties in the calibration of SMT data, we routinely monitored the flux of $^{12}$CO spectral lines of bright carbon stars IRC+10216 and CIT6 from October 2020 through June 2023. These observations were conducted using similar instrument configurations to our science observations. Scans were carried out using position switching and typically consisted of groups of 6 scans of 1 minute duration (2022-2023) or 3 scans of 2 minute duration (2020-2021). Data reduction was performed using our {\sc chaos} pipeline. Each scan was inspected by eye, and scans with significant artifacts were flagged and removed. We then fitted and removed a polynomial baseline from each scan, manually selecting the order of the polynomial that best corrects broad oscillations in the position switched baselines,\footnote{The choice of baseline has little effect on the recovered fluxes, as the lines monitored are very bright relative to these low-order corrections to the baseline level.} and computed the integrated flux over a fixed spectral window set to contain all emission from the CO line.

\subsubsection{Calibration Uncertainty for CO(2--1) Data}

The left panels of Figure~\ref{fig:fluxcal} show the median line flux for each group of CO(2--1) observations for CIT6 and IRC+10216. In all cases, the statistical uncertainty in the flux is very small and the variations between points reflect calibration error.  The scatter around the mean integrated flux can be interpreted as the calibration uncertainty for data taken with the SMT 1.2~mm receiver. The standard deviation in the fluxes amounts to 5.3\% of the integrated flux for IRC+10216 and 6.6\% for CIT6. Using half the spread between the 16th and 84th percentile of the data, which is more robust to outliers, gives a calibration uncertainty of 4.5\% for both sources. Results are consistent between the two sources, despite the factor of 5 difference in brightness, providing further confirmation that calibration errors rather than noise account for the scatter. 

Data taken in Fall 2020 and Spring 2021 show a larger scatter (spread between the 16th and 84th percentile of roughly 10\% of the median) than subsequent semesters (spread between the 16th and 84th percentile of roughly 5\% of the median). This can be attributed to a resurfacing of the SMT tertiary mirror performed in Fall 2021 which resulted in a significant improvement to the optical throughput and more reliable calibration results. 

A handful of points lie well below the typical range. The cause of these deviations is unknown, although many occurred in poorer weather conditions. This phenomenon is rare and appears to be transient, therefore we do not attempt to account for it in the flux calibration of our data. All of our science targets are observed for at least 10 scans, meaning that our final spectra average over 10 or more independent calibration measurements, taken over the course of more than an hour. As a result, transient calibration problems should have little effect on our final flux estimates.

\subsubsection{Calibration Uncertainty for CO(3--2) Data}

The right panels of Figure~\ref{fig:fluxcal} show the results of our CO(3--2) spectral line monitoring with the SMT 0.8~mm receiver. Measurements with this receiver show a larger scatter than at 1.2~mm. The fluxes for IRC+10216 and CIT6 have standard deviations of 11 and 12\% respectively. 

The larger scatter likely reflects a combination of less stable atmospheric conditions at 0.8~mm and standing waves in some 0.8~mm receiver data. The standing waves are narrow -- with typical periods less than the width of the target spectral line -- and not corrected by our low-order polynomial baseline fits. They therefore increase the scatter in our flux measurements. Our science observations are conducted with wobbler switching which is less susceptible to these standing waves. We take these results to be an upper limit on the calibration error. 

There is also a systematic offset between the two polarizations in the 0.8~mm monitoring. However, this effect is present only in spectral line monitoring, and we see no evidence of that our science observations  or continuum beam efficiencies require a correction for the effect (see Section~\ref{ss:errors-jackknife} and Figure~\ref{fig:pol32and10}).

\subsection{Telescope Efficiencies and the Flux Scale}\label{ss:errors-be}

The flux scale of our measurements is set by the point source sensitivity term in Equation~\ref{eq:Snu}. In this section, we describe our determination of this quantity and its uncertainty.

The coupling of flux into the telescope power pattern is quantified in terms of the forward\footnote{Conventions differ about whether the so-called ``main beam efficiency'' $\eta_{\rm mb}$ is defined to be the ratio of power falling in the main beam and the entire telescope power pattern or just the forward hemisphere. The standard procedure at the ARO is to measure and report the latter version. For definiteness, we refer to this as the forward main beam efficiency $\eta_{\rm fmb}$ here. It is equivalent to the main beam efficiency $\eta_{\rm mb}$ as defined in \citet{mangum93}, but differs from the main beam efficiencies reported in, for example, the IRAM 30m documentation by a factor of $F_{\rm eff}$ (the forward efficiency).} main beam efficiency ($\eta_{\rm fmb}$), the fraction of forward directed power falling in the main beam of the telescope. This is related to the point source sensitivity by \begin{equation}\label{eq:pss}
    \chi_{\rm ps} = \frac{2k\Omega_{\rm mb}(\lambda)}{\eta_{\rm fmb}\lambda^2}\,,
\end{equation}
where $k$ is the Boltzmann constant, $\Omega_{\rm mb}$ is the size of the primary beam, and $\lambda$ is the observing wavelength\footnote{$\Omega_{\rm mb}\propto\lambda^2$, so this relation has no wavelength dependence outside of that introduced by the changing value of $\eta_{\rm fmb}$.}.

Measurements of $\eta_{\rm fmb}$ for the ARO telescopes are typically made by collecting continuum observations of planets -- most frequently Jupiter and Mars -- and comparing the measured $T_A^*$ to models that account for the known brightness temperature of the planet and its coupling to the telescope beam. We made routine measurements of $\eta_{\rm fmb}$ during our SMT observing runs and we have also compiled $\eta_{\rm fmb}$ measurements made by observatory staff and other observing projects during the same time period. As the beam efficiency varies as a function of frequency, and can differ between receivers and polarizations due to differences in alignment and illumination of the primary mirror, we determined values of $\eta_{\rm fmb}$ and $\chi_{\rm ps}$ for each receiver system and polarization independently. Furthermore, our SMT observations were conducted over a period of four years from spring 2020 through summer 2023. During summers between observing seasons, maintenance and modifications to the dish, optics, and receiver systems are performed, and these modifications can affect the beam efficiency. In particular, in the summers of 2021 and 2022 adjustments were made to receiver alignment and the surface of the primary mirror. There were discernible differences in $\eta_{\rm fmb}$ after each of these adjustments, and therefore we assessed $\eta_{\rm fmb}$ separately for Spring 2020--Spring 2021, Spring 2022, and Fall 2022--Spring 2023. Our adopted values of $\eta_{\rm fmb}$ and $\chi_{\rm ps}$ are given in Table~\ref{tab:pss}.

The continuum-based efficiency measurements used for the 3~mm and 0.8~mm receivers show a large scatter (with standard deviations up to $\sim13$\% of the mean value for a given receiver and semester). However, the stability of our spectral line standards indicates that the beam efficiency was not varying significantly in periods between dish adjustments. The scatter likely resulted from instabilities in the continuum power detection chain -- which is primarily used for diagnostic purposes -- and the fact that power levels measured in wider bandwidths are intrinsically more unstable relative to their noise. We determined final $\eta_{\rm fmb}$ and $\chi_{\rm ps}$ values for the 3~mm and 0.8~mm receivers by averaging measurements made over the relevant observing seasons with those receivers. 

For the SMT 1~mm receiver, we employed a bootstraping procedure, using our spectral line standard observations to ensure the consistency of our flux calibrations over the full survey. We first determined the average $\eta_{\rm fmb}$, as measured using Mars, for each observing period (Spring 2020--Spring 2021, Spring 2022, and Fall 2022--Spring 2023). We used these efficiencies to flux calibrate our spectral standard data, and from these determined the average observed flux of the IRC+10216 CO(2--1) line in each semester. Treating measurements in the two receiver polarizations separately, this gave us six estimates of the line flux. We took the median of these as our best estimate for the true flux, and used this to derive values of $\eta_{\rm fmb}$ that would make the line fluxes for each observing period and polarization equal the median. We used these as our final beam efficiencies for the SMT. Because the AMISS science observations were carried out using identical hardware and similar observing procedures to the spectral line standards, we believe this procedure produces the most consistent flux calibration across semesters.\footnote{Using $\eta_{\rm fmb}$ derived from IRC+10216 removes semester to semester variations in the median flux of IRC+10216 in our spectral line flux monitoring (Section~\ref{ss:errors-fluxcal}). However, the CIT6 data are unaffected by this choice.}\footnote{Fluxes of some line species in IRC+10216 have been found to vary with a period of around 650 days \citep{he+17,pardo+18}. The low-$J$ CO lines are not among those that show significant variability \citep{cernicharo+14}, and we see no evidence that the IRC+10216 CO fluxes varied systematically relative to CIT6 or our planet flux measurements over the relevant timescales.}

Uncertainty in our flux scale is driven by two factors -- measurement uncertainty for our planet observations, and uncertainty in the planet flux models. We use at least 12 independent beam efficiency measurements in determining the average value for every time period reported in Table~\ref{tab:pss}, and therefore expect the measurement error to be minimal. Uncertainties in the planet fluxes are typically quoted as $\sim 5\%$. Comparison between commonly used models of Mars and Jupiter and Planck observations calibrated on the CMB temperature suggests this estimate is reasonable \citep{akrami+17}. We find similar results comparing independent beam efficiency measurements made on Mars and Jupiter: SMT flux scales calibrated using Mars or Jupiter observations differ by an average of 5\% for a given observing semester and receiver, with a maximum difference of 10\%.

Based on this, we estimate the errors in our flux scale to be $\sim 5\%$. This error appears only as a constant scale factor applied uniformly across our entire dataset, meaning the uncertainty in flux scale can be further reduced if more accurate planet models becomes available.

\subsection{Source-Telescope Coupling}\label{ss:errors-beam}

\begin{figure*}
    \centering
    \includegraphics[width=\textwidth]{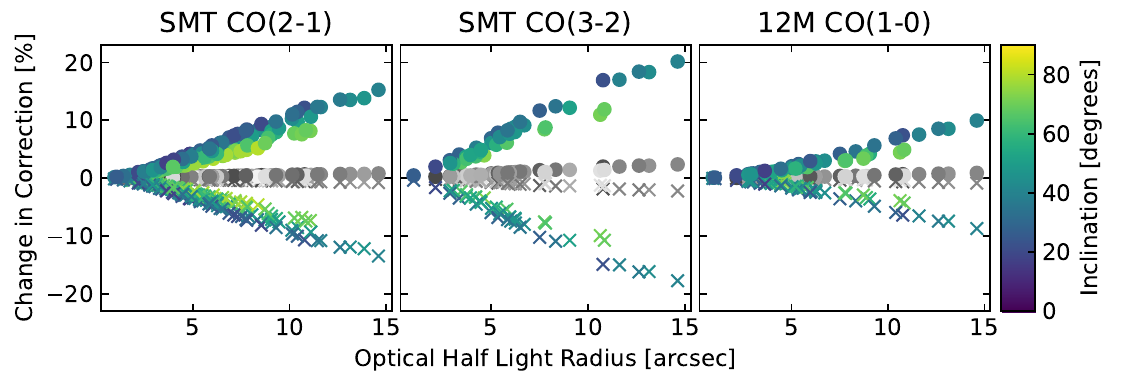}
    \caption{Colored points show the change in aperture correction ($1/K$) caused by a 20\% over- (circles) or under- (crosses) estimation of the CO disk size, plotted as a function of the size of the galaxy, for each target in our CO(2--1) (left), CO(3--2) (center), and CO(1--0) (right) samples. Points are colored according to each target's inclination. Gray points show the effect of a over- or underestimating the beam size by $\pm1\sigma$.}
    \label{fig:sizecorr}
\end{figure*}

The final source of error in our integrated flux measurements is the source-beam coupling term $K$ in Equation~\ref{eq:sdv}. To compute $K$ we define $s(x,y)dxdy$ as the fractional contribution of position $(x,y)$ to the total flux of the galaxy (i.e. $s(x,y)$ integrates to one), where the coordinate system is centered on the galaxy with the $x$ coordinate aligned to the major axis. We define $b(x\prime,y\prime)$ as the power pattern of the telescope beam, normalized such that the response on the beam axis is $b(0,0)=1$. The coupling term is then
\begin{equation} \label{eq:beamcorr}
    K = \int\int b(x\prime-x,y\prime-y) s(x,y) dx dy .
\end{equation}
and the aperture correction is $1/K$.

We model the telescope beam pattern as a Gaussian
\begin{equation}
    b(x,y) = \exp\Big(-4\ln2 \frac{x^2+y^2}{\theta^2}\Big) .
\end{equation}
where $\theta$ is the full width at half power (FWHM). We model each galaxy as thin, exponential disks with half light radius $r_{50}$ and inclination $i$
\begin{equation}\label{eq:fluxprofile}
    s(x,y) = \frac{1}{2\pi r_e^2 \cos i}\exp\Bigg(-\frac{\sqrt{x^2+\big(\frac{y}{\cos i}\big)^2}}{0.596 r_{50}}\Bigg) 
\end{equation}
(the factor of 0.596 is the ratio between the exponential disk scale length and $r_{50}$).

From Equation~\ref{eq:beamcorr} it is clear that three factors can introduce uncertainty in $K$: the flux distribution of the source ($s$), the size and shape of the primary beam ($b$), and the pointing of the telescope ($x\prime-x,y\prime-y)$. In the following, we evaluate the uncertainty in each of these terms and calculate how they translate to uncertainty in our CO line fluxes.

\subsubsection{Source Sizes and Profiles}

\begin{deluxetable*}{ll|cc|cc}
    \tablecaption{Systematic uncertainties introduced by aperture corrections \label{tab:diskuncertainties}}
    
    \tablehead{
        \multicolumn{2}{c}{} & \multicolumn{2}{c}{Disk Size Uncertainty} & \multicolumn{2}{c}{Beam Size Uncertainty} \\
        \colhead{Line or Line Ratio} & \colhead{Telescope} & \colhead{Median} & \colhead{Maximum} & \colhead{Median} & \colhead{Maximum}
    }
    
    \startdata
    CO(1--0) & ARO 12m & $2\%$ & $9\%$ & $<1\%$ & $<1\%$ \\
    CO(1--0) & IRAM 30m & $6\%$ & $19\%$ & $<1\%$ & $2\%$ \\
    CO(2--1) & SMT & $3\%$ & $14\%$ & $<1\%$ & $<1\%$ \\
    CO(3--2) & SMT & $6\%$ & $19\%$ & $<1\%$ & $2\%$ \\
    CO(2--1) / CO(1--0) & SMT / ARO 12m & $2\%$ & $5\%$ & $<1\%$ & $1\%$ \\
    CO(2--1) / CO(1--0) & SMT / IRAM 30m & $3\%$ & $5\%$ & $<1\%$ & $2\%$ \\
    CO(3--2) / CO(1--0) & SMT / ARO 12m & $4\%$ & $10\%$ & $<1\%$ & $2\%$ \\
    CO(3--2) / CO(1--0) & SMT / IRAM 30m & $<1\%$ & $<1\%$ & $<1\%$ & $3\%$ \\
    CO(3--2) / CO(2--1) & SMT / SMT & $2\%$ & $5\%$ & $<1\%$ & $2\%$ \\
    CO(1--0) / CO(1--0) & IRAM 30m / ARO 12m & $4\%$ & $10\%$ & $<1\%$ & $2\%$ \\
    CO(2--1) / CO(2--1) & IRAM 30m / SMT & $10\%$ & $14\%$ & $1\%$ & $2\%$ \\
    \enddata

    \tablecomments{We give the uncertainty in disk-integrated CO luminosity or CO luminosity ratios caused a 20\% uncertainty in the scale length of the CO disk and a $1\sigma$ uncertainty in the telescope beam sizes. For each line/ratio and telescope/telescope-pair considered in our survey, we quantify the systematic uncertainty in CO luminosity or line ratio in terms of the median error for all of our targets, and the error for the target most significantly affected (maximum).}
\end{deluxetable*}

The most significant uncertainty in our aperture corrections can be attributed to the physical model used to describe our sources. Resolved studies of the disks of nearby star forming galaxies show that the distribution of CO emission is globally well described by an exponential profile with a scale length comparable to the optical disk \citep{regan+01,leroy+08,leroy+09,bolatto+17}. A number of works have used resolved maps to demonstrate that aperture corrections similar to Equation~\ref{eq:beamcorr} accurately recover, on average, the integrated CO flux of star forming galaxies \citep{lisenfeld+11,boselli+14,bothwell+14,leroy+21}. However, the correlation between CO and optical disk sizes shows significant scatter, which can lead to substantial uncertainties in the corrections for individual galaxies. \citet{leroy+08} find a roughly 20\% scatter in the ratio of optical and CO scale lengths. \citet{leroy+21} consider numerous methods of aperture correction based on multi-wavelength maps of very nearby disk galaxies. They find that exponential disk models give similar results to more accurate methods using WISE mid-infrared maps (not available for our more distant sources), but have a scatter of $\sim 20$\%.

In Figure~\ref{fig:sizecorr} we show the change in aperture correction caused by a 20\% departure of the true CO disk size from the value assumed when computing $K$. The effect increases for larger galaxies and smaller telescope beams, but is less than 20\% for every source in our sample, and typically less than 10\% for CO(2--1). For ratios between luminosity (or flux) pairs considered in Section~\ref{ss:results-ratios}, the errors in aperture corrections partially cancel resulting in smaller uncertainties. We tabulate the median and maximum uncertainties in CO luminosity and CO line ratios resulting from uncertainty in the galaxy disk sizes in Table~\ref{tab:diskuncertainties}.

\subsubsection{The Primary Beam}

The beam size is set by the illumination pattern of the receivers. SMT and 12m receivers are designed to produce a Gaussian illumination pattern with a $\sim$12~dB taper at the edge of the primary reflector, resulting in a primary beam FWHM of $1.17\lambda/D$ where $\lambda$ is the observing wavelength and $D$ is the diameter of the dish. We mapped the primary beams of the receivers used for AMISS by measuring the response to a bright source as a function of offset from the beam axis. Based on these, we find beam sizes of $(1.18\pm0.01)\lambda/D$ for the SMT 1.2~mm receiver, $(1.24\pm0.03)\lambda/D$ for the SMT 0.8~mm receiver,\footnote{Except for the horizontal polarization in December 2022 and January 2023, which had a beam size of $1.37\lambda/D$.} and $(1.06\pm0.02)\lambda/D$ for the 12m 3~mm receiver. Beam sizes for each receiver at representative frequencies of 230.5, 345.8 and 115.3~GHz are given in Table~\ref{tab:pss}.

The beam sizes and uncertainties for the 1.2 and 3~mm receivers have been verified over multiple independent observations. 

All beam profiles are well described by a two dimensional Gaussian to a radius of at least $\lambda/D$. Our source sizes are typically comparable to or smaller than the beam FWHM for all receivers used in this study, and we do not expect significant flux to enter through non-Gaussian parts of our beam. Therefore we conclude that our modeling of the primary beam introduces minimal error into our estimate of $K$. This provides a considerable calibration advantage over resolved studies, where flux entering through sidelobes and error beams can be a significant source of uncertainty \citep{garcia-burillo+93, denbrok+22}. 

We compute the effects on $K$ of perturbing the beam size by its uncertainty. Figure~\ref{fig:sizecorr} shows the results for each of our sources -- such errors result in changes of less than 2\%. Median and maximum uncertainties in the CO luminosities and line ratios are listed in Table~\ref{tab:diskuncertainties}.

\subsubsection{Telescope Pointing}

\begin{figure*}
    \centering
    \includegraphics[width=\textwidth]{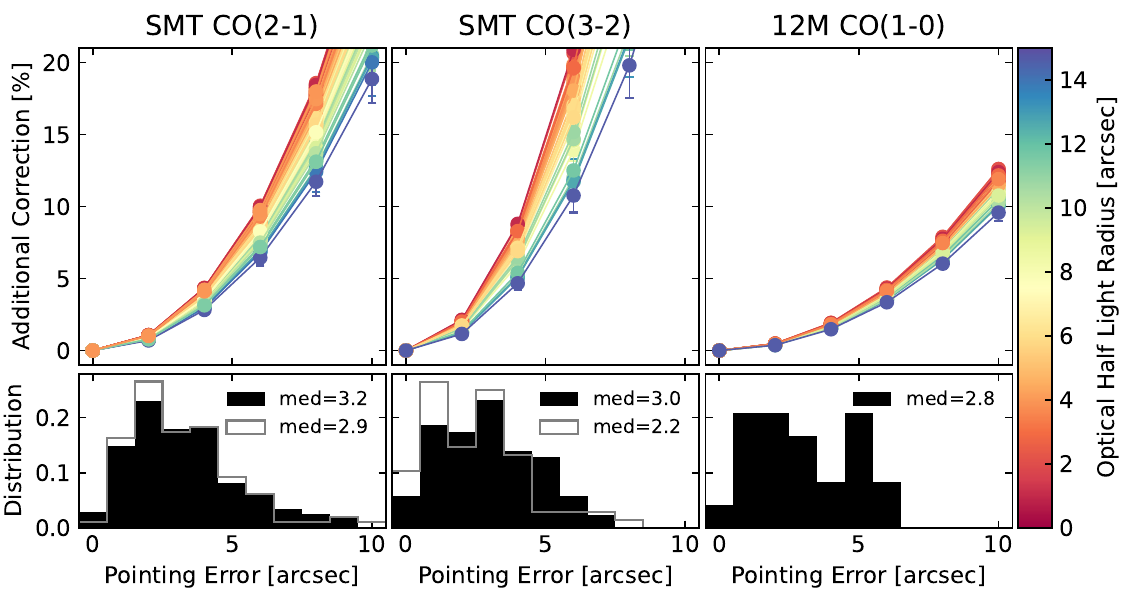}
    \caption{The upper row shows the change in aperture correction factor ($1/K$) resulting from pointing errors of 0-10 arcseconds for each of our sources. Points show the calculation for pointing errors at 45 degrees to the source position angle, while error bars show the results at 0 and 90 degrees. Points are color coded by the optical half light radius of the source.
    The bottom row shows the magnitude of shifts between pointing measurements taken at intervals of less than 1.5 hours (gray open histogram), and 1.5 to 2.5 hours.
    From left to right columns show results for SMT CO(2--1), SMT CO(3--2) observations, and 12m CO(1--0) observations (note that the scaling with pointing error is also applicable to IRAM 30m CO(1--0) observations from xCOLD~GASS).}
    \label{fig:pointing}
\end{figure*}

Errors or drifts in telescope pointing can cause the pointing axis of the telescope to differ from the source coordinates, producing sub-optimal coupling. We checked telescope pointing at approximately two hour intervals throughout our observing program, and found pointing changes on such a timescale were generally small relative to the size of beam.

To quantify the pointing errors we compiled pointing offsets from 456 pointing checks with the SMT 1.2~mm receiver, 194 pointing checks with the SMT 0.8~mm receiver, and 51 pointing checks with the 12m 3~mm receiver. These measurements allow us to compute the magnitude of changes in the pointing offsets between different checks. We grouped the pointing differences into two timescales: differences between checks made within 1.5 hours of one another, which we take to be indicative of the uncertainty in the pointing measurements themselves,\footnote{Pointing measurements taken within 1 hour of one another typically come from observations done for calibration and/or spectral line standards, and are often taken on multiple sources. Differences between offsets from these measurements therefore include effects of both measurement uncertainty and errors in the telescope's all-sky pointing model.} and differences between checks made between 1.5 and 2.5 hours apart, which also capture any drifts in pointing on the typical timescale of our science observations.

Figure~\ref{fig:pointing} shows the increase in aperture correction as a function of pointing error for each of our sources, along with the distribution of pointing shifts for the three receivers. On both timescales we find a median pointing change of $\sim3$ arcseconds. We take this to mean that our pointing uncertainties are dominated by uncertainties in the offset measurement or the pointing model, rather than significant pointing drifts over time. 

The measured pointing changes are the difference of two uncertain measurements, meaning that the pointing uncertainty itself is $ 3/\sqrt{2} \sim 2$ arcseconds. This translates to a decrease in $K$ by less than 1\% at 1.2 and 3~mm, and 1 to 2\% at 0.8~mm. As these pointing losses are small compared to our uncertainty level we do not attempt to correct for them.

\subsection{Comparison to Previous Observations}\label{ss:errors-othertels}

\begin{figure*}
    \centering
    \includegraphics[width=\textwidth]{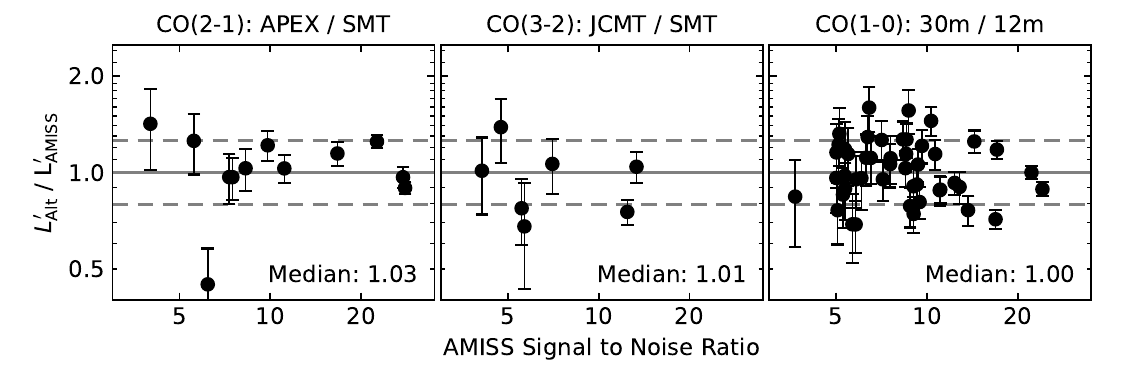}
    \caption{Comparison of CO luminosities measured by AMISS and other studies. Each panel shows the ratio of the beam corrected CO luminosities as a function of the AMISS signal to noise ratio. The solid gray line shows a ratio of 1 and the dashed lines show ratios of $\pm 0.1$~dex. Median ratios for each panel are given in the lower right. Comparison data comes from CO(2--1) observations with APEX \citep[left]{saintonge+17}, CO(3--2) observations with the JCMT \citep[middle]{lamperti+20}, and xCOLD~GASS CO(1--0) observations rescaled as discussed in Appendix~\ref{ap:xcg_flux} (right).}
    \label{fig:othertels}
\end{figure*}

As a further check on our uncertainties, we can compare our measurements with those from other facilities. In the left panel of Figure~\ref{fig:othertels} we show the ratio of AMISS and Atacama Pathfinder Experiment (APEX) CO(2--1) luminosities for 12 sources observed by \citet{saintonge+17}. In the middle panel we show the ratio of AMISS and James Clerk Maxwell Telescope (JCMT) CO(3--2) luminosities for 7 sources shared between our sample and \citet{lamperti+20}. In the right panel we compare ARO 12m and IRAM 30m luminosities for our CO(1--0) targets. In all comparisons, we apply aperture corrections to each luminosity before computing the ratios. We consider only sources where both telescopes detected the line with ${\rm SNR}>3$. For the IRAM CO(1--0) comparison, we use our re-reduction of the xCOLD~GASS data rather than the luminosities originally reported by \citet{saintonge+17}.

The median ratio between the APEX and SMT CO(2--1) luminosities is $1.03\pm0.11$ and the geometric mean\footnote{The geometric mean is most appropriate for this type of comparison. To see this, consider luminosity ratios of 0.5 and 2.0, i.e. one measurement where the telescope A reports twice the luminosity of telescope B and one where telescope B reports twice the luminosity of telescope A. These have a geometric mean of 1.0, but an arithmetic mean of 1.25} is $1.04\pm0.03$. For CO(3--2) the median JCMT to SMT luminosity ratio is $1.01\pm0.11$ and the geometric mean is $0.90\pm0.05$. Median and mean ratios close to unity imply that the flux scale is consistent between different facilities, and suggests that the calibration differs by at most about 10\%.

The scatter in telescope--telescope ratios can be used as a check on our error estimates. In analogy to Equation~\ref{eq:hv}, we write uncertainty-normalized luminosity-differences as
\begin{equation}
    \Delta_{T_1T_2} = \frac{L_{T_1}-L_{T_2}}{\sqrt{\sigma_{L_{T1}}^2+\sigma_{L_{T2}}^2}} .
\end{equation}
When we account only for measurement uncertainties, the distribution of $\Delta_{T_1T_2}$ is 1.3 to 1.8 times wider than expected. We find that including the aperture correction uncertainty (Section~\ref{ss:errors-beam}) and an additional systematic uncertainty of 5-10\% for each telescope is sufficient to produce a distribution of $\Delta_{T_1T_2}$ with the expected variance of 1. This is consistent with our estimates of the relative calibration uncertainties in Section~\ref{ss:errors-fluxcal}. Note that our recalibration forces the median 12m-30m ratio to be unity for the CO(1--0) comparison, but the distribution of $\Delta_{T_1T_2}$ is still a valid check on our errors.

Among all luminosity pairs considered considered, only one differs by more than a factor of $\sim1.5$: the APEX CO(2--1) luminosity for source AMISS.2039 (COLD GASS ID 1977) is lower than the SMT result by a factor of 3. The APEX luminosity would imply that the CO(2--1) / CO(1--0) luminosity ratio for this source is $0.2$ while the CO(3--2) / CO(2--1) ratio is $1.5$. Such a combination of line ratios is incompatible with expectations for the bulk of molecular gas. The AMISS CO(2--1) luminosity gives a more reasonable set of ratios: CO(2--1) / CO(1--0) $\sim$ 0.6 and CO(3--2) / CO(2--1) $\sim$ 0.5. We therefore conclude that an erroneous APEX luminosity is the cause of the large discrepancy.

\subsection{Summary of the AMISS Uncertainty Budget}\label{ss:errors-summarry}

\begin{figure}
    \centering
    \includegraphics[width=.45\textwidth]{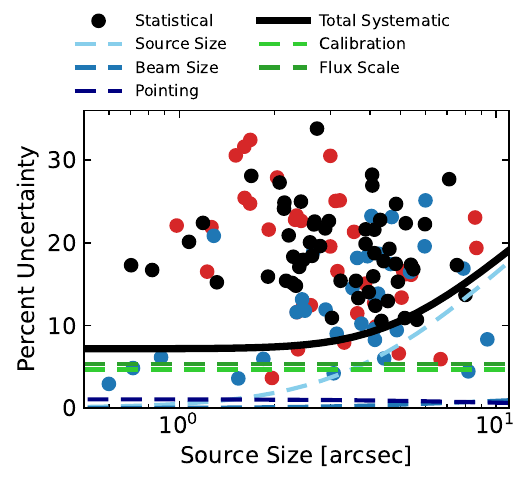}
    \caption{Sources of error on CO(2--1) fluxes for each of our detected CO targets are plotted as a function of optical half light radius. Errors are given as percent uncertainties. Points denote the statistical uncertainty, and are color coded by subsample -- primary (black), star forming (blue), and filler (red). The solid black line shows the total systematic uncertainty. Dashed lines show the various elements contributing to the systematic uncertainty -- calibration errors (light green), flux scale uncertainty (dark green) , pointing errors (dark blue), uncertainty in the source-beam coupling due to measurement in the beam size (medium blue) and unknown source size (light blue).} 
    \label{fig:errors}
\end{figure}

To summarize, for most of our targets, statistical uncertainties dominate over systematic errors in our uncertainty budget. The statistical uncertainty in a flux measurement for a typical line detected in our survey is $\sim 20\%$. We have cross-checked these noise estimates by comparing sets of statistically independent measurements taken in the horizontal and vertical polarizations, and found no evidence of additional errors.

Systematic uncertainties in our analysis can be broken into two categories -- those that add additional scatter between measurements of different sources, and those that scale the luminosity of every source in a constant manner, but do not alter the relative luminosities (or line ratios), when comparing one source to another.

Among the first type of systematic error, monitoring of the CO flux of bright carbon stars shows that there is an additional 5-10\% uncertainty in the flux calibration over time. The scatter between AMISS measurements and those made at other telescopes also suggests the presence of such a scatter. Additional uncertainty is introduced by the unknown spatial profile of CO emission for each source. The uncertainty caused by imperfect aperture corrections can reach 10-20\% for the largest sources in our survey. However, this uncertainty is reduced significantly when comparing the ratios of lines measured with similar beam sizes.

The primary scale factor-like error for our survey is the uncertainty in the flux scale of each telescope. This uncertainty is due to the flux models for our calibration sources, which are accurate to about the 5\% level, and by imperfect measurements of the beam efficiency. Additional uncertainties come from measurement of the primary beam size (at most a 2\% effect), and from errors in telescope pointing which systematically lower the flux for all targets comparable to or smaller than the beam (also a 1-2\% effect).

Figure~\ref{fig:errors} summarizes these various sources of errors, showing the fractional statistical uncertainty for each galaxy detected by AMISS in CO(2--1), along with typical uncertainties or errors from calibration, flux scale, aperture correction, and telescope pointing. On the whole, systematic uncertainties are small, well understood, and uniform across the entire survey. This allows us to utilize these data to precisely measure beam filling factors (Section~\ref{ss:results-disk}) and CO line ratios \citep{keenan+24b} that vary over an intrinsically limited dynamic range.


\section{Results}\label{sec:results}

The final AMISS survey includes CO(2--1) observations of 176 galaxies, of these ${\rm SNR}>3$ (${\rm SNR}>5$) detections are achieved for 130 (87) and $3\sigma$ upper limits of $L_{\rm CO(2--1)}^\prime<10^8$~K~km~s$^{-1}$~pc$^2$ are obtained for another 38. We have obtained CO(3--2) observations of 45 of these galaxies, with detections of 34. We complement these observations with xCOLD~GASS CO(1--0) data for all of our targets, and additional ARO 12m telescope CO(1--0) observations -- covering a larger faction of the gas disks -- for 56 (54 detections). Together, these observations constitute the largest homogeneous sample of unresolved CO(1--0) through CO(3--2) spectra of nearby-universe galaxies of which we are aware. The careful calibration of this dataset makes it possible to carry out detailed studies of CO line ratios \citep{keenan+24b}. While the breadth and depth of the CO(2--1) sample make it possible to study CO(2--1) properties as a function of host galaxy properties over a large range in galaxy properties. 

\begin{figure*}[p]
    \centering
    \includegraphics[width=.97\textwidth]{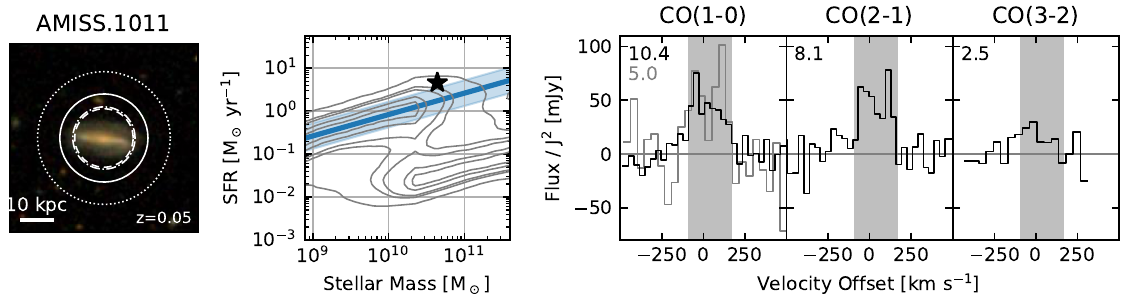}
    \includegraphics[width=.97\textwidth]{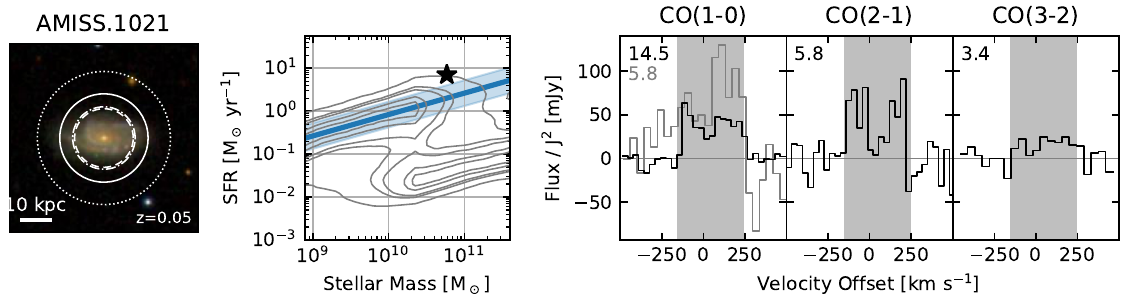}
    \includegraphics[width=.97\textwidth]{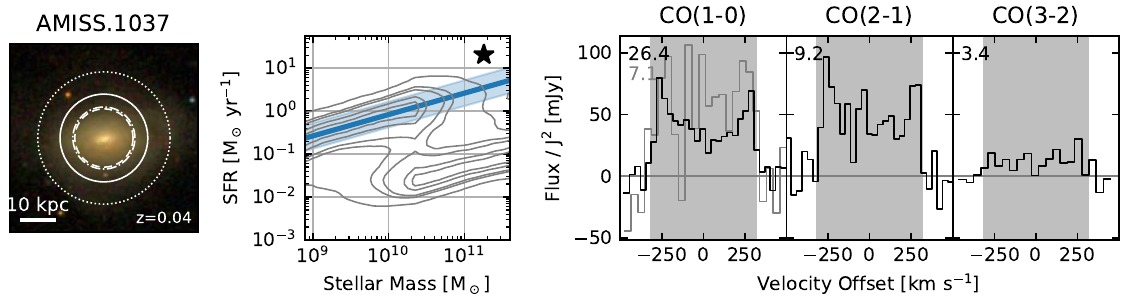}
    \includegraphics[width=.97\textwidth]{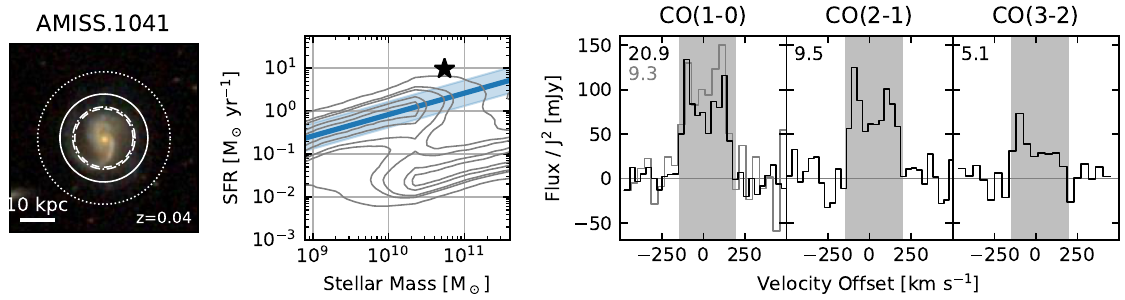}
    \caption{Gallery of AMISS targets and spectra. Select targets are shown here, a complete figure set (177 images) is available in the online journal.
    Left column: SDSS cutouts of each target. Solid, dashed, and dotted lines show the beam sizes of the SMT for CO(2--1), the SMT for CO(3--2) (equal to the IRAM 30m for CO(1--0)) and the 12m for CO(1--0) respectively. The scale bar in the lower left shows 10 kiloparsecs. 
    Middle column: contours show the distribution of star formation rates at a given stellar mass, while the blue line and filled region show the main sequence of star forming galaxies. The stellar mass and star formation rate of the target galaxy is marked by a star. 
    Right column: CO(1--0) spectra from AMISS (gray) and xCOLD~GASS (black), CO(2--1) spectra from AMISS, and CO(3--2) spectra from AMISS. Numbers in the upper right corner give the signal to noise ratio for each line. When a CO line is detected, the gray band indicates the region used to measure the line flux. The scale of the $y$-axis is such that lines would have the same amplitude in each transition for thermalized CO emission. The relative amplitudes of each spectrum give a sense of the luminosity ratios between the different lines.}
    \label{fig:spectra}
\end{figure*}

Figure~\ref{fig:spectra} shows SDSS images and CO spectra of four example galaxies observed in all three CO transitions. Plots for all 177 AMISS targets are available in the online version of this article. The CO line luminosities, along with a range of ancillary information are summarized in Table~\ref{tab:results}, an extended, machine readable version of which is also available online. All 277 spectra observed by AMISS are available online. \emph{[note to reviewer: spectra data will be made available online via Zenodo upon publication of this article]}.

In the remainder of this section, we present some initial scientific results from AMISS. We measure the average size of molecular gas disks relative to galaxy stellar sizes (Section~\ref{ss:results-disk}) and search for excesses of molecular gas in the outskirts of galaxies (Section~\ref{ss:results-amiss10}). We also present typical CO(1--0), CO(2--1), and CO(3--2) line ratios in Section~\ref{ss:results-ratios}.

\figsetstart
\figsetnum{15}
\figsettitle{Gallery of AMISS targets and spectra}

\figsetgrpstart
\figsetgrpnum{15.1}
\figsetgrptitle{AMISS.1000}
\figsetplot{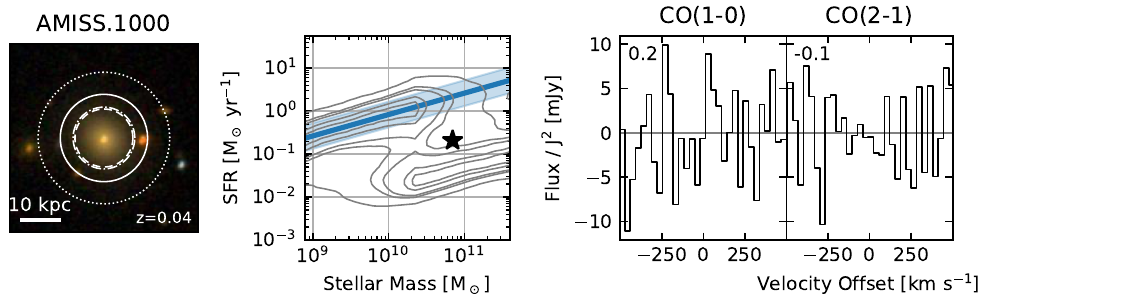}
\figsetgrpnote{Left column: SDSS cutouts of each target. Solid, dashed, and dotted lines show the 
beam sizes of the SMT for CO(2--1), the SMT for CO(3--2) (and the IRAM 30m for CO(1--0)) 
and the 12m for CO(1--0) respectively. The scale bar in the lower left shows 10 
kiloparsecs. 
Middle column: contours show the distribution of star formation rates at a given stellar 
mass, while the blue line and filled region show the main sequence of star forming 
galaxies. The stellar mass and star formation rate of the target galaxy is marked by a 
star. 
Right column: CO(1--0) spectra from AMISS (gray) and xCOLD~GASS (black), CO(2--1) spectra 
from AMISS, and CO(3--2) spectra from AMISS. Numbers in the upper right corner give the 
signal to noise ratio for each line. When a CO line is detected, the gray band indicates 
the region used to measure the line flux. The scale of the $y$-axis is such that lines 
would have the same amplitude in each transition for thermalized CO emission. The relative 
amplitudes of each spectrum give a sense of the luminosity ratios between the different 
lines.}
\figsetgrpend

\figsetgrpstart
\figsetgrpnum{15.2}
\figsetgrptitle{AMISS.1001}
\figsetplot{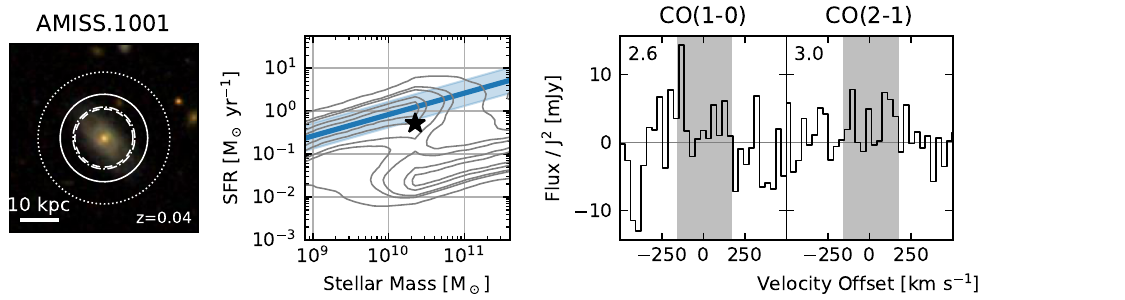}
\figsetgrpnote{Left column: SDSS cutouts of each target. Solid, dashed, and dotted lines show the 
beam sizes of the SMT for CO(2--1), the SMT for CO(3--2) (and the IRAM 30m for CO(1--0)) 
and the 12m for CO(1--0) respectively. The scale bar in the lower left shows 10 
kiloparsecs. 
Middle column: contours show the distribution of star formation rates at a given stellar 
mass, while the blue line and filled region show the main sequence of star forming 
galaxies. The stellar mass and star formation rate of the target galaxy is marked by a 
star. 
Right column: CO(1--0) spectra from AMISS (gray) and xCOLD~GASS (black), CO(2--1) spectra 
from AMISS, and CO(3--2) spectra from AMISS. Numbers in the upper right corner give the 
signal to noise ratio for each line. When a CO line is detected, the gray band indicates 
the region used to measure the line flux. The scale of the $y$-axis is such that lines 
would have the same amplitude in each transition for thermalized CO emission. The relative 
amplitudes of each spectrum give a sense of the luminosity ratios between the different 
lines.}
\figsetgrpend

\figsetgrpstart
\figsetgrpnum{15.3}
\figsetgrptitle{AMISS.1002}
\figsetplot{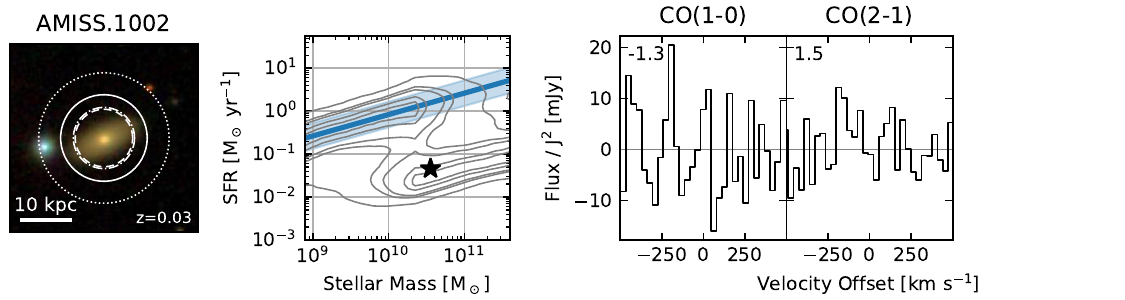}
\figsetgrpnote{Left column: SDSS cutouts of each target. Solid, dashed, and dotted lines show the 
beam sizes of the SMT for CO(2--1), the SMT for CO(3--2) (and the IRAM 30m for CO(1--0)) 
and the 12m for CO(1--0) respectively. The scale bar in the lower left shows 10 
kiloparsecs. 
Middle column: contours show the distribution of star formation rates at a given stellar 
mass, while the blue line and filled region show the main sequence of star forming 
galaxies. The stellar mass and star formation rate of the target galaxy is marked by a 
star. 
Right column: CO(1--0) spectra from AMISS (gray) and xCOLD~GASS (black), CO(2--1) spectra 
from AMISS, and CO(3--2) spectra from AMISS. Numbers in the upper right corner give the 
signal to noise ratio for each line. When a CO line is detected, the gray band indicates 
the region used to measure the line flux. The scale of the $y$-axis is such that lines 
would have the same amplitude in each transition for thermalized CO emission. The relative 
amplitudes of each spectrum give a sense of the luminosity ratios between the different 
lines.}
\figsetgrpend

\figsetgrpstart
\figsetgrpnum{15.4}
\figsetgrptitle{AMISS.1003}
\figsetplot{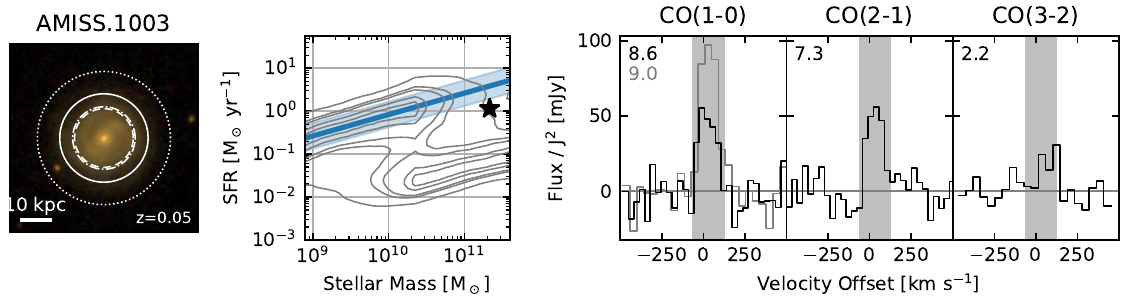}
\figsetgrpnote{Left column: SDSS cutouts of each target. Solid, dashed, and dotted lines show the 
beam sizes of the SMT for CO(2--1), the SMT for CO(3--2) (and the IRAM 30m for CO(1--0)) 
and the 12m for CO(1--0) respectively. The scale bar in the lower left shows 10 
kiloparsecs. 
Middle column: contours show the distribution of star formation rates at a given stellar 
mass, while the blue line and filled region show the main sequence of star forming 
galaxies. The stellar mass and star formation rate of the target galaxy is marked by a 
star. 
Right column: CO(1--0) spectra from AMISS (gray) and xCOLD~GASS (black), CO(2--1) spectra 
from AMISS, and CO(3--2) spectra from AMISS. Numbers in the upper right corner give the 
signal to noise ratio for each line. When a CO line is detected, the gray band indicates 
the region used to measure the line flux. The scale of the $y$-axis is such that lines 
would have the same amplitude in each transition for thermalized CO emission. The relative 
amplitudes of each spectrum give a sense of the luminosity ratios between the different 
lines.}
\figsetgrpend

\figsetgrpstart
\figsetgrpnum{15.5}
\figsetgrptitle{AMISS.1004}
\figsetplot{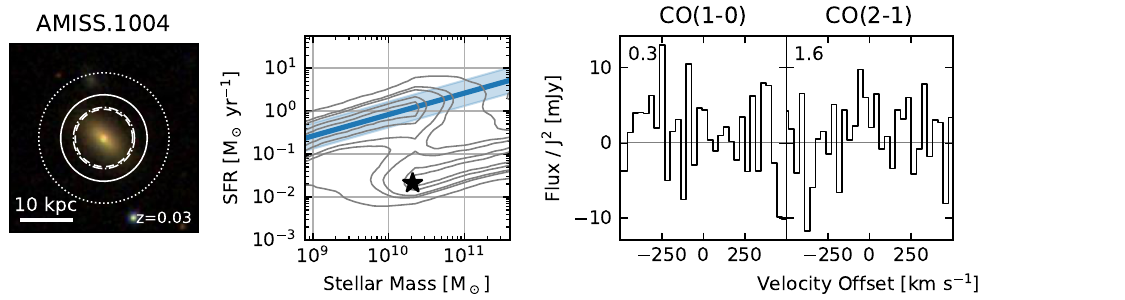}
\figsetgrpnote{Left column: SDSS cutouts of each target. Solid, dashed, and dotted lines show the 
beam sizes of the SMT for CO(2--1), the SMT for CO(3--2) (and the IRAM 30m for CO(1--0)) 
and the 12m for CO(1--0) respectively. The scale bar in the lower left shows 10 
kiloparsecs. 
Middle column: contours show the distribution of star formation rates at a given stellar 
mass, while the blue line and filled region show the main sequence of star forming 
galaxies. The stellar mass and star formation rate of the target galaxy is marked by a 
star. 
Right column: CO(1--0) spectra from AMISS (gray) and xCOLD~GASS (black), CO(2--1) spectra 
from AMISS, and CO(3--2) spectra from AMISS. Numbers in the upper right corner give the 
signal to noise ratio for each line. When a CO line is detected, the gray band indicates 
the region used to measure the line flux. The scale of the $y$-axis is such that lines 
would have the same amplitude in each transition for thermalized CO emission. The relative 
amplitudes of each spectrum give a sense of the luminosity ratios between the different 
lines.}
\figsetgrpend

\figsetgrpstart
\figsetgrpnum{15.6}
\figsetgrptitle{AMISS.1005}
\figsetplot{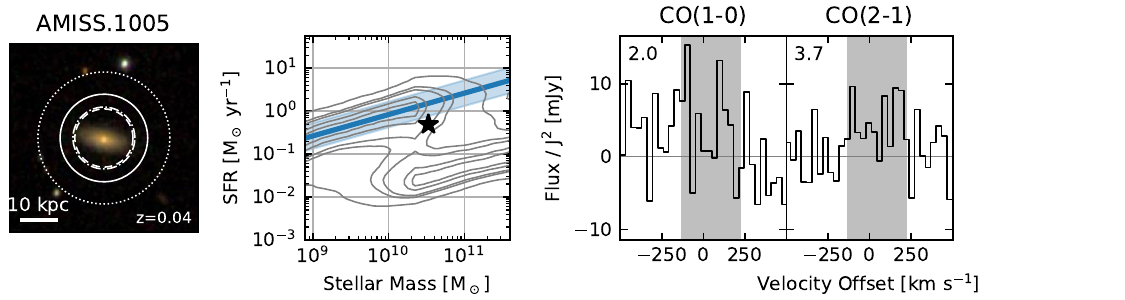}
\figsetgrpnote{Left column: SDSS cutouts of each target. Solid, dashed, and dotted lines show the 
beam sizes of the SMT for CO(2--1), the SMT for CO(3--2) (and the IRAM 30m for CO(1--0)) 
and the 12m for CO(1--0) respectively. The scale bar in the lower left shows 10 
kiloparsecs. 
Middle column: contours show the distribution of star formation rates at a given stellar 
mass, while the blue line and filled region show the main sequence of star forming 
galaxies. The stellar mass and star formation rate of the target galaxy is marked by a 
star. 
Right column: CO(1--0) spectra from AMISS (gray) and xCOLD~GASS (black), CO(2--1) spectra 
from AMISS, and CO(3--2) spectra from AMISS. Numbers in the upper right corner give the 
signal to noise ratio for each line. When a CO line is detected, the gray band indicates 
the region used to measure the line flux. The scale of the $y$-axis is such that lines 
would have the same amplitude in each transition for thermalized CO emission. The relative 
amplitudes of each spectrum give a sense of the luminosity ratios between the different 
lines.}
\figsetgrpend

\figsetgrpstart
\figsetgrpnum{15.7}
\figsetgrptitle{AMISS.1006}
\figsetplot{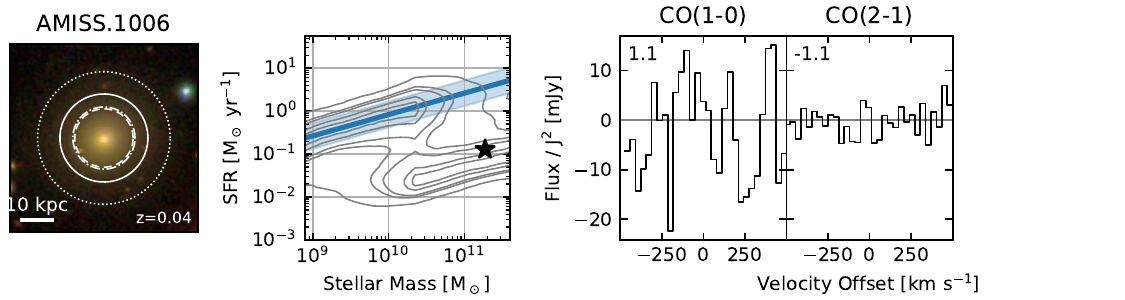}
\figsetgrpnote{Left column: SDSS cutouts of each target. Solid, dashed, and dotted lines show the 
beam sizes of the SMT for CO(2--1), the SMT for CO(3--2) (and the IRAM 30m for CO(1--0)) 
and the 12m for CO(1--0) respectively. The scale bar in the lower left shows 10 
kiloparsecs. 
Middle column: contours show the distribution of star formation rates at a given stellar 
mass, while the blue line and filled region show the main sequence of star forming 
galaxies. The stellar mass and star formation rate of the target galaxy is marked by a 
star. 
Right column: CO(1--0) spectra from AMISS (gray) and xCOLD~GASS (black), CO(2--1) spectra 
from AMISS, and CO(3--2) spectra from AMISS. Numbers in the upper right corner give the 
signal to noise ratio for each line. When a CO line is detected, the gray band indicates 
the region used to measure the line flux. The scale of the $y$-axis is such that lines 
would have the same amplitude in each transition for thermalized CO emission. The relative 
amplitudes of each spectrum give a sense of the luminosity ratios between the different 
lines.}
\figsetgrpend

\figsetgrpstart
\figsetgrpnum{15.8}
\figsetgrptitle{AMISS.1007}
\figsetplot{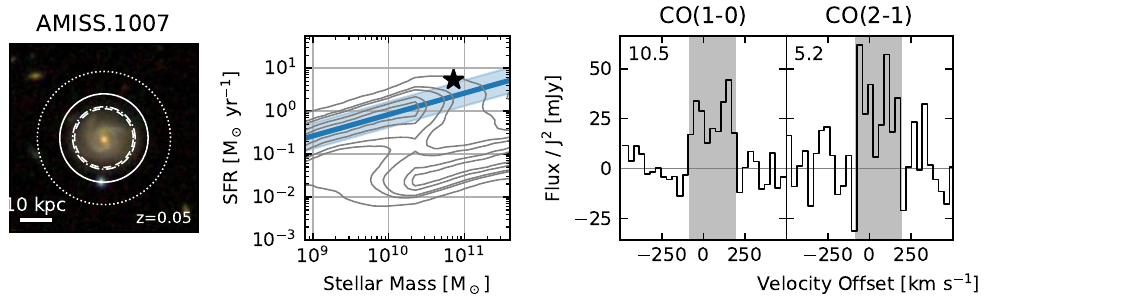}
\figsetgrpnote{Left column: SDSS cutouts of each target. Solid, dashed, and dotted lines show the 
beam sizes of the SMT for CO(2--1), the SMT for CO(3--2) (and the IRAM 30m for CO(1--0)) 
and the 12m for CO(1--0) respectively. The scale bar in the lower left shows 10 
kiloparsecs. 
Middle column: contours show the distribution of star formation rates at a given stellar 
mass, while the blue line and filled region show the main sequence of star forming 
galaxies. The stellar mass and star formation rate of the target galaxy is marked by a 
star. 
Right column: CO(1--0) spectra from AMISS (gray) and xCOLD~GASS (black), CO(2--1) spectra 
from AMISS, and CO(3--2) spectra from AMISS. Numbers in the upper right corner give the 
signal to noise ratio for each line. When a CO line is detected, the gray band indicates 
the region used to measure the line flux. The scale of the $y$-axis is such that lines 
would have the same amplitude in each transition for thermalized CO emission. The relative 
amplitudes of each spectrum give a sense of the luminosity ratios between the different 
lines.}
\figsetgrpend

\figsetgrpstart
\figsetgrpnum{15.9}
\figsetgrptitle{AMISS.1008}
\figsetplot{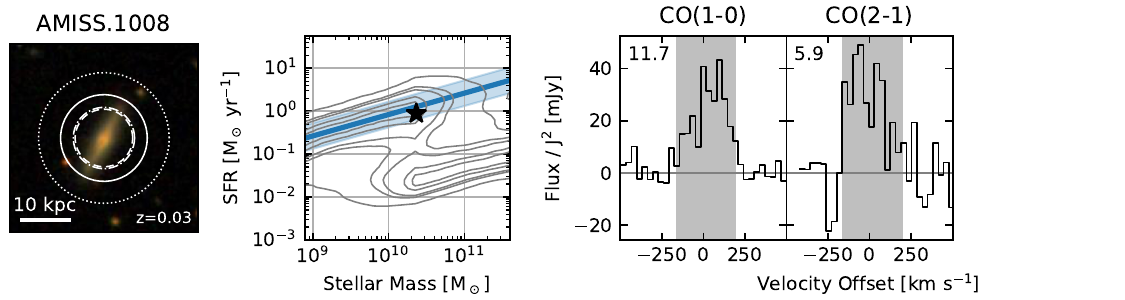}
\figsetgrpnote{Left column: SDSS cutouts of each target. Solid, dashed, and dotted lines show the 
beam sizes of the SMT for CO(2--1), the SMT for CO(3--2) (and the IRAM 30m for CO(1--0)) 
and the 12m for CO(1--0) respectively. The scale bar in the lower left shows 10 
kiloparsecs. 
Middle column: contours show the distribution of star formation rates at a given stellar 
mass, while the blue line and filled region show the main sequence of star forming 
galaxies. The stellar mass and star formation rate of the target galaxy is marked by a 
star. 
Right column: CO(1--0) spectra from AMISS (gray) and xCOLD~GASS (black), CO(2--1) spectra 
from AMISS, and CO(3--2) spectra from AMISS. Numbers in the upper right corner give the 
signal to noise ratio for each line. When a CO line is detected, the gray band indicates 
the region used to measure the line flux. The scale of the $y$-axis is such that lines 
would have the same amplitude in each transition for thermalized CO emission. The relative 
amplitudes of each spectrum give a sense of the luminosity ratios between the different 
lines.}
\figsetgrpend

\figsetgrpstart
\figsetgrpnum{15.10}
\figsetgrptitle{AMISS.1009}
\figsetplot{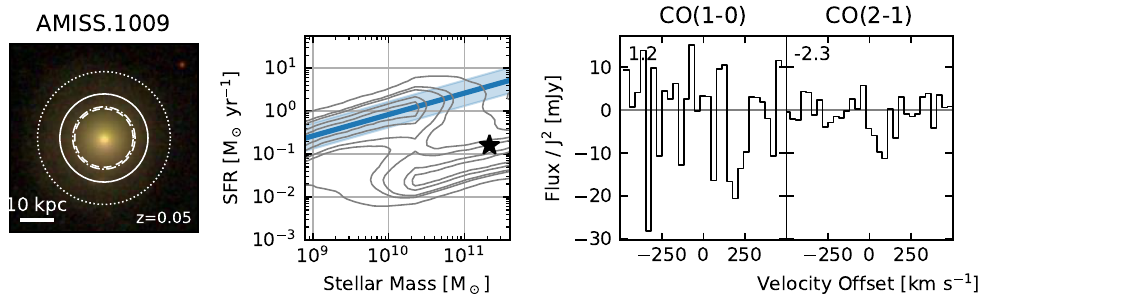}
\figsetgrpnote{Left column: SDSS cutouts of each target. Solid, dashed, and dotted lines show the 
beam sizes of the SMT for CO(2--1), the SMT for CO(3--2) (and the IRAM 30m for CO(1--0)) 
and the 12m for CO(1--0) respectively. The scale bar in the lower left shows 10 
kiloparsecs. 
Middle column: contours show the distribution of star formation rates at a given stellar 
mass, while the blue line and filled region show the main sequence of star forming 
galaxies. The stellar mass and star formation rate of the target galaxy is marked by a 
star. 
Right column: CO(1--0) spectra from AMISS (gray) and xCOLD~GASS (black), CO(2--1) spectra 
from AMISS, and CO(3--2) spectra from AMISS. Numbers in the upper right corner give the 
signal to noise ratio for each line. When a CO line is detected, the gray band indicates 
the region used to measure the line flux. The scale of the $y$-axis is such that lines 
would have the same amplitude in each transition for thermalized CO emission. The relative 
amplitudes of each spectrum give a sense of the luminosity ratios between the different 
lines.}
\figsetgrpend

\figsetgrpstart
\figsetgrpnum{15.11}
\figsetgrptitle{AMISS.1010}
\figsetplot{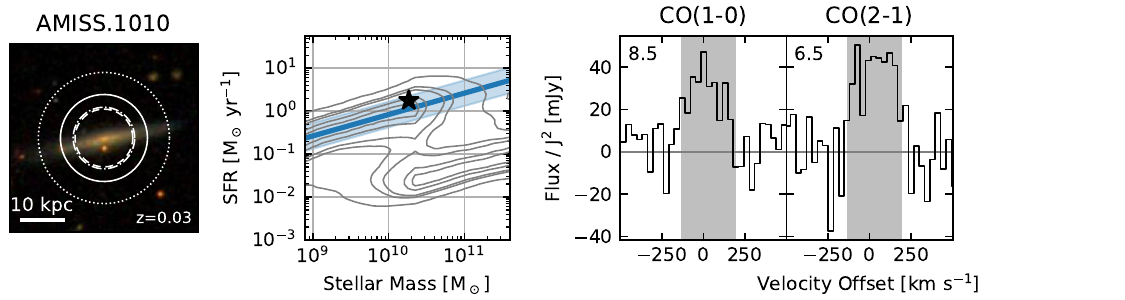}
\figsetgrpnote{Left column: SDSS cutouts of each target. Solid, dashed, and dotted lines show the 
beam sizes of the SMT for CO(2--1), the SMT for CO(3--2) (and the IRAM 30m for CO(1--0)) 
and the 12m for CO(1--0) respectively. The scale bar in the lower left shows 10 
kiloparsecs. 
Middle column: contours show the distribution of star formation rates at a given stellar 
mass, while the blue line and filled region show the main sequence of star forming 
galaxies. The stellar mass and star formation rate of the target galaxy is marked by a 
star. 
Right column: CO(1--0) spectra from AMISS (gray) and xCOLD~GASS (black), CO(2--1) spectra 
from AMISS, and CO(3--2) spectra from AMISS. Numbers in the upper right corner give the 
signal to noise ratio for each line. When a CO line is detected, the gray band indicates 
the region used to measure the line flux. The scale of the $y$-axis is such that lines 
would have the same amplitude in each transition for thermalized CO emission. The relative 
amplitudes of each spectrum give a sense of the luminosity ratios between the different 
lines.}
\figsetgrpend

\figsetgrpstart
\figsetgrpnum{15.12}
\figsetgrptitle{AMISS.1011}
\figsetplot{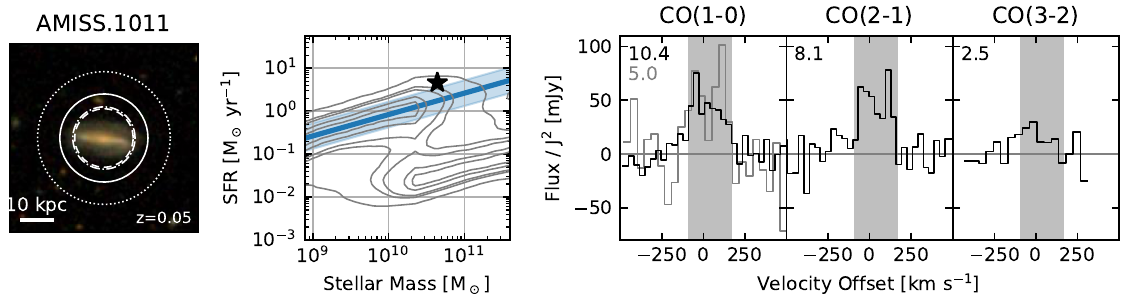}
\figsetgrpnote{Left column: SDSS cutouts of each target. Solid, dashed, and dotted lines show the 
beam sizes of the SMT for CO(2--1), the SMT for CO(3--2) (and the IRAM 30m for CO(1--0)) 
and the 12m for CO(1--0) respectively. The scale bar in the lower left shows 10 
kiloparsecs. 
Middle column: contours show the distribution of star formation rates at a given stellar 
mass, while the blue line and filled region show the main sequence of star forming 
galaxies. The stellar mass and star formation rate of the target galaxy is marked by a 
star. 
Right column: CO(1--0) spectra from AMISS (gray) and xCOLD~GASS (black), CO(2--1) spectra 
from AMISS, and CO(3--2) spectra from AMISS. Numbers in the upper right corner give the 
signal to noise ratio for each line. When a CO line is detected, the gray band indicates 
the region used to measure the line flux. The scale of the $y$-axis is such that lines 
would have the same amplitude in each transition for thermalized CO emission. The relative 
amplitudes of each spectrum give a sense of the luminosity ratios between the different 
lines.}
\figsetgrpend

\figsetgrpstart
\figsetgrpnum{15.13}
\figsetgrptitle{AMISS.1012}
\figsetplot{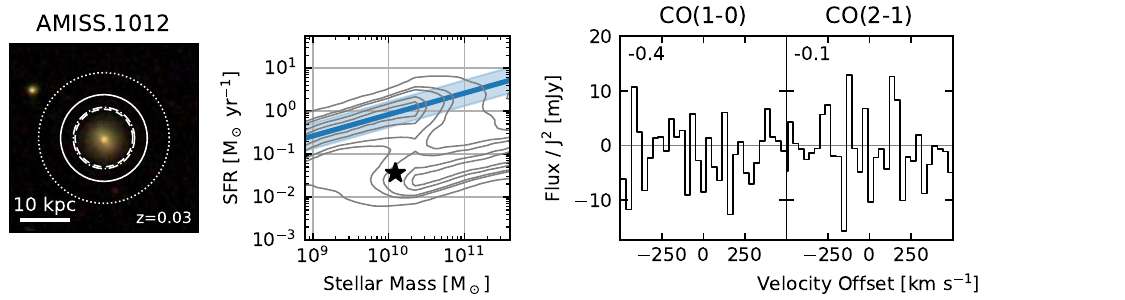}
\figsetgrpnote{Left column: SDSS cutouts of each target. Solid, dashed, and dotted lines show the 
beam sizes of the SMT for CO(2--1), the SMT for CO(3--2) (and the IRAM 30m for CO(1--0)) 
and the 12m for CO(1--0) respectively. The scale bar in the lower left shows 10 
kiloparsecs. 
Middle column: contours show the distribution of star formation rates at a given stellar 
mass, while the blue line and filled region show the main sequence of star forming 
galaxies. The stellar mass and star formation rate of the target galaxy is marked by a 
star. 
Right column: CO(1--0) spectra from AMISS (gray) and xCOLD~GASS (black), CO(2--1) spectra 
from AMISS, and CO(3--2) spectra from AMISS. Numbers in the upper right corner give the 
signal to noise ratio for each line. When a CO line is detected, the gray band indicates 
the region used to measure the line flux. The scale of the $y$-axis is such that lines 
would have the same amplitude in each transition for thermalized CO emission. The relative 
amplitudes of each spectrum give a sense of the luminosity ratios between the different 
lines.}
\figsetgrpend

\figsetgrpstart
\figsetgrpnum{15.14}
\figsetgrptitle{AMISS.1013}
\figsetplot{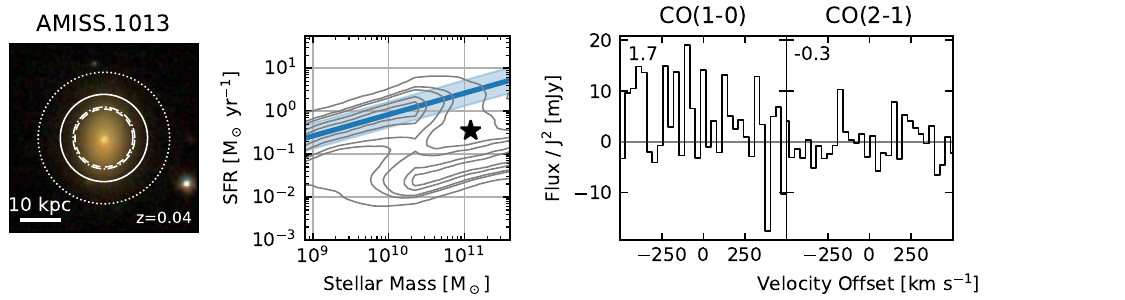}
\figsetgrpnote{Left column: SDSS cutouts of each target. Solid, dashed, and dotted lines show the 
beam sizes of the SMT for CO(2--1), the SMT for CO(3--2) (and the IRAM 30m for CO(1--0)) 
and the 12m for CO(1--0) respectively. The scale bar in the lower left shows 10 
kiloparsecs. 
Middle column: contours show the distribution of star formation rates at a given stellar 
mass, while the blue line and filled region show the main sequence of star forming 
galaxies. The stellar mass and star formation rate of the target galaxy is marked by a 
star. 
Right column: CO(1--0) spectra from AMISS (gray) and xCOLD~GASS (black), CO(2--1) spectra 
from AMISS, and CO(3--2) spectra from AMISS. Numbers in the upper right corner give the 
signal to noise ratio for each line. When a CO line is detected, the gray band indicates 
the region used to measure the line flux. The scale of the $y$-axis is such that lines 
would have the same amplitude in each transition for thermalized CO emission. The relative 
amplitudes of each spectrum give a sense of the luminosity ratios between the different 
lines.}
\figsetgrpend

\figsetgrpstart
\figsetgrpnum{15.15}
\figsetgrptitle{AMISS.1014}
\figsetplot{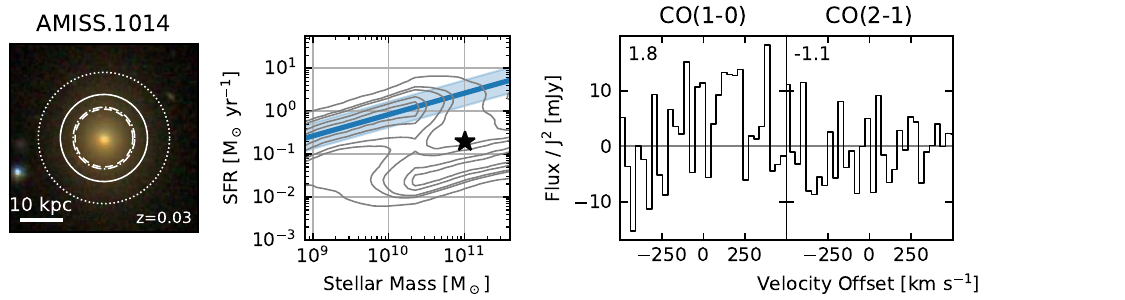}
\figsetgrpnote{Left column: SDSS cutouts of each target. Solid, dashed, and dotted lines show the 
beam sizes of the SMT for CO(2--1), the SMT for CO(3--2) (and the IRAM 30m for CO(1--0)) 
and the 12m for CO(1--0) respectively. The scale bar in the lower left shows 10 
kiloparsecs. 
Middle column: contours show the distribution of star formation rates at a given stellar 
mass, while the blue line and filled region show the main sequence of star forming 
galaxies. The stellar mass and star formation rate of the target galaxy is marked by a 
star. 
Right column: CO(1--0) spectra from AMISS (gray) and xCOLD~GASS (black), CO(2--1) spectra 
from AMISS, and CO(3--2) spectra from AMISS. Numbers in the upper right corner give the 
signal to noise ratio for each line. When a CO line is detected, the gray band indicates 
the region used to measure the line flux. The scale of the $y$-axis is such that lines 
would have the same amplitude in each transition for thermalized CO emission. The relative 
amplitudes of each spectrum give a sense of the luminosity ratios between the different 
lines.}
\figsetgrpend

\figsetgrpstart
\figsetgrpnum{15.16}
\figsetgrptitle{AMISS.1015}
\figsetplot{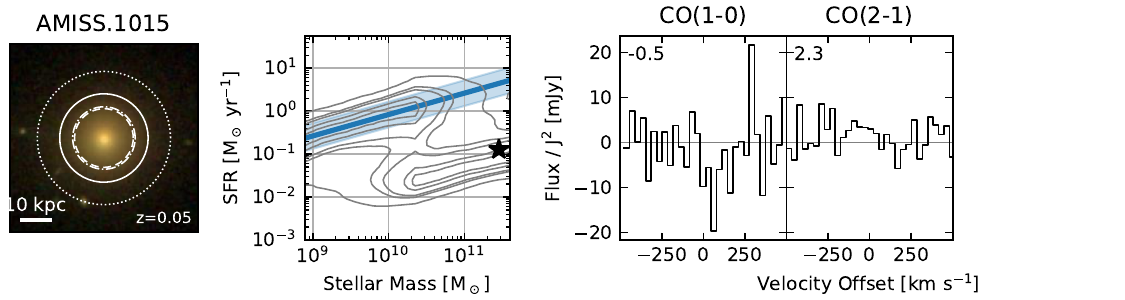}
\figsetgrpnote{Left column: SDSS cutouts of each target. Solid, dashed, and dotted lines show the 
beam sizes of the SMT for CO(2--1), the SMT for CO(3--2) (and the IRAM 30m for CO(1--0)) 
and the 12m for CO(1--0) respectively. The scale bar in the lower left shows 10 
kiloparsecs. 
Middle column: contours show the distribution of star formation rates at a given stellar 
mass, while the blue line and filled region show the main sequence of star forming 
galaxies. The stellar mass and star formation rate of the target galaxy is marked by a 
star. 
Right column: CO(1--0) spectra from AMISS (gray) and xCOLD~GASS (black), CO(2--1) spectra 
from AMISS, and CO(3--2) spectra from AMISS. Numbers in the upper right corner give the 
signal to noise ratio for each line. When a CO line is detected, the gray band indicates 
the region used to measure the line flux. The scale of the $y$-axis is such that lines 
would have the same amplitude in each transition for thermalized CO emission. The relative 
amplitudes of each spectrum give a sense of the luminosity ratios between the different 
lines.}
\figsetgrpend

\figsetgrpstart
\figsetgrpnum{15.17}
\figsetgrptitle{AMISS.1016}
\figsetplot{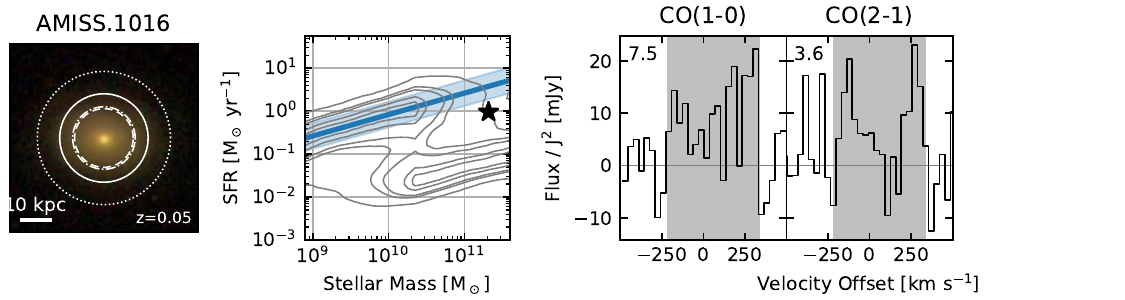}
\figsetgrpnote{Left column: SDSS cutouts of each target. Solid, dashed, and dotted lines show the 
beam sizes of the SMT for CO(2--1), the SMT for CO(3--2) (and the IRAM 30m for CO(1--0)) 
and the 12m for CO(1--0) respectively. The scale bar in the lower left shows 10 
kiloparsecs. 
Middle column: contours show the distribution of star formation rates at a given stellar 
mass, while the blue line and filled region show the main sequence of star forming 
galaxies. The stellar mass and star formation rate of the target galaxy is marked by a 
star. 
Right column: CO(1--0) spectra from AMISS (gray) and xCOLD~GASS (black), CO(2--1) spectra 
from AMISS, and CO(3--2) spectra from AMISS. Numbers in the upper right corner give the 
signal to noise ratio for each line. When a CO line is detected, the gray band indicates 
the region used to measure the line flux. The scale of the $y$-axis is such that lines 
would have the same amplitude in each transition for thermalized CO emission. The relative 
amplitudes of each spectrum give a sense of the luminosity ratios between the different 
lines.}
\figsetgrpend

\figsetgrpstart
\figsetgrpnum{15.18}
\figsetgrptitle{AMISS.1017}
\figsetplot{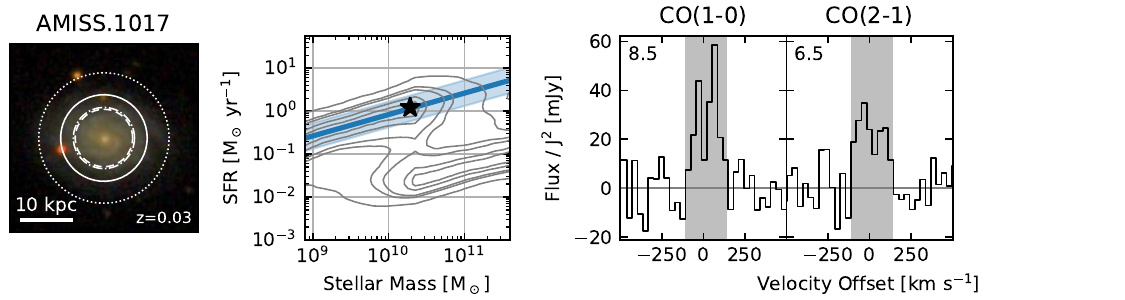}
\figsetgrpnote{Left column: SDSS cutouts of each target. Solid, dashed, and dotted lines show the 
beam sizes of the SMT for CO(2--1), the SMT for CO(3--2) (and the IRAM 30m for CO(1--0)) 
and the 12m for CO(1--0) respectively. The scale bar in the lower left shows 10 
kiloparsecs. 
Middle column: contours show the distribution of star formation rates at a given stellar 
mass, while the blue line and filled region show the main sequence of star forming 
galaxies. The stellar mass and star formation rate of the target galaxy is marked by a 
star. 
Right column: CO(1--0) spectra from AMISS (gray) and xCOLD~GASS (black), CO(2--1) spectra 
from AMISS, and CO(3--2) spectra from AMISS. Numbers in the upper right corner give the 
signal to noise ratio for each line. When a CO line is detected, the gray band indicates 
the region used to measure the line flux. The scale of the $y$-axis is such that lines 
would have the same amplitude in each transition for thermalized CO emission. The relative 
amplitudes of each spectrum give a sense of the luminosity ratios between the different 
lines.}
\figsetgrpend

\figsetgrpstart
\figsetgrpnum{15.19}
\figsetgrptitle{AMISS.1018}
\figsetplot{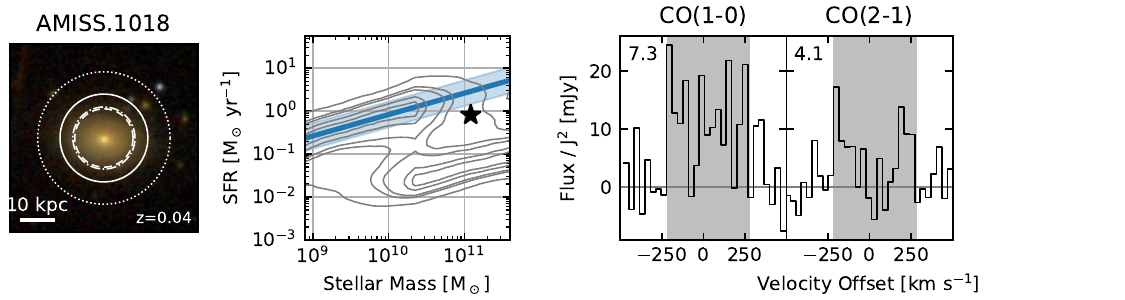}
\figsetgrpnote{Left column: SDSS cutouts of each target. Solid, dashed, and dotted lines show the 
beam sizes of the SMT for CO(2--1), the SMT for CO(3--2) (and the IRAM 30m for CO(1--0)) 
and the 12m for CO(1--0) respectively. The scale bar in the lower left shows 10 
kiloparsecs. 
Middle column: contours show the distribution of star formation rates at a given stellar 
mass, while the blue line and filled region show the main sequence of star forming 
galaxies. The stellar mass and star formation rate of the target galaxy is marked by a 
star. 
Right column: CO(1--0) spectra from AMISS (gray) and xCOLD~GASS (black), CO(2--1) spectra 
from AMISS, and CO(3--2) spectra from AMISS. Numbers in the upper right corner give the 
signal to noise ratio for each line. When a CO line is detected, the gray band indicates 
the region used to measure the line flux. The scale of the $y$-axis is such that lines 
would have the same amplitude in each transition for thermalized CO emission. The relative 
amplitudes of each spectrum give a sense of the luminosity ratios between the different 
lines.}
\figsetgrpend

\figsetgrpstart
\figsetgrpnum{15.20}
\figsetgrptitle{AMISS.1019}
\figsetplot{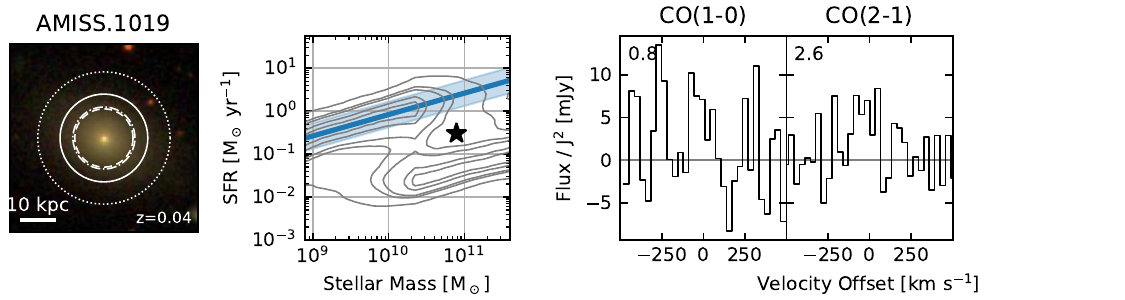}
\figsetgrpnote{Left column: SDSS cutouts of each target. Solid, dashed, and dotted lines show the 
beam sizes of the SMT for CO(2--1), the SMT for CO(3--2) (and the IRAM 30m for CO(1--0)) 
and the 12m for CO(1--0) respectively. The scale bar in the lower left shows 10 
kiloparsecs. 
Middle column: contours show the distribution of star formation rates at a given stellar 
mass, while the blue line and filled region show the main sequence of star forming 
galaxies. The stellar mass and star formation rate of the target galaxy is marked by a 
star. 
Right column: CO(1--0) spectra from AMISS (gray) and xCOLD~GASS (black), CO(2--1) spectra 
from AMISS, and CO(3--2) spectra from AMISS. Numbers in the upper right corner give the 
signal to noise ratio for each line. When a CO line is detected, the gray band indicates 
the region used to measure the line flux. The scale of the $y$-axis is such that lines 
would have the same amplitude in each transition for thermalized CO emission. The relative 
amplitudes of each spectrum give a sense of the luminosity ratios between the different 
lines.}
\figsetgrpend

\figsetgrpstart
\figsetgrpnum{15.21}
\figsetgrptitle{AMISS.1020}
\figsetplot{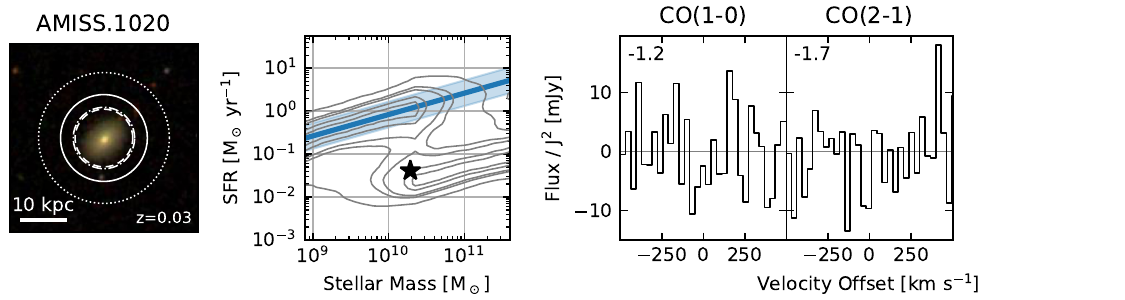}
\figsetgrpnote{Left column: SDSS cutouts of each target. Solid, dashed, and dotted lines show the 
beam sizes of the SMT for CO(2--1), the SMT for CO(3--2) (and the IRAM 30m for CO(1--0)) 
and the 12m for CO(1--0) respectively. The scale bar in the lower left shows 10 
kiloparsecs. 
Middle column: contours show the distribution of star formation rates at a given stellar 
mass, while the blue line and filled region show the main sequence of star forming 
galaxies. The stellar mass and star formation rate of the target galaxy is marked by a 
star. 
Right column: CO(1--0) spectra from AMISS (gray) and xCOLD~GASS (black), CO(2--1) spectra 
from AMISS, and CO(3--2) spectra from AMISS. Numbers in the upper right corner give the 
signal to noise ratio for each line. When a CO line is detected, the gray band indicates 
the region used to measure the line flux. The scale of the $y$-axis is such that lines 
would have the same amplitude in each transition for thermalized CO emission. The relative 
amplitudes of each spectrum give a sense of the luminosity ratios between the different 
lines.}
\figsetgrpend

\figsetgrpstart
\figsetgrpnum{15.22}
\figsetgrptitle{AMISS.1021}
\figsetplot{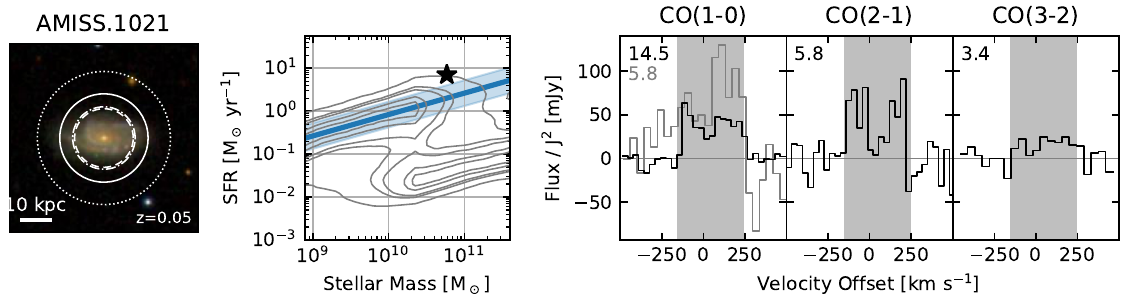}
\figsetgrpnote{Left column: SDSS cutouts of each target. Solid, dashed, and dotted lines show the 
beam sizes of the SMT for CO(2--1), the SMT for CO(3--2) (and the IRAM 30m for CO(1--0)) 
and the 12m for CO(1--0) respectively. The scale bar in the lower left shows 10 
kiloparsecs. 
Middle column: contours show the distribution of star formation rates at a given stellar 
mass, while the blue line and filled region show the main sequence of star forming 
galaxies. The stellar mass and star formation rate of the target galaxy is marked by a 
star. 
Right column: CO(1--0) spectra from AMISS (gray) and xCOLD~GASS (black), CO(2--1) spectra 
from AMISS, and CO(3--2) spectra from AMISS. Numbers in the upper right corner give the 
signal to noise ratio for each line. When a CO line is detected, the gray band indicates 
the region used to measure the line flux. The scale of the $y$-axis is such that lines 
would have the same amplitude in each transition for thermalized CO emission. The relative 
amplitudes of each spectrum give a sense of the luminosity ratios between the different 
lines.}
\figsetgrpend

\figsetgrpstart
\figsetgrpnum{15.23}
\figsetgrptitle{AMISS.1022}
\figsetplot{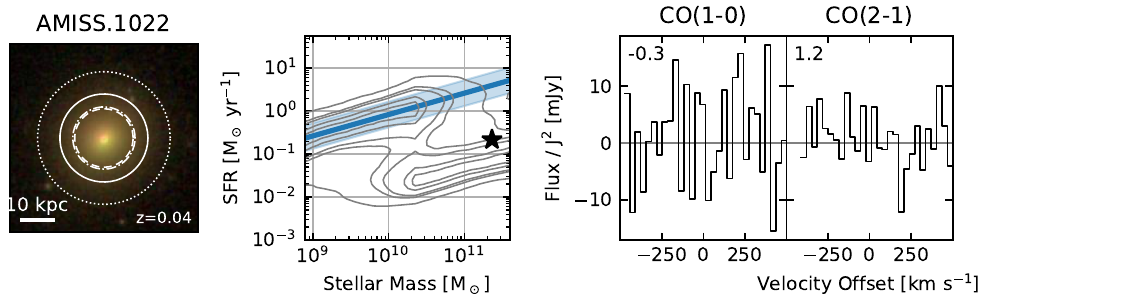}
\figsetgrpnote{Left column: SDSS cutouts of each target. Solid, dashed, and dotted lines show the 
beam sizes of the SMT for CO(2--1), the SMT for CO(3--2) (and the IRAM 30m for CO(1--0)) 
and the 12m for CO(1--0) respectively. The scale bar in the lower left shows 10 
kiloparsecs. 
Middle column: contours show the distribution of star formation rates at a given stellar 
mass, while the blue line and filled region show the main sequence of star forming 
galaxies. The stellar mass and star formation rate of the target galaxy is marked by a 
star. 
Right column: CO(1--0) spectra from AMISS (gray) and xCOLD~GASS (black), CO(2--1) spectra 
from AMISS, and CO(3--2) spectra from AMISS. Numbers in the upper right corner give the 
signal to noise ratio for each line. When a CO line is detected, the gray band indicates 
the region used to measure the line flux. The scale of the $y$-axis is such that lines 
would have the same amplitude in each transition for thermalized CO emission. The relative 
amplitudes of each spectrum give a sense of the luminosity ratios between the different 
lines.}
\figsetgrpend

\figsetgrpstart
\figsetgrpnum{15.24}
\figsetgrptitle{AMISS.1023}
\figsetplot{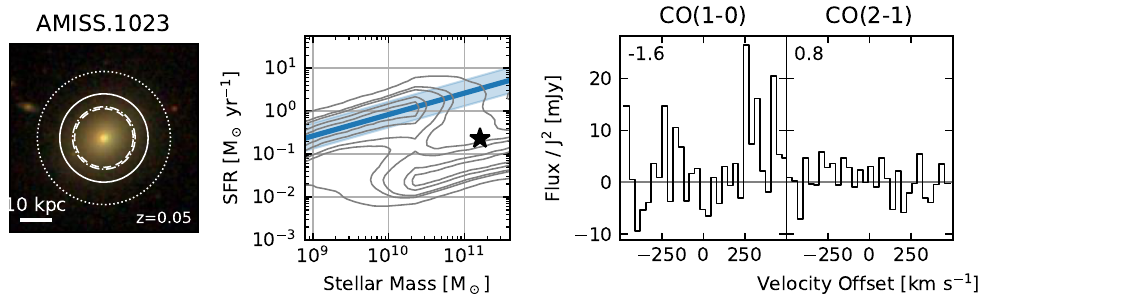}
\figsetgrpnote{Left column: SDSS cutouts of each target. Solid, dashed, and dotted lines show the 
beam sizes of the SMT for CO(2--1), the SMT for CO(3--2) (and the IRAM 30m for CO(1--0)) 
and the 12m for CO(1--0) respectively. The scale bar in the lower left shows 10 
kiloparsecs. 
Middle column: contours show the distribution of star formation rates at a given stellar 
mass, while the blue line and filled region show the main sequence of star forming 
galaxies. The stellar mass and star formation rate of the target galaxy is marked by a 
star. 
Right column: CO(1--0) spectra from AMISS (gray) and xCOLD~GASS (black), CO(2--1) spectra 
from AMISS, and CO(3--2) spectra from AMISS. Numbers in the upper right corner give the 
signal to noise ratio for each line. When a CO line is detected, the gray band indicates 
the region used to measure the line flux. The scale of the $y$-axis is such that lines 
would have the same amplitude in each transition for thermalized CO emission. The relative 
amplitudes of each spectrum give a sense of the luminosity ratios between the different 
lines.}
\figsetgrpend

\figsetgrpstart
\figsetgrpnum{15.25}
\figsetgrptitle{AMISS.1024}
\figsetplot{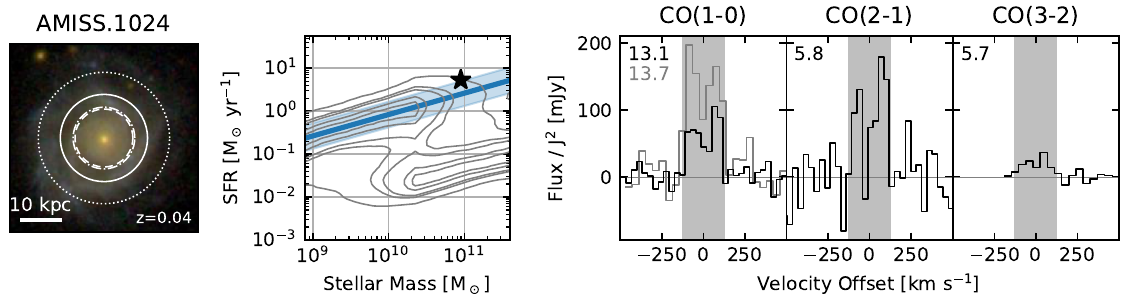}
\figsetgrpnote{Left column: SDSS cutouts of each target. Solid, dashed, and dotted lines show the 
beam sizes of the SMT for CO(2--1), the SMT for CO(3--2) (and the IRAM 30m for CO(1--0)) 
and the 12m for CO(1--0) respectively. The scale bar in the lower left shows 10 
kiloparsecs. 
Middle column: contours show the distribution of star formation rates at a given stellar 
mass, while the blue line and filled region show the main sequence of star forming 
galaxies. The stellar mass and star formation rate of the target galaxy is marked by a 
star. 
Right column: CO(1--0) spectra from AMISS (gray) and xCOLD~GASS (black), CO(2--1) spectra 
from AMISS, and CO(3--2) spectra from AMISS. Numbers in the upper right corner give the 
signal to noise ratio for each line. When a CO line is detected, the gray band indicates 
the region used to measure the line flux. The scale of the $y$-axis is such that lines 
would have the same amplitude in each transition for thermalized CO emission. The relative 
amplitudes of each spectrum give a sense of the luminosity ratios between the different 
lines.}
\figsetgrpend

\figsetgrpstart
\figsetgrpnum{15.26}
\figsetgrptitle{AMISS.1025}
\figsetplot{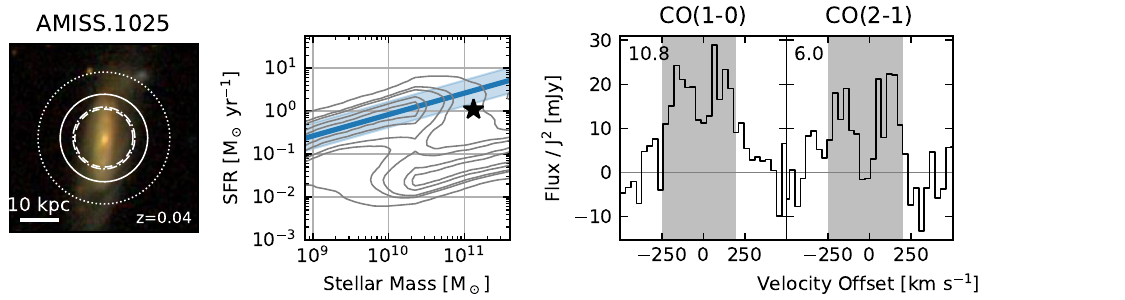}
\figsetgrpnote{Left column: SDSS cutouts of each target. Solid, dashed, and dotted lines show the 
beam sizes of the SMT for CO(2--1), the SMT for CO(3--2) (and the IRAM 30m for CO(1--0)) 
and the 12m for CO(1--0) respectively. The scale bar in the lower left shows 10 
kiloparsecs. 
Middle column: contours show the distribution of star formation rates at a given stellar 
mass, while the blue line and filled region show the main sequence of star forming 
galaxies. The stellar mass and star formation rate of the target galaxy is marked by a 
star. 
Right column: CO(1--0) spectra from AMISS (gray) and xCOLD~GASS (black), CO(2--1) spectra 
from AMISS, and CO(3--2) spectra from AMISS. Numbers in the upper right corner give the 
signal to noise ratio for each line. When a CO line is detected, the gray band indicates 
the region used to measure the line flux. The scale of the $y$-axis is such that lines 
would have the same amplitude in each transition for thermalized CO emission. The relative 
amplitudes of each spectrum give a sense of the luminosity ratios between the different 
lines.}
\figsetgrpend

\figsetgrpstart
\figsetgrpnum{15.27}
\figsetgrptitle{AMISS.1026}
\figsetplot{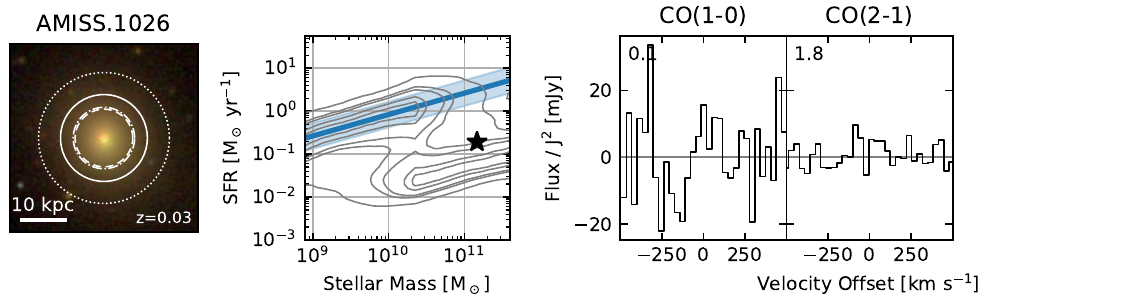}
\figsetgrpnote{Left column: SDSS cutouts of each target. Solid, dashed, and dotted lines show the 
beam sizes of the SMT for CO(2--1), the SMT for CO(3--2) (and the IRAM 30m for CO(1--0)) 
and the 12m for CO(1--0) respectively. The scale bar in the lower left shows 10 
kiloparsecs. 
Middle column: contours show the distribution of star formation rates at a given stellar 
mass, while the blue line and filled region show the main sequence of star forming 
galaxies. The stellar mass and star formation rate of the target galaxy is marked by a 
star. 
Right column: CO(1--0) spectra from AMISS (gray) and xCOLD~GASS (black), CO(2--1) spectra 
from AMISS, and CO(3--2) spectra from AMISS. Numbers in the upper right corner give the 
signal to noise ratio for each line. When a CO line is detected, the gray band indicates 
the region used to measure the line flux. The scale of the $y$-axis is such that lines 
would have the same amplitude in each transition for thermalized CO emission. The relative 
amplitudes of each spectrum give a sense of the luminosity ratios between the different 
lines.}
\figsetgrpend

\figsetgrpstart
\figsetgrpnum{15.28}
\figsetgrptitle{AMISS.1027}
\figsetplot{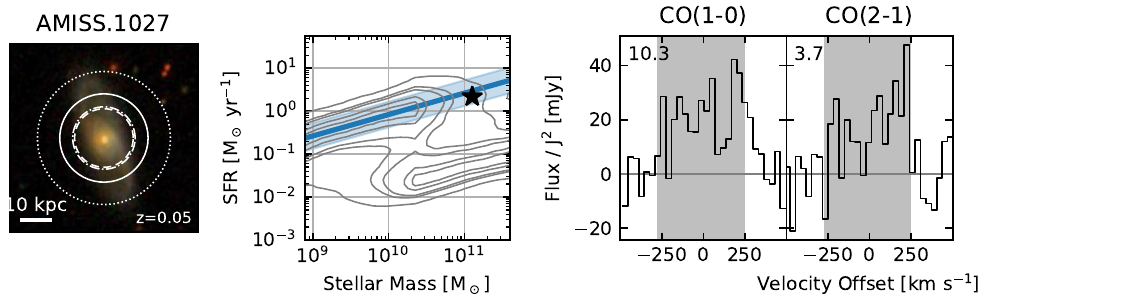}
\figsetgrpnote{Left column: SDSS cutouts of each target. Solid, dashed, and dotted lines show the 
beam sizes of the SMT for CO(2--1), the SMT for CO(3--2) (and the IRAM 30m for CO(1--0)) 
and the 12m for CO(1--0) respectively. The scale bar in the lower left shows 10 
kiloparsecs. 
Middle column: contours show the distribution of star formation rates at a given stellar 
mass, while the blue line and filled region show the main sequence of star forming 
galaxies. The stellar mass and star formation rate of the target galaxy is marked by a 
star. 
Right column: CO(1--0) spectra from AMISS (gray) and xCOLD~GASS (black), CO(2--1) spectra 
from AMISS, and CO(3--2) spectra from AMISS. Numbers in the upper right corner give the 
signal to noise ratio for each line. When a CO line is detected, the gray band indicates 
the region used to measure the line flux. The scale of the $y$-axis is such that lines 
would have the same amplitude in each transition for thermalized CO emission. The relative 
amplitudes of each spectrum give a sense of the luminosity ratios between the different 
lines.}
\figsetgrpend

\figsetgrpstart
\figsetgrpnum{15.29}
\figsetgrptitle{AMISS.1028}
\figsetplot{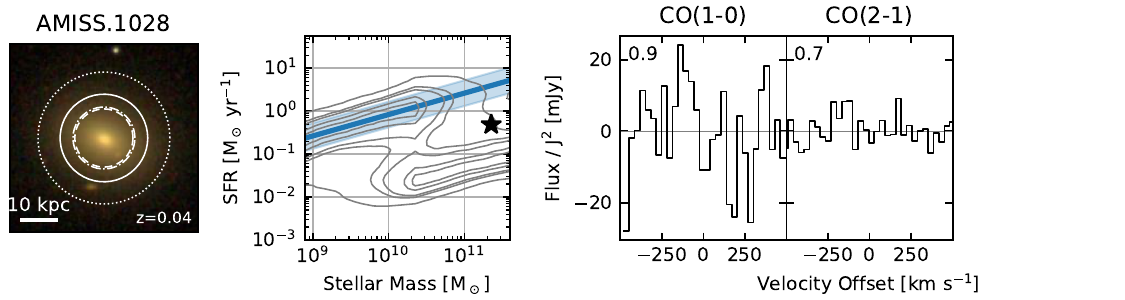}
\figsetgrpnote{Left column: SDSS cutouts of each target. Solid, dashed, and dotted lines show the 
beam sizes of the SMT for CO(2--1), the SMT for CO(3--2) (and the IRAM 30m for CO(1--0)) 
and the 12m for CO(1--0) respectively. The scale bar in the lower left shows 10 
kiloparsecs. 
Middle column: contours show the distribution of star formation rates at a given stellar 
mass, while the blue line and filled region show the main sequence of star forming 
galaxies. The stellar mass and star formation rate of the target galaxy is marked by a 
star. 
Right column: CO(1--0) spectra from AMISS (gray) and xCOLD~GASS (black), CO(2--1) spectra 
from AMISS, and CO(3--2) spectra from AMISS. Numbers in the upper right corner give the 
signal to noise ratio for each line. When a CO line is detected, the gray band indicates 
the region used to measure the line flux. The scale of the $y$-axis is such that lines 
would have the same amplitude in each transition for thermalized CO emission. The relative 
amplitudes of each spectrum give a sense of the luminosity ratios between the different 
lines.}
\figsetgrpend

\figsetgrpstart
\figsetgrpnum{15.30}
\figsetgrptitle{AMISS.1029}
\figsetplot{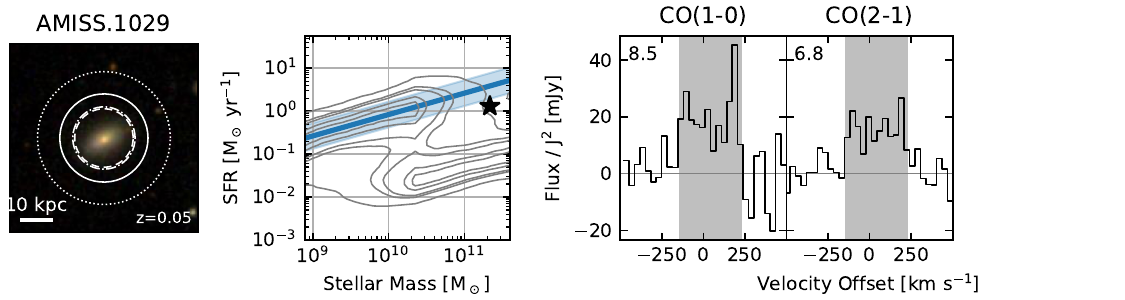}
\figsetgrpnote{Left column: SDSS cutouts of each target. Solid, dashed, and dotted lines show the 
beam sizes of the SMT for CO(2--1), the SMT for CO(3--2) (and the IRAM 30m for CO(1--0)) 
and the 12m for CO(1--0) respectively. The scale bar in the lower left shows 10 
kiloparsecs. 
Middle column: contours show the distribution of star formation rates at a given stellar 
mass, while the blue line and filled region show the main sequence of star forming 
galaxies. The stellar mass and star formation rate of the target galaxy is marked by a 
star. 
Right column: CO(1--0) spectra from AMISS (gray) and xCOLD~GASS (black), CO(2--1) spectra 
from AMISS, and CO(3--2) spectra from AMISS. Numbers in the upper right corner give the 
signal to noise ratio for each line. When a CO line is detected, the gray band indicates 
the region used to measure the line flux. The scale of the $y$-axis is such that lines 
would have the same amplitude in each transition for thermalized CO emission. The relative 
amplitudes of each spectrum give a sense of the luminosity ratios between the different 
lines.}
\figsetgrpend

\figsetgrpstart
\figsetgrpnum{15.31}
\figsetgrptitle{AMISS.1030}
\figsetplot{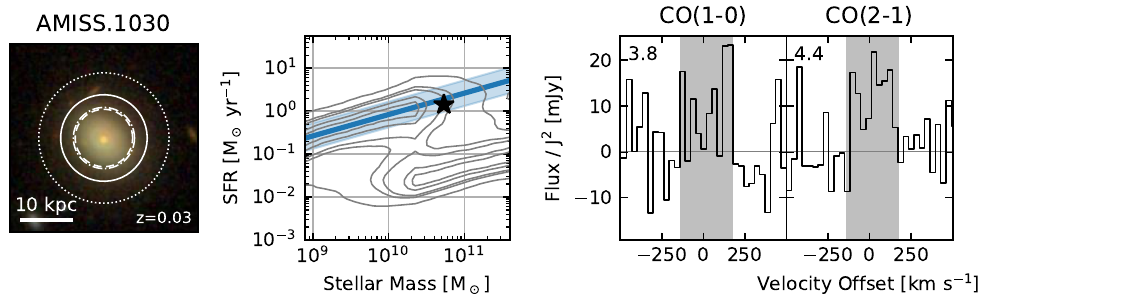}
\figsetgrpnote{Left column: SDSS cutouts of each target. Solid, dashed, and dotted lines show the 
beam sizes of the SMT for CO(2--1), the SMT for CO(3--2) (and the IRAM 30m for CO(1--0)) 
and the 12m for CO(1--0) respectively. The scale bar in the lower left shows 10 
kiloparsecs. 
Middle column: contours show the distribution of star formation rates at a given stellar 
mass, while the blue line and filled region show the main sequence of star forming 
galaxies. The stellar mass and star formation rate of the target galaxy is marked by a 
star. 
Right column: CO(1--0) spectra from AMISS (gray) and xCOLD~GASS (black), CO(2--1) spectra 
from AMISS, and CO(3--2) spectra from AMISS. Numbers in the upper right corner give the 
signal to noise ratio for each line. When a CO line is detected, the gray band indicates 
the region used to measure the line flux. The scale of the $y$-axis is such that lines 
would have the same amplitude in each transition for thermalized CO emission. The relative 
amplitudes of each spectrum give a sense of the luminosity ratios between the different 
lines.}
\figsetgrpend

\figsetgrpstart
\figsetgrpnum{15.32}
\figsetgrptitle{AMISS.1031}
\figsetplot{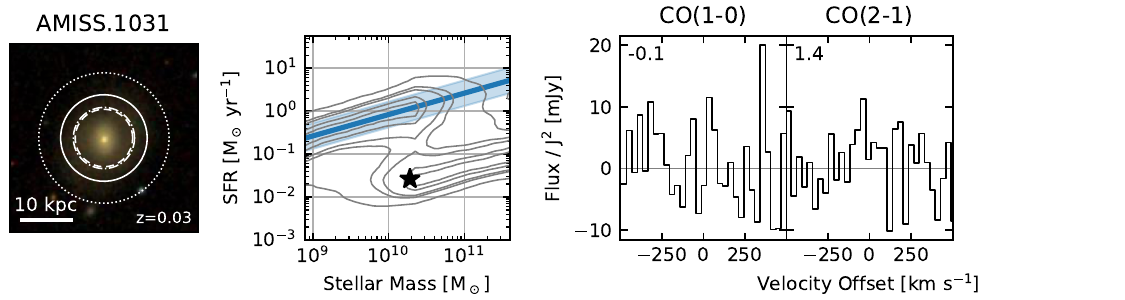}
\figsetgrpnote{Left column: SDSS cutouts of each target. Solid, dashed, and dotted lines show the 
beam sizes of the SMT for CO(2--1), the SMT for CO(3--2) (and the IRAM 30m for CO(1--0)) 
and the 12m for CO(1--0) respectively. The scale bar in the lower left shows 10 
kiloparsecs. 
Middle column: contours show the distribution of star formation rates at a given stellar 
mass, while the blue line and filled region show the main sequence of star forming 
galaxies. The stellar mass and star formation rate of the target galaxy is marked by a 
star. 
Right column: CO(1--0) spectra from AMISS (gray) and xCOLD~GASS (black), CO(2--1) spectra 
from AMISS, and CO(3--2) spectra from AMISS. Numbers in the upper right corner give the 
signal to noise ratio for each line. When a CO line is detected, the gray band indicates 
the region used to measure the line flux. The scale of the $y$-axis is such that lines 
would have the same amplitude in each transition for thermalized CO emission. The relative 
amplitudes of each spectrum give a sense of the luminosity ratios between the different 
lines.}
\figsetgrpend

\figsetgrpstart
\figsetgrpnum{15.33}
\figsetgrptitle{AMISS.1032}
\figsetplot{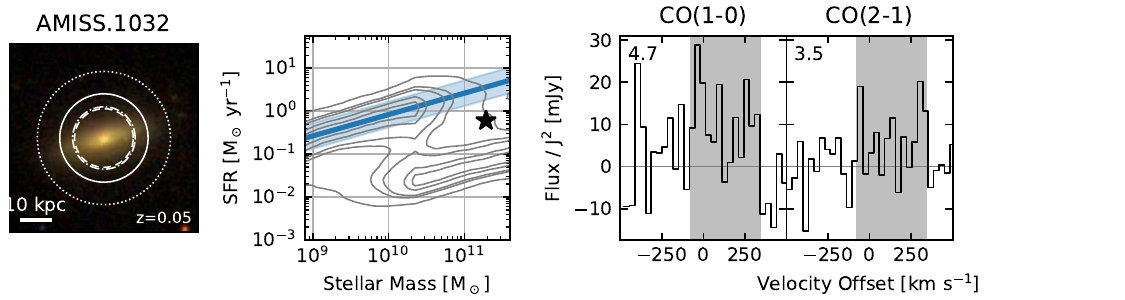}
\figsetgrpnote{Left column: SDSS cutouts of each target. Solid, dashed, and dotted lines show the 
beam sizes of the SMT for CO(2--1), the SMT for CO(3--2) (and the IRAM 30m for CO(1--0)) 
and the 12m for CO(1--0) respectively. The scale bar in the lower left shows 10 
kiloparsecs. 
Middle column: contours show the distribution of star formation rates at a given stellar 
mass, while the blue line and filled region show the main sequence of star forming 
galaxies. The stellar mass and star formation rate of the target galaxy is marked by a 
star. 
Right column: CO(1--0) spectra from AMISS (gray) and xCOLD~GASS (black), CO(2--1) spectra 
from AMISS, and CO(3--2) spectra from AMISS. Numbers in the upper right corner give the 
signal to noise ratio for each line. When a CO line is detected, the gray band indicates 
the region used to measure the line flux. The scale of the $y$-axis is such that lines 
would have the same amplitude in each transition for thermalized CO emission. The relative 
amplitudes of each spectrum give a sense of the luminosity ratios between the different 
lines.}
\figsetgrpend

\figsetgrpstart
\figsetgrpnum{15.34}
\figsetgrptitle{AMISS.1033}
\figsetplot{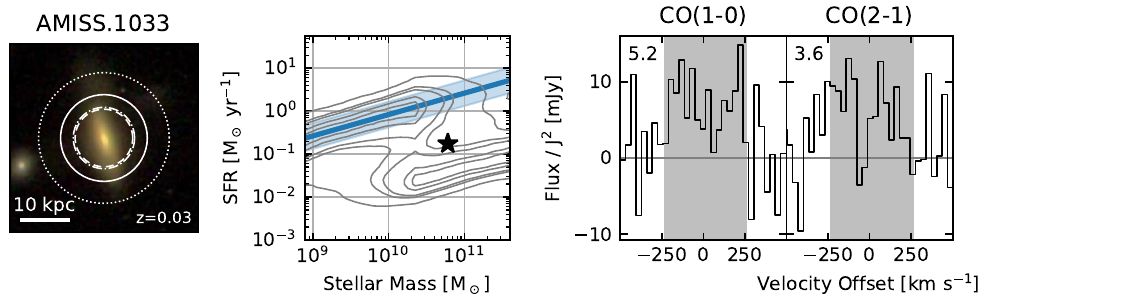}
\figsetgrpnote{Left column: SDSS cutouts of each target. Solid, dashed, and dotted lines show the 
beam sizes of the SMT for CO(2--1), the SMT for CO(3--2) (and the IRAM 30m for CO(1--0)) 
and the 12m for CO(1--0) respectively. The scale bar in the lower left shows 10 
kiloparsecs. 
Middle column: contours show the distribution of star formation rates at a given stellar 
mass, while the blue line and filled region show the main sequence of star forming 
galaxies. The stellar mass and star formation rate of the target galaxy is marked by a 
star. 
Right column: CO(1--0) spectra from AMISS (gray) and xCOLD~GASS (black), CO(2--1) spectra 
from AMISS, and CO(3--2) spectra from AMISS. Numbers in the upper right corner give the 
signal to noise ratio for each line. When a CO line is detected, the gray band indicates 
the region used to measure the line flux. The scale of the $y$-axis is such that lines 
would have the same amplitude in each transition for thermalized CO emission. The relative 
amplitudes of each spectrum give a sense of the luminosity ratios between the different 
lines.}
\figsetgrpend

\figsetgrpstart
\figsetgrpnum{15.35}
\figsetgrptitle{AMISS.1034}
\figsetplot{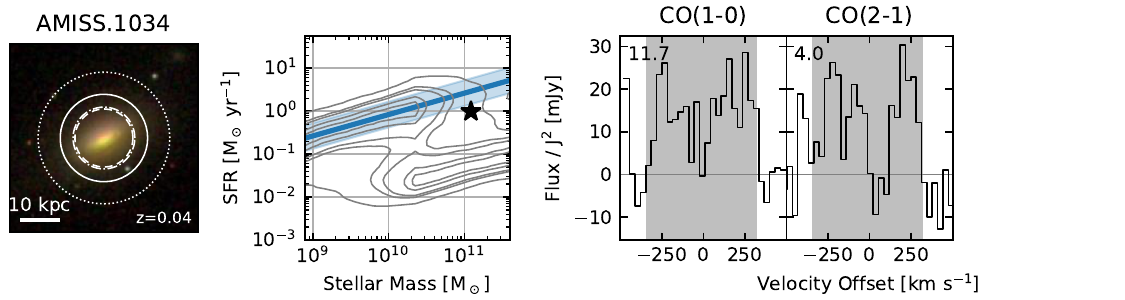}
\figsetgrpnote{Left column: SDSS cutouts of each target. Solid, dashed, and dotted lines show the 
beam sizes of the SMT for CO(2--1), the SMT for CO(3--2) (and the IRAM 30m for CO(1--0)) 
and the 12m for CO(1--0) respectively. The scale bar in the lower left shows 10 
kiloparsecs. 
Middle column: contours show the distribution of star formation rates at a given stellar 
mass, while the blue line and filled region show the main sequence of star forming 
galaxies. The stellar mass and star formation rate of the target galaxy is marked by a 
star. 
Right column: CO(1--0) spectra from AMISS (gray) and xCOLD~GASS (black), CO(2--1) spectra 
from AMISS, and CO(3--2) spectra from AMISS. Numbers in the upper right corner give the 
signal to noise ratio for each line. When a CO line is detected, the gray band indicates 
the region used to measure the line flux. The scale of the $y$-axis is such that lines 
would have the same amplitude in each transition for thermalized CO emission. The relative 
amplitudes of each spectrum give a sense of the luminosity ratios between the different 
lines.}
\figsetgrpend

\figsetgrpstart
\figsetgrpnum{15.36}
\figsetgrptitle{AMISS.1035}
\figsetplot{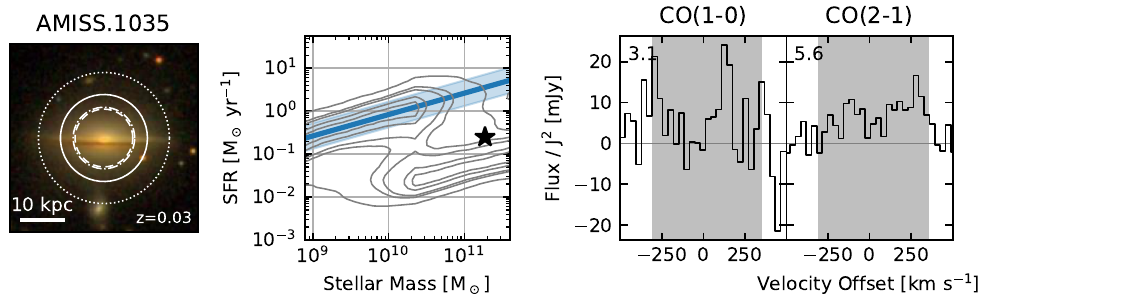}
\figsetgrpnote{Left column: SDSS cutouts of each target. Solid, dashed, and dotted lines show the 
beam sizes of the SMT for CO(2--1), the SMT for CO(3--2) (and the IRAM 30m for CO(1--0)) 
and the 12m for CO(1--0) respectively. The scale bar in the lower left shows 10 
kiloparsecs. 
Middle column: contours show the distribution of star formation rates at a given stellar 
mass, while the blue line and filled region show the main sequence of star forming 
galaxies. The stellar mass and star formation rate of the target galaxy is marked by a 
star. 
Right column: CO(1--0) spectra from AMISS (gray) and xCOLD~GASS (black), CO(2--1) spectra 
from AMISS, and CO(3--2) spectra from AMISS. Numbers in the upper right corner give the 
signal to noise ratio for each line. When a CO line is detected, the gray band indicates 
the region used to measure the line flux. The scale of the $y$-axis is such that lines 
would have the same amplitude in each transition for thermalized CO emission. The relative 
amplitudes of each spectrum give a sense of the luminosity ratios between the different 
lines.}
\figsetgrpend

\figsetgrpstart
\figsetgrpnum{15.37}
\figsetgrptitle{AMISS.1036}
\figsetplot{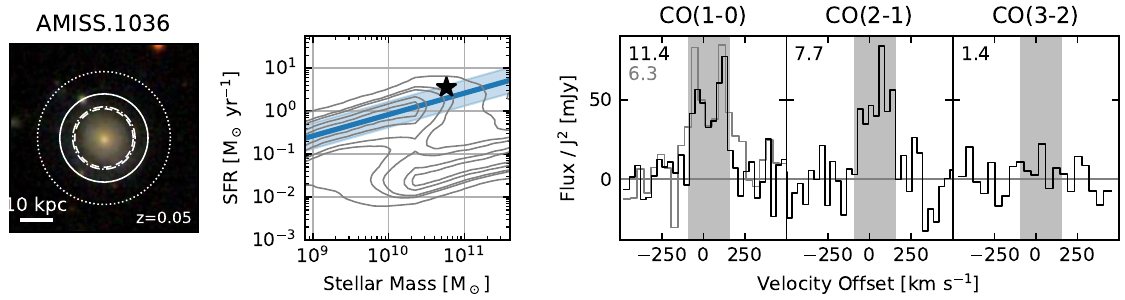}
\figsetgrpnote{Left column: SDSS cutouts of each target. Solid, dashed, and dotted lines show the 
beam sizes of the SMT for CO(2--1), the SMT for CO(3--2) (and the IRAM 30m for CO(1--0)) 
and the 12m for CO(1--0) respectively. The scale bar in the lower left shows 10 
kiloparsecs. 
Middle column: contours show the distribution of star formation rates at a given stellar 
mass, while the blue line and filled region show the main sequence of star forming 
galaxies. The stellar mass and star formation rate of the target galaxy is marked by a 
star. 
Right column: CO(1--0) spectra from AMISS (gray) and xCOLD~GASS (black), CO(2--1) spectra 
from AMISS, and CO(3--2) spectra from AMISS. Numbers in the upper right corner give the 
signal to noise ratio for each line. When a CO line is detected, the gray band indicates 
the region used to measure the line flux. The scale of the $y$-axis is such that lines 
would have the same amplitude in each transition for thermalized CO emission. The relative 
amplitudes of each spectrum give a sense of the luminosity ratios between the different 
lines.}
\figsetgrpend

\figsetgrpstart
\figsetgrpnum{15.38}
\figsetgrptitle{AMISS.1037}
\figsetplot{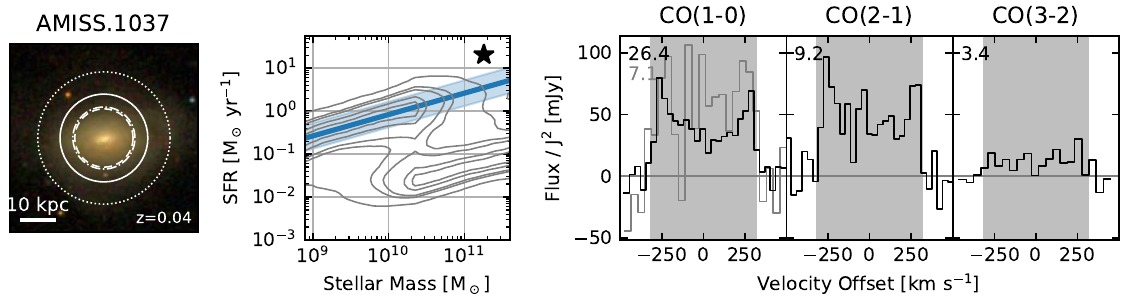}
\figsetgrpnote{Left column: SDSS cutouts of each target. Solid, dashed, and dotted lines show the 
beam sizes of the SMT for CO(2--1), the SMT for CO(3--2) (and the IRAM 30m for CO(1--0)) 
and the 12m for CO(1--0) respectively. The scale bar in the lower left shows 10 
kiloparsecs. 
Middle column: contours show the distribution of star formation rates at a given stellar 
mass, while the blue line and filled region show the main sequence of star forming 
galaxies. The stellar mass and star formation rate of the target galaxy is marked by a 
star. 
Right column: CO(1--0) spectra from AMISS (gray) and xCOLD~GASS (black), CO(2--1) spectra 
from AMISS, and CO(3--2) spectra from AMISS. Numbers in the upper right corner give the 
signal to noise ratio for each line. When a CO line is detected, the gray band indicates 
the region used to measure the line flux. The scale of the $y$-axis is such that lines 
would have the same amplitude in each transition for thermalized CO emission. The relative 
amplitudes of each spectrum give a sense of the luminosity ratios between the different 
lines.}
\figsetgrpend

\figsetgrpstart
\figsetgrpnum{15.39}
\figsetgrptitle{AMISS.1038}
\figsetplot{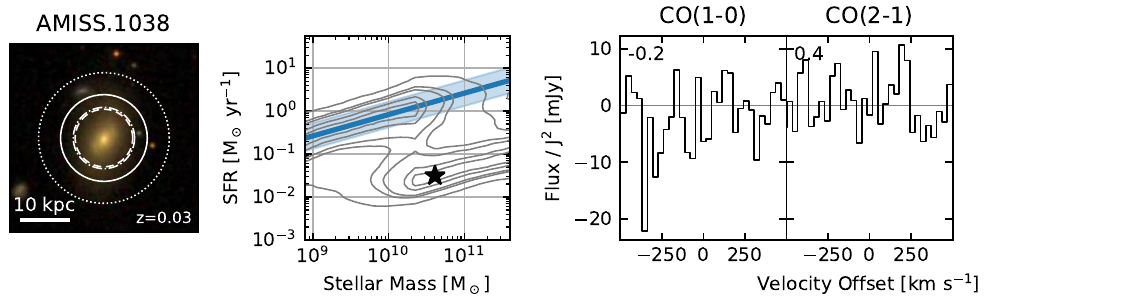}
\figsetgrpnote{Left column: SDSS cutouts of each target. Solid, dashed, and dotted lines show the 
beam sizes of the SMT for CO(2--1), the SMT for CO(3--2) (and the IRAM 30m for CO(1--0)) 
and the 12m for CO(1--0) respectively. The scale bar in the lower left shows 10 
kiloparsecs. 
Middle column: contours show the distribution of star formation rates at a given stellar 
mass, while the blue line and filled region show the main sequence of star forming 
galaxies. The stellar mass and star formation rate of the target galaxy is marked by a 
star. 
Right column: CO(1--0) spectra from AMISS (gray) and xCOLD~GASS (black), CO(2--1) spectra 
from AMISS, and CO(3--2) spectra from AMISS. Numbers in the upper right corner give the 
signal to noise ratio for each line. When a CO line is detected, the gray band indicates 
the region used to measure the line flux. The scale of the $y$-axis is such that lines 
would have the same amplitude in each transition for thermalized CO emission. The relative 
amplitudes of each spectrum give a sense of the luminosity ratios between the different 
lines.}
\figsetgrpend

\figsetgrpstart
\figsetgrpnum{15.40}
\figsetgrptitle{AMISS.1039}
\figsetplot{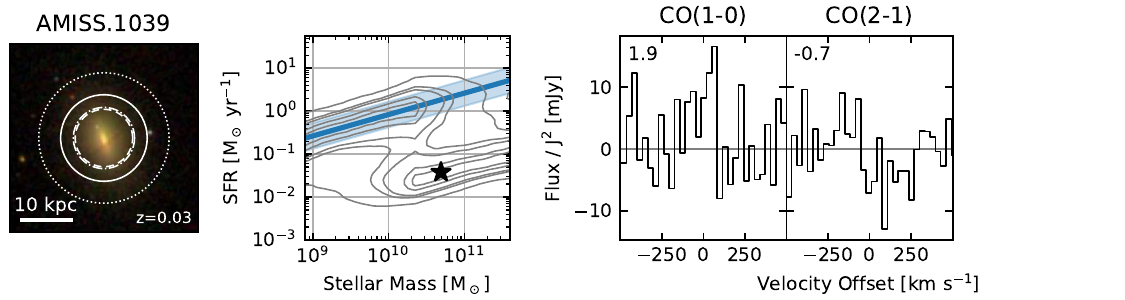}
\figsetgrpnote{Left column: SDSS cutouts of each target. Solid, dashed, and dotted lines show the 
beam sizes of the SMT for CO(2--1), the SMT for CO(3--2) (and the IRAM 30m for CO(1--0)) 
and the 12m for CO(1--0) respectively. The scale bar in the lower left shows 10 
kiloparsecs. 
Middle column: contours show the distribution of star formation rates at a given stellar 
mass, while the blue line and filled region show the main sequence of star forming 
galaxies. The stellar mass and star formation rate of the target galaxy is marked by a 
star. 
Right column: CO(1--0) spectra from AMISS (gray) and xCOLD~GASS (black), CO(2--1) spectra 
from AMISS, and CO(3--2) spectra from AMISS. Numbers in the upper right corner give the 
signal to noise ratio for each line. When a CO line is detected, the gray band indicates 
the region used to measure the line flux. The scale of the $y$-axis is such that lines 
would have the same amplitude in each transition for thermalized CO emission. The relative 
amplitudes of each spectrum give a sense of the luminosity ratios between the different 
lines.}
\figsetgrpend

\figsetgrpstart
\figsetgrpnum{15.41}
\figsetgrptitle{AMISS.1040}
\figsetplot{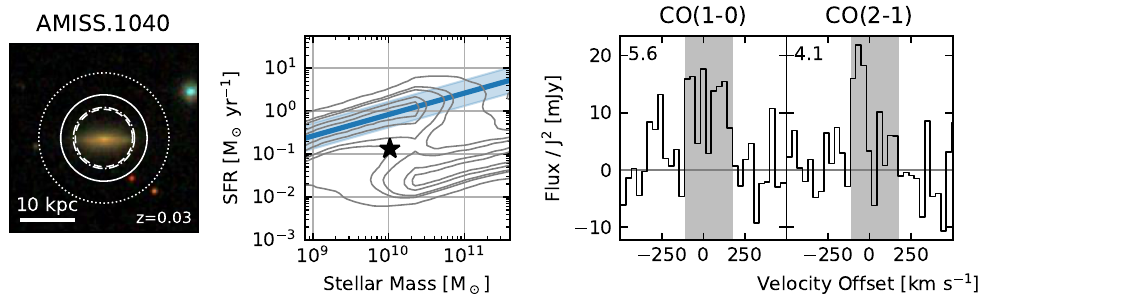}
\figsetgrpnote{Left column: SDSS cutouts of each target. Solid, dashed, and dotted lines show the 
beam sizes of the SMT for CO(2--1), the SMT for CO(3--2) (and the IRAM 30m for CO(1--0)) 
and the 12m for CO(1--0) respectively. The scale bar in the lower left shows 10 
kiloparsecs. 
Middle column: contours show the distribution of star formation rates at a given stellar 
mass, while the blue line and filled region show the main sequence of star forming 
galaxies. The stellar mass and star formation rate of the target galaxy is marked by a 
star. 
Right column: CO(1--0) spectra from AMISS (gray) and xCOLD~GASS (black), CO(2--1) spectra 
from AMISS, and CO(3--2) spectra from AMISS. Numbers in the upper right corner give the 
signal to noise ratio for each line. When a CO line is detected, the gray band indicates 
the region used to measure the line flux. The scale of the $y$-axis is such that lines 
would have the same amplitude in each transition for thermalized CO emission. The relative 
amplitudes of each spectrum give a sense of the luminosity ratios between the different 
lines.}
\figsetgrpend

\figsetgrpstart
\figsetgrpnum{15.42}
\figsetgrptitle{AMISS.1041}
\figsetplot{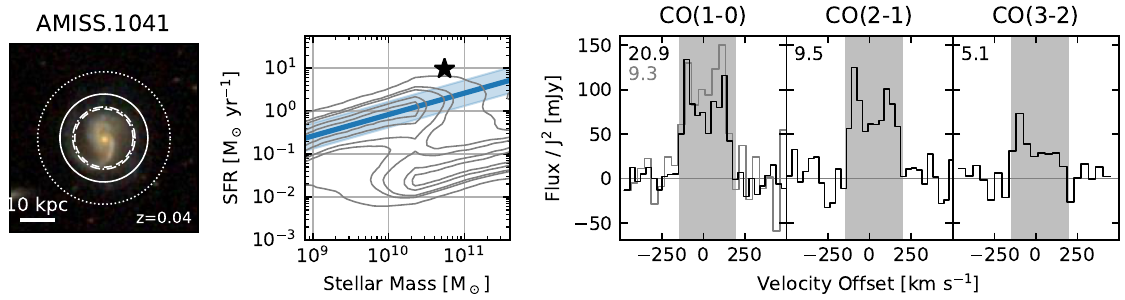}
\figsetgrpnote{Left column: SDSS cutouts of each target. Solid, dashed, and dotted lines show the 
beam sizes of the SMT for CO(2--1), the SMT for CO(3--2) (and the IRAM 30m for CO(1--0)) 
and the 12m for CO(1--0) respectively. The scale bar in the lower left shows 10 
kiloparsecs. 
Middle column: contours show the distribution of star formation rates at a given stellar 
mass, while the blue line and filled region show the main sequence of star forming 
galaxies. The stellar mass and star formation rate of the target galaxy is marked by a 
star. 
Right column: CO(1--0) spectra from AMISS (gray) and xCOLD~GASS (black), CO(2--1) spectra 
from AMISS, and CO(3--2) spectra from AMISS. Numbers in the upper right corner give the 
signal to noise ratio for each line. When a CO line is detected, the gray band indicates 
the region used to measure the line flux. The scale of the $y$-axis is such that lines 
would have the same amplitude in each transition for thermalized CO emission. The relative 
amplitudes of each spectrum give a sense of the luminosity ratios between the different 
lines.}
\figsetgrpend

\figsetgrpstart
\figsetgrpnum{15.43}
\figsetgrptitle{AMISS.1042}
\figsetplot{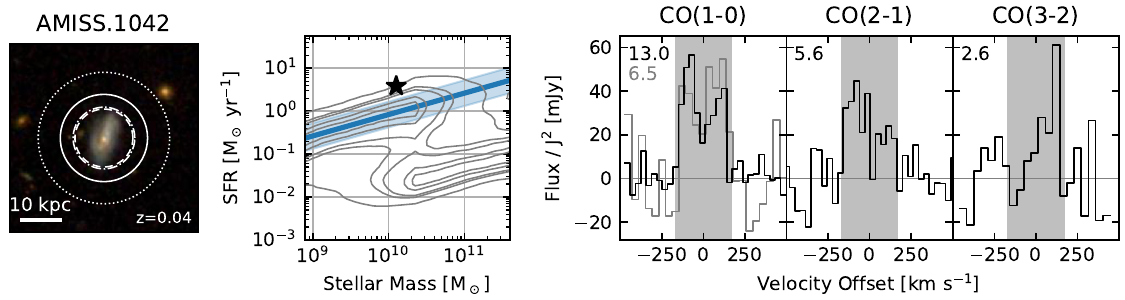}
\figsetgrpnote{Left column: SDSS cutouts of each target. Solid, dashed, and dotted lines show the 
beam sizes of the SMT for CO(2--1), the SMT for CO(3--2) (and the IRAM 30m for CO(1--0)) 
and the 12m for CO(1--0) respectively. The scale bar in the lower left shows 10 
kiloparsecs. 
Middle column: contours show the distribution of star formation rates at a given stellar 
mass, while the blue line and filled region show the main sequence of star forming 
galaxies. The stellar mass and star formation rate of the target galaxy is marked by a 
star. 
Right column: CO(1--0) spectra from AMISS (gray) and xCOLD~GASS (black), CO(2--1) spectra 
from AMISS, and CO(3--2) spectra from AMISS. Numbers in the upper right corner give the 
signal to noise ratio for each line. When a CO line is detected, the gray band indicates 
the region used to measure the line flux. The scale of the $y$-axis is such that lines 
would have the same amplitude in each transition for thermalized CO emission. The relative 
amplitudes of each spectrum give a sense of the luminosity ratios between the different 
lines.}
\figsetgrpend

\figsetgrpstart
\figsetgrpnum{15.44}
\figsetgrptitle{AMISS.1043}
\figsetplot{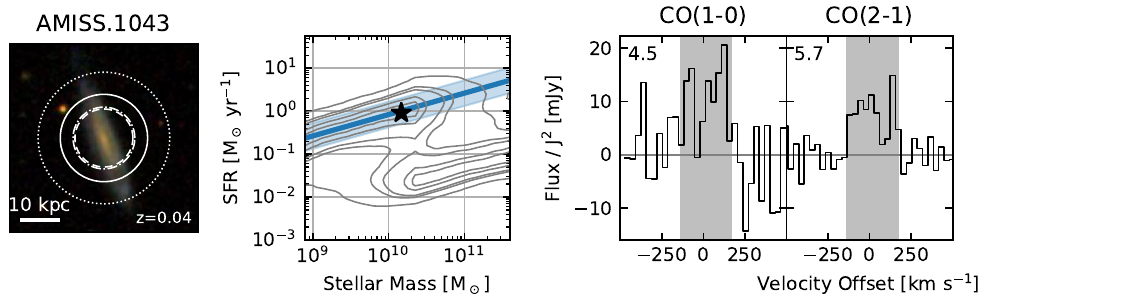}
\figsetgrpnote{Left column: SDSS cutouts of each target. Solid, dashed, and dotted lines show the 
beam sizes of the SMT for CO(2--1), the SMT for CO(3--2) (and the IRAM 30m for CO(1--0)) 
and the 12m for CO(1--0) respectively. The scale bar in the lower left shows 10 
kiloparsecs. 
Middle column: contours show the distribution of star formation rates at a given stellar 
mass, while the blue line and filled region show the main sequence of star forming 
galaxies. The stellar mass and star formation rate of the target galaxy is marked by a 
star. 
Right column: CO(1--0) spectra from AMISS (gray) and xCOLD~GASS (black), CO(2--1) spectra 
from AMISS, and CO(3--2) spectra from AMISS. Numbers in the upper right corner give the 
signal to noise ratio for each line. When a CO line is detected, the gray band indicates 
the region used to measure the line flux. The scale of the $y$-axis is such that lines 
would have the same amplitude in each transition for thermalized CO emission. The relative 
amplitudes of each spectrum give a sense of the luminosity ratios between the different 
lines.}
\figsetgrpend

\figsetgrpstart
\figsetgrpnum{15.45}
\figsetgrptitle{AMISS.1044}
\figsetplot{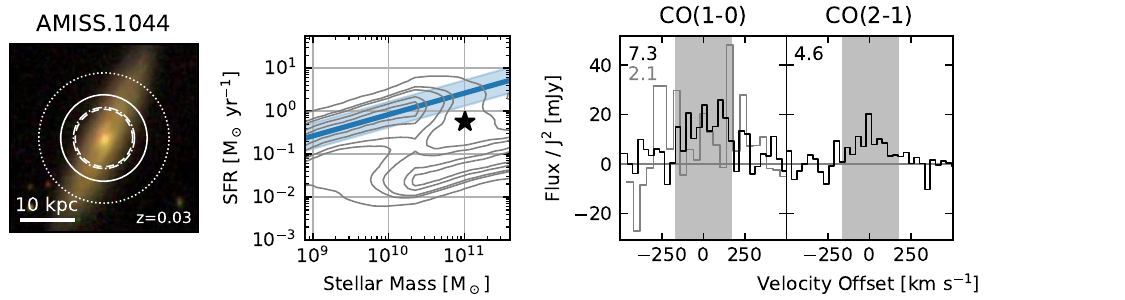}
\figsetgrpnote{Left column: SDSS cutouts of each target. Solid, dashed, and dotted lines show the 
beam sizes of the SMT for CO(2--1), the SMT for CO(3--2) (and the IRAM 30m for CO(1--0)) 
and the 12m for CO(1--0) respectively. The scale bar in the lower left shows 10 
kiloparsecs. 
Middle column: contours show the distribution of star formation rates at a given stellar 
mass, while the blue line and filled region show the main sequence of star forming 
galaxies. The stellar mass and star formation rate of the target galaxy is marked by a 
star. 
Right column: CO(1--0) spectra from AMISS (gray) and xCOLD~GASS (black), CO(2--1) spectra 
from AMISS, and CO(3--2) spectra from AMISS. Numbers in the upper right corner give the 
signal to noise ratio for each line. When a CO line is detected, the gray band indicates 
the region used to measure the line flux. The scale of the $y$-axis is such that lines 
would have the same amplitude in each transition for thermalized CO emission. The relative 
amplitudes of each spectrum give a sense of the luminosity ratios between the different 
lines.}
\figsetgrpend

\figsetgrpstart
\figsetgrpnum{15.46}
\figsetgrptitle{AMISS.1045}
\figsetplot{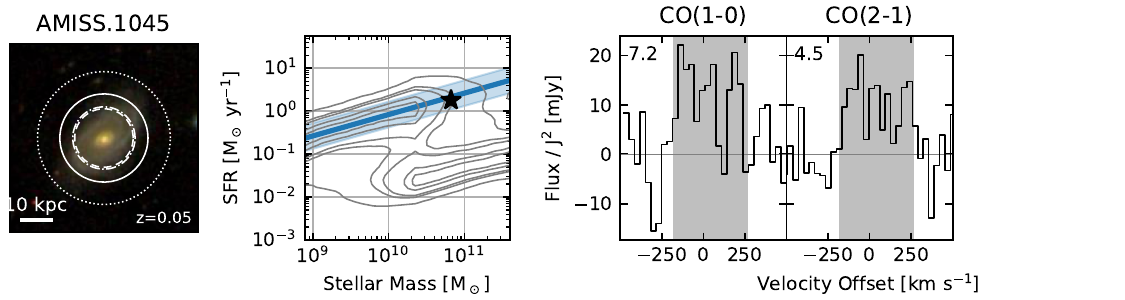}
\figsetgrpnote{Left column: SDSS cutouts of each target. Solid, dashed, and dotted lines show the 
beam sizes of the SMT for CO(2--1), the SMT for CO(3--2) (and the IRAM 30m for CO(1--0)) 
and the 12m for CO(1--0) respectively. The scale bar in the lower left shows 10 
kiloparsecs. 
Middle column: contours show the distribution of star formation rates at a given stellar 
mass, while the blue line and filled region show the main sequence of star forming 
galaxies. The stellar mass and star formation rate of the target galaxy is marked by a 
star. 
Right column: CO(1--0) spectra from AMISS (gray) and xCOLD~GASS (black), CO(2--1) spectra 
from AMISS, and CO(3--2) spectra from AMISS. Numbers in the upper right corner give the 
signal to noise ratio for each line. When a CO line is detected, the gray band indicates 
the region used to measure the line flux. The scale of the $y$-axis is such that lines 
would have the same amplitude in each transition for thermalized CO emission. The relative 
amplitudes of each spectrum give a sense of the luminosity ratios between the different 
lines.}
\figsetgrpend

\figsetgrpstart
\figsetgrpnum{15.47}
\figsetgrptitle{AMISS.1046}
\figsetplot{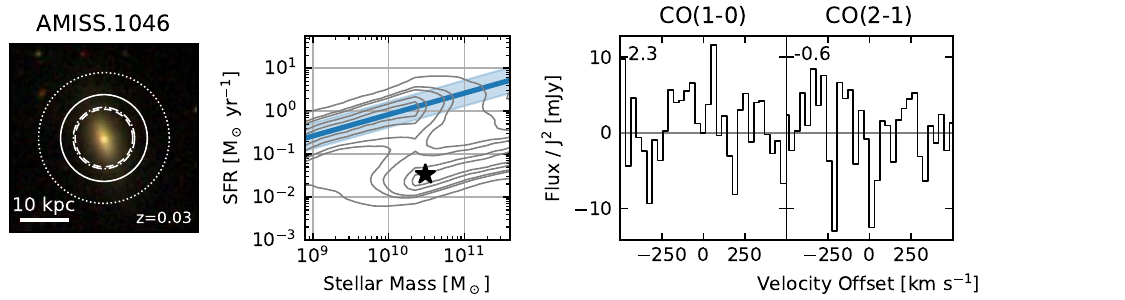}
\figsetgrpnote{Left column: SDSS cutouts of each target. Solid, dashed, and dotted lines show the 
beam sizes of the SMT for CO(2--1), the SMT for CO(3--2) (and the IRAM 30m for CO(1--0)) 
and the 12m for CO(1--0) respectively. The scale bar in the lower left shows 10 
kiloparsecs. 
Middle column: contours show the distribution of star formation rates at a given stellar 
mass, while the blue line and filled region show the main sequence of star forming 
galaxies. The stellar mass and star formation rate of the target galaxy is marked by a 
star. 
Right column: CO(1--0) spectra from AMISS (gray) and xCOLD~GASS (black), CO(2--1) spectra 
from AMISS, and CO(3--2) spectra from AMISS. Numbers in the upper right corner give the 
signal to noise ratio for each line. When a CO line is detected, the gray band indicates 
the region used to measure the line flux. The scale of the $y$-axis is such that lines 
would have the same amplitude in each transition for thermalized CO emission. The relative 
amplitudes of each spectrum give a sense of the luminosity ratios between the different 
lines.}
\figsetgrpend

\figsetgrpstart
\figsetgrpnum{15.48}
\figsetgrptitle{AMISS.1047}
\figsetplot{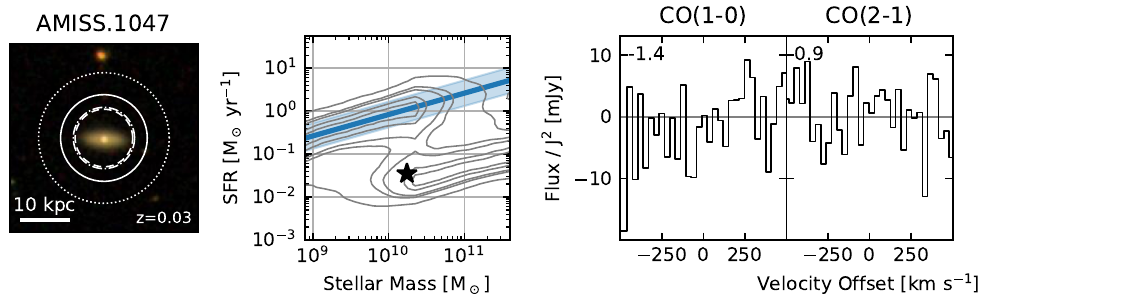}
\figsetgrpnote{Left column: SDSS cutouts of each target. Solid, dashed, and dotted lines show the 
beam sizes of the SMT for CO(2--1), the SMT for CO(3--2) (and the IRAM 30m for CO(1--0)) 
and the 12m for CO(1--0) respectively. The scale bar in the lower left shows 10 
kiloparsecs. 
Middle column: contours show the distribution of star formation rates at a given stellar 
mass, while the blue line and filled region show the main sequence of star forming 
galaxies. The stellar mass and star formation rate of the target galaxy is marked by a 
star. 
Right column: CO(1--0) spectra from AMISS (gray) and xCOLD~GASS (black), CO(2--1) spectra 
from AMISS, and CO(3--2) spectra from AMISS. Numbers in the upper right corner give the 
signal to noise ratio for each line. When a CO line is detected, the gray band indicates 
the region used to measure the line flux. The scale of the $y$-axis is such that lines 
would have the same amplitude in each transition for thermalized CO emission. The relative 
amplitudes of each spectrum give a sense of the luminosity ratios between the different 
lines.}
\figsetgrpend

\figsetgrpstart
\figsetgrpnum{15.49}
\figsetgrptitle{AMISS.1048}
\figsetplot{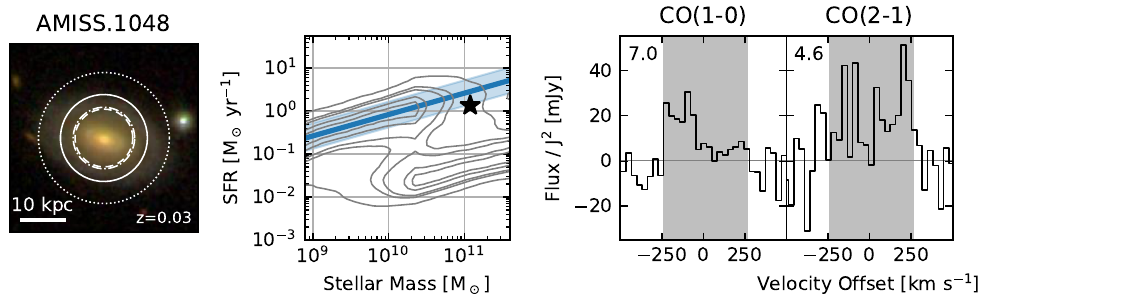}
\figsetgrpnote{Left column: SDSS cutouts of each target. Solid, dashed, and dotted lines show the 
beam sizes of the SMT for CO(2--1), the SMT for CO(3--2) (and the IRAM 30m for CO(1--0)) 
and the 12m for CO(1--0) respectively. The scale bar in the lower left shows 10 
kiloparsecs. 
Middle column: contours show the distribution of star formation rates at a given stellar 
mass, while the blue line and filled region show the main sequence of star forming 
galaxies. The stellar mass and star formation rate of the target galaxy is marked by a 
star. 
Right column: CO(1--0) spectra from AMISS (gray) and xCOLD~GASS (black), CO(2--1) spectra 
from AMISS, and CO(3--2) spectra from AMISS. Numbers in the upper right corner give the 
signal to noise ratio for each line. When a CO line is detected, the gray band indicates 
the region used to measure the line flux. The scale of the $y$-axis is such that lines 
would have the same amplitude in each transition for thermalized CO emission. The relative 
amplitudes of each spectrum give a sense of the luminosity ratios between the different 
lines.}
\figsetgrpend

\figsetgrpstart
\figsetgrpnum{15.50}
\figsetgrptitle{AMISS.1049}
\figsetplot{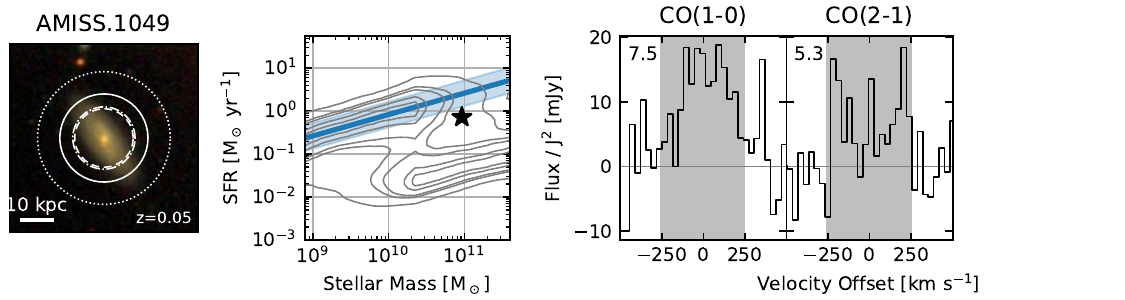}
\figsetgrpnote{Left column: SDSS cutouts of each target. Solid, dashed, and dotted lines show the 
beam sizes of the SMT for CO(2--1), the SMT for CO(3--2) (and the IRAM 30m for CO(1--0)) 
and the 12m for CO(1--0) respectively. The scale bar in the lower left shows 10 
kiloparsecs. 
Middle column: contours show the distribution of star formation rates at a given stellar 
mass, while the blue line and filled region show the main sequence of star forming 
galaxies. The stellar mass and star formation rate of the target galaxy is marked by a 
star. 
Right column: CO(1--0) spectra from AMISS (gray) and xCOLD~GASS (black), CO(2--1) spectra 
from AMISS, and CO(3--2) spectra from AMISS. Numbers in the upper right corner give the 
signal to noise ratio for each line. When a CO line is detected, the gray band indicates 
the region used to measure the line flux. The scale of the $y$-axis is such that lines 
would have the same amplitude in each transition for thermalized CO emission. The relative 
amplitudes of each spectrum give a sense of the luminosity ratios between the different 
lines.}
\figsetgrpend

\figsetgrpstart
\figsetgrpnum{15.51}
\figsetgrptitle{AMISS.1050}
\figsetplot{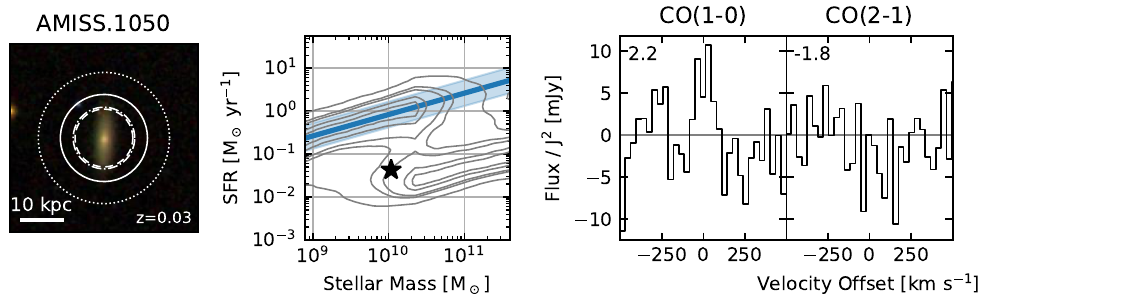}
\figsetgrpnote{Left column: SDSS cutouts of each target. Solid, dashed, and dotted lines show the 
beam sizes of the SMT for CO(2--1), the SMT for CO(3--2) (and the IRAM 30m for CO(1--0)) 
and the 12m for CO(1--0) respectively. The scale bar in the lower left shows 10 
kiloparsecs. 
Middle column: contours show the distribution of star formation rates at a given stellar 
mass, while the blue line and filled region show the main sequence of star forming 
galaxies. The stellar mass and star formation rate of the target galaxy is marked by a 
star. 
Right column: CO(1--0) spectra from AMISS (gray) and xCOLD~GASS (black), CO(2--1) spectra 
from AMISS, and CO(3--2) spectra from AMISS. Numbers in the upper right corner give the 
signal to noise ratio for each line. When a CO line is detected, the gray band indicates 
the region used to measure the line flux. The scale of the $y$-axis is such that lines 
would have the same amplitude in each transition for thermalized CO emission. The relative 
amplitudes of each spectrum give a sense of the luminosity ratios between the different 
lines.}
\figsetgrpend

\figsetgrpstart
\figsetgrpnum{15.52}
\figsetgrptitle{AMISS.1051}
\figsetplot{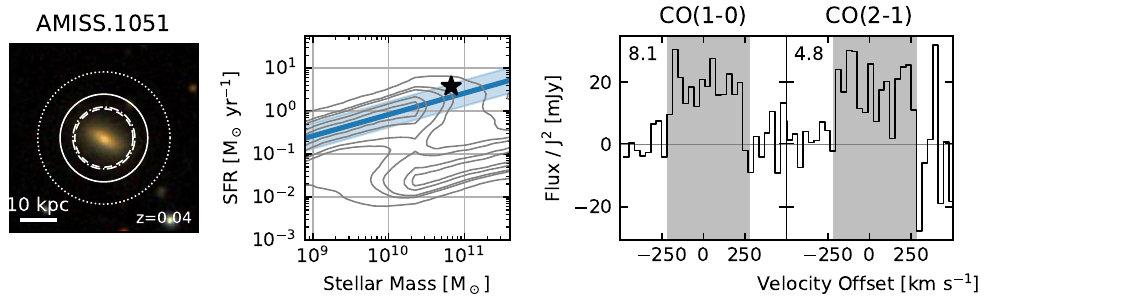}
\figsetgrpnote{Left column: SDSS cutouts of each target. Solid, dashed, and dotted lines show the 
beam sizes of the SMT for CO(2--1), the SMT for CO(3--2) (and the IRAM 30m for CO(1--0)) 
and the 12m for CO(1--0) respectively. The scale bar in the lower left shows 10 
kiloparsecs. 
Middle column: contours show the distribution of star formation rates at a given stellar 
mass, while the blue line and filled region show the main sequence of star forming 
galaxies. The stellar mass and star formation rate of the target galaxy is marked by a 
star. 
Right column: CO(1--0) spectra from AMISS (gray) and xCOLD~GASS (black), CO(2--1) spectra 
from AMISS, and CO(3--2) spectra from AMISS. Numbers in the upper right corner give the 
signal to noise ratio for each line. When a CO line is detected, the gray band indicates 
the region used to measure the line flux. The scale of the $y$-axis is such that lines 
would have the same amplitude in each transition for thermalized CO emission. The relative 
amplitudes of each spectrum give a sense of the luminosity ratios between the different 
lines.}
\figsetgrpend

\figsetgrpstart
\figsetgrpnum{15.53}
\figsetgrptitle{AMISS.1052}
\figsetplot{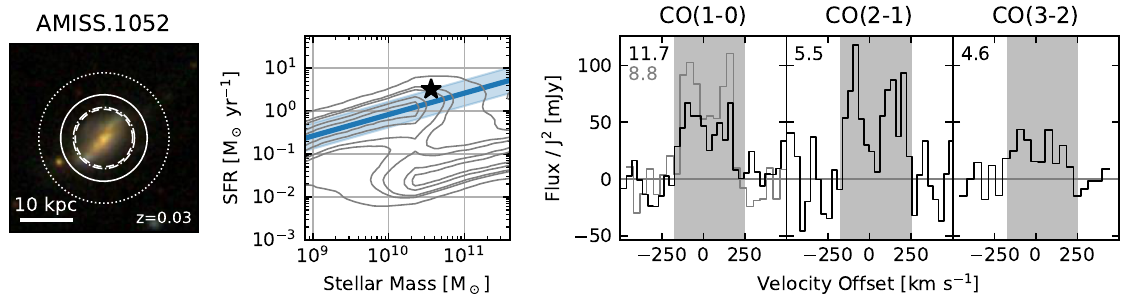}
\figsetgrpnote{Left column: SDSS cutouts of each target. Solid, dashed, and dotted lines show the 
beam sizes of the SMT for CO(2--1), the SMT for CO(3--2) (and the IRAM 30m for CO(1--0)) 
and the 12m for CO(1--0) respectively. The scale bar in the lower left shows 10 
kiloparsecs. 
Middle column: contours show the distribution of star formation rates at a given stellar 
mass, while the blue line and filled region show the main sequence of star forming 
galaxies. The stellar mass and star formation rate of the target galaxy is marked by a 
star. 
Right column: CO(1--0) spectra from AMISS (gray) and xCOLD~GASS (black), CO(2--1) spectra 
from AMISS, and CO(3--2) spectra from AMISS. Numbers in the upper right corner give the 
signal to noise ratio for each line. When a CO line is detected, the gray band indicates 
the region used to measure the line flux. The scale of the $y$-axis is such that lines 
would have the same amplitude in each transition for thermalized CO emission. The relative 
amplitudes of each spectrum give a sense of the luminosity ratios between the different 
lines.}
\figsetgrpend

\figsetgrpstart
\figsetgrpnum{15.54}
\figsetgrptitle{AMISS.1053}
\figsetplot{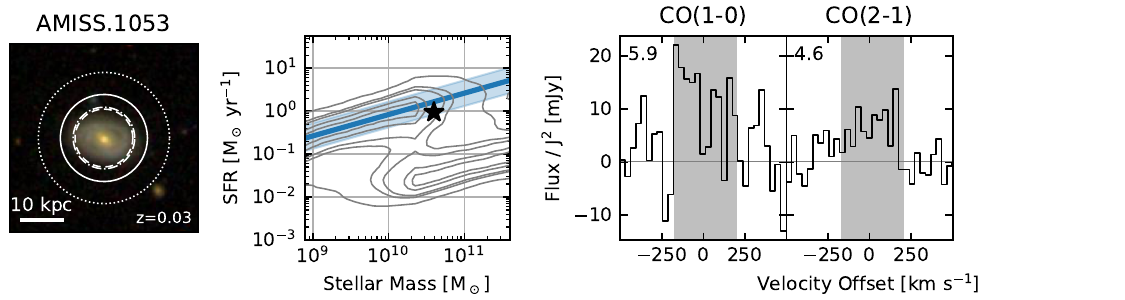}
\figsetgrpnote{Left column: SDSS cutouts of each target. Solid, dashed, and dotted lines show the 
beam sizes of the SMT for CO(2--1), the SMT for CO(3--2) (and the IRAM 30m for CO(1--0)) 
and the 12m for CO(1--0) respectively. The scale bar in the lower left shows 10 
kiloparsecs. 
Middle column: contours show the distribution of star formation rates at a given stellar 
mass, while the blue line and filled region show the main sequence of star forming 
galaxies. The stellar mass and star formation rate of the target galaxy is marked by a 
star. 
Right column: CO(1--0) spectra from AMISS (gray) and xCOLD~GASS (black), CO(2--1) spectra 
from AMISS, and CO(3--2) spectra from AMISS. Numbers in the upper right corner give the 
signal to noise ratio for each line. When a CO line is detected, the gray band indicates 
the region used to measure the line flux. The scale of the $y$-axis is such that lines 
would have the same amplitude in each transition for thermalized CO emission. The relative 
amplitudes of each spectrum give a sense of the luminosity ratios between the different 
lines.}
\figsetgrpend

\figsetgrpstart
\figsetgrpnum{15.55}
\figsetgrptitle{AMISS.1054}
\figsetplot{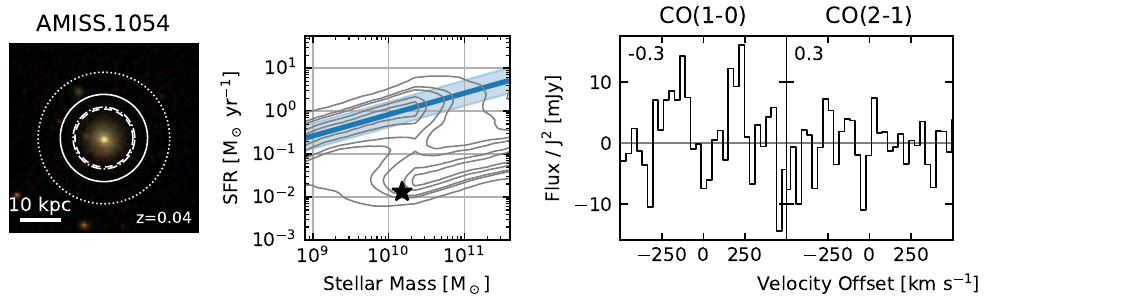}
\figsetgrpnote{Left column: SDSS cutouts of each target. Solid, dashed, and dotted lines show the 
beam sizes of the SMT for CO(2--1), the SMT for CO(3--2) (and the IRAM 30m for CO(1--0)) 
and the 12m for CO(1--0) respectively. The scale bar in the lower left shows 10 
kiloparsecs. 
Middle column: contours show the distribution of star formation rates at a given stellar 
mass, while the blue line and filled region show the main sequence of star forming 
galaxies. The stellar mass and star formation rate of the target galaxy is marked by a 
star. 
Right column: CO(1--0) spectra from AMISS (gray) and xCOLD~GASS (black), CO(2--1) spectra 
from AMISS, and CO(3--2) spectra from AMISS. Numbers in the upper right corner give the 
signal to noise ratio for each line. When a CO line is detected, the gray band indicates 
the region used to measure the line flux. The scale of the $y$-axis is such that lines 
would have the same amplitude in each transition for thermalized CO emission. The relative 
amplitudes of each spectrum give a sense of the luminosity ratios between the different 
lines.}
\figsetgrpend

\figsetgrpstart
\figsetgrpnum{15.56}
\figsetgrptitle{AMISS.1055}
\figsetplot{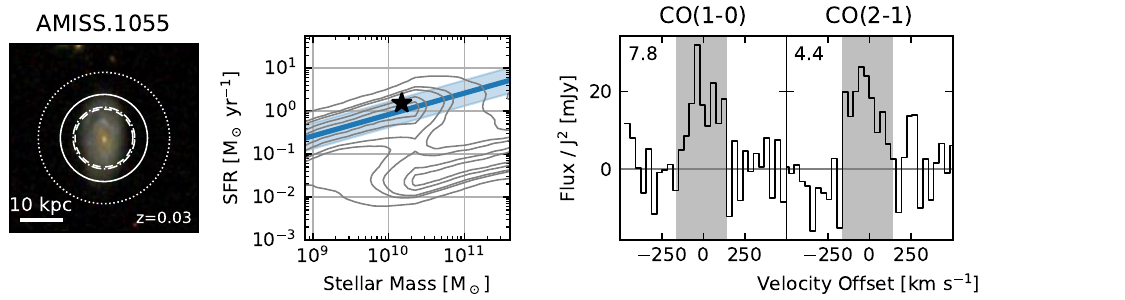}
\figsetgrpnote{Left column: SDSS cutouts of each target. Solid, dashed, and dotted lines show the 
beam sizes of the SMT for CO(2--1), the SMT for CO(3--2) (and the IRAM 30m for CO(1--0)) 
and the 12m for CO(1--0) respectively. The scale bar in the lower left shows 10 
kiloparsecs. 
Middle column: contours show the distribution of star formation rates at a given stellar 
mass, while the blue line and filled region show the main sequence of star forming 
galaxies. The stellar mass and star formation rate of the target galaxy is marked by a 
star. 
Right column: CO(1--0) spectra from AMISS (gray) and xCOLD~GASS (black), CO(2--1) spectra 
from AMISS, and CO(3--2) spectra from AMISS. Numbers in the upper right corner give the 
signal to noise ratio for each line. When a CO line is detected, the gray band indicates 
the region used to measure the line flux. The scale of the $y$-axis is such that lines 
would have the same amplitude in each transition for thermalized CO emission. The relative 
amplitudes of each spectrum give a sense of the luminosity ratios between the different 
lines.}
\figsetgrpend

\figsetgrpstart
\figsetgrpnum{15.57}
\figsetgrptitle{AMISS.1056}
\figsetplot{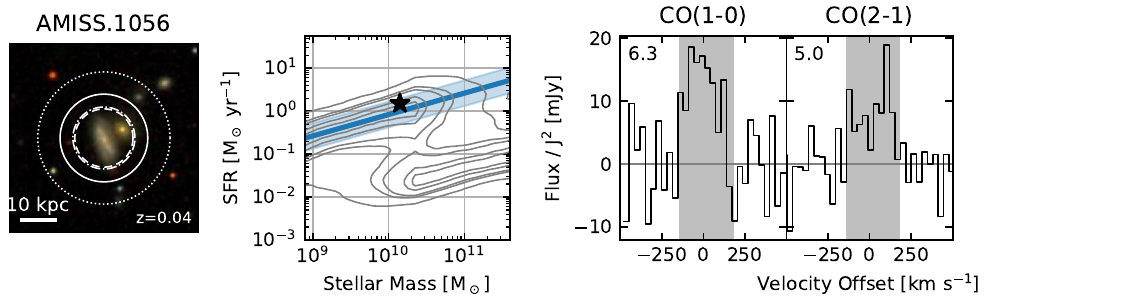}
\figsetgrpnote{Left column: SDSS cutouts of each target. Solid, dashed, and dotted lines show the 
beam sizes of the SMT for CO(2--1), the SMT for CO(3--2) (and the IRAM 30m for CO(1--0)) 
and the 12m for CO(1--0) respectively. The scale bar in the lower left shows 10 
kiloparsecs. 
Middle column: contours show the distribution of star formation rates at a given stellar 
mass, while the blue line and filled region show the main sequence of star forming 
galaxies. The stellar mass and star formation rate of the target galaxy is marked by a 
star. 
Right column: CO(1--0) spectra from AMISS (gray) and xCOLD~GASS (black), CO(2--1) spectra 
from AMISS, and CO(3--2) spectra from AMISS. Numbers in the upper right corner give the 
signal to noise ratio for each line. When a CO line is detected, the gray band indicates 
the region used to measure the line flux. The scale of the $y$-axis is such that lines 
would have the same amplitude in each transition for thermalized CO emission. The relative 
amplitudes of each spectrum give a sense of the luminosity ratios between the different 
lines.}
\figsetgrpend

\figsetgrpstart
\figsetgrpnum{15.58}
\figsetgrptitle{AMISS.1057}
\figsetplot{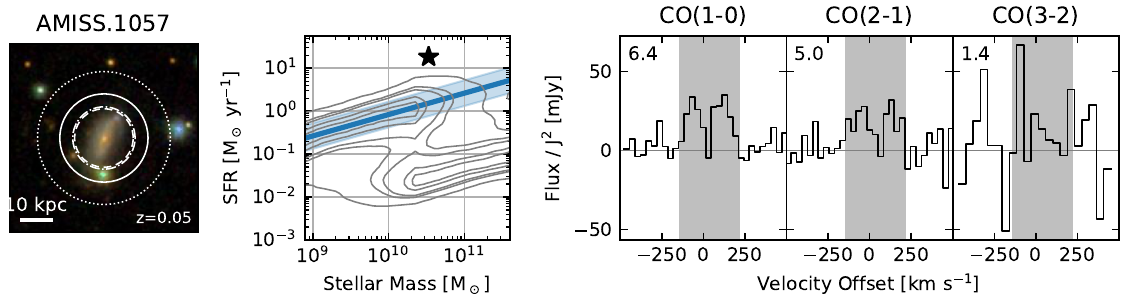}
\figsetgrpnote{Left column: SDSS cutouts of each target. Solid, dashed, and dotted lines show the 
beam sizes of the SMT for CO(2--1), the SMT for CO(3--2) (and the IRAM 30m for CO(1--0)) 
and the 12m for CO(1--0) respectively. The scale bar in the lower left shows 10 
kiloparsecs. 
Middle column: contours show the distribution of star formation rates at a given stellar 
mass, while the blue line and filled region show the main sequence of star forming 
galaxies. The stellar mass and star formation rate of the target galaxy is marked by a 
star. 
Right column: CO(1--0) spectra from AMISS (gray) and xCOLD~GASS (black), CO(2--1) spectra 
from AMISS, and CO(3--2) spectra from AMISS. Numbers in the upper right corner give the 
signal to noise ratio for each line. When a CO line is detected, the gray band indicates 
the region used to measure the line flux. The scale of the $y$-axis is such that lines 
would have the same amplitude in each transition for thermalized CO emission. The relative 
amplitudes of each spectrum give a sense of the luminosity ratios between the different 
lines.}
\figsetgrpend

\figsetgrpstart
\figsetgrpnum{15.59}
\figsetgrptitle{AMISS.1058}
\figsetplot{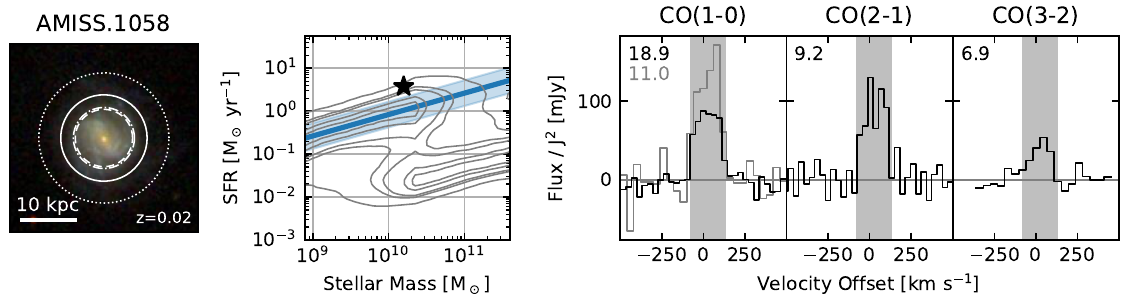}
\figsetgrpnote{Left column: SDSS cutouts of each target. Solid, dashed, and dotted lines show the 
beam sizes of the SMT for CO(2--1), the SMT for CO(3--2) (and the IRAM 30m for CO(1--0)) 
and the 12m for CO(1--0) respectively. The scale bar in the lower left shows 10 
kiloparsecs. 
Middle column: contours show the distribution of star formation rates at a given stellar 
mass, while the blue line and filled region show the main sequence of star forming 
galaxies. The stellar mass and star formation rate of the target galaxy is marked by a 
star. 
Right column: CO(1--0) spectra from AMISS (gray) and xCOLD~GASS (black), CO(2--1) spectra 
from AMISS, and CO(3--2) spectra from AMISS. Numbers in the upper right corner give the 
signal to noise ratio for each line. When a CO line is detected, the gray band indicates 
the region used to measure the line flux. The scale of the $y$-axis is such that lines 
would have the same amplitude in each transition for thermalized CO emission. The relative 
amplitudes of each spectrum give a sense of the luminosity ratios between the different 
lines.}
\figsetgrpend

\figsetgrpstart
\figsetgrpnum{15.60}
\figsetgrptitle{AMISS.1059}
\figsetplot{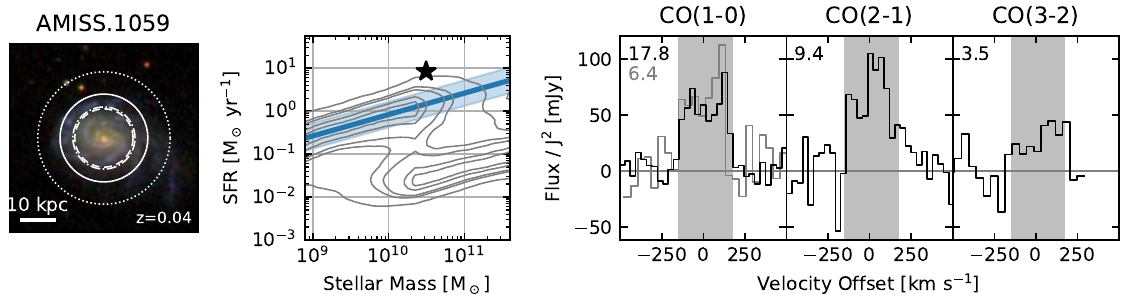}
\figsetgrpnote{Left column: SDSS cutouts of each target. Solid, dashed, and dotted lines show the 
beam sizes of the SMT for CO(2--1), the SMT for CO(3--2) (and the IRAM 30m for CO(1--0)) 
and the 12m for CO(1--0) respectively. The scale bar in the lower left shows 10 
kiloparsecs. 
Middle column: contours show the distribution of star formation rates at a given stellar 
mass, while the blue line and filled region show the main sequence of star forming 
galaxies. The stellar mass and star formation rate of the target galaxy is marked by a 
star. 
Right column: CO(1--0) spectra from AMISS (gray) and xCOLD~GASS (black), CO(2--1) spectra 
from AMISS, and CO(3--2) spectra from AMISS. Numbers in the upper right corner give the 
signal to noise ratio for each line. When a CO line is detected, the gray band indicates 
the region used to measure the line flux. The scale of the $y$-axis is such that lines 
would have the same amplitude in each transition for thermalized CO emission. The relative 
amplitudes of each spectrum give a sense of the luminosity ratios between the different 
lines.}
\figsetgrpend

\figsetgrpstart
\figsetgrpnum{15.61}
\figsetgrptitle{AMISS.1060}
\figsetplot{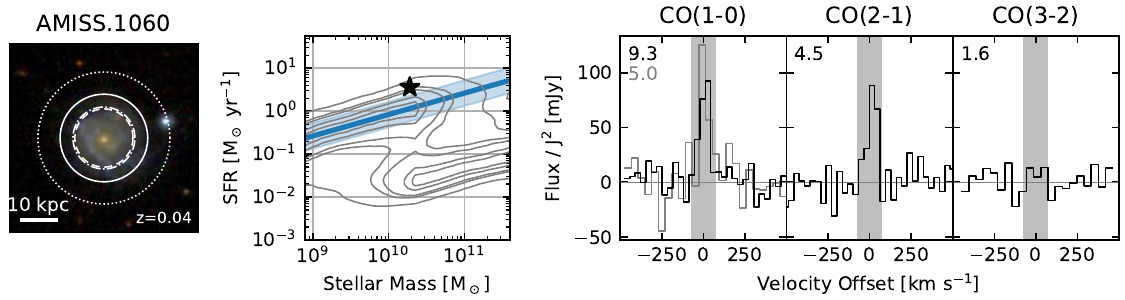}
\figsetgrpnote{Left column: SDSS cutouts of each target. Solid, dashed, and dotted lines show the 
beam sizes of the SMT for CO(2--1), the SMT for CO(3--2) (and the IRAM 30m for CO(1--0)) 
and the 12m for CO(1--0) respectively. The scale bar in the lower left shows 10 
kiloparsecs. 
Middle column: contours show the distribution of star formation rates at a given stellar 
mass, while the blue line and filled region show the main sequence of star forming 
galaxies. The stellar mass and star formation rate of the target galaxy is marked by a 
star. 
Right column: CO(1--0) spectra from AMISS (gray) and xCOLD~GASS (black), CO(2--1) spectra 
from AMISS, and CO(3--2) spectra from AMISS. Numbers in the upper right corner give the 
signal to noise ratio for each line. When a CO line is detected, the gray band indicates 
the region used to measure the line flux. The scale of the $y$-axis is such that lines 
would have the same amplitude in each transition for thermalized CO emission. The relative 
amplitudes of each spectrum give a sense of the luminosity ratios between the different 
lines.}
\figsetgrpend

\figsetgrpstart
\figsetgrpnum{15.62}
\figsetgrptitle{AMISS.1061}
\figsetplot{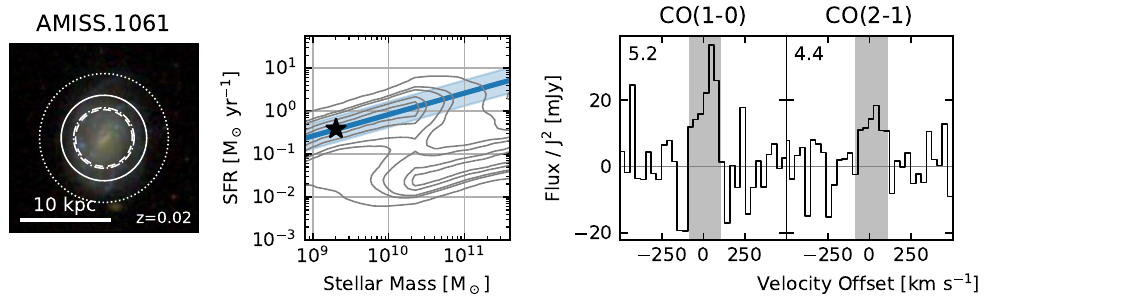}
\figsetgrpnote{Left column: SDSS cutouts of each target. Solid, dashed, and dotted lines show the 
beam sizes of the SMT for CO(2--1), the SMT for CO(3--2) (and the IRAM 30m for CO(1--0)) 
and the 12m for CO(1--0) respectively. The scale bar in the lower left shows 10 
kiloparsecs. 
Middle column: contours show the distribution of star formation rates at a given stellar 
mass, while the blue line and filled region show the main sequence of star forming 
galaxies. The stellar mass and star formation rate of the target galaxy is marked by a 
star. 
Right column: CO(1--0) spectra from AMISS (gray) and xCOLD~GASS (black), CO(2--1) spectra 
from AMISS, and CO(3--2) spectra from AMISS. Numbers in the upper right corner give the 
signal to noise ratio for each line. When a CO line is detected, the gray band indicates 
the region used to measure the line flux. The scale of the $y$-axis is such that lines 
would have the same amplitude in each transition for thermalized CO emission. The relative 
amplitudes of each spectrum give a sense of the luminosity ratios between the different 
lines.}
\figsetgrpend

\figsetgrpstart
\figsetgrpnum{15.63}
\figsetgrptitle{AMISS.1062}
\figsetplot{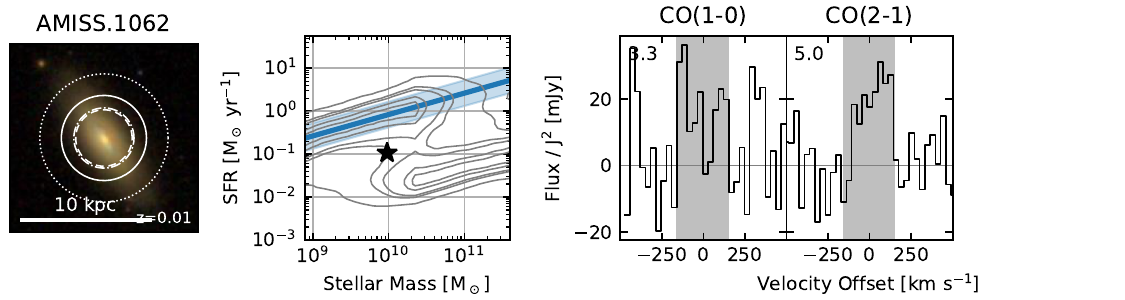}
\figsetgrpnote{Left column: SDSS cutouts of each target. Solid, dashed, and dotted lines show the 
beam sizes of the SMT for CO(2--1), the SMT for CO(3--2) (and the IRAM 30m for CO(1--0)) 
and the 12m for CO(1--0) respectively. The scale bar in the lower left shows 10 
kiloparsecs. 
Middle column: contours show the distribution of star formation rates at a given stellar 
mass, while the blue line and filled region show the main sequence of star forming 
galaxies. The stellar mass and star formation rate of the target galaxy is marked by a 
star. 
Right column: CO(1--0) spectra from AMISS (gray) and xCOLD~GASS (black), CO(2--1) spectra 
from AMISS, and CO(3--2) spectra from AMISS. Numbers in the upper right corner give the 
signal to noise ratio for each line. When a CO line is detected, the gray band indicates 
the region used to measure the line flux. The scale of the $y$-axis is such that lines 
would have the same amplitude in each transition for thermalized CO emission. The relative 
amplitudes of each spectrum give a sense of the luminosity ratios between the different 
lines.}
\figsetgrpend

\figsetgrpstart
\figsetgrpnum{15.64}
\figsetgrptitle{AMISS.1063}
\figsetplot{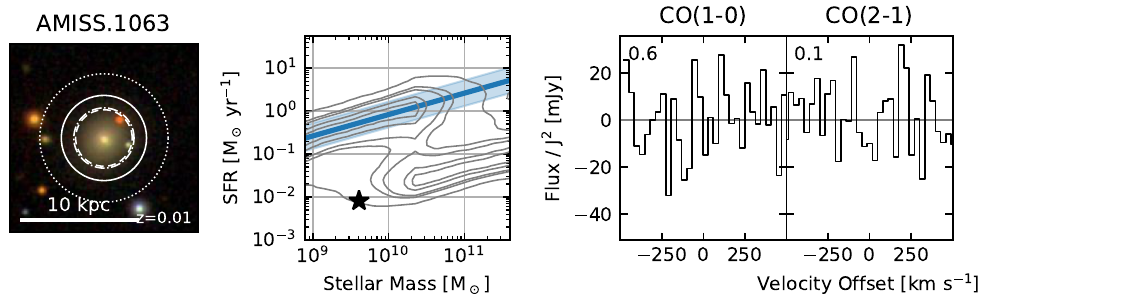}
\figsetgrpnote{Left column: SDSS cutouts of each target. Solid, dashed, and dotted lines show the 
beam sizes of the SMT for CO(2--1), the SMT for CO(3--2) (and the IRAM 30m for CO(1--0)) 
and the 12m for CO(1--0) respectively. The scale bar in the lower left shows 10 
kiloparsecs. 
Middle column: contours show the distribution of star formation rates at a given stellar 
mass, while the blue line and filled region show the main sequence of star forming 
galaxies. The stellar mass and star formation rate of the target galaxy is marked by a 
star. 
Right column: CO(1--0) spectra from AMISS (gray) and xCOLD~GASS (black), CO(2--1) spectra 
from AMISS, and CO(3--2) spectra from AMISS. Numbers in the upper right corner give the 
signal to noise ratio for each line. When a CO line is detected, the gray band indicates 
the region used to measure the line flux. The scale of the $y$-axis is such that lines 
would have the same amplitude in each transition for thermalized CO emission. The relative 
amplitudes of each spectrum give a sense of the luminosity ratios between the different 
lines.}
\figsetgrpend

\figsetgrpstart
\figsetgrpnum{15.65}
\figsetgrptitle{AMISS.1064}
\figsetplot{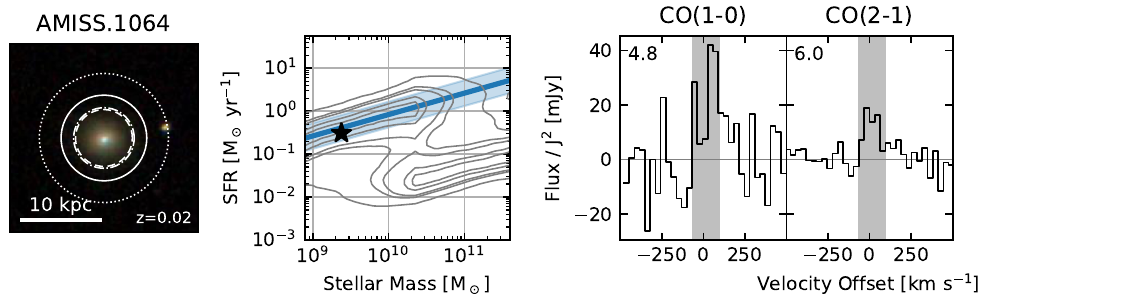}
\figsetgrpnote{Left column: SDSS cutouts of each target. Solid, dashed, and dotted lines show the 
beam sizes of the SMT for CO(2--1), the SMT for CO(3--2) (and the IRAM 30m for CO(1--0)) 
and the 12m for CO(1--0) respectively. The scale bar in the lower left shows 10 
kiloparsecs. 
Middle column: contours show the distribution of star formation rates at a given stellar 
mass, while the blue line and filled region show the main sequence of star forming 
galaxies. The stellar mass and star formation rate of the target galaxy is marked by a 
star. 
Right column: CO(1--0) spectra from AMISS (gray) and xCOLD~GASS (black), CO(2--1) spectra 
from AMISS, and CO(3--2) spectra from AMISS. Numbers in the upper right corner give the 
signal to noise ratio for each line. When a CO line is detected, the gray band indicates 
the region used to measure the line flux. The scale of the $y$-axis is such that lines 
would have the same amplitude in each transition for thermalized CO emission. The relative 
amplitudes of each spectrum give a sense of the luminosity ratios between the different 
lines.}
\figsetgrpend

\figsetgrpstart
\figsetgrpnum{15.66}
\figsetgrptitle{AMISS.1065}
\figsetplot{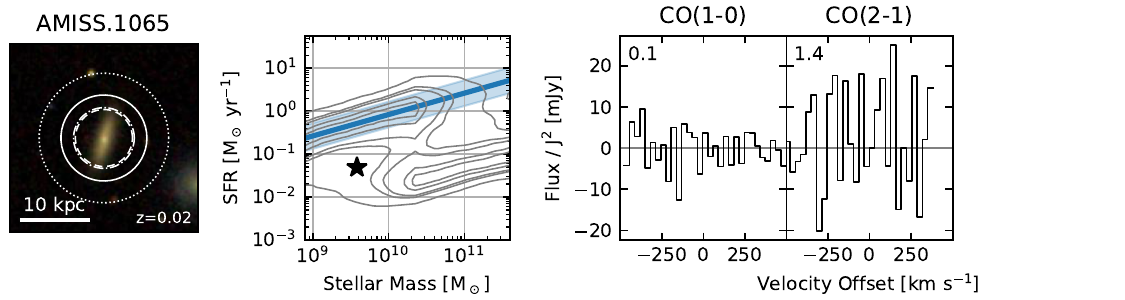}
\figsetgrpnote{Left column: SDSS cutouts of each target. Solid, dashed, and dotted lines show the 
beam sizes of the SMT for CO(2--1), the SMT for CO(3--2) (and the IRAM 30m for CO(1--0)) 
and the 12m for CO(1--0) respectively. The scale bar in the lower left shows 10 
kiloparsecs. 
Middle column: contours show the distribution of star formation rates at a given stellar 
mass, while the blue line and filled region show the main sequence of star forming 
galaxies. The stellar mass and star formation rate of the target galaxy is marked by a 
star. 
Right column: CO(1--0) spectra from AMISS (gray) and xCOLD~GASS (black), CO(2--1) spectra 
from AMISS, and CO(3--2) spectra from AMISS. Numbers in the upper right corner give the 
signal to noise ratio for each line. When a CO line is detected, the gray band indicates 
the region used to measure the line flux. The scale of the $y$-axis is such that lines 
would have the same amplitude in each transition for thermalized CO emission. The relative 
amplitudes of each spectrum give a sense of the luminosity ratios between the different 
lines.}
\figsetgrpend

\figsetgrpstart
\figsetgrpnum{15.67}
\figsetgrptitle{AMISS.1066}
\figsetplot{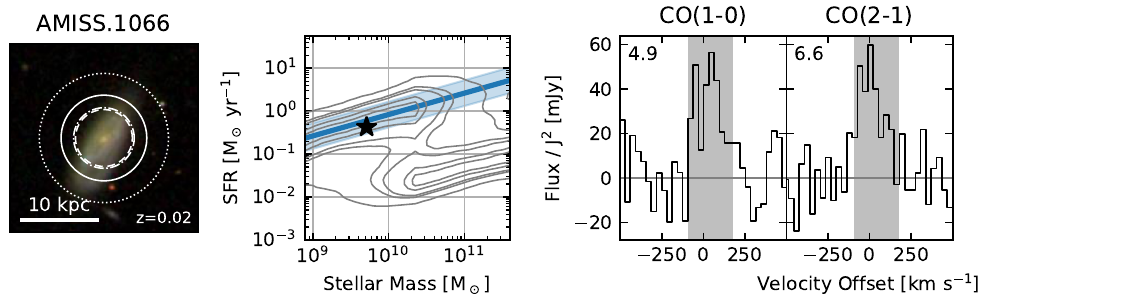}
\figsetgrpnote{Left column: SDSS cutouts of each target. Solid, dashed, and dotted lines show the 
beam sizes of the SMT for CO(2--1), the SMT for CO(3--2) (and the IRAM 30m for CO(1--0)) 
and the 12m for CO(1--0) respectively. The scale bar in the lower left shows 10 
kiloparsecs. 
Middle column: contours show the distribution of star formation rates at a given stellar 
mass, while the blue line and filled region show the main sequence of star forming 
galaxies. The stellar mass and star formation rate of the target galaxy is marked by a 
star. 
Right column: CO(1--0) spectra from AMISS (gray) and xCOLD~GASS (black), CO(2--1) spectra 
from AMISS, and CO(3--2) spectra from AMISS. Numbers in the upper right corner give the 
signal to noise ratio for each line. When a CO line is detected, the gray band indicates 
the region used to measure the line flux. The scale of the $y$-axis is such that lines 
would have the same amplitude in each transition for thermalized CO emission. The relative 
amplitudes of each spectrum give a sense of the luminosity ratios between the different 
lines.}
\figsetgrpend

\figsetgrpstart
\figsetgrpnum{15.68}
\figsetgrptitle{AMISS.1067}
\figsetplot{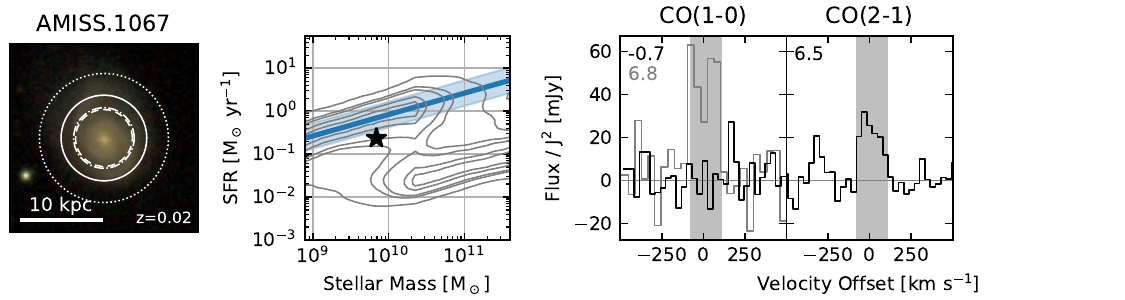}
\figsetgrpnote{Left column: SDSS cutouts of each target. Solid, dashed, and dotted lines show the 
beam sizes of the SMT for CO(2--1), the SMT for CO(3--2) (and the IRAM 30m for CO(1--0)) 
and the 12m for CO(1--0) respectively. The scale bar in the lower left shows 10 
kiloparsecs. 
Middle column: contours show the distribution of star formation rates at a given stellar 
mass, while the blue line and filled region show the main sequence of star forming 
galaxies. The stellar mass and star formation rate of the target galaxy is marked by a 
star. 
Right column: CO(1--0) spectra from AMISS (gray) and xCOLD~GASS (black), CO(2--1) spectra 
from AMISS, and CO(3--2) spectra from AMISS. Numbers in the upper right corner give the 
signal to noise ratio for each line. When a CO line is detected, the gray band indicates 
the region used to measure the line flux. The scale of the $y$-axis is such that lines 
would have the same amplitude in each transition for thermalized CO emission. The relative 
amplitudes of each spectrum give a sense of the luminosity ratios between the different 
lines.}
\figsetgrpend

\figsetgrpstart
\figsetgrpnum{15.69}
\figsetgrptitle{AMISS.1068}
\figsetplot{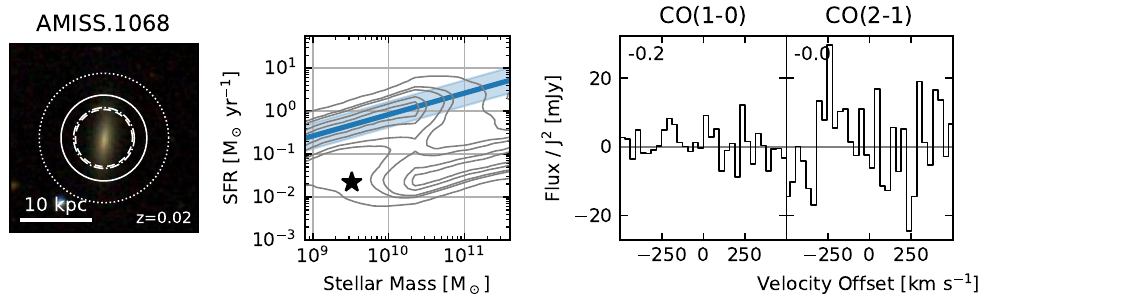}
\figsetgrpnote{Left column: SDSS cutouts of each target. Solid, dashed, and dotted lines show the 
beam sizes of the SMT for CO(2--1), the SMT for CO(3--2) (and the IRAM 30m for CO(1--0)) 
and the 12m for CO(1--0) respectively. The scale bar in the lower left shows 10 
kiloparsecs. 
Middle column: contours show the distribution of star formation rates at a given stellar 
mass, while the blue line and filled region show the main sequence of star forming 
galaxies. The stellar mass and star formation rate of the target galaxy is marked by a 
star. 
Right column: CO(1--0) spectra from AMISS (gray) and xCOLD~GASS (black), CO(2--1) spectra 
from AMISS, and CO(3--2) spectra from AMISS. Numbers in the upper right corner give the 
signal to noise ratio for each line. When a CO line is detected, the gray band indicates 
the region used to measure the line flux. The scale of the $y$-axis is such that lines 
would have the same amplitude in each transition for thermalized CO emission. The relative 
amplitudes of each spectrum give a sense of the luminosity ratios between the different 
lines.}
\figsetgrpend

\figsetgrpstart
\figsetgrpnum{15.70}
\figsetgrptitle{AMISS.1069}
\figsetplot{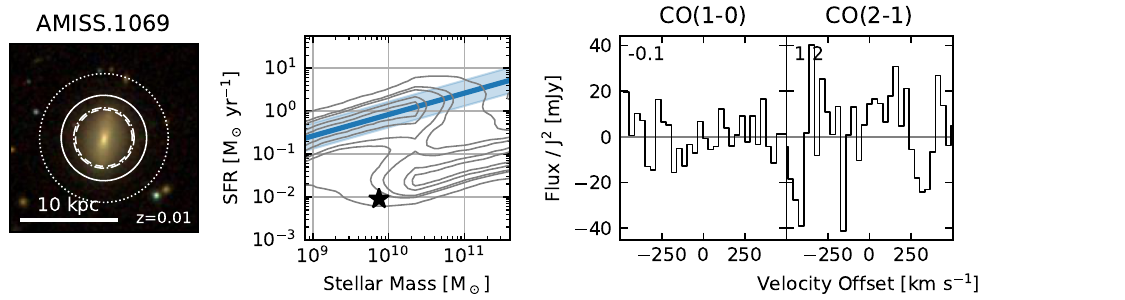}
\figsetgrpnote{Left column: SDSS cutouts of each target. Solid, dashed, and dotted lines show the 
beam sizes of the SMT for CO(2--1), the SMT for CO(3--2) (and the IRAM 30m for CO(1--0)) 
and the 12m for CO(1--0) respectively. The scale bar in the lower left shows 10 
kiloparsecs. 
Middle column: contours show the distribution of star formation rates at a given stellar 
mass, while the blue line and filled region show the main sequence of star forming 
galaxies. The stellar mass and star formation rate of the target galaxy is marked by a 
star. 
Right column: CO(1--0) spectra from AMISS (gray) and xCOLD~GASS (black), CO(2--1) spectra 
from AMISS, and CO(3--2) spectra from AMISS. Numbers in the upper right corner give the 
signal to noise ratio for each line. When a CO line is detected, the gray band indicates 
the region used to measure the line flux. The scale of the $y$-axis is such that lines 
would have the same amplitude in each transition for thermalized CO emission. The relative 
amplitudes of each spectrum give a sense of the luminosity ratios between the different 
lines.}
\figsetgrpend

\figsetgrpstart
\figsetgrpnum{15.71}
\figsetgrptitle{AMISS.1070}
\figsetplot{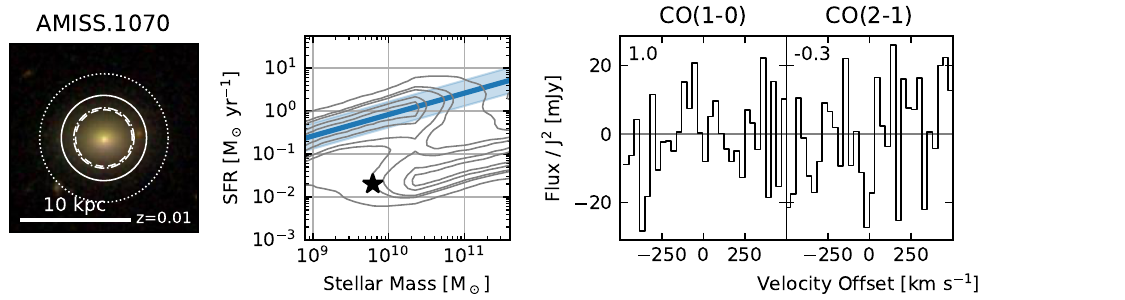}
\figsetgrpnote{Left column: SDSS cutouts of each target. Solid, dashed, and dotted lines show the 
beam sizes of the SMT for CO(2--1), the SMT for CO(3--2) (and the IRAM 30m for CO(1--0)) 
and the 12m for CO(1--0) respectively. The scale bar in the lower left shows 10 
kiloparsecs. 
Middle column: contours show the distribution of star formation rates at a given stellar 
mass, while the blue line and filled region show the main sequence of star forming 
galaxies. The stellar mass and star formation rate of the target galaxy is marked by a 
star. 
Right column: CO(1--0) spectra from AMISS (gray) and xCOLD~GASS (black), CO(2--1) spectra 
from AMISS, and CO(3--2) spectra from AMISS. Numbers in the upper right corner give the 
signal to noise ratio for each line. When a CO line is detected, the gray band indicates 
the region used to measure the line flux. The scale of the $y$-axis is such that lines 
would have the same amplitude in each transition for thermalized CO emission. The relative 
amplitudes of each spectrum give a sense of the luminosity ratios between the different 
lines.}
\figsetgrpend

\figsetgrpstart
\figsetgrpnum{15.72}
\figsetgrptitle{AMISS.1071}
\figsetplot{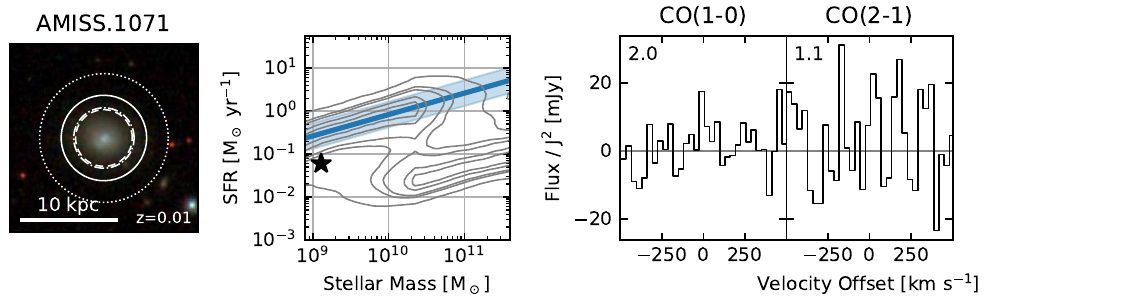}
\figsetgrpnote{Left column: SDSS cutouts of each target. Solid, dashed, and dotted lines show the 
beam sizes of the SMT for CO(2--1), the SMT for CO(3--2) (and the IRAM 30m for CO(1--0)) 
and the 12m for CO(1--0) respectively. The scale bar in the lower left shows 10 
kiloparsecs. 
Middle column: contours show the distribution of star formation rates at a given stellar 
mass, while the blue line and filled region show the main sequence of star forming 
galaxies. The stellar mass and star formation rate of the target galaxy is marked by a 
star. 
Right column: CO(1--0) spectra from AMISS (gray) and xCOLD~GASS (black), CO(2--1) spectra 
from AMISS, and CO(3--2) spectra from AMISS. Numbers in the upper right corner give the 
signal to noise ratio for each line. When a CO line is detected, the gray band indicates 
the region used to measure the line flux. The scale of the $y$-axis is such that lines 
would have the same amplitude in each transition for thermalized CO emission. The relative 
amplitudes of each spectrum give a sense of the luminosity ratios between the different 
lines.}
\figsetgrpend

\figsetgrpstart
\figsetgrpnum{15.73}
\figsetgrptitle{AMISS.1072}
\figsetplot{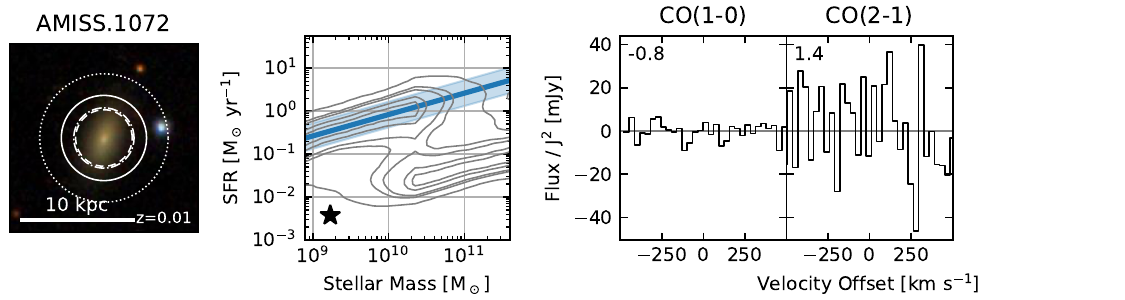}
\figsetgrpnote{Left column: SDSS cutouts of each target. Solid, dashed, and dotted lines show the 
beam sizes of the SMT for CO(2--1), the SMT for CO(3--2) (and the IRAM 30m for CO(1--0)) 
and the 12m for CO(1--0) respectively. The scale bar in the lower left shows 10 
kiloparsecs. 
Middle column: contours show the distribution of star formation rates at a given stellar 
mass, while the blue line and filled region show the main sequence of star forming 
galaxies. The stellar mass and star formation rate of the target galaxy is marked by a 
star. 
Right column: CO(1--0) spectra from AMISS (gray) and xCOLD~GASS (black), CO(2--1) spectra 
from AMISS, and CO(3--2) spectra from AMISS. Numbers in the upper right corner give the 
signal to noise ratio for each line. When a CO line is detected, the gray band indicates 
the region used to measure the line flux. The scale of the $y$-axis is such that lines 
would have the same amplitude in each transition for thermalized CO emission. The relative 
amplitudes of each spectrum give a sense of the luminosity ratios between the different 
lines.}
\figsetgrpend

\figsetgrpstart
\figsetgrpnum{15.74}
\figsetgrptitle{AMISS.1073}
\figsetplot{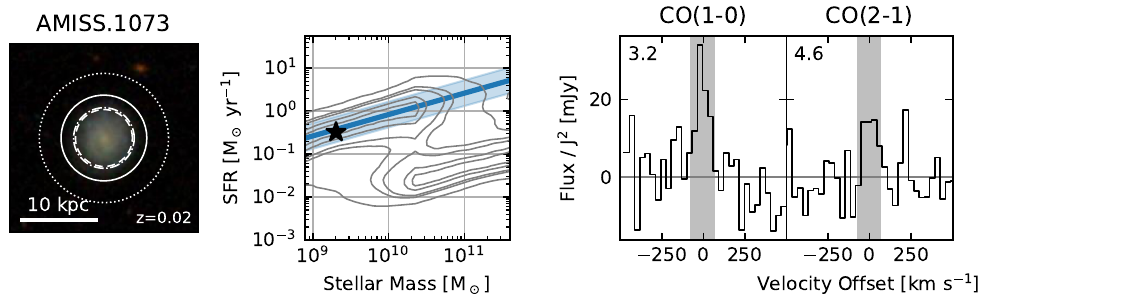}
\figsetgrpnote{Left column: SDSS cutouts of each target. Solid, dashed, and dotted lines show the 
beam sizes of the SMT for CO(2--1), the SMT for CO(3--2) (and the IRAM 30m for CO(1--0)) 
and the 12m for CO(1--0) respectively. The scale bar in the lower left shows 10 
kiloparsecs. 
Middle column: contours show the distribution of star formation rates at a given stellar 
mass, while the blue line and filled region show the main sequence of star forming 
galaxies. The stellar mass and star formation rate of the target galaxy is marked by a 
star. 
Right column: CO(1--0) spectra from AMISS (gray) and xCOLD~GASS (black), CO(2--1) spectra 
from AMISS, and CO(3--2) spectra from AMISS. Numbers in the upper right corner give the 
signal to noise ratio for each line. When a CO line is detected, the gray band indicates 
the region used to measure the line flux. The scale of the $y$-axis is such that lines 
would have the same amplitude in each transition for thermalized CO emission. The relative 
amplitudes of each spectrum give a sense of the luminosity ratios between the different 
lines.}
\figsetgrpend

\figsetgrpstart
\figsetgrpnum{15.75}
\figsetgrptitle{AMISS.1074}
\figsetplot{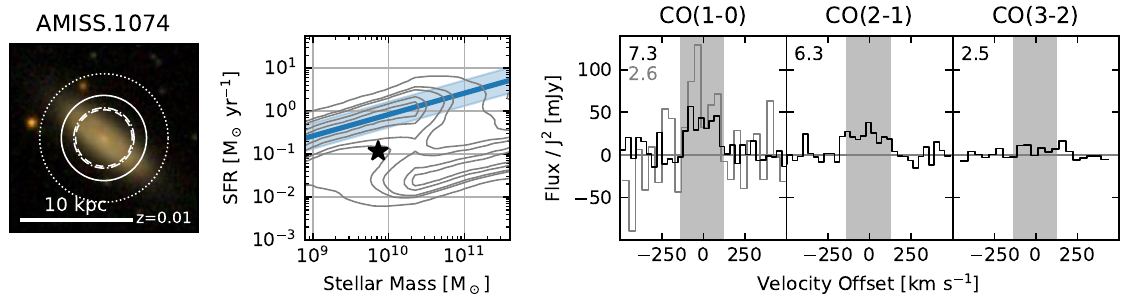}
\figsetgrpnote{Left column: SDSS cutouts of each target. Solid, dashed, and dotted lines show the 
beam sizes of the SMT for CO(2--1), the SMT for CO(3--2) (and the IRAM 30m for CO(1--0)) 
and the 12m for CO(1--0) respectively. The scale bar in the lower left shows 10 
kiloparsecs. 
Middle column: contours show the distribution of star formation rates at a given stellar 
mass, while the blue line and filled region show the main sequence of star forming 
galaxies. The stellar mass and star formation rate of the target galaxy is marked by a 
star. 
Right column: CO(1--0) spectra from AMISS (gray) and xCOLD~GASS (black), CO(2--1) spectra 
from AMISS, and CO(3--2) spectra from AMISS. Numbers in the upper right corner give the 
signal to noise ratio for each line. When a CO line is detected, the gray band indicates 
the region used to measure the line flux. The scale of the $y$-axis is such that lines 
would have the same amplitude in each transition for thermalized CO emission. The relative 
amplitudes of each spectrum give a sense of the luminosity ratios between the different 
lines.}
\figsetgrpend

\figsetgrpstart
\figsetgrpnum{15.76}
\figsetgrptitle{AMISS.1075}
\figsetplot{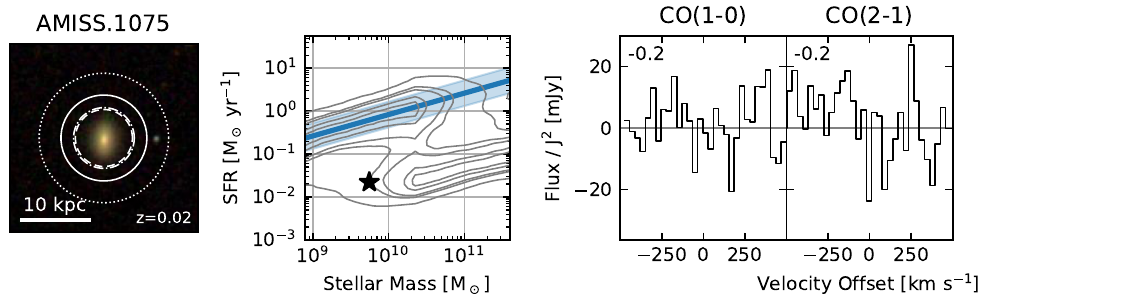}
\figsetgrpnote{Left column: SDSS cutouts of each target. Solid, dashed, and dotted lines show the 
beam sizes of the SMT for CO(2--1), the SMT for CO(3--2) (and the IRAM 30m for CO(1--0)) 
and the 12m for CO(1--0) respectively. The scale bar in the lower left shows 10 
kiloparsecs. 
Middle column: contours show the distribution of star formation rates at a given stellar 
mass, while the blue line and filled region show the main sequence of star forming 
galaxies. The stellar mass and star formation rate of the target galaxy is marked by a 
star. 
Right column: CO(1--0) spectra from AMISS (gray) and xCOLD~GASS (black), CO(2--1) spectra 
from AMISS, and CO(3--2) spectra from AMISS. Numbers in the upper right corner give the 
signal to noise ratio for each line. When a CO line is detected, the gray band indicates 
the region used to measure the line flux. The scale of the $y$-axis is such that lines 
would have the same amplitude in each transition for thermalized CO emission. The relative 
amplitudes of each spectrum give a sense of the luminosity ratios between the different 
lines.}
\figsetgrpend

\figsetgrpstart
\figsetgrpnum{15.77}
\figsetgrptitle{AMISS.1076}
\figsetplot{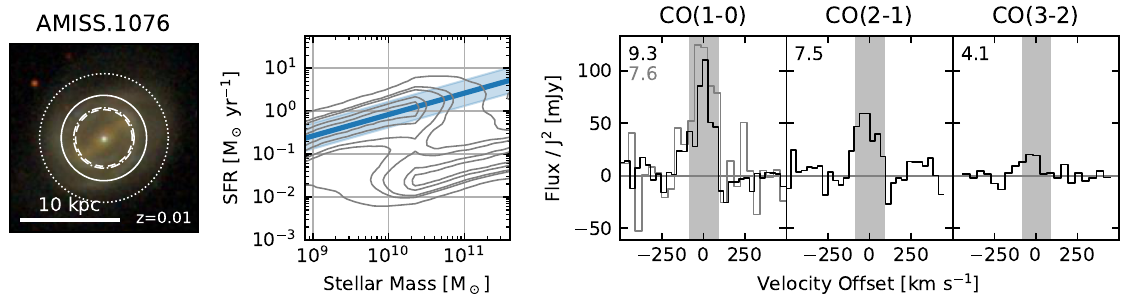}
\figsetgrpnote{Left column: SDSS cutouts of each target. Solid, dashed, and dotted lines show the 
beam sizes of the SMT for CO(2--1), the SMT for CO(3--2) (and the IRAM 30m for CO(1--0)) 
and the 12m for CO(1--0) respectively. The scale bar in the lower left shows 10 
kiloparsecs. 
Middle column: contours show the distribution of star formation rates at a given stellar 
mass, while the blue line and filled region show the main sequence of star forming 
galaxies. The stellar mass and star formation rate of the target galaxy is marked by a 
star. 
Right column: CO(1--0) spectra from AMISS (gray) and xCOLD~GASS (black), CO(2--1) spectra 
from AMISS, and CO(3--2) spectra from AMISS. Numbers in the upper right corner give the 
signal to noise ratio for each line. When a CO line is detected, the gray band indicates 
the region used to measure the line flux. The scale of the $y$-axis is such that lines 
would have the same amplitude in each transition for thermalized CO emission. The relative 
amplitudes of each spectrum give a sense of the luminosity ratios between the different 
lines.}
\figsetgrpend

\figsetgrpstart
\figsetgrpnum{15.78}
\figsetgrptitle{AMISS.1077}
\figsetplot{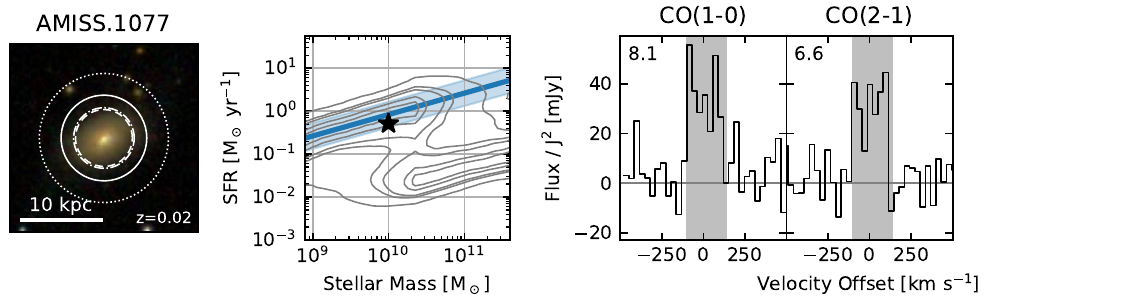}
\figsetgrpnote{Left column: SDSS cutouts of each target. Solid, dashed, and dotted lines show the 
beam sizes of the SMT for CO(2--1), the SMT for CO(3--2) (and the IRAM 30m for CO(1--0)) 
and the 12m for CO(1--0) respectively. The scale bar in the lower left shows 10 
kiloparsecs. 
Middle column: contours show the distribution of star formation rates at a given stellar 
mass, while the blue line and filled region show the main sequence of star forming 
galaxies. The stellar mass and star formation rate of the target galaxy is marked by a 
star. 
Right column: CO(1--0) spectra from AMISS (gray) and xCOLD~GASS (black), CO(2--1) spectra 
from AMISS, and CO(3--2) spectra from AMISS. Numbers in the upper right corner give the 
signal to noise ratio for each line. When a CO line is detected, the gray band indicates 
the region used to measure the line flux. The scale of the $y$-axis is such that lines 
would have the same amplitude in each transition for thermalized CO emission. The relative 
amplitudes of each spectrum give a sense of the luminosity ratios between the different 
lines.}
\figsetgrpend

\figsetgrpstart
\figsetgrpnum{15.79}
\figsetgrptitle{AMISS.1078}
\figsetplot{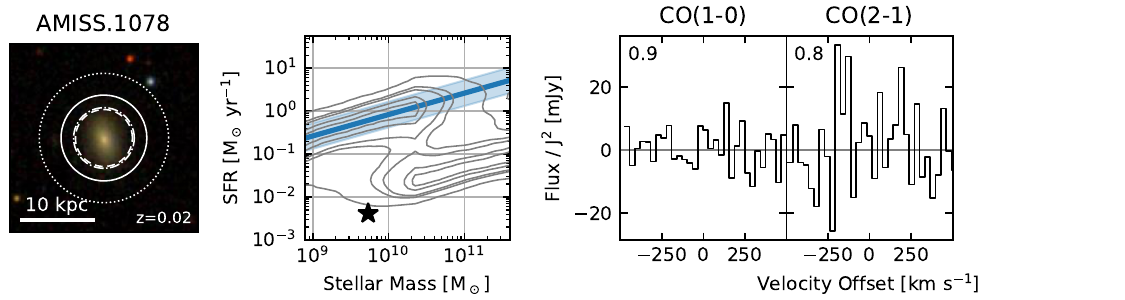}
\figsetgrpnote{Left column: SDSS cutouts of each target. Solid, dashed, and dotted lines show the 
beam sizes of the SMT for CO(2--1), the SMT for CO(3--2) (and the IRAM 30m for CO(1--0)) 
and the 12m for CO(1--0) respectively. The scale bar in the lower left shows 10 
kiloparsecs. 
Middle column: contours show the distribution of star formation rates at a given stellar 
mass, while the blue line and filled region show the main sequence of star forming 
galaxies. The stellar mass and star formation rate of the target galaxy is marked by a 
star. 
Right column: CO(1--0) spectra from AMISS (gray) and xCOLD~GASS (black), CO(2--1) spectra 
from AMISS, and CO(3--2) spectra from AMISS. Numbers in the upper right corner give the 
signal to noise ratio for each line. When a CO line is detected, the gray band indicates 
the region used to measure the line flux. The scale of the $y$-axis is such that lines 
would have the same amplitude in each transition for thermalized CO emission. The relative 
amplitudes of each spectrum give a sense of the luminosity ratios between the different 
lines.}
\figsetgrpend

\figsetgrpstart
\figsetgrpnum{15.80}
\figsetgrptitle{AMISS.1079}
\figsetplot{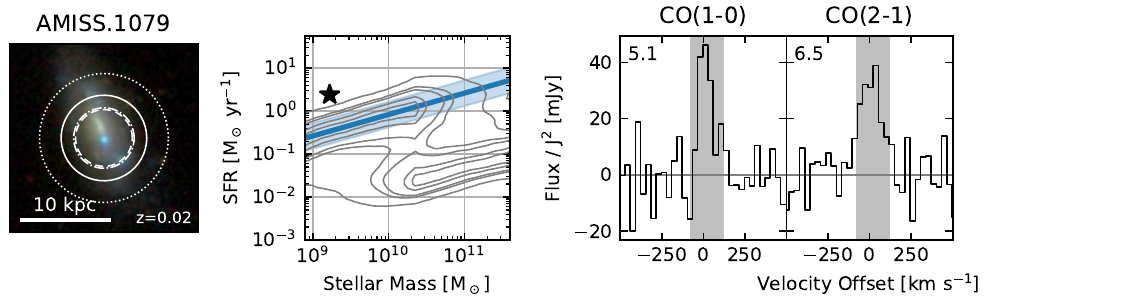}
\figsetgrpnote{Left column: SDSS cutouts of each target. Solid, dashed, and dotted lines show the 
beam sizes of the SMT for CO(2--1), the SMT for CO(3--2) (and the IRAM 30m for CO(1--0)) 
and the 12m for CO(1--0) respectively. The scale bar in the lower left shows 10 
kiloparsecs. 
Middle column: contours show the distribution of star formation rates at a given stellar 
mass, while the blue line and filled region show the main sequence of star forming 
galaxies. The stellar mass and star formation rate of the target galaxy is marked by a 
star. 
Right column: CO(1--0) spectra from AMISS (gray) and xCOLD~GASS (black), CO(2--1) spectra 
from AMISS, and CO(3--2) spectra from AMISS. Numbers in the upper right corner give the 
signal to noise ratio for each line. When a CO line is detected, the gray band indicates 
the region used to measure the line flux. The scale of the $y$-axis is such that lines 
would have the same amplitude in each transition for thermalized CO emission. The relative 
amplitudes of each spectrum give a sense of the luminosity ratios between the different 
lines.}
\figsetgrpend

\figsetgrpstart
\figsetgrpnum{15.81}
\figsetgrptitle{AMISS.1080}
\figsetplot{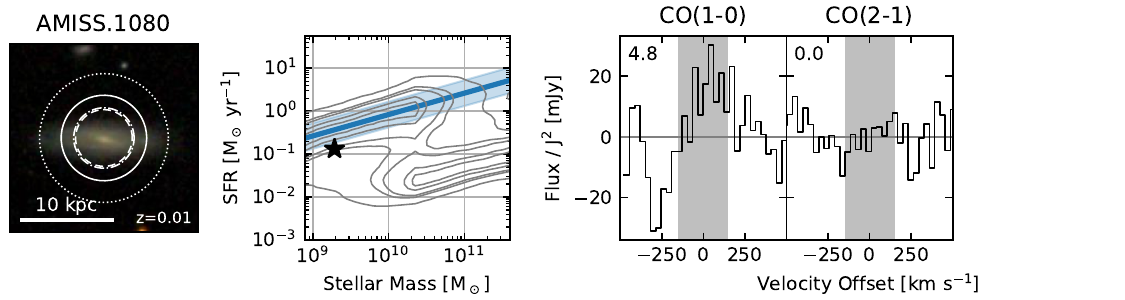}
\figsetgrpnote{Left column: SDSS cutouts of each target. Solid, dashed, and dotted lines show the 
beam sizes of the SMT for CO(2--1), the SMT for CO(3--2) (and the IRAM 30m for CO(1--0)) 
and the 12m for CO(1--0) respectively. The scale bar in the lower left shows 10 
kiloparsecs. 
Middle column: contours show the distribution of star formation rates at a given stellar 
mass, while the blue line and filled region show the main sequence of star forming 
galaxies. The stellar mass and star formation rate of the target galaxy is marked by a 
star. 
Right column: CO(1--0) spectra from AMISS (gray) and xCOLD~GASS (black), CO(2--1) spectra 
from AMISS, and CO(3--2) spectra from AMISS. Numbers in the upper right corner give the 
signal to noise ratio for each line. When a CO line is detected, the gray band indicates 
the region used to measure the line flux. The scale of the $y$-axis is such that lines 
would have the same amplitude in each transition for thermalized CO emission. The relative 
amplitudes of each spectrum give a sense of the luminosity ratios between the different 
lines.}
\figsetgrpend

\figsetgrpstart
\figsetgrpnum{15.82}
\figsetgrptitle{AMISS.1081}
\figsetplot{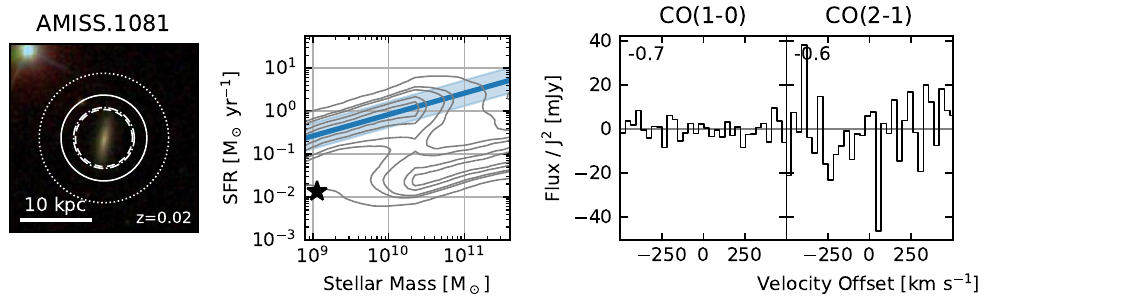}
\figsetgrpnote{Left column: SDSS cutouts of each target. Solid, dashed, and dotted lines show the 
beam sizes of the SMT for CO(2--1), the SMT for CO(3--2) (and the IRAM 30m for CO(1--0)) 
and the 12m for CO(1--0) respectively. The scale bar in the lower left shows 10 
kiloparsecs. 
Middle column: contours show the distribution of star formation rates at a given stellar 
mass, while the blue line and filled region show the main sequence of star forming 
galaxies. The stellar mass and star formation rate of the target galaxy is marked by a 
star. 
Right column: CO(1--0) spectra from AMISS (gray) and xCOLD~GASS (black), CO(2--1) spectra 
from AMISS, and CO(3--2) spectra from AMISS. Numbers in the upper right corner give the 
signal to noise ratio for each line. When a CO line is detected, the gray band indicates 
the region used to measure the line flux. The scale of the $y$-axis is such that lines 
would have the same amplitude in each transition for thermalized CO emission. The relative 
amplitudes of each spectrum give a sense of the luminosity ratios between the different 
lines.}
\figsetgrpend

\figsetgrpstart
\figsetgrpnum{15.83}
\figsetgrptitle{AMISS.1082}
\figsetplot{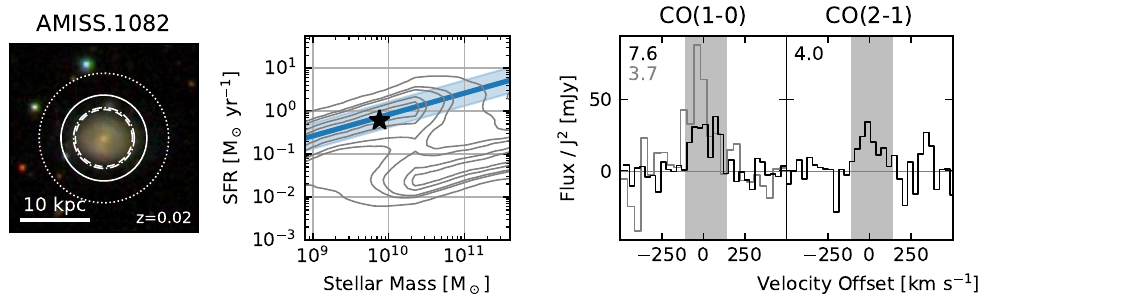}
\figsetgrpnote{Left column: SDSS cutouts of each target. Solid, dashed, and dotted lines show the 
beam sizes of the SMT for CO(2--1), the SMT for CO(3--2) (and the IRAM 30m for CO(1--0)) 
and the 12m for CO(1--0) respectively. The scale bar in the lower left shows 10 
kiloparsecs. 
Middle column: contours show the distribution of star formation rates at a given stellar 
mass, while the blue line and filled region show the main sequence of star forming 
galaxies. The stellar mass and star formation rate of the target galaxy is marked by a 
star. 
Right column: CO(1--0) spectra from AMISS (gray) and xCOLD~GASS (black), CO(2--1) spectra 
from AMISS, and CO(3--2) spectra from AMISS. Numbers in the upper right corner give the 
signal to noise ratio for each line. When a CO line is detected, the gray band indicates 
the region used to measure the line flux. The scale of the $y$-axis is such that lines 
would have the same amplitude in each transition for thermalized CO emission. The relative 
amplitudes of each spectrum give a sense of the luminosity ratios between the different 
lines.}
\figsetgrpend

\figsetgrpstart
\figsetgrpnum{15.84}
\figsetgrptitle{AMISS.1083}
\figsetplot{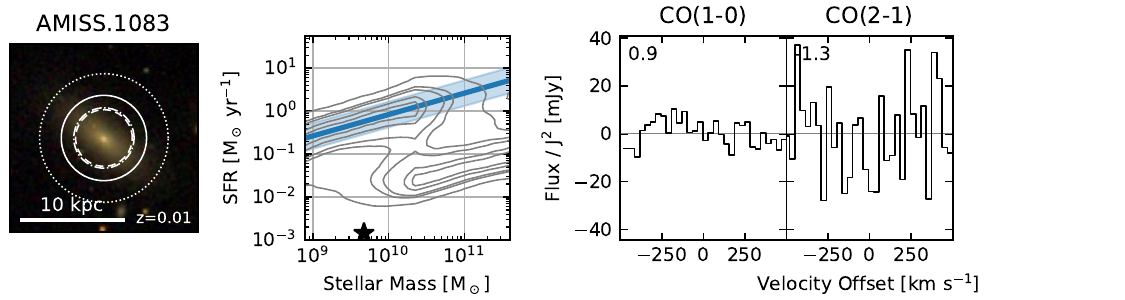}
\figsetgrpnote{Left column: SDSS cutouts of each target. Solid, dashed, and dotted lines show the 
beam sizes of the SMT for CO(2--1), the SMT for CO(3--2) (and the IRAM 30m for CO(1--0)) 
and the 12m for CO(1--0) respectively. The scale bar in the lower left shows 10 
kiloparsecs. 
Middle column: contours show the distribution of star formation rates at a given stellar 
mass, while the blue line and filled region show the main sequence of star forming 
galaxies. The stellar mass and star formation rate of the target galaxy is marked by a 
star. 
Right column: CO(1--0) spectra from AMISS (gray) and xCOLD~GASS (black), CO(2--1) spectra 
from AMISS, and CO(3--2) spectra from AMISS. Numbers in the upper right corner give the 
signal to noise ratio for each line. When a CO line is detected, the gray band indicates 
the region used to measure the line flux. The scale of the $y$-axis is such that lines 
would have the same amplitude in each transition for thermalized CO emission. The relative 
amplitudes of each spectrum give a sense of the luminosity ratios between the different 
lines.}
\figsetgrpend

\figsetgrpstart
\figsetgrpnum{15.85}
\figsetgrptitle{AMISS.1084}
\figsetplot{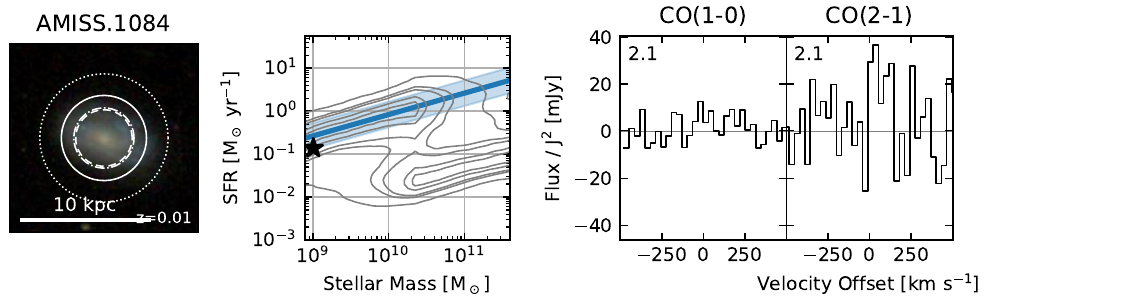}
\figsetgrpnote{Left column: SDSS cutouts of each target. Solid, dashed, and dotted lines show the 
beam sizes of the SMT for CO(2--1), the SMT for CO(3--2) (and the IRAM 30m for CO(1--0)) 
and the 12m for CO(1--0) respectively. The scale bar in the lower left shows 10 
kiloparsecs. 
Middle column: contours show the distribution of star formation rates at a given stellar 
mass, while the blue line and filled region show the main sequence of star forming 
galaxies. The stellar mass and star formation rate of the target galaxy is marked by a 
star. 
Right column: CO(1--0) spectra from AMISS (gray) and xCOLD~GASS (black), CO(2--1) spectra 
from AMISS, and CO(3--2) spectra from AMISS. Numbers in the upper right corner give the 
signal to noise ratio for each line. When a CO line is detected, the gray band indicates 
the region used to measure the line flux. The scale of the $y$-axis is such that lines 
would have the same amplitude in each transition for thermalized CO emission. The relative 
amplitudes of each spectrum give a sense of the luminosity ratios between the different 
lines.}
\figsetgrpend

\figsetgrpstart
\figsetgrpnum{15.86}
\figsetgrptitle{AMISS.1085}
\figsetplot{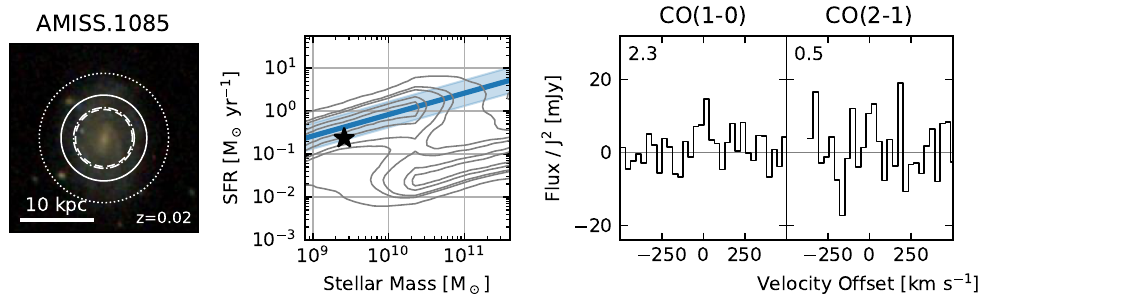}
\figsetgrpnote{Left column: SDSS cutouts of each target. Solid, dashed, and dotted lines show the 
beam sizes of the SMT for CO(2--1), the SMT for CO(3--2) (and the IRAM 30m for CO(1--0)) 
and the 12m for CO(1--0) respectively. The scale bar in the lower left shows 10 
kiloparsecs. 
Middle column: contours show the distribution of star formation rates at a given stellar 
mass, while the blue line and filled region show the main sequence of star forming 
galaxies. The stellar mass and star formation rate of the target galaxy is marked by a 
star. 
Right column: CO(1--0) spectra from AMISS (gray) and xCOLD~GASS (black), CO(2--1) spectra 
from AMISS, and CO(3--2) spectra from AMISS. Numbers in the upper right corner give the 
signal to noise ratio for each line. When a CO line is detected, the gray band indicates 
the region used to measure the line flux. The scale of the $y$-axis is such that lines 
would have the same amplitude in each transition for thermalized CO emission. The relative 
amplitudes of each spectrum give a sense of the luminosity ratios between the different 
lines.}
\figsetgrpend

\figsetgrpstart
\figsetgrpnum{15.87}
\figsetgrptitle{AMISS.1086}
\figsetplot{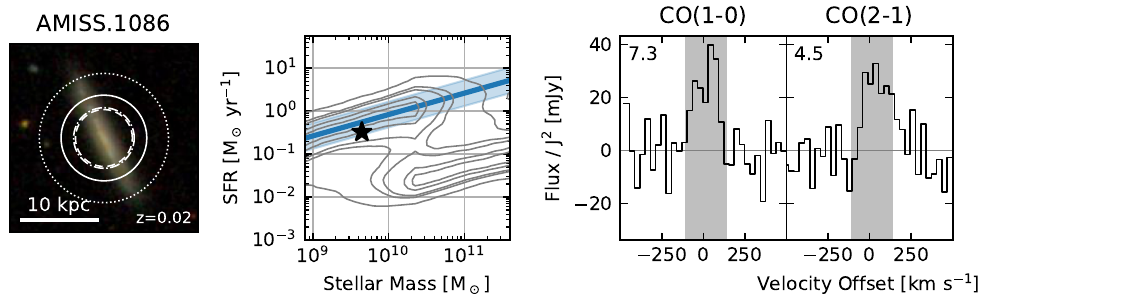}
\figsetgrpnote{Left column: SDSS cutouts of each target. Solid, dashed, and dotted lines show the 
beam sizes of the SMT for CO(2--1), the SMT for CO(3--2) (and the IRAM 30m for CO(1--0)) 
and the 12m for CO(1--0) respectively. The scale bar in the lower left shows 10 
kiloparsecs. 
Middle column: contours show the distribution of star formation rates at a given stellar 
mass, while the blue line and filled region show the main sequence of star forming 
galaxies. The stellar mass and star formation rate of the target galaxy is marked by a 
star. 
Right column: CO(1--0) spectra from AMISS (gray) and xCOLD~GASS (black), CO(2--1) spectra 
from AMISS, and CO(3--2) spectra from AMISS. Numbers in the upper right corner give the 
signal to noise ratio for each line. When a CO line is detected, the gray band indicates 
the region used to measure the line flux. The scale of the $y$-axis is such that lines 
would have the same amplitude in each transition for thermalized CO emission. The relative 
amplitudes of each spectrum give a sense of the luminosity ratios between the different 
lines.}
\figsetgrpend

\figsetgrpstart
\figsetgrpnum{15.88}
\figsetgrptitle{AMISS.1087}
\figsetplot{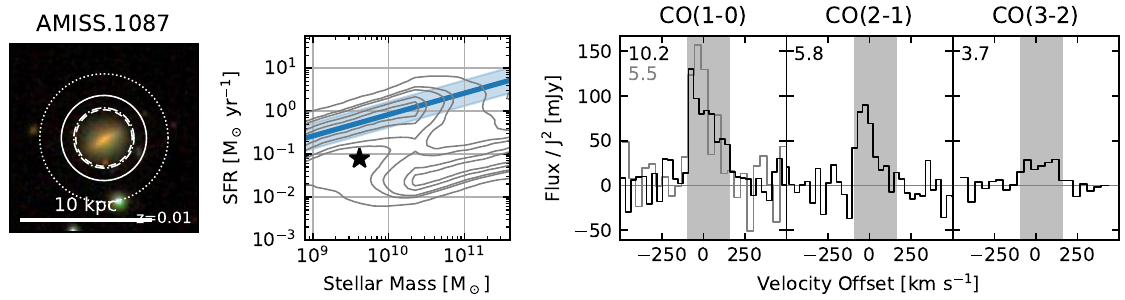}
\figsetgrpnote{Left column: SDSS cutouts of each target. Solid, dashed, and dotted lines show the 
beam sizes of the SMT for CO(2--1), the SMT for CO(3--2) (and the IRAM 30m for CO(1--0)) 
and the 12m for CO(1--0) respectively. The scale bar in the lower left shows 10 
kiloparsecs. 
Middle column: contours show the distribution of star formation rates at a given stellar 
mass, while the blue line and filled region show the main sequence of star forming 
galaxies. The stellar mass and star formation rate of the target galaxy is marked by a 
star. 
Right column: CO(1--0) spectra from AMISS (gray) and xCOLD~GASS (black), CO(2--1) spectra 
from AMISS, and CO(3--2) spectra from AMISS. Numbers in the upper right corner give the 
signal to noise ratio for each line. When a CO line is detected, the gray band indicates 
the region used to measure the line flux. The scale of the $y$-axis is such that lines 
would have the same amplitude in each transition for thermalized CO emission. The relative 
amplitudes of each spectrum give a sense of the luminosity ratios between the different 
lines.}
\figsetgrpend

\figsetgrpstart
\figsetgrpnum{15.89}
\figsetgrptitle{AMISS.1088}
\figsetplot{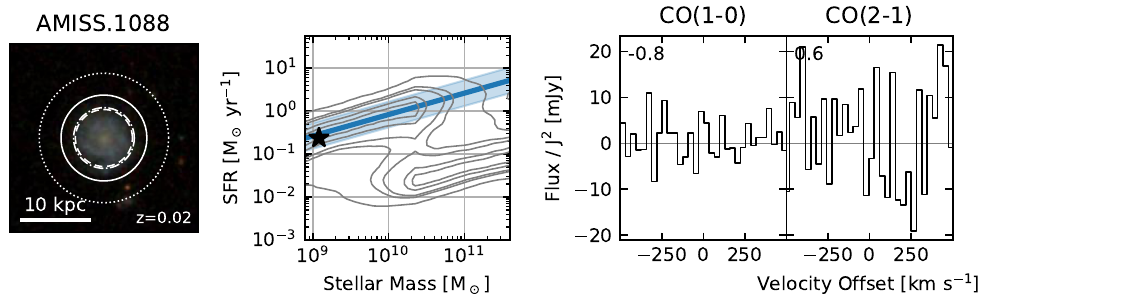}
\figsetgrpnote{Left column: SDSS cutouts of each target. Solid, dashed, and dotted lines show the 
beam sizes of the SMT for CO(2--1), the SMT for CO(3--2) (and the IRAM 30m for CO(1--0)) 
and the 12m for CO(1--0) respectively. The scale bar in the lower left shows 10 
kiloparsecs. 
Middle column: contours show the distribution of star formation rates at a given stellar 
mass, while the blue line and filled region show the main sequence of star forming 
galaxies. The stellar mass and star formation rate of the target galaxy is marked by a 
star. 
Right column: CO(1--0) spectra from AMISS (gray) and xCOLD~GASS (black), CO(2--1) spectra 
from AMISS, and CO(3--2) spectra from AMISS. Numbers in the upper right corner give the 
signal to noise ratio for each line. When a CO line is detected, the gray band indicates 
the region used to measure the line flux. The scale of the $y$-axis is such that lines 
would have the same amplitude in each transition for thermalized CO emission. The relative 
amplitudes of each spectrum give a sense of the luminosity ratios between the different 
lines.}
\figsetgrpend

\figsetgrpstart
\figsetgrpnum{15.90}
\figsetgrptitle{AMISS.1089}
\figsetplot{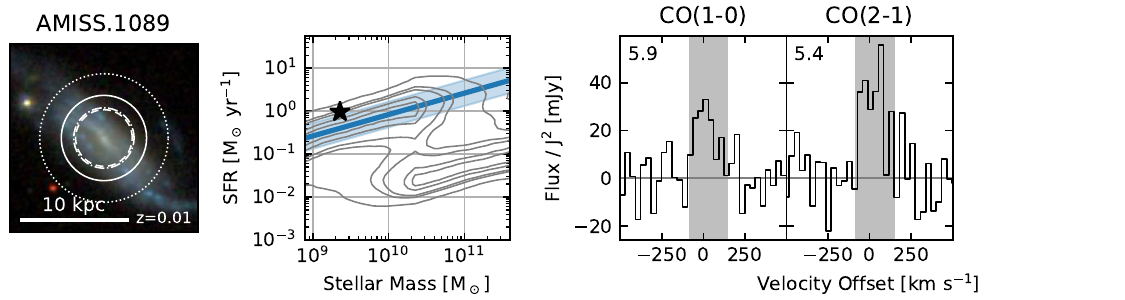}
\figsetgrpnote{Left column: SDSS cutouts of each target. Solid, dashed, and dotted lines show the 
beam sizes of the SMT for CO(2--1), the SMT for CO(3--2) (and the IRAM 30m for CO(1--0)) 
and the 12m for CO(1--0) respectively. The scale bar in the lower left shows 10 
kiloparsecs. 
Middle column: contours show the distribution of star formation rates at a given stellar 
mass, while the blue line and filled region show the main sequence of star forming 
galaxies. The stellar mass and star formation rate of the target galaxy is marked by a 
star. 
Right column: CO(1--0) spectra from AMISS (gray) and xCOLD~GASS (black), CO(2--1) spectra 
from AMISS, and CO(3--2) spectra from AMISS. Numbers in the upper right corner give the 
signal to noise ratio for each line. When a CO line is detected, the gray band indicates 
the region used to measure the line flux. The scale of the $y$-axis is such that lines 
would have the same amplitude in each transition for thermalized CO emission. The relative 
amplitudes of each spectrum give a sense of the luminosity ratios between the different 
lines.}
\figsetgrpend

\figsetgrpstart
\figsetgrpnum{15.91}
\figsetgrptitle{AMISS.1090}
\figsetplot{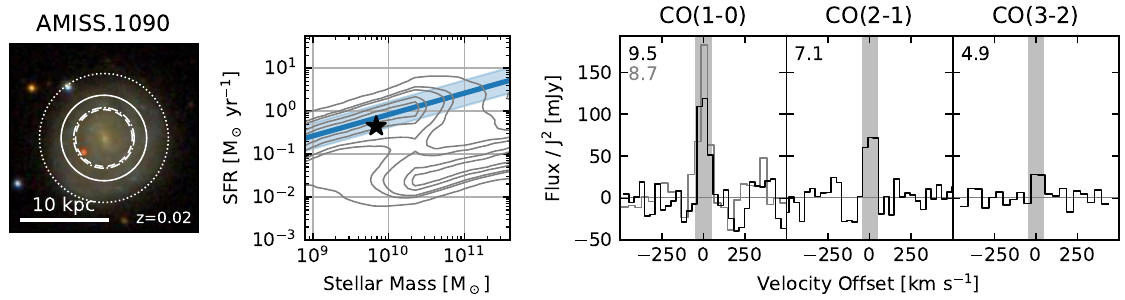}
\figsetgrpnote{Left column: SDSS cutouts of each target. Solid, dashed, and dotted lines show the 
beam sizes of the SMT for CO(2--1), the SMT for CO(3--2) (and the IRAM 30m for CO(1--0)) 
and the 12m for CO(1--0) respectively. The scale bar in the lower left shows 10 
kiloparsecs. 
Middle column: contours show the distribution of star formation rates at a given stellar 
mass, while the blue line and filled region show the main sequence of star forming 
galaxies. The stellar mass and star formation rate of the target galaxy is marked by a 
star. 
Right column: CO(1--0) spectra from AMISS (gray) and xCOLD~GASS (black), CO(2--1) spectra 
from AMISS, and CO(3--2) spectra from AMISS. Numbers in the upper right corner give the 
signal to noise ratio for each line. When a CO line is detected, the gray band indicates 
the region used to measure the line flux. The scale of the $y$-axis is such that lines 
would have the same amplitude in each transition for thermalized CO emission. The relative 
amplitudes of each spectrum give a sense of the luminosity ratios between the different 
lines.}
\figsetgrpend

\figsetgrpstart
\figsetgrpnum{15.92}
\figsetgrptitle{AMISS.1091}
\figsetplot{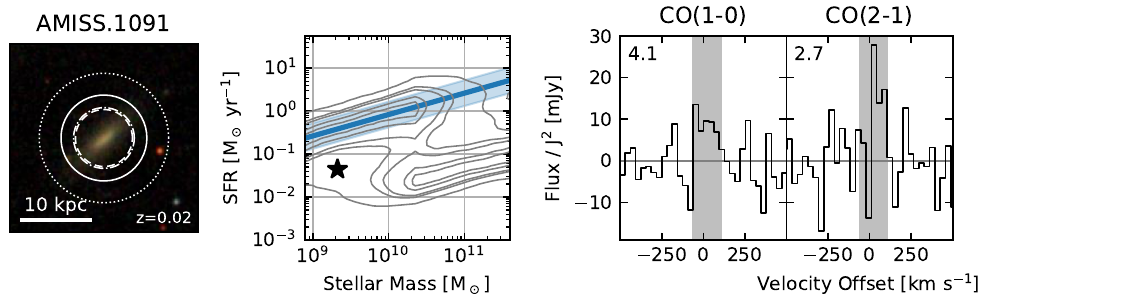}
\figsetgrpnote{Left column: SDSS cutouts of each target. Solid, dashed, and dotted lines show the 
beam sizes of the SMT for CO(2--1), the SMT for CO(3--2) (and the IRAM 30m for CO(1--0)) 
and the 12m for CO(1--0) respectively. The scale bar in the lower left shows 10 
kiloparsecs. 
Middle column: contours show the distribution of star formation rates at a given stellar 
mass, while the blue line and filled region show the main sequence of star forming 
galaxies. The stellar mass and star formation rate of the target galaxy is marked by a 
star. 
Right column: CO(1--0) spectra from AMISS (gray) and xCOLD~GASS (black), CO(2--1) spectra 
from AMISS, and CO(3--2) spectra from AMISS. Numbers in the upper right corner give the 
signal to noise ratio for each line. When a CO line is detected, the gray band indicates 
the region used to measure the line flux. The scale of the $y$-axis is such that lines 
would have the same amplitude in each transition for thermalized CO emission. The relative 
amplitudes of each spectrum give a sense of the luminosity ratios between the different 
lines.}
\figsetgrpend

\figsetgrpstart
\figsetgrpnum{15.93}
\figsetgrptitle{AMISS.1092}
\figsetplot{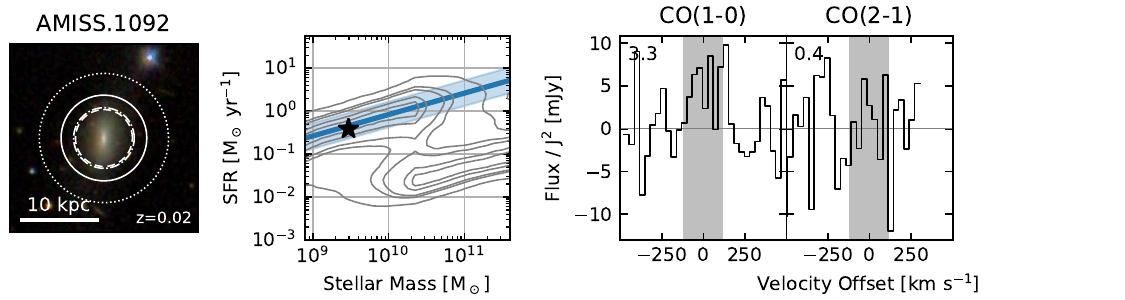}
\figsetgrpnote{Left column: SDSS cutouts of each target. Solid, dashed, and dotted lines show the 
beam sizes of the SMT for CO(2--1), the SMT for CO(3--2) (and the IRAM 30m for CO(1--0)) 
and the 12m for CO(1--0) respectively. The scale bar in the lower left shows 10 
kiloparsecs. 
Middle column: contours show the distribution of star formation rates at a given stellar 
mass, while the blue line and filled region show the main sequence of star forming 
galaxies. The stellar mass and star formation rate of the target galaxy is marked by a 
star. 
Right column: CO(1--0) spectra from AMISS (gray) and xCOLD~GASS (black), CO(2--1) spectra 
from AMISS, and CO(3--2) spectra from AMISS. Numbers in the upper right corner give the 
signal to noise ratio for each line. When a CO line is detected, the gray band indicates 
the region used to measure the line flux. The scale of the $y$-axis is such that lines 
would have the same amplitude in each transition for thermalized CO emission. The relative 
amplitudes of each spectrum give a sense of the luminosity ratios between the different 
lines.}
\figsetgrpend

\figsetgrpstart
\figsetgrpnum{15.94}
\figsetgrptitle{AMISS.1093}
\figsetplot{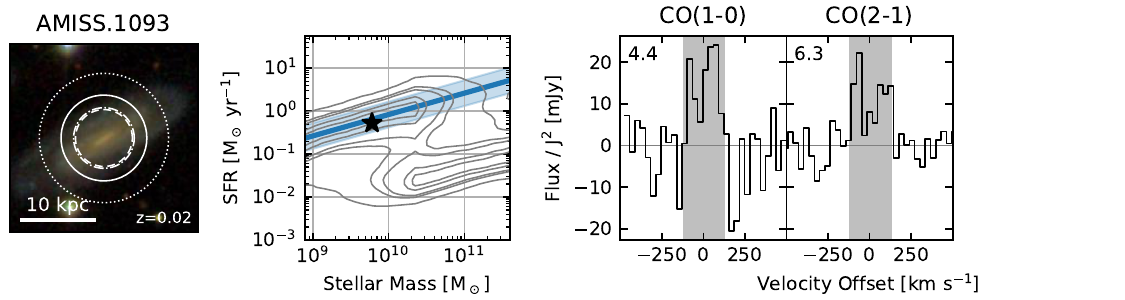}
\figsetgrpnote{Left column: SDSS cutouts of each target. Solid, dashed, and dotted lines show the 
beam sizes of the SMT for CO(2--1), the SMT for CO(3--2) (and the IRAM 30m for CO(1--0)) 
and the 12m for CO(1--0) respectively. The scale bar in the lower left shows 10 
kiloparsecs. 
Middle column: contours show the distribution of star formation rates at a given stellar 
mass, while the blue line and filled region show the main sequence of star forming 
galaxies. The stellar mass and star formation rate of the target galaxy is marked by a 
star. 
Right column: CO(1--0) spectra from AMISS (gray) and xCOLD~GASS (black), CO(2--1) spectra 
from AMISS, and CO(3--2) spectra from AMISS. Numbers in the upper right corner give the 
signal to noise ratio for each line. When a CO line is detected, the gray band indicates 
the region used to measure the line flux. The scale of the $y$-axis is such that lines 
would have the same amplitude in each transition for thermalized CO emission. The relative 
amplitudes of each spectrum give a sense of the luminosity ratios between the different 
lines.}
\figsetgrpend

\figsetgrpstart
\figsetgrpnum{15.95}
\figsetgrptitle{AMISS.1094}
\figsetplot{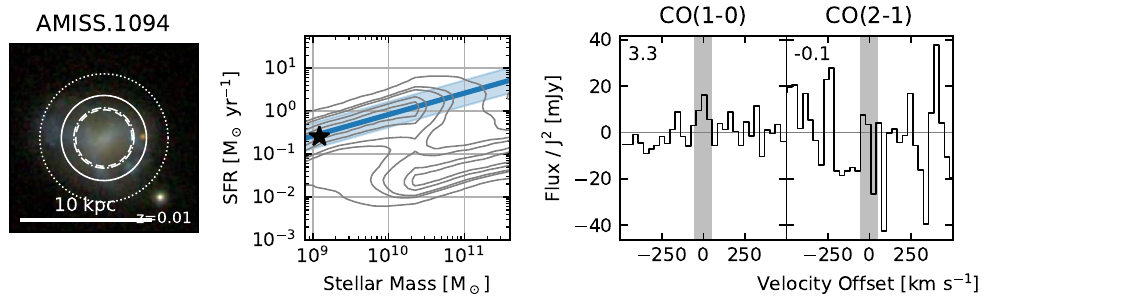}
\figsetgrpnote{Left column: SDSS cutouts of each target. Solid, dashed, and dotted lines show the 
beam sizes of the SMT for CO(2--1), the SMT for CO(3--2) (and the IRAM 30m for CO(1--0)) 
and the 12m for CO(1--0) respectively. The scale bar in the lower left shows 10 
kiloparsecs. 
Middle column: contours show the distribution of star formation rates at a given stellar 
mass, while the blue line and filled region show the main sequence of star forming 
galaxies. The stellar mass and star formation rate of the target galaxy is marked by a 
star. 
Right column: CO(1--0) spectra from AMISS (gray) and xCOLD~GASS (black), CO(2--1) spectra 
from AMISS, and CO(3--2) spectra from AMISS. Numbers in the upper right corner give the 
signal to noise ratio for each line. When a CO line is detected, the gray band indicates 
the region used to measure the line flux. The scale of the $y$-axis is such that lines 
would have the same amplitude in each transition for thermalized CO emission. The relative 
amplitudes of each spectrum give a sense of the luminosity ratios between the different 
lines.}
\figsetgrpend

\figsetgrpstart
\figsetgrpnum{15.96}
\figsetgrptitle{AMISS.1095}
\figsetplot{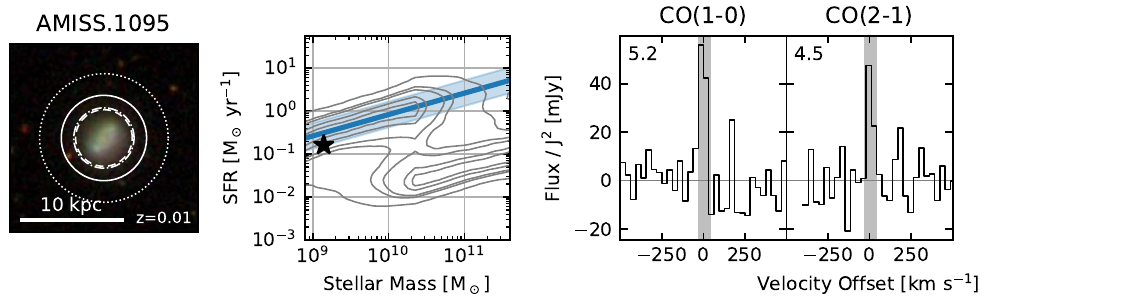}
\figsetgrpnote{Left column: SDSS cutouts of each target. Solid, dashed, and dotted lines show the 
beam sizes of the SMT for CO(2--1), the SMT for CO(3--2) (and the IRAM 30m for CO(1--0)) 
and the 12m for CO(1--0) respectively. The scale bar in the lower left shows 10 
kiloparsecs. 
Middle column: contours show the distribution of star formation rates at a given stellar 
mass, while the blue line and filled region show the main sequence of star forming 
galaxies. The stellar mass and star formation rate of the target galaxy is marked by a 
star. 
Right column: CO(1--0) spectra from AMISS (gray) and xCOLD~GASS (black), CO(2--1) spectra 
from AMISS, and CO(3--2) spectra from AMISS. Numbers in the upper right corner give the 
signal to noise ratio for each line. When a CO line is detected, the gray band indicates 
the region used to measure the line flux. The scale of the $y$-axis is such that lines 
would have the same amplitude in each transition for thermalized CO emission. The relative 
amplitudes of each spectrum give a sense of the luminosity ratios between the different 
lines.}
\figsetgrpend

\figsetgrpstart
\figsetgrpnum{15.97}
\figsetgrptitle{AMISS.1096}
\figsetplot{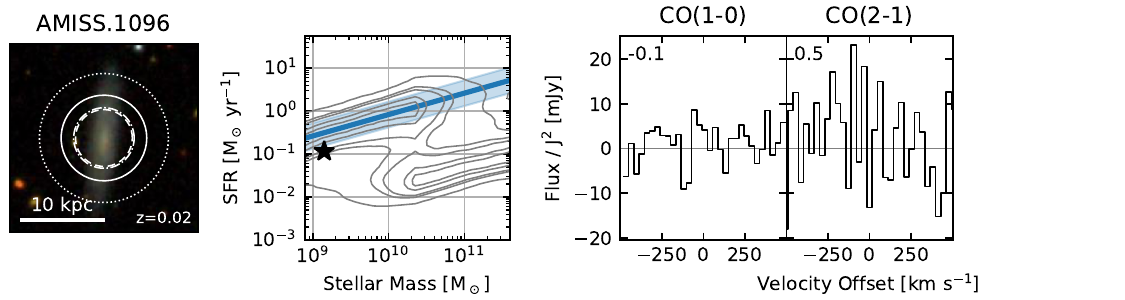}
\figsetgrpnote{Left column: SDSS cutouts of each target. Solid, dashed, and dotted lines show the 
beam sizes of the SMT for CO(2--1), the SMT for CO(3--2) (and the IRAM 30m for CO(1--0)) 
and the 12m for CO(1--0) respectively. The scale bar in the lower left shows 10 
kiloparsecs. 
Middle column: contours show the distribution of star formation rates at a given stellar 
mass, while the blue line and filled region show the main sequence of star forming 
galaxies. The stellar mass and star formation rate of the target galaxy is marked by a 
star. 
Right column: CO(1--0) spectra from AMISS (gray) and xCOLD~GASS (black), CO(2--1) spectra 
from AMISS, and CO(3--2) spectra from AMISS. Numbers in the upper right corner give the 
signal to noise ratio for each line. When a CO line is detected, the gray band indicates 
the region used to measure the line flux. The scale of the $y$-axis is such that lines 
would have the same amplitude in each transition for thermalized CO emission. The relative 
amplitudes of each spectrum give a sense of the luminosity ratios between the different 
lines.}
\figsetgrpend

\figsetgrpstart
\figsetgrpnum{15.98}
\figsetgrptitle{AMISS.1097}
\figsetplot{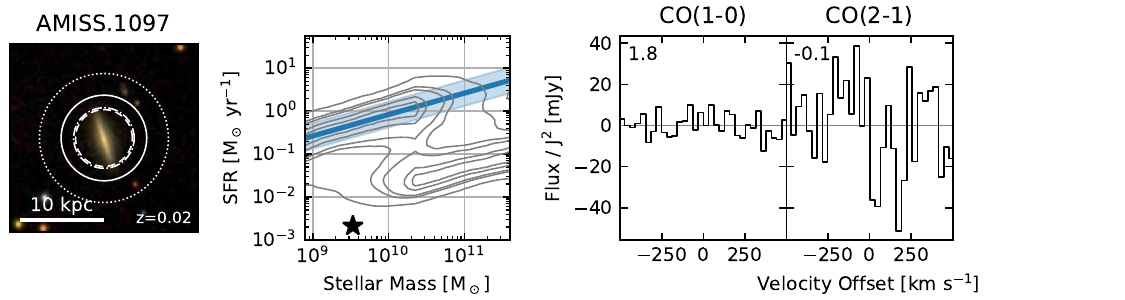}
\figsetgrpnote{Left column: SDSS cutouts of each target. Solid, dashed, and dotted lines show the 
beam sizes of the SMT for CO(2--1), the SMT for CO(3--2) (and the IRAM 30m for CO(1--0)) 
and the 12m for CO(1--0) respectively. The scale bar in the lower left shows 10 
kiloparsecs. 
Middle column: contours show the distribution of star formation rates at a given stellar 
mass, while the blue line and filled region show the main sequence of star forming 
galaxies. The stellar mass and star formation rate of the target galaxy is marked by a 
star. 
Right column: CO(1--0) spectra from AMISS (gray) and xCOLD~GASS (black), CO(2--1) spectra 
from AMISS, and CO(3--2) spectra from AMISS. Numbers in the upper right corner give the 
signal to noise ratio for each line. When a CO line is detected, the gray band indicates 
the region used to measure the line flux. The scale of the $y$-axis is such that lines 
would have the same amplitude in each transition for thermalized CO emission. The relative 
amplitudes of each spectrum give a sense of the luminosity ratios between the different 
lines.}
\figsetgrpend

\figsetgrpstart
\figsetgrpnum{15.99}
\figsetgrptitle{AMISS.1098}
\figsetplot{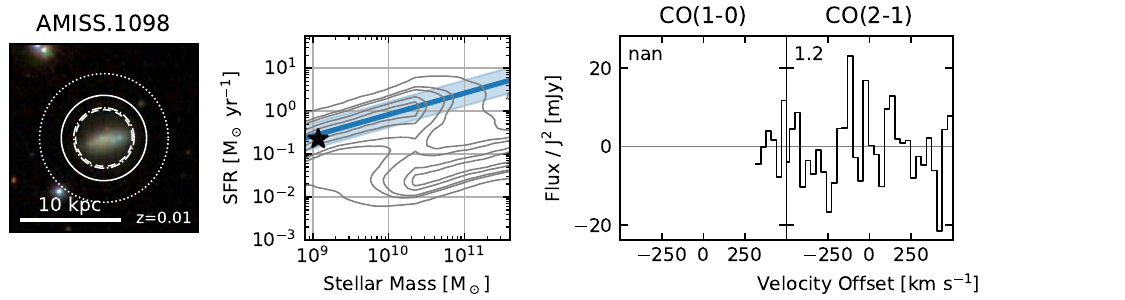}
\figsetgrpnote{Left column: SDSS cutouts of each target. Solid, dashed, and dotted lines show the 
beam sizes of the SMT for CO(2--1), the SMT for CO(3--2) (and the IRAM 30m for CO(1--0)) 
and the 12m for CO(1--0) respectively. The scale bar in the lower left shows 10 
kiloparsecs. 
Middle column: contours show the distribution of star formation rates at a given stellar 
mass, while the blue line and filled region show the main sequence of star forming 
galaxies. The stellar mass and star formation rate of the target galaxy is marked by a 
star. 
Right column: CO(1--0) spectra from AMISS (gray) and xCOLD~GASS (black), CO(2--1) spectra 
from AMISS, and CO(3--2) spectra from AMISS. Numbers in the upper right corner give the 
signal to noise ratio for each line. When a CO line is detected, the gray band indicates 
the region used to measure the line flux. The scale of the $y$-axis is such that lines 
would have the same amplitude in each transition for thermalized CO emission. The relative 
amplitudes of each spectrum give a sense of the luminosity ratios between the different 
lines.}
\figsetgrpend

\figsetgrpstart
\figsetgrpnum{15.100}
\figsetgrptitle{AMISS.1099}
\figsetplot{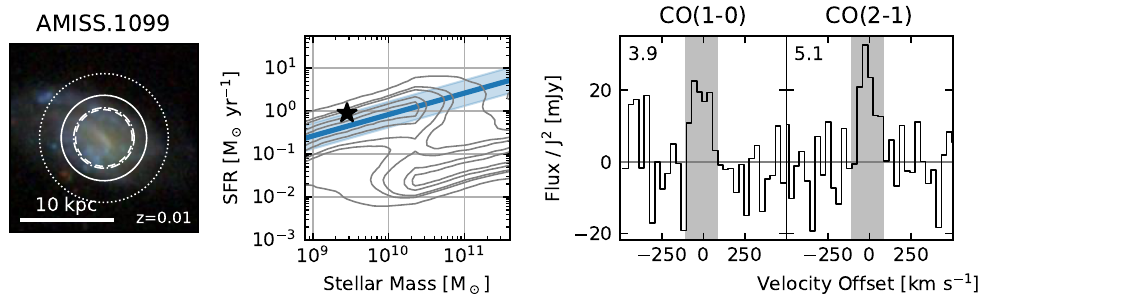}
\figsetgrpnote{Left column: SDSS cutouts of each target. Solid, dashed, and dotted lines show the 
beam sizes of the SMT for CO(2--1), the SMT for CO(3--2) (and the IRAM 30m for CO(1--0)) 
and the 12m for CO(1--0) respectively. The scale bar in the lower left shows 10 
kiloparsecs. 
Middle column: contours show the distribution of star formation rates at a given stellar 
mass, while the blue line and filled region show the main sequence of star forming 
galaxies. The stellar mass and star formation rate of the target galaxy is marked by a 
star. 
Right column: CO(1--0) spectra from AMISS (gray) and xCOLD~GASS (black), CO(2--1) spectra 
from AMISS, and CO(3--2) spectra from AMISS. Numbers in the upper right corner give the 
signal to noise ratio for each line. When a CO line is detected, the gray band indicates 
the region used to measure the line flux. The scale of the $y$-axis is such that lines 
would have the same amplitude in each transition for thermalized CO emission. The relative 
amplitudes of each spectrum give a sense of the luminosity ratios between the different 
lines.}
\figsetgrpend

\figsetgrpstart
\figsetgrpnum{15.101}
\figsetgrptitle{AMISS.1100}
\figsetplot{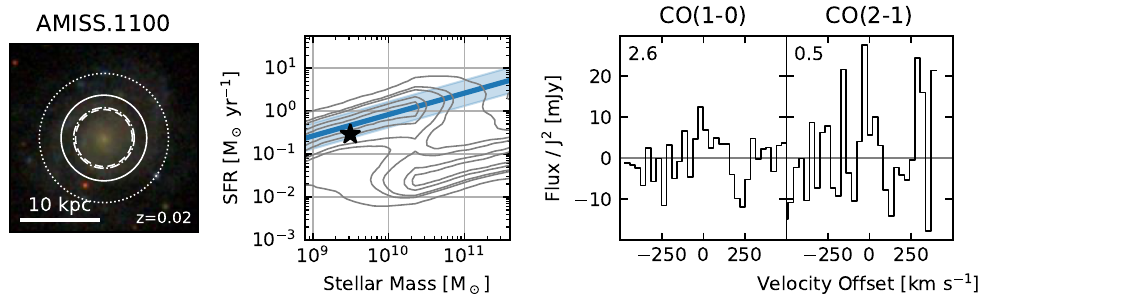}
\figsetgrpnote{Left column: SDSS cutouts of each target. Solid, dashed, and dotted lines show the 
beam sizes of the SMT for CO(2--1), the SMT for CO(3--2) (and the IRAM 30m for CO(1--0)) 
and the 12m for CO(1--0) respectively. The scale bar in the lower left shows 10 
kiloparsecs. 
Middle column: contours show the distribution of star formation rates at a given stellar 
mass, while the blue line and filled region show the main sequence of star forming 
galaxies. The stellar mass and star formation rate of the target galaxy is marked by a 
star. 
Right column: CO(1--0) spectra from AMISS (gray) and xCOLD~GASS (black), CO(2--1) spectra 
from AMISS, and CO(3--2) spectra from AMISS. Numbers in the upper right corner give the 
signal to noise ratio for each line. When a CO line is detected, the gray band indicates 
the region used to measure the line flux. The scale of the $y$-axis is such that lines 
would have the same amplitude in each transition for thermalized CO emission. The relative 
amplitudes of each spectrum give a sense of the luminosity ratios between the different 
lines.}
\figsetgrpend

\figsetgrpstart
\figsetgrpnum{15.102}
\figsetgrptitle{AMISS.2002}
\figsetplot{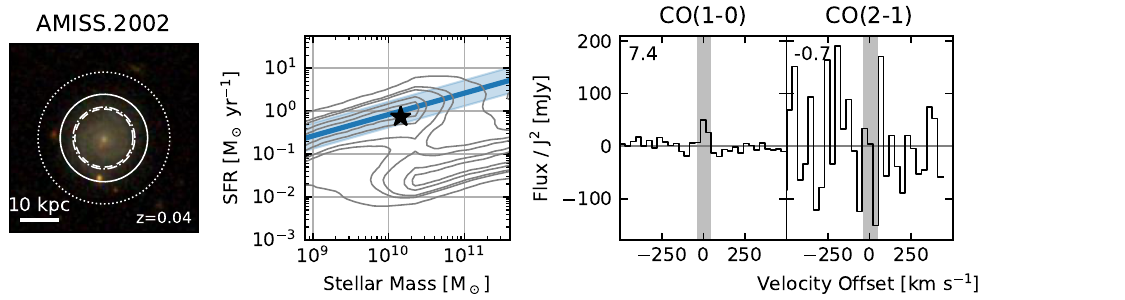}
\figsetgrpnote{Left column: SDSS cutouts of each target. Solid, dashed, and dotted lines show the 
beam sizes of the SMT for CO(2--1), the SMT for CO(3--2) (and the IRAM 30m for CO(1--0)) 
and the 12m for CO(1--0) respectively. The scale bar in the lower left shows 10 
kiloparsecs. 
Middle column: contours show the distribution of star formation rates at a given stellar 
mass, while the blue line and filled region show the main sequence of star forming 
galaxies. The stellar mass and star formation rate of the target galaxy is marked by a 
star. 
Right column: CO(1--0) spectra from AMISS (gray) and xCOLD~GASS (black), CO(2--1) spectra 
from AMISS, and CO(3--2) spectra from AMISS. Numbers in the upper right corner give the 
signal to noise ratio for each line. When a CO line is detected, the gray band indicates 
the region used to measure the line flux. The scale of the $y$-axis is such that lines 
would have the same amplitude in each transition for thermalized CO emission. The relative 
amplitudes of each spectrum give a sense of the luminosity ratios between the different 
lines.}
\figsetgrpend

\figsetgrpstart
\figsetgrpnum{15.103}
\figsetgrptitle{AMISS.2003}
\figsetplot{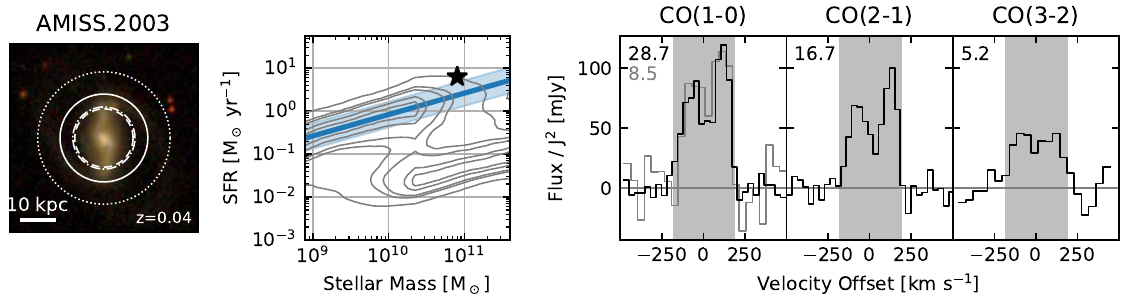}
\figsetgrpnote{Left column: SDSS cutouts of each target. Solid, dashed, and dotted lines show the 
beam sizes of the SMT for CO(2--1), the SMT for CO(3--2) (and the IRAM 30m for CO(1--0)) 
and the 12m for CO(1--0) respectively. The scale bar in the lower left shows 10 
kiloparsecs. 
Middle column: contours show the distribution of star formation rates at a given stellar 
mass, while the blue line and filled region show the main sequence of star forming 
galaxies. The stellar mass and star formation rate of the target galaxy is marked by a 
star. 
Right column: CO(1--0) spectra from AMISS (gray) and xCOLD~GASS (black), CO(2--1) spectra 
from AMISS, and CO(3--2) spectra from AMISS. Numbers in the upper right corner give the 
signal to noise ratio for each line. When a CO line is detected, the gray band indicates 
the region used to measure the line flux. The scale of the $y$-axis is such that lines 
would have the same amplitude in each transition for thermalized CO emission. The relative 
amplitudes of each spectrum give a sense of the luminosity ratios between the different 
lines.}
\figsetgrpend

\figsetgrpstart
\figsetgrpnum{15.104}
\figsetgrptitle{AMISS.2004}
\figsetplot{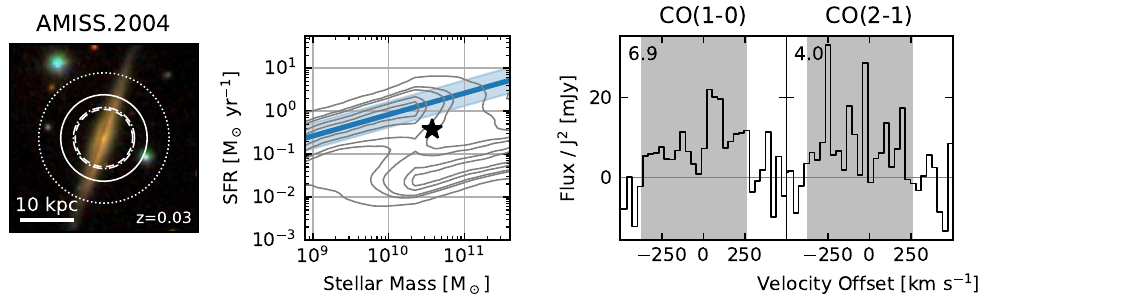}
\figsetgrpnote{Left column: SDSS cutouts of each target. Solid, dashed, and dotted lines show the 
beam sizes of the SMT for CO(2--1), the SMT for CO(3--2) (and the IRAM 30m for CO(1--0)) 
and the 12m for CO(1--0) respectively. The scale bar in the lower left shows 10 
kiloparsecs. 
Middle column: contours show the distribution of star formation rates at a given stellar 
mass, while the blue line and filled region show the main sequence of star forming 
galaxies. The stellar mass and star formation rate of the target galaxy is marked by a 
star. 
Right column: CO(1--0) spectra from AMISS (gray) and xCOLD~GASS (black), CO(2--1) spectra 
from AMISS, and CO(3--2) spectra from AMISS. Numbers in the upper right corner give the 
signal to noise ratio for each line. When a CO line is detected, the gray band indicates 
the region used to measure the line flux. The scale of the $y$-axis is such that lines 
would have the same amplitude in each transition for thermalized CO emission. The relative 
amplitudes of each spectrum give a sense of the luminosity ratios between the different 
lines.}
\figsetgrpend

\figsetgrpstart
\figsetgrpnum{15.105}
\figsetgrptitle{AMISS.2011}
\figsetplot{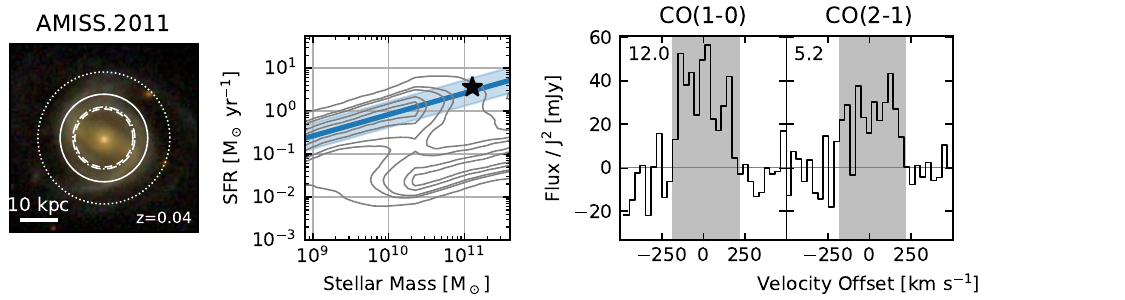}
\figsetgrpnote{Left column: SDSS cutouts of each target. Solid, dashed, and dotted lines show the 
beam sizes of the SMT for CO(2--1), the SMT for CO(3--2) (and the IRAM 30m for CO(1--0)) 
and the 12m for CO(1--0) respectively. The scale bar in the lower left shows 10 
kiloparsecs. 
Middle column: contours show the distribution of star formation rates at a given stellar 
mass, while the blue line and filled region show the main sequence of star forming 
galaxies. The stellar mass and star formation rate of the target galaxy is marked by a 
star. 
Right column: CO(1--0) spectra from AMISS (gray) and xCOLD~GASS (black), CO(2--1) spectra 
from AMISS, and CO(3--2) spectra from AMISS. Numbers in the upper right corner give the 
signal to noise ratio for each line. When a CO line is detected, the gray band indicates 
the region used to measure the line flux. The scale of the $y$-axis is such that lines 
would have the same amplitude in each transition for thermalized CO emission. The relative 
amplitudes of each spectrum give a sense of the luminosity ratios between the different 
lines.}
\figsetgrpend

\figsetgrpstart
\figsetgrpnum{15.106}
\figsetgrptitle{AMISS.2013}
\figsetplot{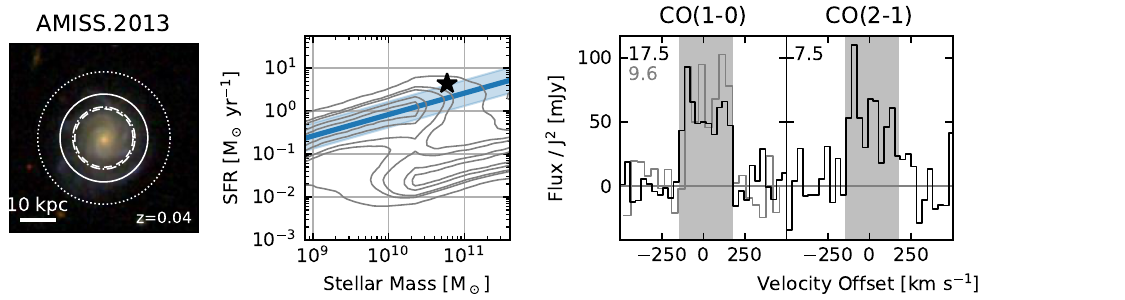}
\figsetgrpnote{Left column: SDSS cutouts of each target. Solid, dashed, and dotted lines show the 
beam sizes of the SMT for CO(2--1), the SMT for CO(3--2) (and the IRAM 30m for CO(1--0)) 
and the 12m for CO(1--0) respectively. The scale bar in the lower left shows 10 
kiloparsecs. 
Middle column: contours show the distribution of star formation rates at a given stellar 
mass, while the blue line and filled region show the main sequence of star forming 
galaxies. The stellar mass and star formation rate of the target galaxy is marked by a 
star. 
Right column: CO(1--0) spectra from AMISS (gray) and xCOLD~GASS (black), CO(2--1) spectra 
from AMISS, and CO(3--2) spectra from AMISS. Numbers in the upper right corner give the 
signal to noise ratio for each line. When a CO line is detected, the gray band indicates 
the region used to measure the line flux. The scale of the $y$-axis is such that lines 
would have the same amplitude in each transition for thermalized CO emission. The relative 
amplitudes of each spectrum give a sense of the luminosity ratios between the different 
lines.}
\figsetgrpend

\figsetgrpstart
\figsetgrpnum{15.107}
\figsetgrptitle{AMISS.2024}
\figsetplot{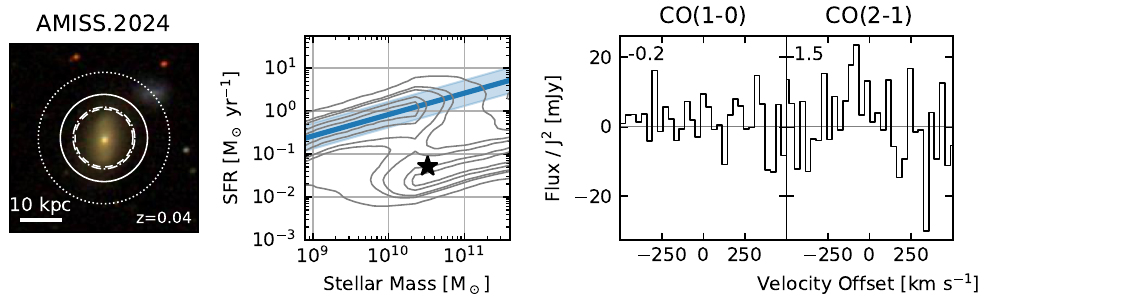}
\figsetgrpnote{Left column: SDSS cutouts of each target. Solid, dashed, and dotted lines show the 
beam sizes of the SMT for CO(2--1), the SMT for CO(3--2) (and the IRAM 30m for CO(1--0)) 
and the 12m for CO(1--0) respectively. The scale bar in the lower left shows 10 
kiloparsecs. 
Middle column: contours show the distribution of star formation rates at a given stellar 
mass, while the blue line and filled region show the main sequence of star forming 
galaxies. The stellar mass and star formation rate of the target galaxy is marked by a 
star. 
Right column: CO(1--0) spectra from AMISS (gray) and xCOLD~GASS (black), CO(2--1) spectra 
from AMISS, and CO(3--2) spectra from AMISS. Numbers in the upper right corner give the 
signal to noise ratio for each line. When a CO line is detected, the gray band indicates 
the region used to measure the line flux. The scale of the $y$-axis is such that lines 
would have the same amplitude in each transition for thermalized CO emission. The relative 
amplitudes of each spectrum give a sense of the luminosity ratios between the different 
lines.}
\figsetgrpend

\figsetgrpstart
\figsetgrpnum{15.108}
\figsetgrptitle{AMISS.2027}
\figsetplot{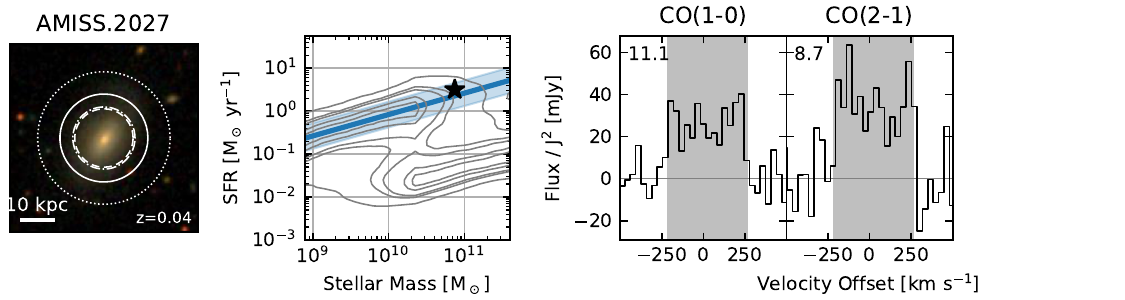}
\figsetgrpnote{Left column: SDSS cutouts of each target. Solid, dashed, and dotted lines show the 
beam sizes of the SMT for CO(2--1), the SMT for CO(3--2) (and the IRAM 30m for CO(1--0)) 
and the 12m for CO(1--0) respectively. The scale bar in the lower left shows 10 
kiloparsecs. 
Middle column: contours show the distribution of star formation rates at a given stellar 
mass, while the blue line and filled region show the main sequence of star forming 
galaxies. The stellar mass and star formation rate of the target galaxy is marked by a 
star. 
Right column: CO(1--0) spectra from AMISS (gray) and xCOLD~GASS (black), CO(2--1) spectra 
from AMISS, and CO(3--2) spectra from AMISS. Numbers in the upper right corner give the 
signal to noise ratio for each line. When a CO line is detected, the gray band indicates 
the region used to measure the line flux. The scale of the $y$-axis is such that lines 
would have the same amplitude in each transition for thermalized CO emission. The relative 
amplitudes of each spectrum give a sense of the luminosity ratios between the different 
lines.}
\figsetgrpend

\figsetgrpstart
\figsetgrpnum{15.109}
\figsetgrptitle{AMISS.2033}
\figsetplot{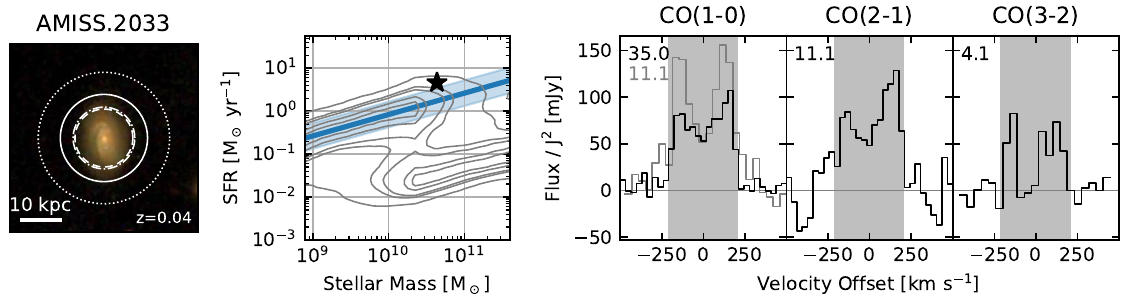}
\figsetgrpnote{Left column: SDSS cutouts of each target. Solid, dashed, and dotted lines show the 
beam sizes of the SMT for CO(2--1), the SMT for CO(3--2) (and the IRAM 30m for CO(1--0)) 
and the 12m for CO(1--0) respectively. The scale bar in the lower left shows 10 
kiloparsecs. 
Middle column: contours show the distribution of star formation rates at a given stellar 
mass, while the blue line and filled region show the main sequence of star forming 
galaxies. The stellar mass and star formation rate of the target galaxy is marked by a 
star. 
Right column: CO(1--0) spectra from AMISS (gray) and xCOLD~GASS (black), CO(2--1) spectra 
from AMISS, and CO(3--2) spectra from AMISS. Numbers in the upper right corner give the 
signal to noise ratio for each line. When a CO line is detected, the gray band indicates 
the region used to measure the line flux. The scale of the $y$-axis is such that lines 
would have the same amplitude in each transition for thermalized CO emission. The relative 
amplitudes of each spectrum give a sense of the luminosity ratios between the different 
lines.}
\figsetgrpend

\figsetgrpstart
\figsetgrpnum{15.110}
\figsetgrptitle{AMISS.2034}
\figsetplot{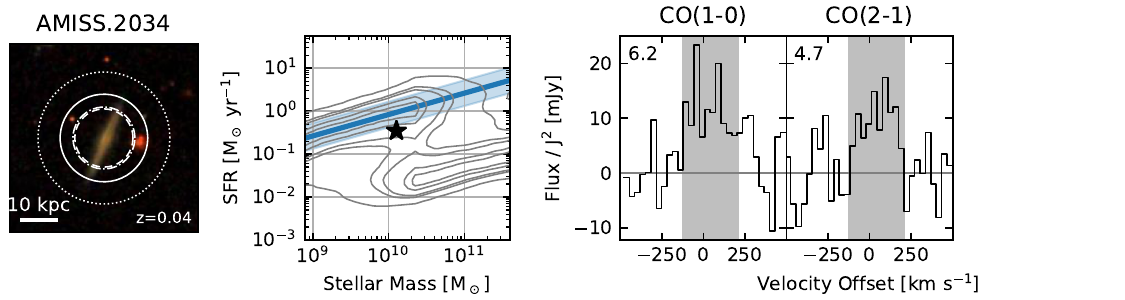}
\figsetgrpnote{Left column: SDSS cutouts of each target. Solid, dashed, and dotted lines show the 
beam sizes of the SMT for CO(2--1), the SMT for CO(3--2) (and the IRAM 30m for CO(1--0)) 
and the 12m for CO(1--0) respectively. The scale bar in the lower left shows 10 
kiloparsecs. 
Middle column: contours show the distribution of star formation rates at a given stellar 
mass, while the blue line and filled region show the main sequence of star forming 
galaxies. The stellar mass and star formation rate of the target galaxy is marked by a 
star. 
Right column: CO(1--0) spectra from AMISS (gray) and xCOLD~GASS (black), CO(2--1) spectra 
from AMISS, and CO(3--2) spectra from AMISS. Numbers in the upper right corner give the 
signal to noise ratio for each line. When a CO line is detected, the gray band indicates 
the region used to measure the line flux. The scale of the $y$-axis is such that lines 
would have the same amplitude in each transition for thermalized CO emission. The relative 
amplitudes of each spectrum give a sense of the luminosity ratios between the different 
lines.}
\figsetgrpend

\figsetgrpstart
\figsetgrpnum{15.111}
\figsetgrptitle{AMISS.2036}
\figsetplot{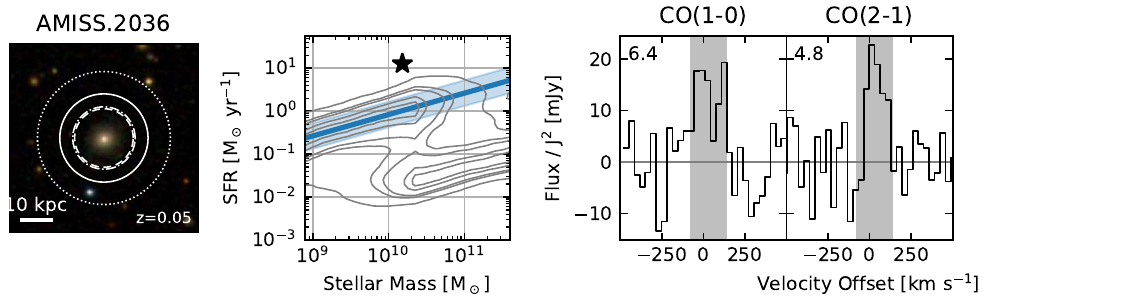}
\figsetgrpnote{Left column: SDSS cutouts of each target. Solid, dashed, and dotted lines show the 
beam sizes of the SMT for CO(2--1), the SMT for CO(3--2) (and the IRAM 30m for CO(1--0)) 
and the 12m for CO(1--0) respectively. The scale bar in the lower left shows 10 
kiloparsecs. 
Middle column: contours show the distribution of star formation rates at a given stellar 
mass, while the blue line and filled region show the main sequence of star forming 
galaxies. The stellar mass and star formation rate of the target galaxy is marked by a 
star. 
Right column: CO(1--0) spectra from AMISS (gray) and xCOLD~GASS (black), CO(2--1) spectra 
from AMISS, and CO(3--2) spectra from AMISS. Numbers in the upper right corner give the 
signal to noise ratio for each line. When a CO line is detected, the gray band indicates 
the region used to measure the line flux. The scale of the $y$-axis is such that lines 
would have the same amplitude in each transition for thermalized CO emission. The relative 
amplitudes of each spectrum give a sense of the luminosity ratios between the different 
lines.}
\figsetgrpend

\figsetgrpstart
\figsetgrpnum{15.112}
\figsetgrptitle{AMISS.2039}
\figsetplot{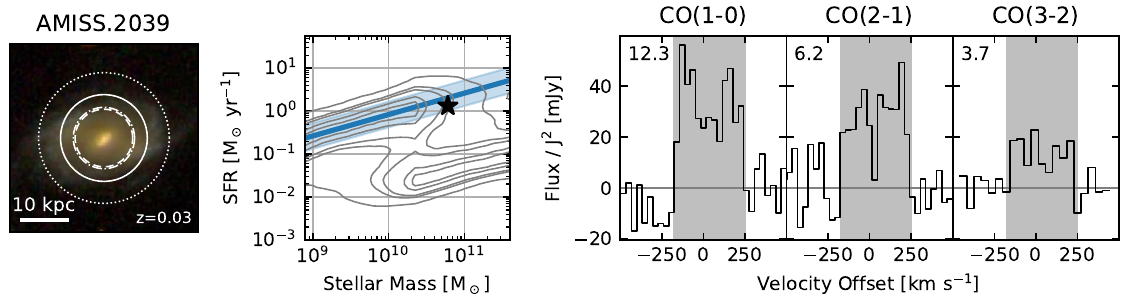}
\figsetgrpnote{Left column: SDSS cutouts of each target. Solid, dashed, and dotted lines show the 
beam sizes of the SMT for CO(2--1), the SMT for CO(3--2) (and the IRAM 30m for CO(1--0)) 
and the 12m for CO(1--0) respectively. The scale bar in the lower left shows 10 
kiloparsecs. 
Middle column: contours show the distribution of star formation rates at a given stellar 
mass, while the blue line and filled region show the main sequence of star forming 
galaxies. The stellar mass and star formation rate of the target galaxy is marked by a 
star. 
Right column: CO(1--0) spectra from AMISS (gray) and xCOLD~GASS (black), CO(2--1) spectra 
from AMISS, and CO(3--2) spectra from AMISS. Numbers in the upper right corner give the 
signal to noise ratio for each line. When a CO line is detected, the gray band indicates 
the region used to measure the line flux. The scale of the $y$-axis is such that lines 
would have the same amplitude in each transition for thermalized CO emission. The relative 
amplitudes of each spectrum give a sense of the luminosity ratios between the different 
lines.}
\figsetgrpend

\figsetgrpstart
\figsetgrpnum{15.113}
\figsetgrptitle{AMISS.2047}
\figsetplot{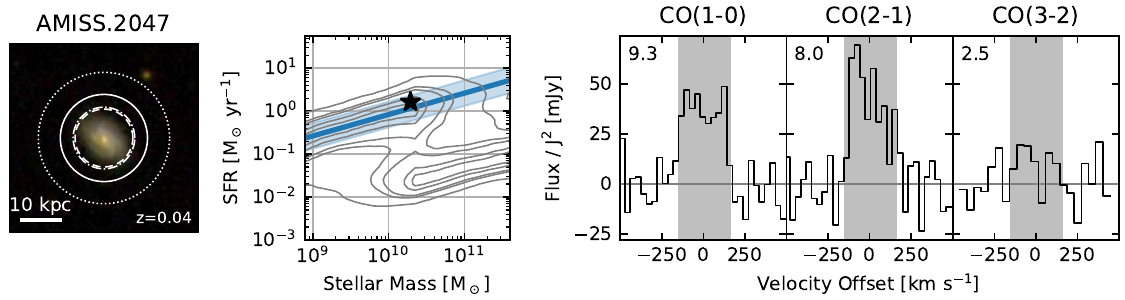}
\figsetgrpnote{Left column: SDSS cutouts of each target. Solid, dashed, and dotted lines show the 
beam sizes of the SMT for CO(2--1), the SMT for CO(3--2) (and the IRAM 30m for CO(1--0)) 
and the 12m for CO(1--0) respectively. The scale bar in the lower left shows 10 
kiloparsecs. 
Middle column: contours show the distribution of star formation rates at a given stellar 
mass, while the blue line and filled region show the main sequence of star forming 
galaxies. The stellar mass and star formation rate of the target galaxy is marked by a 
star. 
Right column: CO(1--0) spectra from AMISS (gray) and xCOLD~GASS (black), CO(2--1) spectra 
from AMISS, and CO(3--2) spectra from AMISS. Numbers in the upper right corner give the 
signal to noise ratio for each line. When a CO line is detected, the gray band indicates 
the region used to measure the line flux. The scale of the $y$-axis is such that lines 
would have the same amplitude in each transition for thermalized CO emission. The relative 
amplitudes of each spectrum give a sense of the luminosity ratios between the different 
lines.}
\figsetgrpend

\figsetgrpstart
\figsetgrpnum{15.114}
\figsetgrptitle{AMISS.2049}
\figsetplot{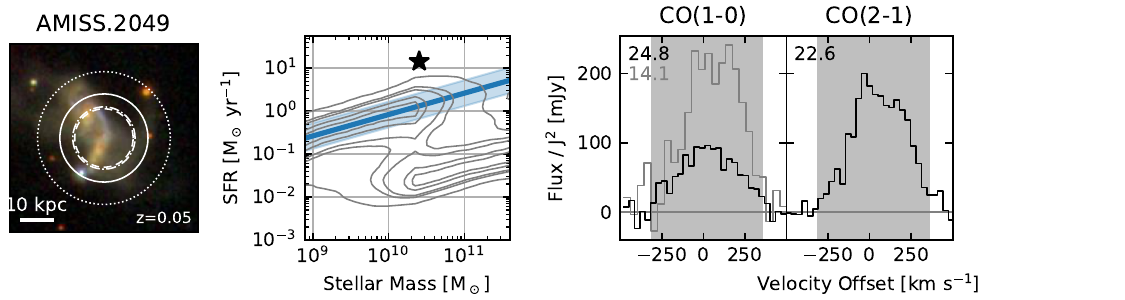}
\figsetgrpnote{Left column: SDSS cutouts of each target. Solid, dashed, and dotted lines show the 
beam sizes of the SMT for CO(2--1), the SMT for CO(3--2) (and the IRAM 30m for CO(1--0)) 
and the 12m for CO(1--0) respectively. The scale bar in the lower left shows 10 
kiloparsecs. 
Middle column: contours show the distribution of star formation rates at a given stellar 
mass, while the blue line and filled region show the main sequence of star forming 
galaxies. The stellar mass and star formation rate of the target galaxy is marked by a 
star. 
Right column: CO(1--0) spectra from AMISS (gray) and xCOLD~GASS (black), CO(2--1) spectra 
from AMISS, and CO(3--2) spectra from AMISS. Numbers in the upper right corner give the 
signal to noise ratio for each line. When a CO line is detected, the gray band indicates 
the region used to measure the line flux. The scale of the $y$-axis is such that lines 
would have the same amplitude in each transition for thermalized CO emission. The relative 
amplitudes of each spectrum give a sense of the luminosity ratios between the different 
lines.}
\figsetgrpend

\figsetgrpstart
\figsetgrpnum{15.115}
\figsetgrptitle{AMISS.2050}
\figsetplot{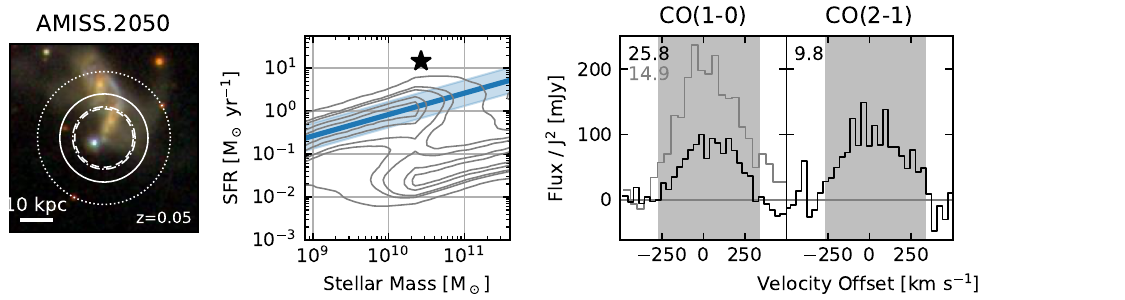}
\figsetgrpnote{Left column: SDSS cutouts of each target. Solid, dashed, and dotted lines show the 
beam sizes of the SMT for CO(2--1), the SMT for CO(3--2) (and the IRAM 30m for CO(1--0)) 
and the 12m for CO(1--0) respectively. The scale bar in the lower left shows 10 
kiloparsecs. 
Middle column: contours show the distribution of star formation rates at a given stellar 
mass, while the blue line and filled region show the main sequence of star forming 
galaxies. The stellar mass and star formation rate of the target galaxy is marked by a 
star. 
Right column: CO(1--0) spectra from AMISS (gray) and xCOLD~GASS (black), CO(2--1) spectra 
from AMISS, and CO(3--2) spectra from AMISS. Numbers in the upper right corner give the 
signal to noise ratio for each line. When a CO line is detected, the gray band indicates 
the region used to measure the line flux. The scale of the $y$-axis is such that lines 
would have the same amplitude in each transition for thermalized CO emission. The relative 
amplitudes of each spectrum give a sense of the luminosity ratios between the different 
lines.}
\figsetgrpend

\figsetgrpstart
\figsetgrpnum{15.116}
\figsetgrptitle{AMISS.2051}
\figsetplot{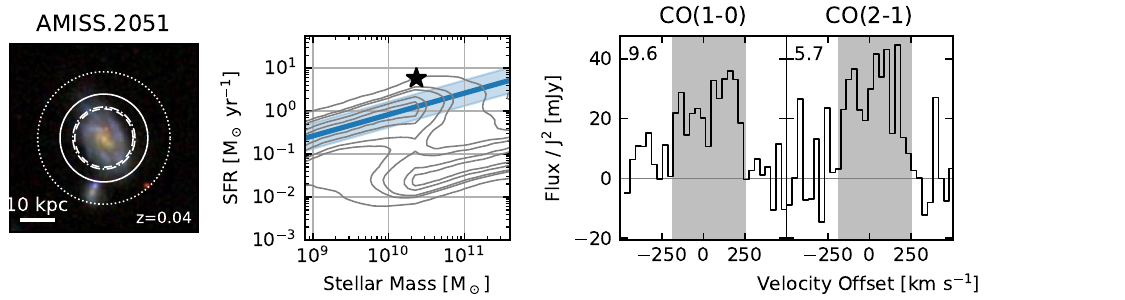}
\figsetgrpnote{Left column: SDSS cutouts of each target. Solid, dashed, and dotted lines show the 
beam sizes of the SMT for CO(2--1), the SMT for CO(3--2) (and the IRAM 30m for CO(1--0)) 
and the 12m for CO(1--0) respectively. The scale bar in the lower left shows 10 
kiloparsecs. 
Middle column: contours show the distribution of star formation rates at a given stellar 
mass, while the blue line and filled region show the main sequence of star forming 
galaxies. The stellar mass and star formation rate of the target galaxy is marked by a 
star. 
Right column: CO(1--0) spectra from AMISS (gray) and xCOLD~GASS (black), CO(2--1) spectra 
from AMISS, and CO(3--2) spectra from AMISS. Numbers in the upper right corner give the 
signal to noise ratio for each line. When a CO line is detected, the gray band indicates 
the region used to measure the line flux. The scale of the $y$-axis is such that lines 
would have the same amplitude in each transition for thermalized CO emission. The relative 
amplitudes of each spectrum give a sense of the luminosity ratios between the different 
lines.}
\figsetgrpend

\figsetgrpstart
\figsetgrpnum{15.117}
\figsetgrptitle{AMISS.2052}
\figsetplot{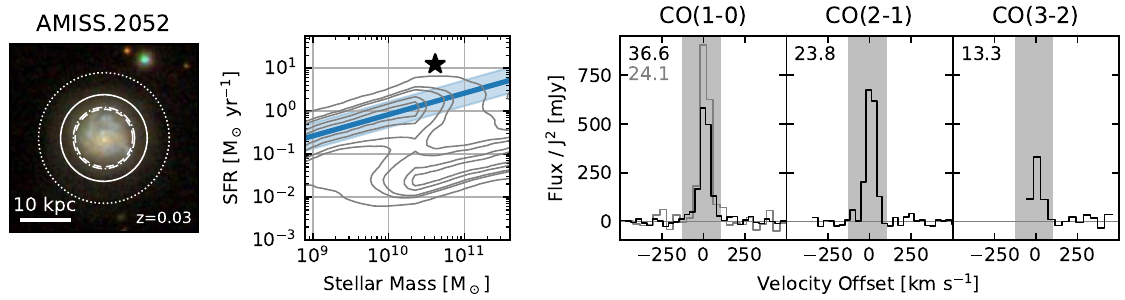}
\figsetgrpnote{Left column: SDSS cutouts of each target. Solid, dashed, and dotted lines show the 
beam sizes of the SMT for CO(2--1), the SMT for CO(3--2) (and the IRAM 30m for CO(1--0)) 
and the 12m for CO(1--0) respectively. The scale bar in the lower left shows 10 
kiloparsecs. 
Middle column: contours show the distribution of star formation rates at a given stellar 
mass, while the blue line and filled region show the main sequence of star forming 
galaxies. The stellar mass and star formation rate of the target galaxy is marked by a 
star. 
Right column: CO(1--0) spectra from AMISS (gray) and xCOLD~GASS (black), CO(2--1) spectra 
from AMISS, and CO(3--2) spectra from AMISS. Numbers in the upper right corner give the 
signal to noise ratio for each line. When a CO line is detected, the gray band indicates 
the region used to measure the line flux. The scale of the $y$-axis is such that lines 
would have the same amplitude in each transition for thermalized CO emission. The relative 
amplitudes of each spectrum give a sense of the luminosity ratios between the different 
lines.}
\figsetgrpend

\figsetgrpstart
\figsetgrpnum{15.118}
\figsetgrptitle{AMISS.2053}
\figsetplot{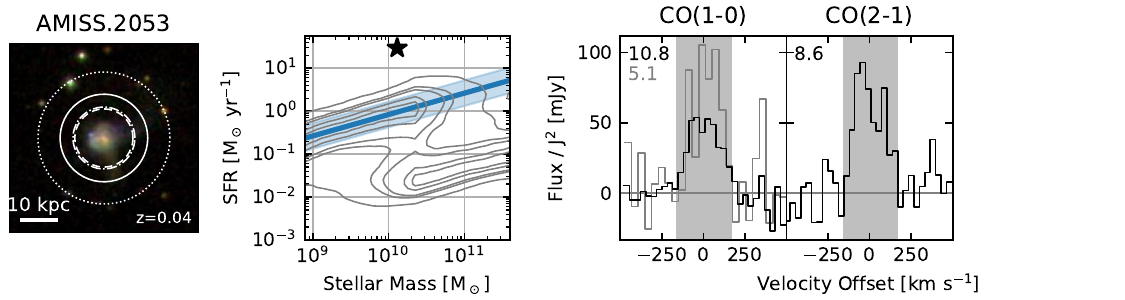}
\figsetgrpnote{Left column: SDSS cutouts of each target. Solid, dashed, and dotted lines show the 
beam sizes of the SMT for CO(2--1), the SMT for CO(3--2) (and the IRAM 30m for CO(1--0)) 
and the 12m for CO(1--0) respectively. The scale bar in the lower left shows 10 
kiloparsecs. 
Middle column: contours show the distribution of star formation rates at a given stellar 
mass, while the blue line and filled region show the main sequence of star forming 
galaxies. The stellar mass and star formation rate of the target galaxy is marked by a 
star. 
Right column: CO(1--0) spectra from AMISS (gray) and xCOLD~GASS (black), CO(2--1) spectra 
from AMISS, and CO(3--2) spectra from AMISS. Numbers in the upper right corner give the 
signal to noise ratio for each line. When a CO line is detected, the gray band indicates 
the region used to measure the line flux. The scale of the $y$-axis is such that lines 
would have the same amplitude in each transition for thermalized CO emission. The relative 
amplitudes of each spectrum give a sense of the luminosity ratios between the different 
lines.}
\figsetgrpend

\figsetgrpstart
\figsetgrpnum{15.119}
\figsetgrptitle{AMISS.2054}
\figsetplot{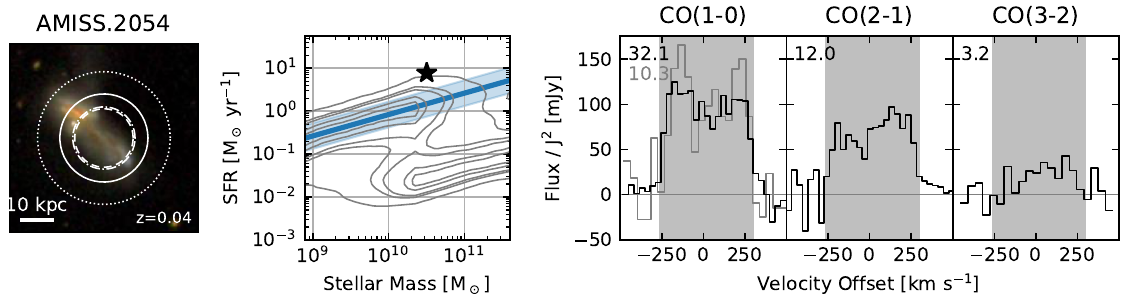}
\figsetgrpnote{Left column: SDSS cutouts of each target. Solid, dashed, and dotted lines show the 
beam sizes of the SMT for CO(2--1), the SMT for CO(3--2) (and the IRAM 30m for CO(1--0)) 
and the 12m for CO(1--0) respectively. The scale bar in the lower left shows 10 
kiloparsecs. 
Middle column: contours show the distribution of star formation rates at a given stellar 
mass, while the blue line and filled region show the main sequence of star forming 
galaxies. The stellar mass and star formation rate of the target galaxy is marked by a 
star. 
Right column: CO(1--0) spectra from AMISS (gray) and xCOLD~GASS (black), CO(2--1) spectra 
from AMISS, and CO(3--2) spectra from AMISS. Numbers in the upper right corner give the 
signal to noise ratio for each line. When a CO line is detected, the gray band indicates 
the region used to measure the line flux. The scale of the $y$-axis is such that lines 
would have the same amplitude in each transition for thermalized CO emission. The relative 
amplitudes of each spectrum give a sense of the luminosity ratios between the different 
lines.}
\figsetgrpend

\figsetgrpstart
\figsetgrpnum{15.120}
\figsetgrptitle{AMISS.2059}
\figsetplot{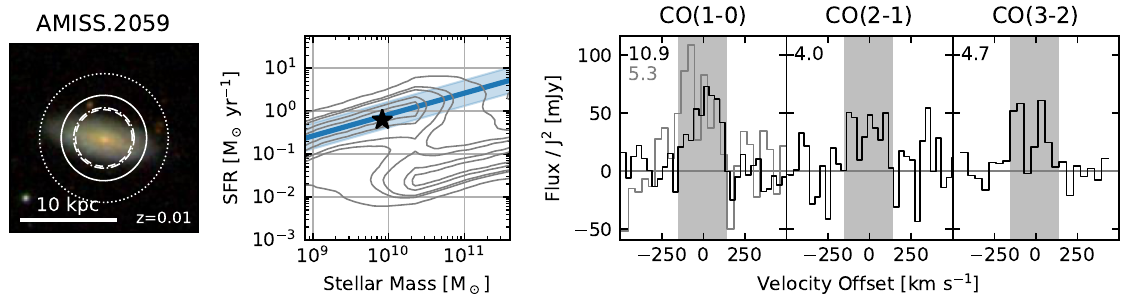}
\figsetgrpnote{Left column: SDSS cutouts of each target. Solid, dashed, and dotted lines show the 
beam sizes of the SMT for CO(2--1), the SMT for CO(3--2) (and the IRAM 30m for CO(1--0)) 
and the 12m for CO(1--0) respectively. The scale bar in the lower left shows 10 
kiloparsecs. 
Middle column: contours show the distribution of star formation rates at a given stellar 
mass, while the blue line and filled region show the main sequence of star forming 
galaxies. The stellar mass and star formation rate of the target galaxy is marked by a 
star. 
Right column: CO(1--0) spectra from AMISS (gray) and xCOLD~GASS (black), CO(2--1) spectra 
from AMISS, and CO(3--2) spectra from AMISS. Numbers in the upper right corner give the 
signal to noise ratio for each line. When a CO line is detected, the gray band indicates 
the region used to measure the line flux. The scale of the $y$-axis is such that lines 
would have the same amplitude in each transition for thermalized CO emission. The relative 
amplitudes of each spectrum give a sense of the luminosity ratios between the different 
lines.}
\figsetgrpend

\figsetgrpstart
\figsetgrpnum{15.121}
\figsetgrptitle{AMISS.2063}
\figsetplot{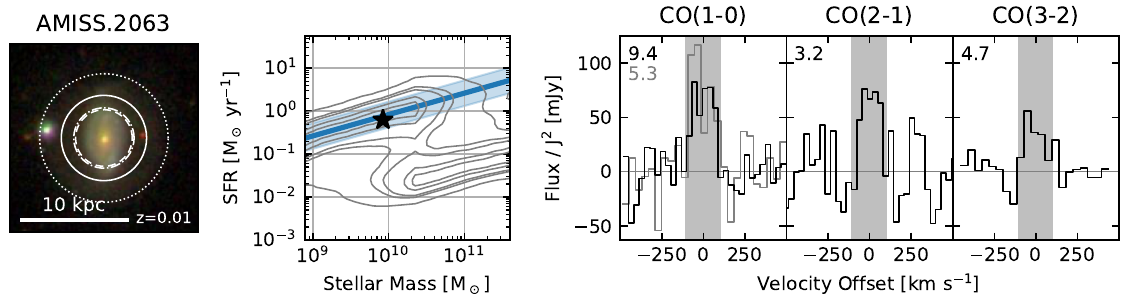}
\figsetgrpnote{Left column: SDSS cutouts of each target. Solid, dashed, and dotted lines show the 
beam sizes of the SMT for CO(2--1), the SMT for CO(3--2) (and the IRAM 30m for CO(1--0)) 
and the 12m for CO(1--0) respectively. The scale bar in the lower left shows 10 
kiloparsecs. 
Middle column: contours show the distribution of star formation rates at a given stellar 
mass, while the blue line and filled region show the main sequence of star forming 
galaxies. The stellar mass and star formation rate of the target galaxy is marked by a 
star. 
Right column: CO(1--0) spectra from AMISS (gray) and xCOLD~GASS (black), CO(2--1) spectra 
from AMISS, and CO(3--2) spectra from AMISS. Numbers in the upper right corner give the 
signal to noise ratio for each line. When a CO line is detected, the gray band indicates 
the region used to measure the line flux. The scale of the $y$-axis is such that lines 
would have the same amplitude in each transition for thermalized CO emission. The relative 
amplitudes of each spectrum give a sense of the luminosity ratios between the different 
lines.}
\figsetgrpend

\figsetgrpstart
\figsetgrpnum{15.122}
\figsetgrptitle{AMISS.2067}
\figsetplot{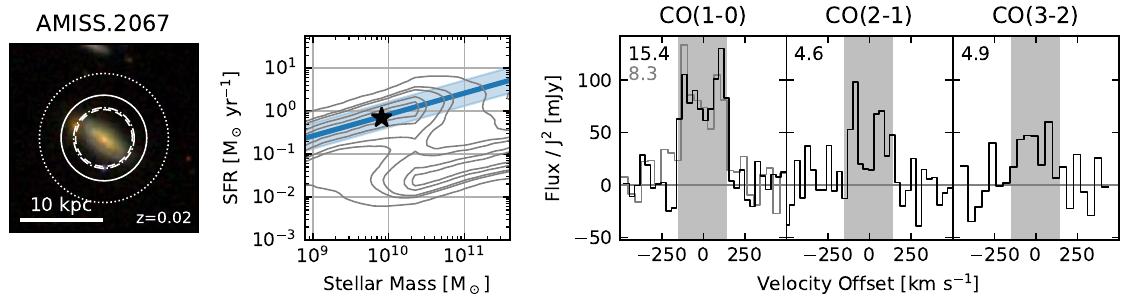}
\figsetgrpnote{Left column: SDSS cutouts of each target. Solid, dashed, and dotted lines show the 
beam sizes of the SMT for CO(2--1), the SMT for CO(3--2) (and the IRAM 30m for CO(1--0)) 
and the 12m for CO(1--0) respectively. The scale bar in the lower left shows 10 
kiloparsecs. 
Middle column: contours show the distribution of star formation rates at a given stellar 
mass, while the blue line and filled region show the main sequence of star forming 
galaxies. The stellar mass and star formation rate of the target galaxy is marked by a 
star. 
Right column: CO(1--0) spectra from AMISS (gray) and xCOLD~GASS (black), CO(2--1) spectra 
from AMISS, and CO(3--2) spectra from AMISS. Numbers in the upper right corner give the 
signal to noise ratio for each line. When a CO line is detected, the gray band indicates 
the region used to measure the line flux. The scale of the $y$-axis is such that lines 
would have the same amplitude in each transition for thermalized CO emission. The relative 
amplitudes of each spectrum give a sense of the luminosity ratios between the different 
lines.}
\figsetgrpend

\figsetgrpstart
\figsetgrpnum{15.123}
\figsetgrptitle{AMISS.2072}
\figsetplot{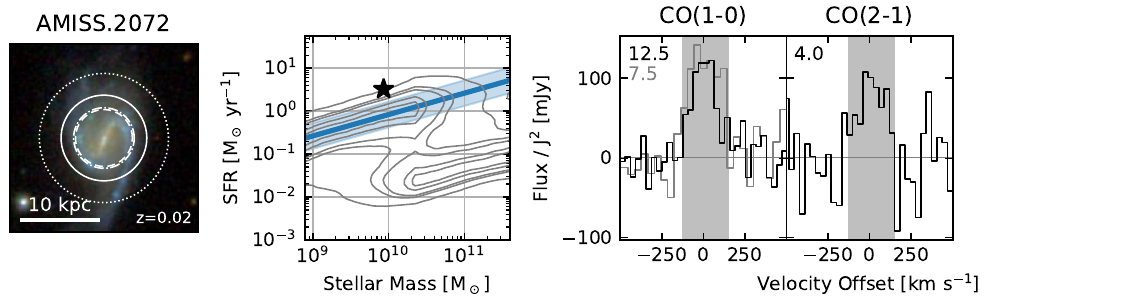}
\figsetgrpnote{Left column: SDSS cutouts of each target. Solid, dashed, and dotted lines show the 
beam sizes of the SMT for CO(2--1), the SMT for CO(3--2) (and the IRAM 30m for CO(1--0)) 
and the 12m for CO(1--0) respectively. The scale bar in the lower left shows 10 
kiloparsecs. 
Middle column: contours show the distribution of star formation rates at a given stellar 
mass, while the blue line and filled region show the main sequence of star forming 
galaxies. The stellar mass and star formation rate of the target galaxy is marked by a 
star. 
Right column: CO(1--0) spectra from AMISS (gray) and xCOLD~GASS (black), CO(2--1) spectra 
from AMISS, and CO(3--2) spectra from AMISS. Numbers in the upper right corner give the 
signal to noise ratio for each line. When a CO line is detected, the gray band indicates 
the region used to measure the line flux. The scale of the $y$-axis is such that lines 
would have the same amplitude in each transition for thermalized CO emission. The relative 
amplitudes of each spectrum give a sense of the luminosity ratios between the different 
lines.}
\figsetgrpend

\figsetgrpstart
\figsetgrpnum{15.124}
\figsetgrptitle{AMISS.2073}
\figsetplot{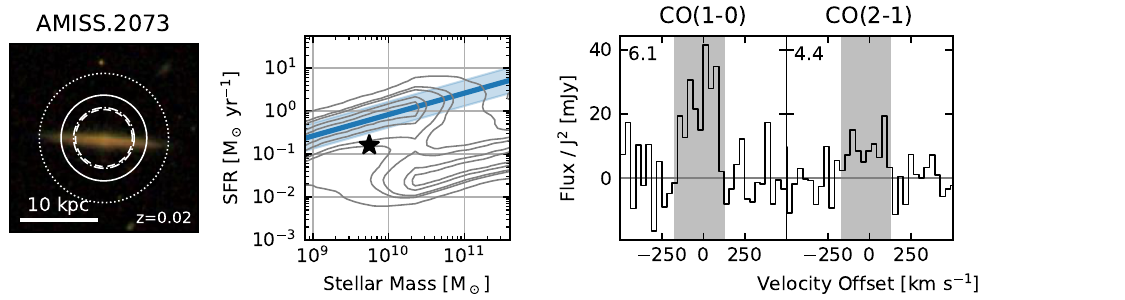}
\figsetgrpnote{Left column: SDSS cutouts of each target. Solid, dashed, and dotted lines show the 
beam sizes of the SMT for CO(2--1), the SMT for CO(3--2) (and the IRAM 30m for CO(1--0)) 
and the 12m for CO(1--0) respectively. The scale bar in the lower left shows 10 
kiloparsecs. 
Middle column: contours show the distribution of star formation rates at a given stellar 
mass, while the blue line and filled region show the main sequence of star forming 
galaxies. The stellar mass and star formation rate of the target galaxy is marked by a 
star. 
Right column: CO(1--0) spectra from AMISS (gray) and xCOLD~GASS (black), CO(2--1) spectra 
from AMISS, and CO(3--2) spectra from AMISS. Numbers in the upper right corner give the 
signal to noise ratio for each line. When a CO line is detected, the gray band indicates 
the region used to measure the line flux. The scale of the $y$-axis is such that lines 
would have the same amplitude in each transition for thermalized CO emission. The relative 
amplitudes of each spectrum give a sense of the luminosity ratios between the different 
lines.}
\figsetgrpend

\figsetgrpstart
\figsetgrpnum{15.125}
\figsetgrptitle{AMISS.2087}
\figsetplot{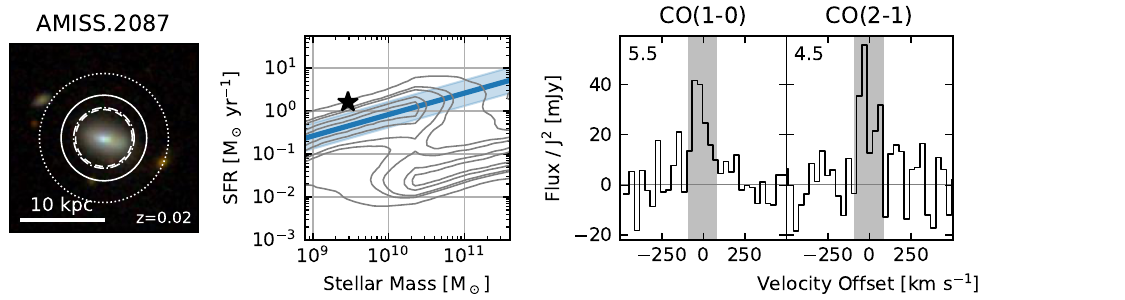}
\figsetgrpnote{Left column: SDSS cutouts of each target. Solid, dashed, and dotted lines show the 
beam sizes of the SMT for CO(2--1), the SMT for CO(3--2) (and the IRAM 30m for CO(1--0)) 
and the 12m for CO(1--0) respectively. The scale bar in the lower left shows 10 
kiloparsecs. 
Middle column: contours show the distribution of star formation rates at a given stellar 
mass, while the blue line and filled region show the main sequence of star forming 
galaxies. The stellar mass and star formation rate of the target galaxy is marked by a 
star. 
Right column: CO(1--0) spectra from AMISS (gray) and xCOLD~GASS (black), CO(2--1) spectra 
from AMISS, and CO(3--2) spectra from AMISS. Numbers in the upper right corner give the 
signal to noise ratio for each line. When a CO line is detected, the gray band indicates 
the region used to measure the line flux. The scale of the $y$-axis is such that lines 
would have the same amplitude in each transition for thermalized CO emission. The relative 
amplitudes of each spectrum give a sense of the luminosity ratios between the different 
lines.}
\figsetgrpend

\figsetgrpstart
\figsetgrpnum{15.126}
\figsetgrptitle{AMISS.2089}
\figsetplot{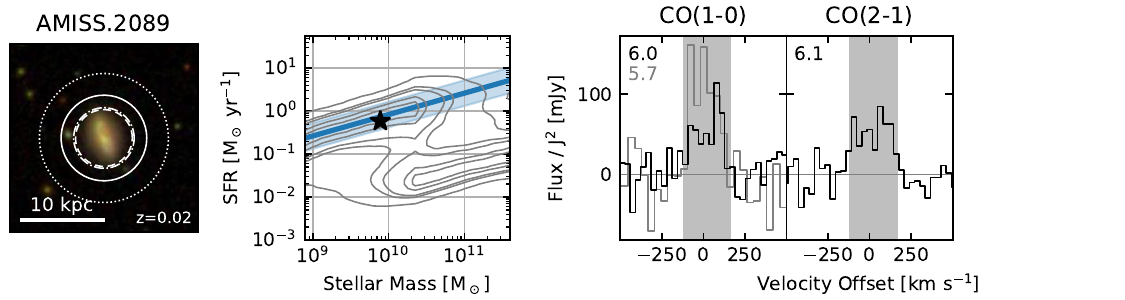}
\figsetgrpnote{Left column: SDSS cutouts of each target. Solid, dashed, and dotted lines show the 
beam sizes of the SMT for CO(2--1), the SMT for CO(3--2) (and the IRAM 30m for CO(1--0)) 
and the 12m for CO(1--0) respectively. The scale bar in the lower left shows 10 
kiloparsecs. 
Middle column: contours show the distribution of star formation rates at a given stellar 
mass, while the blue line and filled region show the main sequence of star forming 
galaxies. The stellar mass and star formation rate of the target galaxy is marked by a 
star. 
Right column: CO(1--0) spectra from AMISS (gray) and xCOLD~GASS (black), CO(2--1) spectra 
from AMISS, and CO(3--2) spectra from AMISS. Numbers in the upper right corner give the 
signal to noise ratio for each line. When a CO line is detected, the gray band indicates 
the region used to measure the line flux. The scale of the $y$-axis is such that lines 
would have the same amplitude in each transition for thermalized CO emission. The relative 
amplitudes of each spectrum give a sense of the luminosity ratios between the different 
lines.}
\figsetgrpend

\figsetgrpstart
\figsetgrpnum{15.127}
\figsetgrptitle{AMISS.2090}
\figsetplot{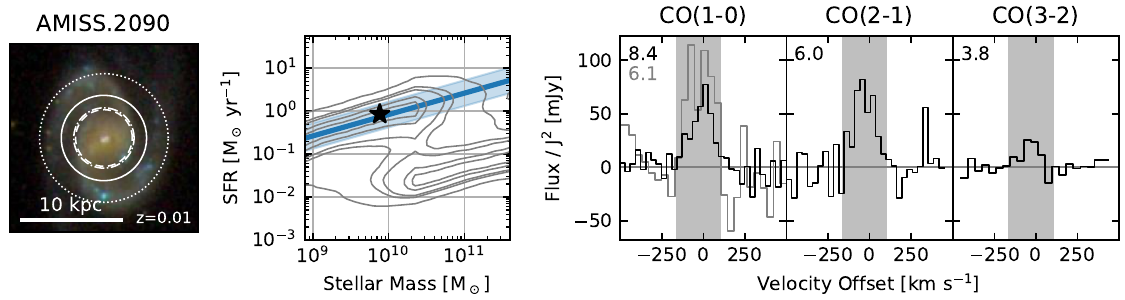}
\figsetgrpnote{Left column: SDSS cutouts of each target. Solid, dashed, and dotted lines show the 
beam sizes of the SMT for CO(2--1), the SMT for CO(3--2) (and the IRAM 30m for CO(1--0)) 
and the 12m for CO(1--0) respectively. The scale bar in the lower left shows 10 
kiloparsecs. 
Middle column: contours show the distribution of star formation rates at a given stellar 
mass, while the blue line and filled region show the main sequence of star forming 
galaxies. The stellar mass and star formation rate of the target galaxy is marked by a 
star. 
Right column: CO(1--0) spectra from AMISS (gray) and xCOLD~GASS (black), CO(2--1) spectra 
from AMISS, and CO(3--2) spectra from AMISS. Numbers in the upper right corner give the 
signal to noise ratio for each line. When a CO line is detected, the gray band indicates 
the region used to measure the line flux. The scale of the $y$-axis is such that lines 
would have the same amplitude in each transition for thermalized CO emission. The relative 
amplitudes of each spectrum give a sense of the luminosity ratios between the different 
lines.}
\figsetgrpend

\figsetgrpstart
\figsetgrpnum{15.128}
\figsetgrptitle{AMISS.2096}
\figsetplot{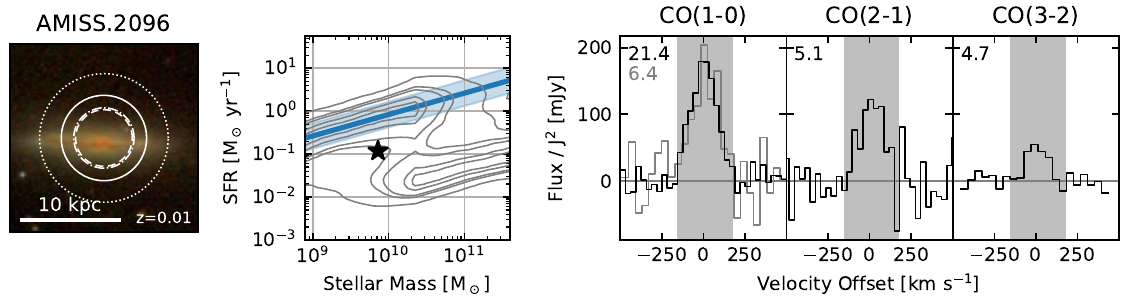}
\figsetgrpnote{Left column: SDSS cutouts of each target. Solid, dashed, and dotted lines show the 
beam sizes of the SMT for CO(2--1), the SMT for CO(3--2) (and the IRAM 30m for CO(1--0)) 
and the 12m for CO(1--0) respectively. The scale bar in the lower left shows 10 
kiloparsecs. 
Middle column: contours show the distribution of star formation rates at a given stellar 
mass, while the blue line and filled region show the main sequence of star forming 
galaxies. The stellar mass and star formation rate of the target galaxy is marked by a 
star. 
Right column: CO(1--0) spectra from AMISS (gray) and xCOLD~GASS (black), CO(2--1) spectra 
from AMISS, and CO(3--2) spectra from AMISS. Numbers in the upper right corner give the 
signal to noise ratio for each line. When a CO line is detected, the gray band indicates 
the region used to measure the line flux. The scale of the $y$-axis is such that lines 
would have the same amplitude in each transition for thermalized CO emission. The relative 
amplitudes of each spectrum give a sense of the luminosity ratios between the different 
lines.}
\figsetgrpend

\figsetgrpstart
\figsetgrpnum{15.129}
\figsetgrptitle{AMISS.3002}
\figsetplot{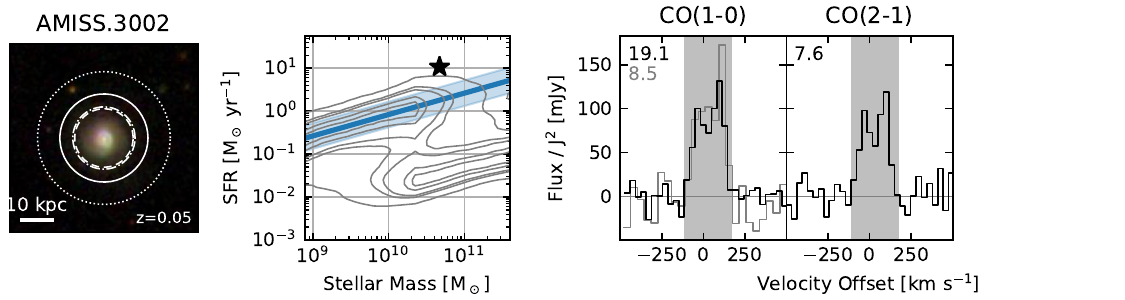}
\figsetgrpnote{Left column: SDSS cutouts of each target. Solid, dashed, and dotted lines show the 
beam sizes of the SMT for CO(2--1), the SMT for CO(3--2) (and the IRAM 30m for CO(1--0)) 
and the 12m for CO(1--0) respectively. The scale bar in the lower left shows 10 
kiloparsecs. 
Middle column: contours show the distribution of star formation rates at a given stellar 
mass, while the blue line and filled region show the main sequence of star forming 
galaxies. The stellar mass and star formation rate of the target galaxy is marked by a 
star. 
Right column: CO(1--0) spectra from AMISS (gray) and xCOLD~GASS (black), CO(2--1) spectra 
from AMISS, and CO(3--2) spectra from AMISS. Numbers in the upper right corner give the 
signal to noise ratio for each line. When a CO line is detected, the gray band indicates 
the region used to measure the line flux. The scale of the $y$-axis is such that lines 
would have the same amplitude in each transition for thermalized CO emission. The relative 
amplitudes of each spectrum give a sense of the luminosity ratios between the different 
lines.}
\figsetgrpend

\figsetgrpstart
\figsetgrpnum{15.130}
\figsetgrptitle{AMISS.3004}
\figsetplot{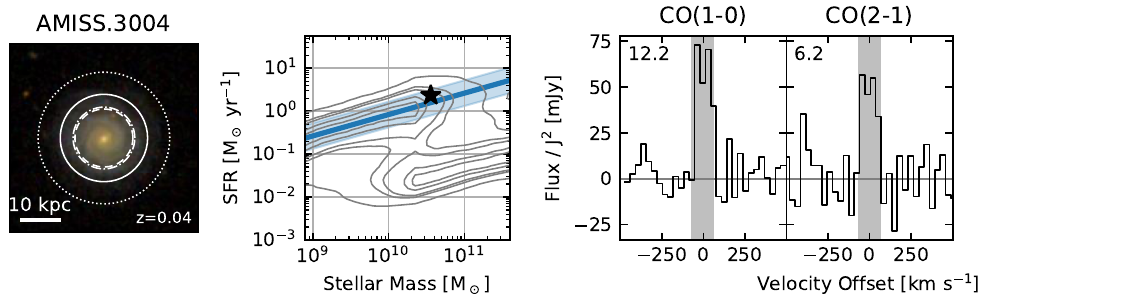}
\figsetgrpnote{Left column: SDSS cutouts of each target. Solid, dashed, and dotted lines show the 
beam sizes of the SMT for CO(2--1), the SMT for CO(3--2) (and the IRAM 30m for CO(1--0)) 
and the 12m for CO(1--0) respectively. The scale bar in the lower left shows 10 
kiloparsecs. 
Middle column: contours show the distribution of star formation rates at a given stellar 
mass, while the blue line and filled region show the main sequence of star forming 
galaxies. The stellar mass and star formation rate of the target galaxy is marked by a 
star. 
Right column: CO(1--0) spectra from AMISS (gray) and xCOLD~GASS (black), CO(2--1) spectra 
from AMISS, and CO(3--2) spectra from AMISS. Numbers in the upper right corner give the 
signal to noise ratio for each line. When a CO line is detected, the gray band indicates 
the region used to measure the line flux. The scale of the $y$-axis is such that lines 
would have the same amplitude in each transition for thermalized CO emission. The relative 
amplitudes of each spectrum give a sense of the luminosity ratios between the different 
lines.}
\figsetgrpend

\figsetgrpstart
\figsetgrpnum{15.131}
\figsetgrptitle{AMISS.3027}
\figsetplot{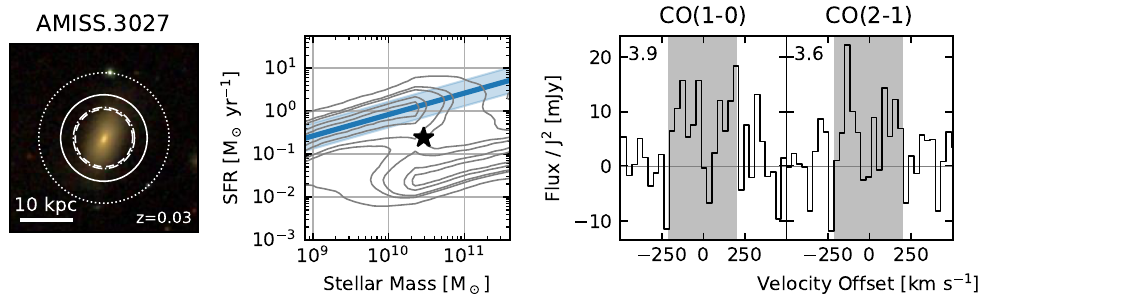}
\figsetgrpnote{Left column: SDSS cutouts of each target. Solid, dashed, and dotted lines show the 
beam sizes of the SMT for CO(2--1), the SMT for CO(3--2) (and the IRAM 30m for CO(1--0)) 
and the 12m for CO(1--0) respectively. The scale bar in the lower left shows 10 
kiloparsecs. 
Middle column: contours show the distribution of star formation rates at a given stellar 
mass, while the blue line and filled region show the main sequence of star forming 
galaxies. The stellar mass and star formation rate of the target galaxy is marked by a 
star. 
Right column: CO(1--0) spectra from AMISS (gray) and xCOLD~GASS (black), CO(2--1) spectra 
from AMISS, and CO(3--2) spectra from AMISS. Numbers in the upper right corner give the 
signal to noise ratio for each line. When a CO line is detected, the gray band indicates 
the region used to measure the line flux. The scale of the $y$-axis is such that lines 
would have the same amplitude in each transition for thermalized CO emission. The relative 
amplitudes of each spectrum give a sense of the luminosity ratios between the different 
lines.}
\figsetgrpend

\figsetgrpstart
\figsetgrpnum{15.132}
\figsetgrptitle{AMISS.3048}
\figsetplot{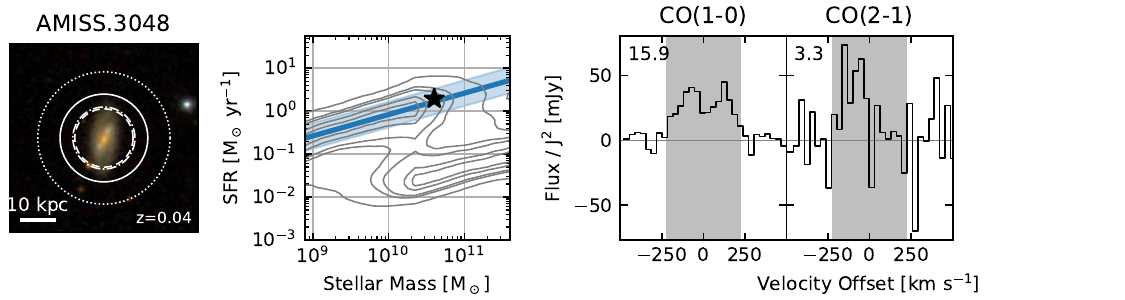}
\figsetgrpnote{Left column: SDSS cutouts of each target. Solid, dashed, and dotted lines show the 
beam sizes of the SMT for CO(2--1), the SMT for CO(3--2) (and the IRAM 30m for CO(1--0)) 
and the 12m for CO(1--0) respectively. The scale bar in the lower left shows 10 
kiloparsecs. 
Middle column: contours show the distribution of star formation rates at a given stellar 
mass, while the blue line and filled region show the main sequence of star forming 
galaxies. The stellar mass and star formation rate of the target galaxy is marked by a 
star. 
Right column: CO(1--0) spectra from AMISS (gray) and xCOLD~GASS (black), CO(2--1) spectra 
from AMISS, and CO(3--2) spectra from AMISS. Numbers in the upper right corner give the 
signal to noise ratio for each line. When a CO line is detected, the gray band indicates 
the region used to measure the line flux. The scale of the $y$-axis is such that lines 
would have the same amplitude in each transition for thermalized CO emission. The relative 
amplitudes of each spectrum give a sense of the luminosity ratios between the different 
lines.}
\figsetgrpend

\figsetgrpstart
\figsetgrpnum{15.133}
\figsetgrptitle{AMISS.3052}
\figsetplot{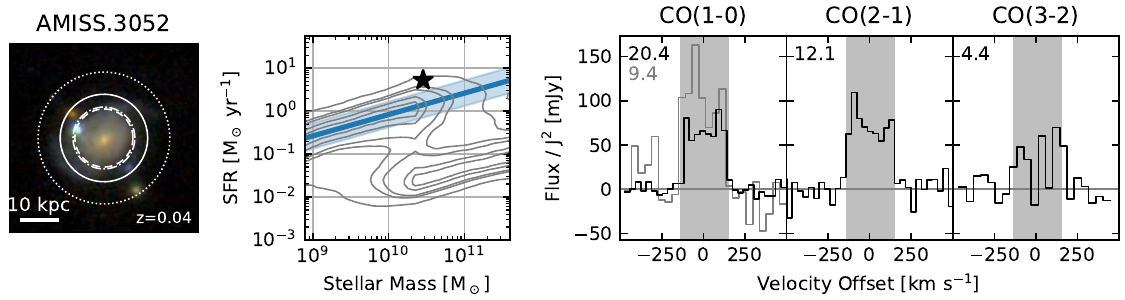}
\figsetgrpnote{Left column: SDSS cutouts of each target. Solid, dashed, and dotted lines show the 
beam sizes of the SMT for CO(2--1), the SMT for CO(3--2) (and the IRAM 30m for CO(1--0)) 
and the 12m for CO(1--0) respectively. The scale bar in the lower left shows 10 
kiloparsecs. 
Middle column: contours show the distribution of star formation rates at a given stellar 
mass, while the blue line and filled region show the main sequence of star forming 
galaxies. The stellar mass and star formation rate of the target galaxy is marked by a 
star. 
Right column: CO(1--0) spectra from AMISS (gray) and xCOLD~GASS (black), CO(2--1) spectra 
from AMISS, and CO(3--2) spectra from AMISS. Numbers in the upper right corner give the 
signal to noise ratio for each line. When a CO line is detected, the gray band indicates 
the region used to measure the line flux. The scale of the $y$-axis is such that lines 
would have the same amplitude in each transition for thermalized CO emission. The relative 
amplitudes of each spectrum give a sense of the luminosity ratios between the different 
lines.}
\figsetgrpend

\figsetgrpstart
\figsetgrpnum{15.134}
\figsetgrptitle{AMISS.3053}
\figsetplot{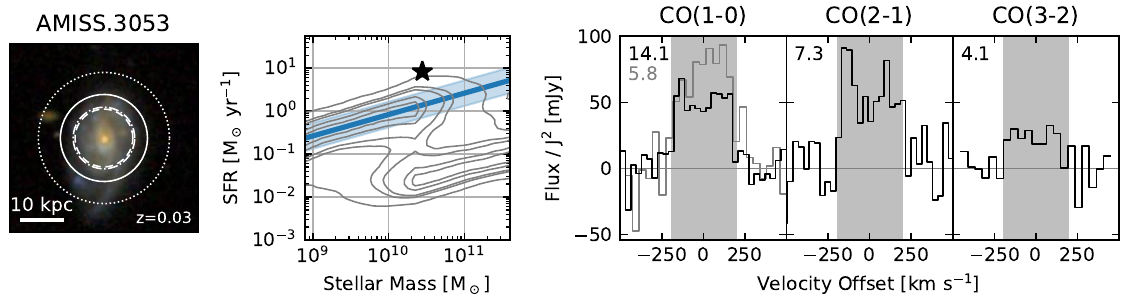}
\figsetgrpnote{Left column: SDSS cutouts of each target. Solid, dashed, and dotted lines show the 
beam sizes of the SMT for CO(2--1), the SMT for CO(3--2) (and the IRAM 30m for CO(1--0)) 
and the 12m for CO(1--0) respectively. The scale bar in the lower left shows 10 
kiloparsecs. 
Middle column: contours show the distribution of star formation rates at a given stellar 
mass, while the blue line and filled region show the main sequence of star forming 
galaxies. The stellar mass and star formation rate of the target galaxy is marked by a 
star. 
Right column: CO(1--0) spectra from AMISS (gray) and xCOLD~GASS (black), CO(2--1) spectra 
from AMISS, and CO(3--2) spectra from AMISS. Numbers in the upper right corner give the 
signal to noise ratio for each line. When a CO line is detected, the gray band indicates 
the region used to measure the line flux. The scale of the $y$-axis is such that lines 
would have the same amplitude in each transition for thermalized CO emission. The relative 
amplitudes of each spectrum give a sense of the luminosity ratios between the different 
lines.}
\figsetgrpend

\figsetgrpstart
\figsetgrpnum{15.135}
\figsetgrptitle{AMISS.3078}
\figsetplot{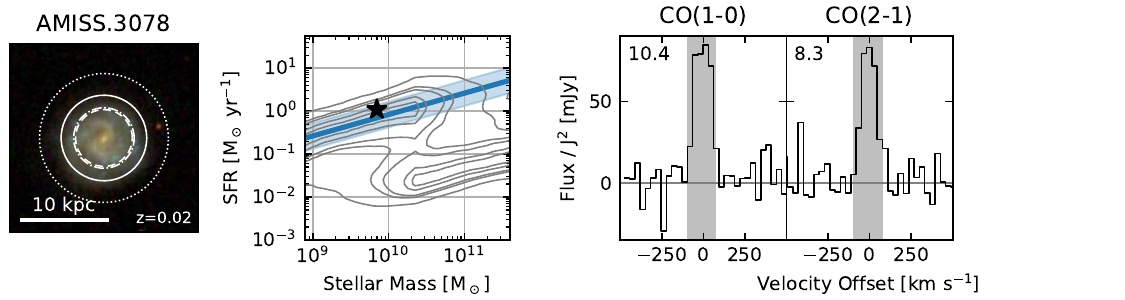}
\figsetgrpnote{Left column: SDSS cutouts of each target. Solid, dashed, and dotted lines show the 
beam sizes of the SMT for CO(2--1), the SMT for CO(3--2) (and the IRAM 30m for CO(1--0)) 
and the 12m for CO(1--0) respectively. The scale bar in the lower left shows 10 
kiloparsecs. 
Middle column: contours show the distribution of star formation rates at a given stellar 
mass, while the blue line and filled region show the main sequence of star forming 
galaxies. The stellar mass and star formation rate of the target galaxy is marked by a 
star. 
Right column: CO(1--0) spectra from AMISS (gray) and xCOLD~GASS (black), CO(2--1) spectra 
from AMISS, and CO(3--2) spectra from AMISS. Numbers in the upper right corner give the 
signal to noise ratio for each line. When a CO line is detected, the gray band indicates 
the region used to measure the line flux. The scale of the $y$-axis is such that lines 
would have the same amplitude in each transition for thermalized CO emission. The relative 
amplitudes of each spectrum give a sense of the luminosity ratios between the different 
lines.}
\figsetgrpend

\figsetgrpstart
\figsetgrpnum{15.136}
\figsetgrptitle{AMISS.3081}
\figsetplot{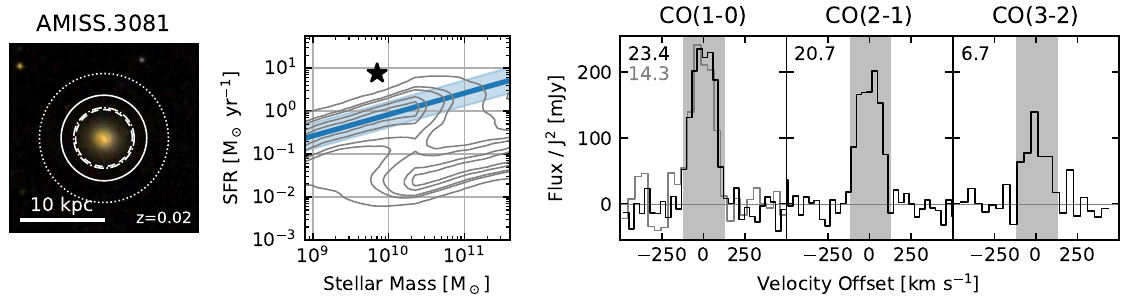}
\figsetgrpnote{Left column: SDSS cutouts of each target. Solid, dashed, and dotted lines show the 
beam sizes of the SMT for CO(2--1), the SMT for CO(3--2) (and the IRAM 30m for CO(1--0)) 
and the 12m for CO(1--0) respectively. The scale bar in the lower left shows 10 
kiloparsecs. 
Middle column: contours show the distribution of star formation rates at a given stellar 
mass, while the blue line and filled region show the main sequence of star forming 
galaxies. The stellar mass and star formation rate of the target galaxy is marked by a 
star. 
Right column: CO(1--0) spectra from AMISS (gray) and xCOLD~GASS (black), CO(2--1) spectra 
from AMISS, and CO(3--2) spectra from AMISS. Numbers in the upper right corner give the 
signal to noise ratio for each line. When a CO line is detected, the gray band indicates 
the region used to measure the line flux. The scale of the $y$-axis is such that lines 
would have the same amplitude in each transition for thermalized CO emission. The relative 
amplitudes of each spectrum give a sense of the luminosity ratios between the different 
lines.}
\figsetgrpend

\figsetgrpstart
\figsetgrpnum{15.137}
\figsetgrptitle{AMISS.3083}
\figsetplot{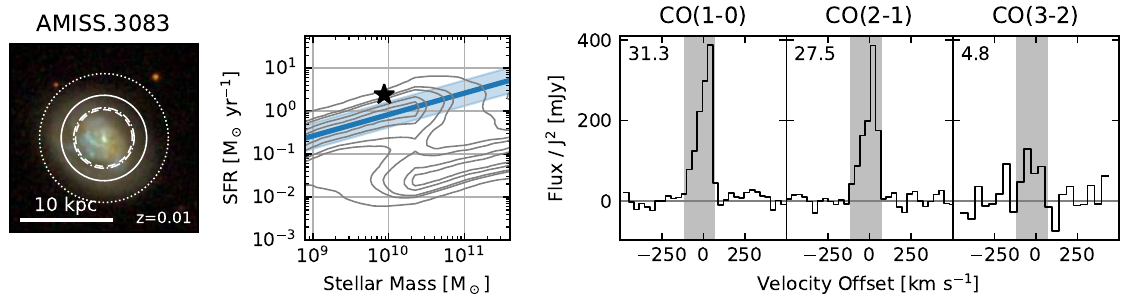}
\figsetgrpnote{Left column: SDSS cutouts of each target. Solid, dashed, and dotted lines show the 
beam sizes of the SMT for CO(2--1), the SMT for CO(3--2) (and the IRAM 30m for CO(1--0)) 
and the 12m for CO(1--0) respectively. The scale bar in the lower left shows 10 
kiloparsecs. 
Middle column: contours show the distribution of star formation rates at a given stellar 
mass, while the blue line and filled region show the main sequence of star forming 
galaxies. The stellar mass and star formation rate of the target galaxy is marked by a 
star. 
Right column: CO(1--0) spectra from AMISS (gray) and xCOLD~GASS (black), CO(2--1) spectra 
from AMISS, and CO(3--2) spectra from AMISS. Numbers in the upper right corner give the 
signal to noise ratio for each line. When a CO line is detected, the gray band indicates 
the region used to measure the line flux. The scale of the $y$-axis is such that lines 
would have the same amplitude in each transition for thermalized CO emission. The relative 
amplitudes of each spectrum give a sense of the luminosity ratios between the different 
lines.}
\figsetgrpend

\figsetgrpstart
\figsetgrpnum{15.138}
\figsetgrptitle{AMISS.9019}
\figsetplot{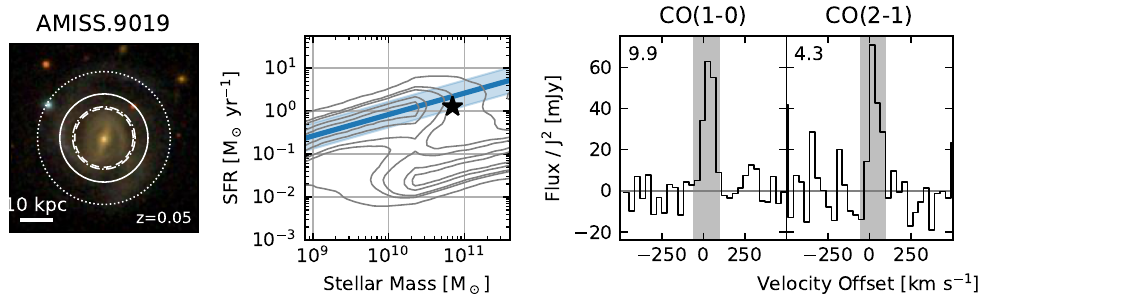}
\figsetgrpnote{Left column: SDSS cutouts of each target. Solid, dashed, and dotted lines show the 
beam sizes of the SMT for CO(2--1), the SMT for CO(3--2) (and the IRAM 30m for CO(1--0)) 
and the 12m for CO(1--0) respectively. The scale bar in the lower left shows 10 
kiloparsecs. 
Middle column: contours show the distribution of star formation rates at a given stellar 
mass, while the blue line and filled region show the main sequence of star forming 
galaxies. The stellar mass and star formation rate of the target galaxy is marked by a 
star. 
Right column: CO(1--0) spectra from AMISS (gray) and xCOLD~GASS (black), CO(2--1) spectra 
from AMISS, and CO(3--2) spectra from AMISS. Numbers in the upper right corner give the 
signal to noise ratio for each line. When a CO line is detected, the gray band indicates 
the region used to measure the line flux. The scale of the $y$-axis is such that lines 
would have the same amplitude in each transition for thermalized CO emission. The relative 
amplitudes of each spectrum give a sense of the luminosity ratios between the different 
lines.}
\figsetgrpend

\figsetgrpstart
\figsetgrpnum{15.139}
\figsetgrptitle{AMISS.9031}
\figsetplot{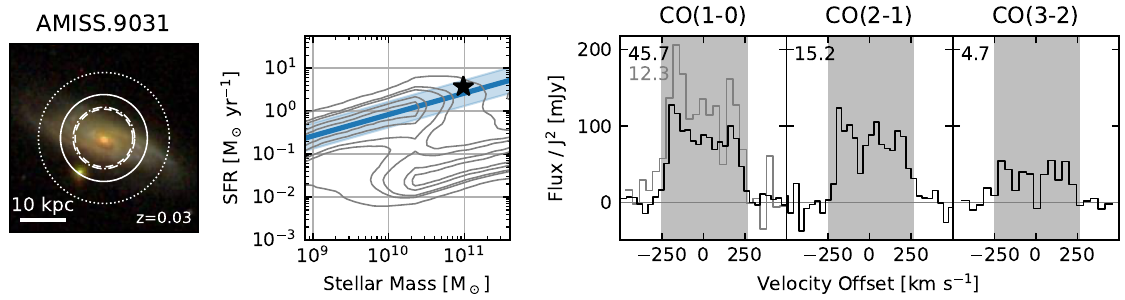}
\figsetgrpnote{Left column: SDSS cutouts of each target. Solid, dashed, and dotted lines show the 
beam sizes of the SMT for CO(2--1), the SMT for CO(3--2) (and the IRAM 30m for CO(1--0)) 
and the 12m for CO(1--0) respectively. The scale bar in the lower left shows 10 
kiloparsecs. 
Middle column: contours show the distribution of star formation rates at a given stellar 
mass, while the blue line and filled region show the main sequence of star forming 
galaxies. The stellar mass and star formation rate of the target galaxy is marked by a 
star. 
Right column: CO(1--0) spectra from AMISS (gray) and xCOLD~GASS (black), CO(2--1) spectra 
from AMISS, and CO(3--2) spectra from AMISS. Numbers in the upper right corner give the 
signal to noise ratio for each line. When a CO line is detected, the gray band indicates 
the region used to measure the line flux. The scale of the $y$-axis is such that lines 
would have the same amplitude in each transition for thermalized CO emission. The relative 
amplitudes of each spectrum give a sense of the luminosity ratios between the different 
lines.}
\figsetgrpend

\figsetgrpstart
\figsetgrpnum{15.140}
\figsetgrptitle{AMISS.9034}
\figsetplot{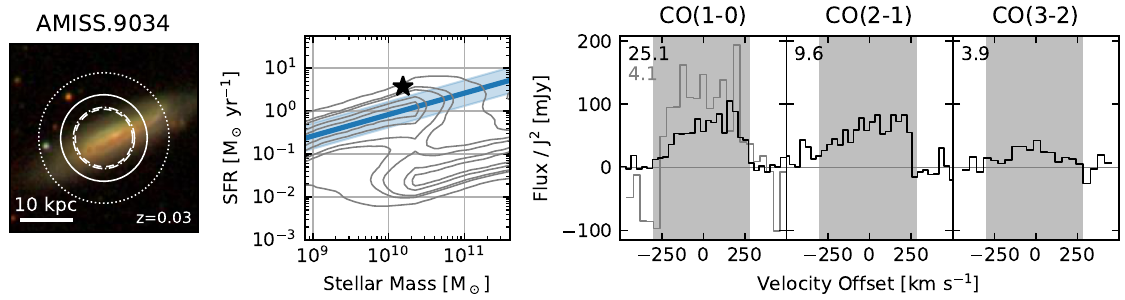}
\figsetgrpnote{Left column: SDSS cutouts of each target. Solid, dashed, and dotted lines show the 
beam sizes of the SMT for CO(2--1), the SMT for CO(3--2) (and the IRAM 30m for CO(1--0)) 
and the 12m for CO(1--0) respectively. The scale bar in the lower left shows 10 
kiloparsecs. 
Middle column: contours show the distribution of star formation rates at a given stellar 
mass, while the blue line and filled region show the main sequence of star forming 
galaxies. The stellar mass and star formation rate of the target galaxy is marked by a 
star. 
Right column: CO(1--0) spectra from AMISS (gray) and xCOLD~GASS (black), CO(2--1) spectra 
from AMISS, and CO(3--2) spectra from AMISS. Numbers in the upper right corner give the 
signal to noise ratio for each line. When a CO line is detected, the gray band indicates 
the region used to measure the line flux. The scale of the $y$-axis is such that lines 
would have the same amplitude in each transition for thermalized CO emission. The relative 
amplitudes of each spectrum give a sense of the luminosity ratios between the different 
lines.}
\figsetgrpend

\figsetgrpstart
\figsetgrpnum{15.141}
\figsetgrptitle{AMISS.9050}
\figsetplot{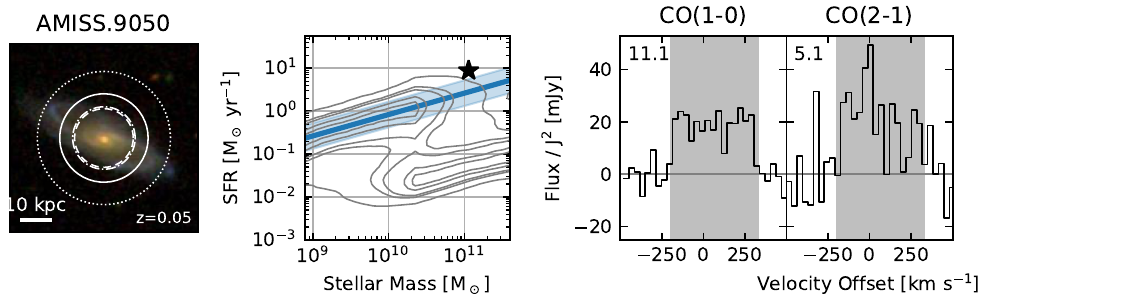}
\figsetgrpnote{Left column: SDSS cutouts of each target. Solid, dashed, and dotted lines show the 
beam sizes of the SMT for CO(2--1), the SMT for CO(3--2) (and the IRAM 30m for CO(1--0)) 
and the 12m for CO(1--0) respectively. The scale bar in the lower left shows 10 
kiloparsecs. 
Middle column: contours show the distribution of star formation rates at a given stellar 
mass, while the blue line and filled region show the main sequence of star forming 
galaxies. The stellar mass and star formation rate of the target galaxy is marked by a 
star. 
Right column: CO(1--0) spectra from AMISS (gray) and xCOLD~GASS (black), CO(2--1) spectra 
from AMISS, and CO(3--2) spectra from AMISS. Numbers in the upper right corner give the 
signal to noise ratio for each line. When a CO line is detected, the gray band indicates 
the region used to measure the line flux. The scale of the $y$-axis is such that lines 
would have the same amplitude in each transition for thermalized CO emission. The relative 
amplitudes of each spectrum give a sense of the luminosity ratios between the different 
lines.}
\figsetgrpend

\figsetgrpstart
\figsetgrpnum{15.142}
\figsetgrptitle{AMISS.9060}
\figsetplot{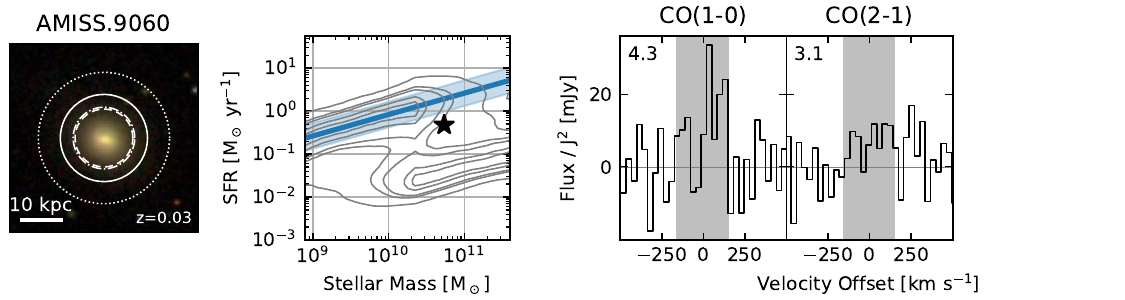}
\figsetgrpnote{Left column: SDSS cutouts of each target. Solid, dashed, and dotted lines show the 
beam sizes of the SMT for CO(2--1), the SMT for CO(3--2) (and the IRAM 30m for CO(1--0)) 
and the 12m for CO(1--0) respectively. The scale bar in the lower left shows 10 
kiloparsecs. 
Middle column: contours show the distribution of star formation rates at a given stellar 
mass, while the blue line and filled region show the main sequence of star forming 
galaxies. The stellar mass and star formation rate of the target galaxy is marked by a 
star. 
Right column: CO(1--0) spectra from AMISS (gray) and xCOLD~GASS (black), CO(2--1) spectra 
from AMISS, and CO(3--2) spectra from AMISS. Numbers in the upper right corner give the 
signal to noise ratio for each line. When a CO line is detected, the gray band indicates 
the region used to measure the line flux. The scale of the $y$-axis is such that lines 
would have the same amplitude in each transition for thermalized CO emission. The relative 
amplitudes of each spectrum give a sense of the luminosity ratios between the different 
lines.}
\figsetgrpend

\figsetgrpstart
\figsetgrpnum{15.143}
\figsetgrptitle{AMISS.9080}
\figsetplot{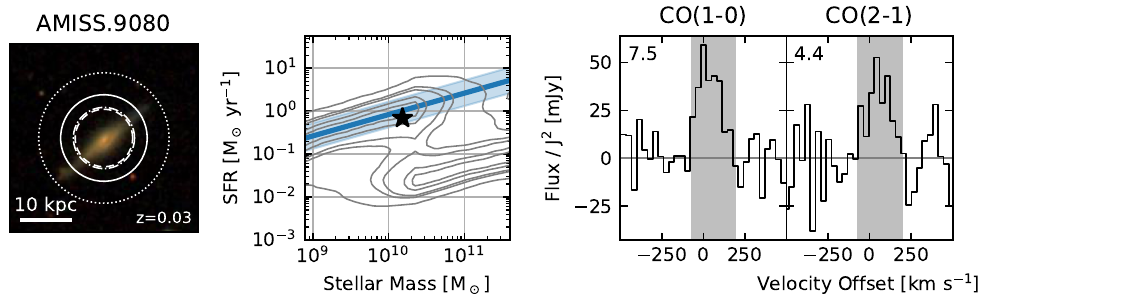}
\figsetgrpnote{Left column: SDSS cutouts of each target. Solid, dashed, and dotted lines show the 
beam sizes of the SMT for CO(2--1), the SMT for CO(3--2) (and the IRAM 30m for CO(1--0)) 
and the 12m for CO(1--0) respectively. The scale bar in the lower left shows 10 
kiloparsecs. 
Middle column: contours show the distribution of star formation rates at a given stellar 
mass, while the blue line and filled region show the main sequence of star forming 
galaxies. The stellar mass and star formation rate of the target galaxy is marked by a 
star. 
Right column: CO(1--0) spectra from AMISS (gray) and xCOLD~GASS (black), CO(2--1) spectra 
from AMISS, and CO(3--2) spectra from AMISS. Numbers in the upper right corner give the 
signal to noise ratio for each line. When a CO line is detected, the gray band indicates 
the region used to measure the line flux. The scale of the $y$-axis is such that lines 
would have the same amplitude in each transition for thermalized CO emission. The relative 
amplitudes of each spectrum give a sense of the luminosity ratios between the different 
lines.}
\figsetgrpend

\figsetgrpstart
\figsetgrpnum{15.144}
\figsetgrptitle{AMISS.9083}
\figsetplot{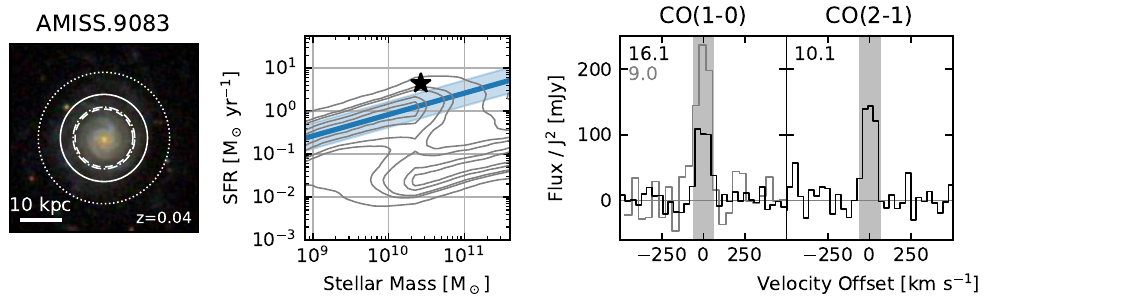}
\figsetgrpnote{Left column: SDSS cutouts of each target. Solid, dashed, and dotted lines show the 
beam sizes of the SMT for CO(2--1), the SMT for CO(3--2) (and the IRAM 30m for CO(1--0)) 
and the 12m for CO(1--0) respectively. The scale bar in the lower left shows 10 
kiloparsecs. 
Middle column: contours show the distribution of star formation rates at a given stellar 
mass, while the blue line and filled region show the main sequence of star forming 
galaxies. The stellar mass and star formation rate of the target galaxy is marked by a 
star. 
Right column: CO(1--0) spectra from AMISS (gray) and xCOLD~GASS (black), CO(2--1) spectra 
from AMISS, and CO(3--2) spectra from AMISS. Numbers in the upper right corner give the 
signal to noise ratio for each line. When a CO line is detected, the gray band indicates 
the region used to measure the line flux. The scale of the $y$-axis is such that lines 
would have the same amplitude in each transition for thermalized CO emission. The relative 
amplitudes of each spectrum give a sense of the luminosity ratios between the different 
lines.}
\figsetgrpend

\figsetgrpstart
\figsetgrpnum{15.145}
\figsetgrptitle{AMISS.9092}
\figsetplot{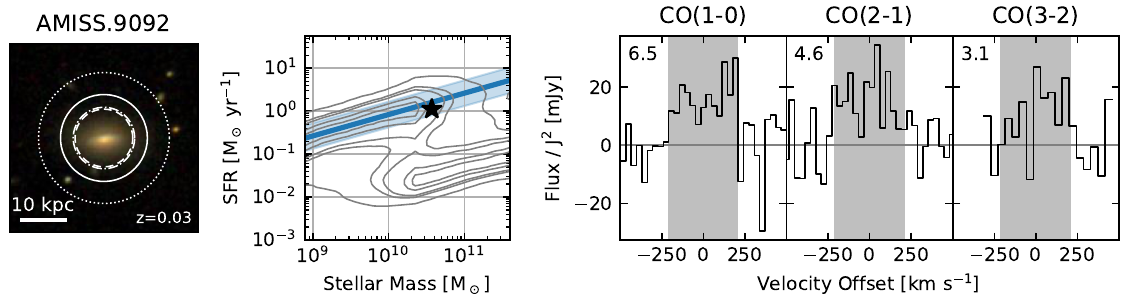}
\figsetgrpnote{Left column: SDSS cutouts of each target. Solid, dashed, and dotted lines show the 
beam sizes of the SMT for CO(2--1), the SMT for CO(3--2) (and the IRAM 30m for CO(1--0)) 
and the 12m for CO(1--0) respectively. The scale bar in the lower left shows 10 
kiloparsecs. 
Middle column: contours show the distribution of star formation rates at a given stellar 
mass, while the blue line and filled region show the main sequence of star forming 
galaxies. The stellar mass and star formation rate of the target galaxy is marked by a 
star. 
Right column: CO(1--0) spectra from AMISS (gray) and xCOLD~GASS (black), CO(2--1) spectra 
from AMISS, and CO(3--2) spectra from AMISS. Numbers in the upper right corner give the 
signal to noise ratio for each line. When a CO line is detected, the gray band indicates 
the region used to measure the line flux. The scale of the $y$-axis is such that lines 
would have the same amplitude in each transition for thermalized CO emission. The relative 
amplitudes of each spectrum give a sense of the luminosity ratios between the different 
lines.}
\figsetgrpend

\figsetgrpstart
\figsetgrpnum{15.146}
\figsetgrptitle{AMISS.9098}
\figsetplot{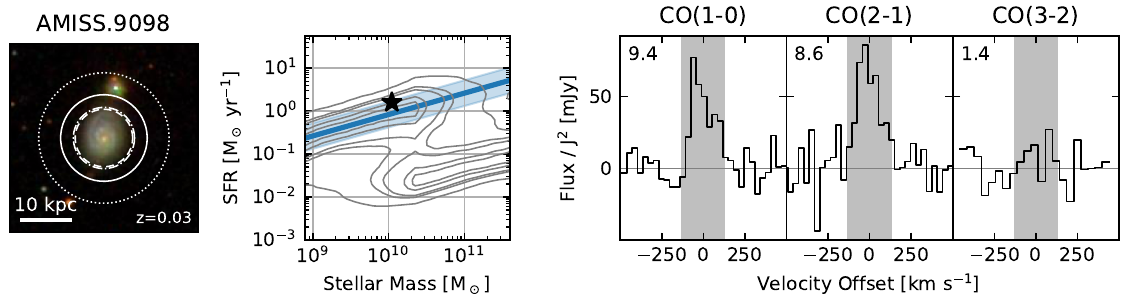}
\figsetgrpnote{Left column: SDSS cutouts of each target. Solid, dashed, and dotted lines show the 
beam sizes of the SMT for CO(2--1), the SMT for CO(3--2) (and the IRAM 30m for CO(1--0)) 
and the 12m for CO(1--0) respectively. The scale bar in the lower left shows 10 
kiloparsecs. 
Middle column: contours show the distribution of star formation rates at a given stellar 
mass, while the blue line and filled region show the main sequence of star forming 
galaxies. The stellar mass and star formation rate of the target galaxy is marked by a 
star. 
Right column: CO(1--0) spectra from AMISS (gray) and xCOLD~GASS (black), CO(2--1) spectra 
from AMISS, and CO(3--2) spectra from AMISS. Numbers in the upper right corner give the 
signal to noise ratio for each line. When a CO line is detected, the gray band indicates 
the region used to measure the line flux. The scale of the $y$-axis is such that lines 
would have the same amplitude in each transition for thermalized CO emission. The relative 
amplitudes of each spectrum give a sense of the luminosity ratios between the different 
lines.}
\figsetgrpend

\figsetgrpstart
\figsetgrpnum{15.147}
\figsetgrptitle{AMISS.9103}
\figsetplot{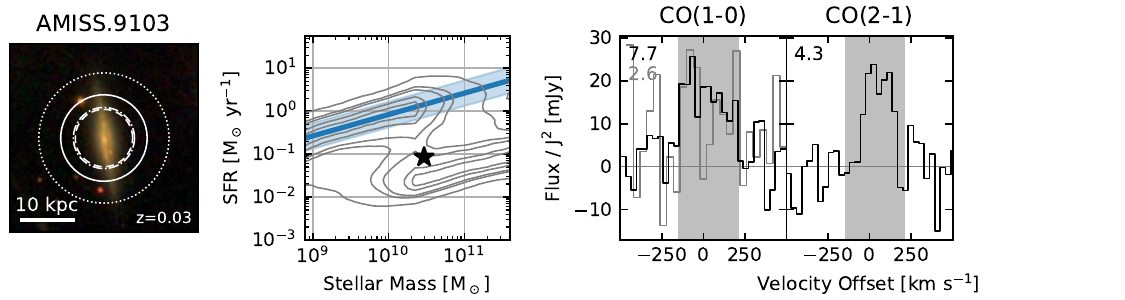}
\figsetgrpnote{Left column: SDSS cutouts of each target. Solid, dashed, and dotted lines show the 
beam sizes of the SMT for CO(2--1), the SMT for CO(3--2) (and the IRAM 30m for CO(1--0)) 
and the 12m for CO(1--0) respectively. The scale bar in the lower left shows 10 
kiloparsecs. 
Middle column: contours show the distribution of star formation rates at a given stellar 
mass, while the blue line and filled region show the main sequence of star forming 
galaxies. The stellar mass and star formation rate of the target galaxy is marked by a 
star. 
Right column: CO(1--0) spectra from AMISS (gray) and xCOLD~GASS (black), CO(2--1) spectra 
from AMISS, and CO(3--2) spectra from AMISS. Numbers in the upper right corner give the 
signal to noise ratio for each line. When a CO line is detected, the gray band indicates 
the region used to measure the line flux. The scale of the $y$-axis is such that lines 
would have the same amplitude in each transition for thermalized CO emission. The relative 
amplitudes of each spectrum give a sense of the luminosity ratios between the different 
lines.}
\figsetgrpend

\figsetgrpstart
\figsetgrpnum{15.148}
\figsetgrptitle{AMISS.9121}
\figsetplot{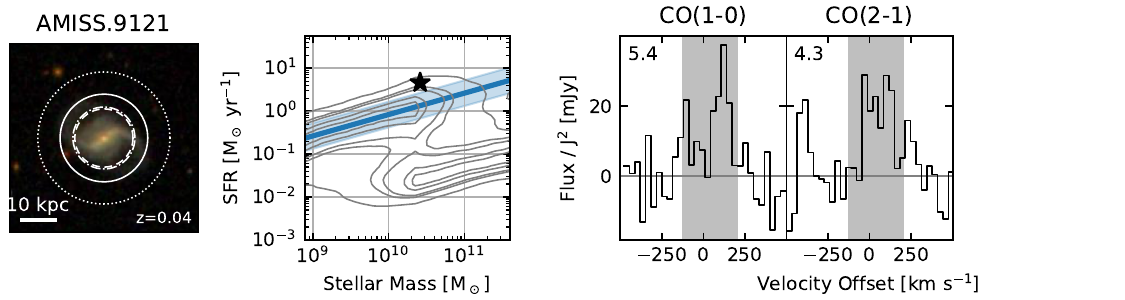}
\figsetgrpnote{Left column: SDSS cutouts of each target. Solid, dashed, and dotted lines show the 
beam sizes of the SMT for CO(2--1), the SMT for CO(3--2) (and the IRAM 30m for CO(1--0)) 
and the 12m for CO(1--0) respectively. The scale bar in the lower left shows 10 
kiloparsecs. 
Middle column: contours show the distribution of star formation rates at a given stellar 
mass, while the blue line and filled region show the main sequence of star forming 
galaxies. The stellar mass and star formation rate of the target galaxy is marked by a 
star. 
Right column: CO(1--0) spectra from AMISS (gray) and xCOLD~GASS (black), CO(2--1) spectra 
from AMISS, and CO(3--2) spectra from AMISS. Numbers in the upper right corner give the 
signal to noise ratio for each line. When a CO line is detected, the gray band indicates 
the region used to measure the line flux. The scale of the $y$-axis is such that lines 
would have the same amplitude in each transition for thermalized CO emission. The relative 
amplitudes of each spectrum give a sense of the luminosity ratios between the different 
lines.}
\figsetgrpend

\figsetgrpstart
\figsetgrpnum{15.149}
\figsetgrptitle{AMISS.9122}
\figsetplot{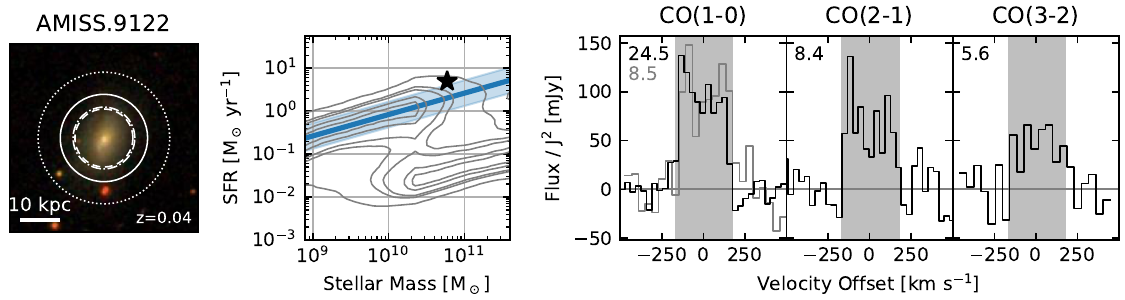}
\figsetgrpnote{Left column: SDSS cutouts of each target. Solid, dashed, and dotted lines show the 
beam sizes of the SMT for CO(2--1), the SMT for CO(3--2) (and the IRAM 30m for CO(1--0)) 
and the 12m for CO(1--0) respectively. The scale bar in the lower left shows 10 
kiloparsecs. 
Middle column: contours show the distribution of star formation rates at a given stellar 
mass, while the blue line and filled region show the main sequence of star forming 
galaxies. The stellar mass and star formation rate of the target galaxy is marked by a 
star. 
Right column: CO(1--0) spectra from AMISS (gray) and xCOLD~GASS (black), CO(2--1) spectra 
from AMISS, and CO(3--2) spectra from AMISS. Numbers in the upper right corner give the 
signal to noise ratio for each line. When a CO line is detected, the gray band indicates 
the region used to measure the line flux. The scale of the $y$-axis is such that lines 
would have the same amplitude in each transition for thermalized CO emission. The relative 
amplitudes of each spectrum give a sense of the luminosity ratios between the different 
lines.}
\figsetgrpend

\figsetgrpstart
\figsetgrpnum{15.150}
\figsetgrptitle{AMISS.9123}
\figsetplot{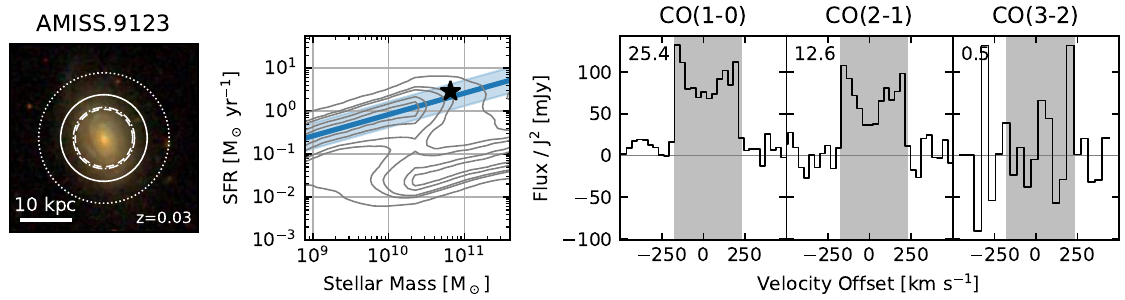}
\figsetgrpnote{Left column: SDSS cutouts of each target. Solid, dashed, and dotted lines show the 
beam sizes of the SMT for CO(2--1), the SMT for CO(3--2) (and the IRAM 30m for CO(1--0)) 
and the 12m for CO(1--0) respectively. The scale bar in the lower left shows 10 
kiloparsecs. 
Middle column: contours show the distribution of star formation rates at a given stellar 
mass, while the blue line and filled region show the main sequence of star forming 
galaxies. The stellar mass and star formation rate of the target galaxy is marked by a 
star. 
Right column: CO(1--0) spectra from AMISS (gray) and xCOLD~GASS (black), CO(2--1) spectra 
from AMISS, and CO(3--2) spectra from AMISS. Numbers in the upper right corner give the 
signal to noise ratio for each line. When a CO line is detected, the gray band indicates 
the region used to measure the line flux. The scale of the $y$-axis is such that lines 
would have the same amplitude in each transition for thermalized CO emission. The relative 
amplitudes of each spectrum give a sense of the luminosity ratios between the different 
lines.}
\figsetgrpend

\figsetgrpstart
\figsetgrpnum{15.151}
\figsetgrptitle{AMISS.9127}
\figsetplot{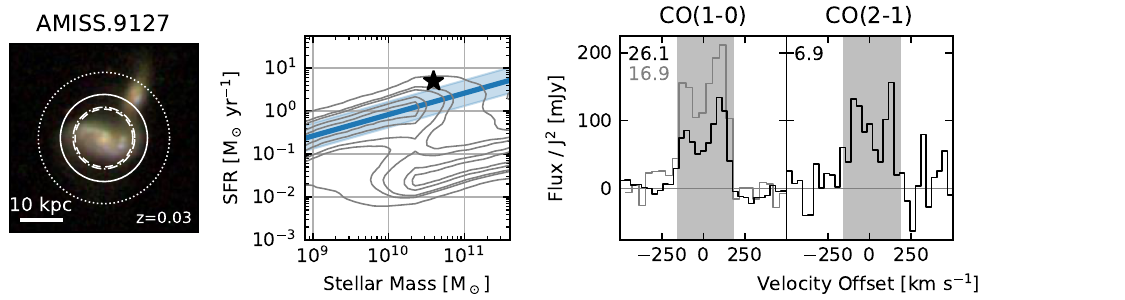}
\figsetgrpnote{Left column: SDSS cutouts of each target. Solid, dashed, and dotted lines show the 
beam sizes of the SMT for CO(2--1), the SMT for CO(3--2) (and the IRAM 30m for CO(1--0)) 
and the 12m for CO(1--0) respectively. The scale bar in the lower left shows 10 
kiloparsecs. 
Middle column: contours show the distribution of star formation rates at a given stellar 
mass, while the blue line and filled region show the main sequence of star forming 
galaxies. The stellar mass and star formation rate of the target galaxy is marked by a 
star. 
Right column: CO(1--0) spectra from AMISS (gray) and xCOLD~GASS (black), CO(2--1) spectra 
from AMISS, and CO(3--2) spectra from AMISS. Numbers in the upper right corner give the 
signal to noise ratio for each line. When a CO line is detected, the gray band indicates 
the region used to measure the line flux. The scale of the $y$-axis is such that lines 
would have the same amplitude in each transition for thermalized CO emission. The relative 
amplitudes of each spectrum give a sense of the luminosity ratios between the different 
lines.}
\figsetgrpend

\figsetgrpstart
\figsetgrpnum{15.152}
\figsetgrptitle{AMISS.9129}
\figsetplot{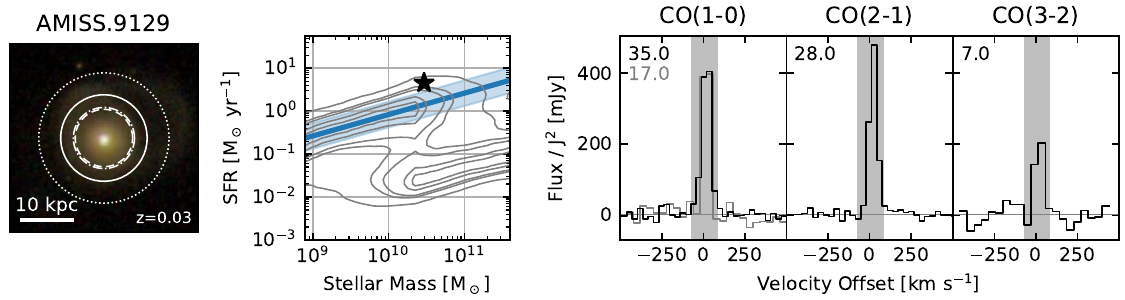}
\figsetgrpnote{Left column: SDSS cutouts of each target. Solid, dashed, and dotted lines show the 
beam sizes of the SMT for CO(2--1), the SMT for CO(3--2) (and the IRAM 30m for CO(1--0)) 
and the 12m for CO(1--0) respectively. The scale bar in the lower left shows 10 
kiloparsecs. 
Middle column: contours show the distribution of star formation rates at a given stellar 
mass, while the blue line and filled region show the main sequence of star forming 
galaxies. The stellar mass and star formation rate of the target galaxy is marked by a 
star. 
Right column: CO(1--0) spectra from AMISS (gray) and xCOLD~GASS (black), CO(2--1) spectra 
from AMISS, and CO(3--2) spectra from AMISS. Numbers in the upper right corner give the 
signal to noise ratio for each line. When a CO line is detected, the gray band indicates 
the region used to measure the line flux. The scale of the $y$-axis is such that lines 
would have the same amplitude in each transition for thermalized CO emission. The relative 
amplitudes of each spectrum give a sense of the luminosity ratios between the different 
lines.}
\figsetgrpend

\figsetgrpstart
\figsetgrpnum{15.153}
\figsetgrptitle{AMISS.9164}
\figsetplot{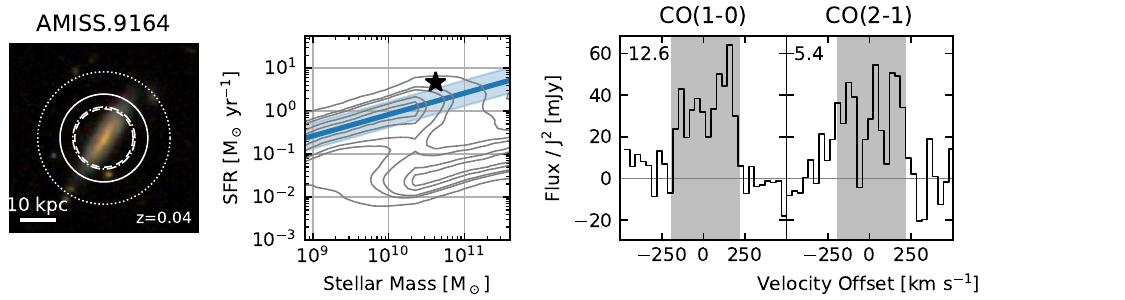}
\figsetgrpnote{Left column: SDSS cutouts of each target. Solid, dashed, and dotted lines show the 
beam sizes of the SMT for CO(2--1), the SMT for CO(3--2) (and the IRAM 30m for CO(1--0)) 
and the 12m for CO(1--0) respectively. The scale bar in the lower left shows 10 
kiloparsecs. 
Middle column: contours show the distribution of star formation rates at a given stellar 
mass, while the blue line and filled region show the main sequence of star forming 
galaxies. The stellar mass and star formation rate of the target galaxy is marked by a 
star. 
Right column: CO(1--0) spectra from AMISS (gray) and xCOLD~GASS (black), CO(2--1) spectra 
from AMISS, and CO(3--2) spectra from AMISS. Numbers in the upper right corner give the 
signal to noise ratio for each line. When a CO line is detected, the gray band indicates 
the region used to measure the line flux. The scale of the $y$-axis is such that lines 
would have the same amplitude in each transition for thermalized CO emission. The relative 
amplitudes of each spectrum give a sense of the luminosity ratios between the different 
lines.}
\figsetgrpend

\figsetgrpstart
\figsetgrpnum{15.154}
\figsetgrptitle{AMISS.9169}
\figsetplot{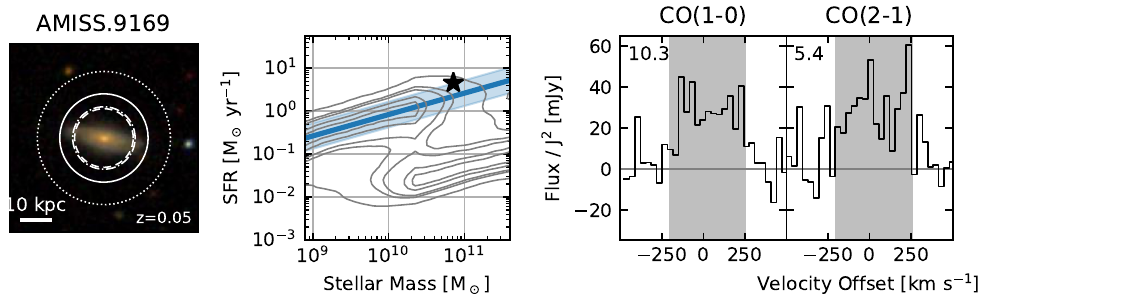}
\figsetgrpnote{Left column: SDSS cutouts of each target. Solid, dashed, and dotted lines show the 
beam sizes of the SMT for CO(2--1), the SMT for CO(3--2) (and the IRAM 30m for CO(1--0)) 
and the 12m for CO(1--0) respectively. The scale bar in the lower left shows 10 
kiloparsecs. 
Middle column: contours show the distribution of star formation rates at a given stellar 
mass, while the blue line and filled region show the main sequence of star forming 
galaxies. The stellar mass and star formation rate of the target galaxy is marked by a 
star. 
Right column: CO(1--0) spectra from AMISS (gray) and xCOLD~GASS (black), CO(2--1) spectra 
from AMISS, and CO(3--2) spectra from AMISS. Numbers in the upper right corner give the 
signal to noise ratio for each line. When a CO line is detected, the gray band indicates 
the region used to measure the line flux. The scale of the $y$-axis is such that lines 
would have the same amplitude in each transition for thermalized CO emission. The relative 
amplitudes of each spectrum give a sense of the luminosity ratios between the different 
lines.}
\figsetgrpend

\figsetgrpstart
\figsetgrpnum{15.155}
\figsetgrptitle{AMISS.9174}
\figsetplot{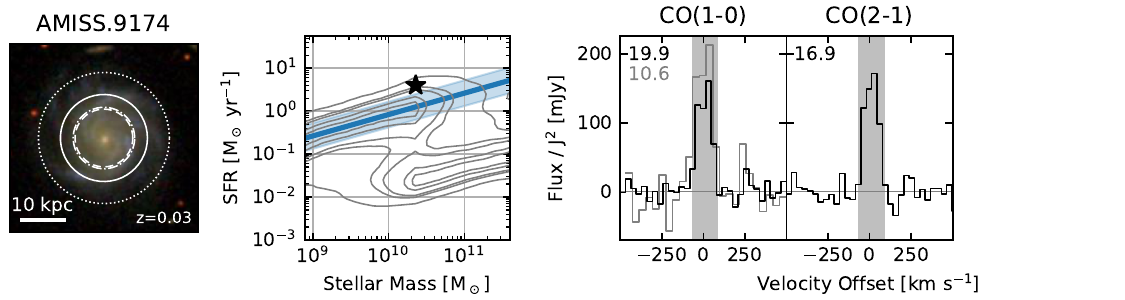}
\figsetgrpnote{Left column: SDSS cutouts of each target. Solid, dashed, and dotted lines show the 
beam sizes of the SMT for CO(2--1), the SMT for CO(3--2) (and the IRAM 30m for CO(1--0)) 
and the 12m for CO(1--0) respectively. The scale bar in the lower left shows 10 
kiloparsecs. 
Middle column: contours show the distribution of star formation rates at a given stellar 
mass, while the blue line and filled region show the main sequence of star forming 
galaxies. The stellar mass and star formation rate of the target galaxy is marked by a 
star. 
Right column: CO(1--0) spectra from AMISS (gray) and xCOLD~GASS (black), CO(2--1) spectra 
from AMISS, and CO(3--2) spectra from AMISS. Numbers in the upper right corner give the 
signal to noise ratio for each line. When a CO line is detected, the gray band indicates 
the region used to measure the line flux. The scale of the $y$-axis is such that lines 
would have the same amplitude in each transition for thermalized CO emission. The relative 
amplitudes of each spectrum give a sense of the luminosity ratios between the different 
lines.}
\figsetgrpend

\figsetgrpstart
\figsetgrpnum{15.156}
\figsetgrptitle{AMISS.9175}
\figsetplot{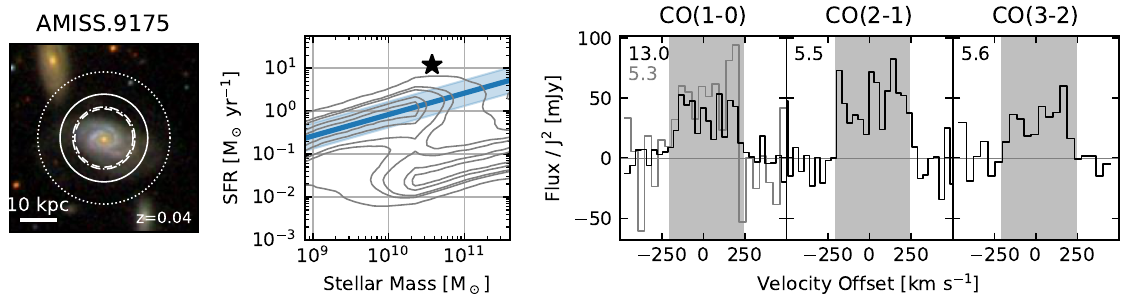}
\figsetgrpnote{Left column: SDSS cutouts of each target. Solid, dashed, and dotted lines show the 
beam sizes of the SMT for CO(2--1), the SMT for CO(3--2) (and the IRAM 30m for CO(1--0)) 
and the 12m for CO(1--0) respectively. The scale bar in the lower left shows 10 
kiloparsecs. 
Middle column: contours show the distribution of star formation rates at a given stellar 
mass, while the blue line and filled region show the main sequence of star forming 
galaxies. The stellar mass and star formation rate of the target galaxy is marked by a 
star. 
Right column: CO(1--0) spectra from AMISS (gray) and xCOLD~GASS (black), CO(2--1) spectra 
from AMISS, and CO(3--2) spectra from AMISS. Numbers in the upper right corner give the 
signal to noise ratio for each line. When a CO line is detected, the gray band indicates 
the region used to measure the line flux. The scale of the $y$-axis is such that lines 
would have the same amplitude in each transition for thermalized CO emission. The relative 
amplitudes of each spectrum give a sense of the luminosity ratios between the different 
lines.}
\figsetgrpend

\figsetgrpstart
\figsetgrpnum{15.157}
\figsetgrptitle{AMISS.9176}
\figsetplot{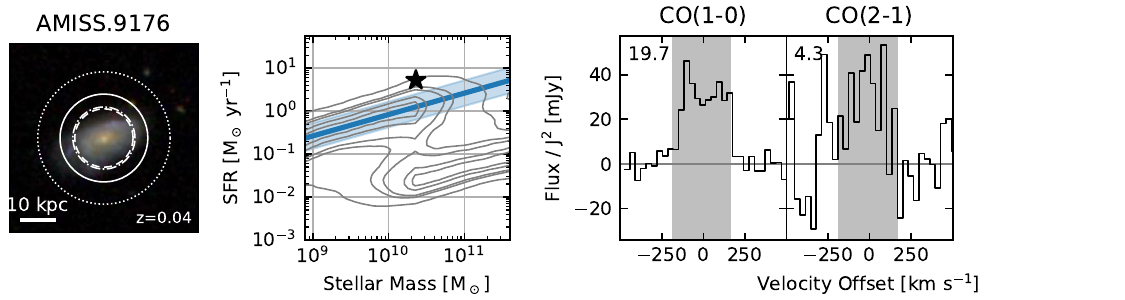}
\figsetgrpnote{Left column: SDSS cutouts of each target. Solid, dashed, and dotted lines show the 
beam sizes of the SMT for CO(2--1), the SMT for CO(3--2) (and the IRAM 30m for CO(1--0)) 
and the 12m for CO(1--0) respectively. The scale bar in the lower left shows 10 
kiloparsecs. 
Middle column: contours show the distribution of star formation rates at a given stellar 
mass, while the blue line and filled region show the main sequence of star forming 
galaxies. The stellar mass and star formation rate of the target galaxy is marked by a 
star. 
Right column: CO(1--0) spectra from AMISS (gray) and xCOLD~GASS (black), CO(2--1) spectra 
from AMISS, and CO(3--2) spectra from AMISS. Numbers in the upper right corner give the 
signal to noise ratio for each line. When a CO line is detected, the gray band indicates 
the region used to measure the line flux. The scale of the $y$-axis is such that lines 
would have the same amplitude in each transition for thermalized CO emission. The relative 
amplitudes of each spectrum give a sense of the luminosity ratios between the different 
lines.}
\figsetgrpend

\figsetgrpstart
\figsetgrpnum{15.158}
\figsetgrptitle{AMISS.9177}
\figsetplot{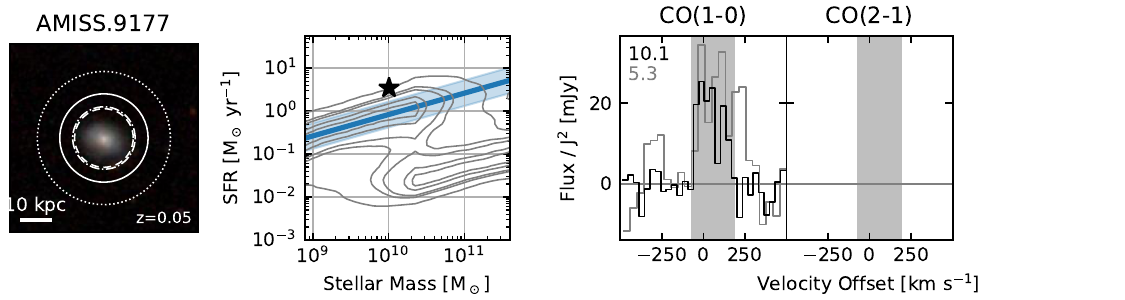}
\figsetgrpnote{Left column: SDSS cutouts of each target. Solid, dashed, and dotted lines show the 
beam sizes of the SMT for CO(2--1), the SMT for CO(3--2) (and the IRAM 30m for CO(1--0)) 
and the 12m for CO(1--0) respectively. The scale bar in the lower left shows 10 
kiloparsecs. 
Middle column: contours show the distribution of star formation rates at a given stellar 
mass, while the blue line and filled region show the main sequence of star forming 
galaxies. The stellar mass and star formation rate of the target galaxy is marked by a 
star. 
Right column: CO(1--0) spectra from AMISS (gray) and xCOLD~GASS (black), CO(2--1) spectra 
from AMISS, and CO(3--2) spectra from AMISS. Numbers in the upper right corner give the 
signal to noise ratio for each line. When a CO line is detected, the gray band indicates 
the region used to measure the line flux. The scale of the $y$-axis is such that lines 
would have the same amplitude in each transition for thermalized CO emission. The relative 
amplitudes of each spectrum give a sense of the luminosity ratios between the different 
lines.}
\figsetgrpend

\figsetgrpstart
\figsetgrpnum{15.159}
\figsetgrptitle{AMISS.9178}
\figsetplot{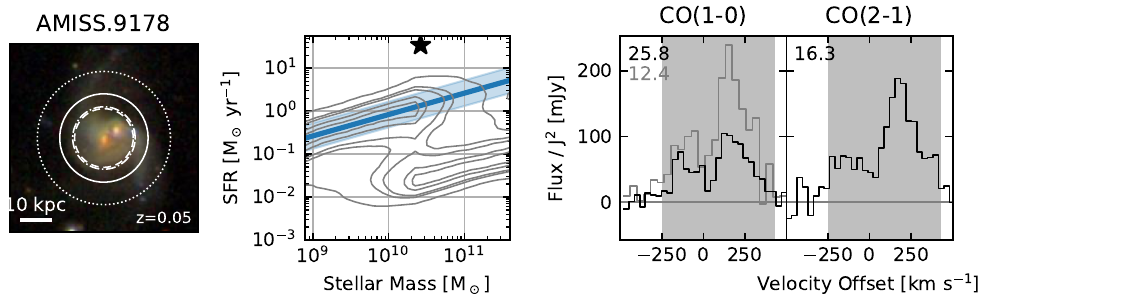}
\figsetgrpnote{Left column: SDSS cutouts of each target. Solid, dashed, and dotted lines show the 
beam sizes of the SMT for CO(2--1), the SMT for CO(3--2) (and the IRAM 30m for CO(1--0)) 
and the 12m for CO(1--0) respectively. The scale bar in the lower left shows 10 
kiloparsecs. 
Middle column: contours show the distribution of star formation rates at a given stellar 
mass, while the blue line and filled region show the main sequence of star forming 
galaxies. The stellar mass and star formation rate of the target galaxy is marked by a 
star. 
Right column: CO(1--0) spectra from AMISS (gray) and xCOLD~GASS (black), CO(2--1) spectra 
from AMISS, and CO(3--2) spectra from AMISS. Numbers in the upper right corner give the 
signal to noise ratio for each line. When a CO line is detected, the gray band indicates 
the region used to measure the line flux. The scale of the $y$-axis is such that lines 
would have the same amplitude in each transition for thermalized CO emission. The relative 
amplitudes of each spectrum give a sense of the luminosity ratios between the different 
lines.}
\figsetgrpend

\figsetgrpstart
\figsetgrpnum{15.160}
\figsetgrptitle{AMISS.9179}
\figsetplot{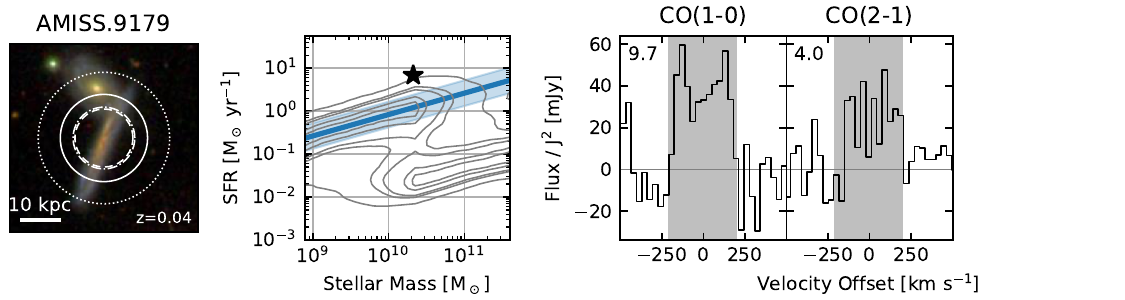}
\figsetgrpnote{Left column: SDSS cutouts of each target. Solid, dashed, and dotted lines show the 
beam sizes of the SMT for CO(2--1), the SMT for CO(3--2) (and the IRAM 30m for CO(1--0)) 
and the 12m for CO(1--0) respectively. The scale bar in the lower left shows 10 
kiloparsecs. 
Middle column: contours show the distribution of star formation rates at a given stellar 
mass, while the blue line and filled region show the main sequence of star forming 
galaxies. The stellar mass and star formation rate of the target galaxy is marked by a 
star. 
Right column: CO(1--0) spectra from AMISS (gray) and xCOLD~GASS (black), CO(2--1) spectra 
from AMISS, and CO(3--2) spectra from AMISS. Numbers in the upper right corner give the 
signal to noise ratio for each line. When a CO line is detected, the gray band indicates 
the region used to measure the line flux. The scale of the $y$-axis is such that lines 
would have the same amplitude in each transition for thermalized CO emission. The relative 
amplitudes of each spectrum give a sense of the luminosity ratios between the different 
lines.}
\figsetgrpend

\figsetgrpstart
\figsetgrpnum{15.161}
\figsetgrptitle{AMISS.9180}
\figsetplot{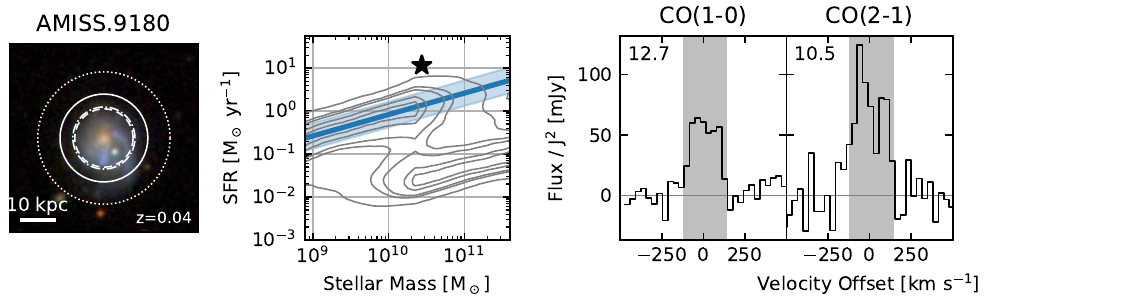}
\figsetgrpnote{Left column: SDSS cutouts of each target. Solid, dashed, and dotted lines show the 
beam sizes of the SMT for CO(2--1), the SMT for CO(3--2) (and the IRAM 30m for CO(1--0)) 
and the 12m for CO(1--0) respectively. The scale bar in the lower left shows 10 
kiloparsecs. 
Middle column: contours show the distribution of star formation rates at a given stellar 
mass, while the blue line and filled region show the main sequence of star forming 
galaxies. The stellar mass and star formation rate of the target galaxy is marked by a 
star. 
Right column: CO(1--0) spectra from AMISS (gray) and xCOLD~GASS (black), CO(2--1) spectra 
from AMISS, and CO(3--2) spectra from AMISS. Numbers in the upper right corner give the 
signal to noise ratio for each line. When a CO line is detected, the gray band indicates 
the region used to measure the line flux. The scale of the $y$-axis is such that lines 
would have the same amplitude in each transition for thermalized CO emission. The relative 
amplitudes of each spectrum give a sense of the luminosity ratios between the different 
lines.}
\figsetgrpend

\figsetgrpstart
\figsetgrpnum{15.162}
\figsetgrptitle{AMISS.9181}
\figsetplot{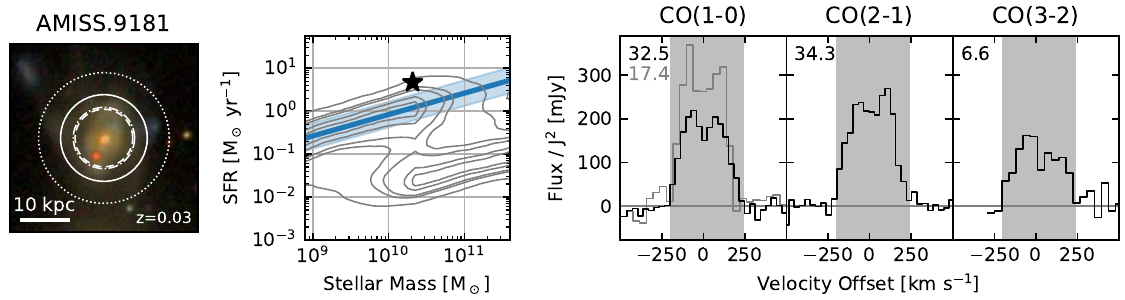}
\figsetgrpnote{Left column: SDSS cutouts of each target. Solid, dashed, and dotted lines show the 
beam sizes of the SMT for CO(2--1), the SMT for CO(3--2) (and the IRAM 30m for CO(1--0)) 
and the 12m for CO(1--0) respectively. The scale bar in the lower left shows 10 
kiloparsecs. 
Middle column: contours show the distribution of star formation rates at a given stellar 
mass, while the blue line and filled region show the main sequence of star forming 
galaxies. The stellar mass and star formation rate of the target galaxy is marked by a 
star. 
Right column: CO(1--0) spectra from AMISS (gray) and xCOLD~GASS (black), CO(2--1) spectra 
from AMISS, and CO(3--2) spectra from AMISS. Numbers in the upper right corner give the 
signal to noise ratio for each line. When a CO line is detected, the gray band indicates 
the region used to measure the line flux. The scale of the $y$-axis is such that lines 
would have the same amplitude in each transition for thermalized CO emission. The relative 
amplitudes of each spectrum give a sense of the luminosity ratios between the different 
lines.}
\figsetgrpend

\figsetgrpstart
\figsetgrpnum{15.163}
\figsetgrptitle{AMISS.9184}
\figsetplot{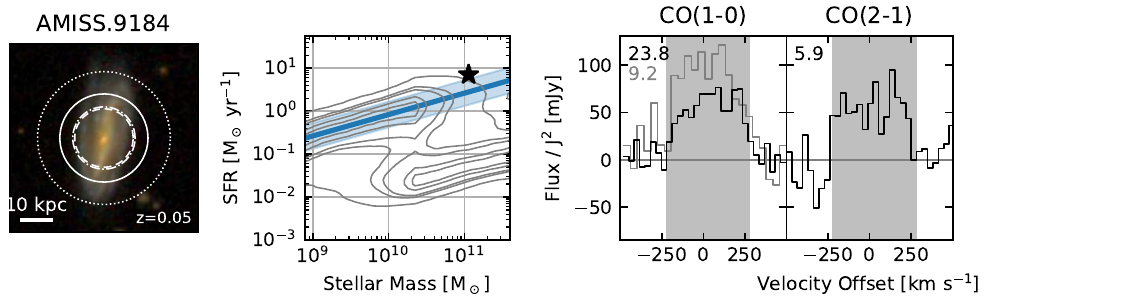}
\figsetgrpnote{Left column: SDSS cutouts of each target. Solid, dashed, and dotted lines show the 
beam sizes of the SMT for CO(2--1), the SMT for CO(3--2) (and the IRAM 30m for CO(1--0)) 
and the 12m for CO(1--0) respectively. The scale bar in the lower left shows 10 
kiloparsecs. 
Middle column: contours show the distribution of star formation rates at a given stellar 
mass, while the blue line and filled region show the main sequence of star forming 
galaxies. The stellar mass and star formation rate of the target galaxy is marked by a 
star. 
Right column: CO(1--0) spectra from AMISS (gray) and xCOLD~GASS (black), CO(2--1) spectra 
from AMISS, and CO(3--2) spectra from AMISS. Numbers in the upper right corner give the 
signal to noise ratio for each line. When a CO line is detected, the gray band indicates 
the region used to measure the line flux. The scale of the $y$-axis is such that lines 
would have the same amplitude in each transition for thermalized CO emission. The relative 
amplitudes of each spectrum give a sense of the luminosity ratios between the different 
lines.}
\figsetgrpend

\figsetgrpstart
\figsetgrpnum{15.164}
\figsetgrptitle{AMISS.9185}
\figsetplot{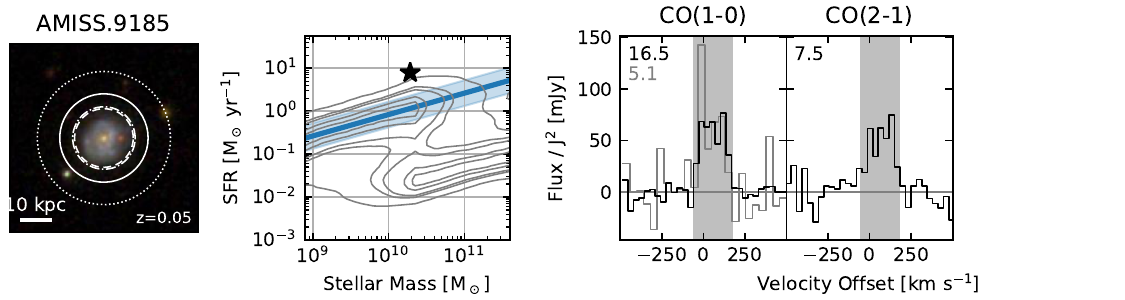}
\figsetgrpnote{Left column: SDSS cutouts of each target. Solid, dashed, and dotted lines show the 
beam sizes of the SMT for CO(2--1), the SMT for CO(3--2) (and the IRAM 30m for CO(1--0)) 
and the 12m for CO(1--0) respectively. The scale bar in the lower left shows 10 
kiloparsecs. 
Middle column: contours show the distribution of star formation rates at a given stellar 
mass, while the blue line and filled region show the main sequence of star forming 
galaxies. The stellar mass and star formation rate of the target galaxy is marked by a 
star. 
Right column: CO(1--0) spectra from AMISS (gray) and xCOLD~GASS (black), CO(2--1) spectra 
from AMISS, and CO(3--2) spectra from AMISS. Numbers in the upper right corner give the 
signal to noise ratio for each line. When a CO line is detected, the gray band indicates 
the region used to measure the line flux. The scale of the $y$-axis is such that lines 
would have the same amplitude in each transition for thermalized CO emission. The relative 
amplitudes of each spectrum give a sense of the luminosity ratios between the different 
lines.}
\figsetgrpend

\figsetgrpstart
\figsetgrpnum{15.165}
\figsetgrptitle{AMISS.9186}
\figsetplot{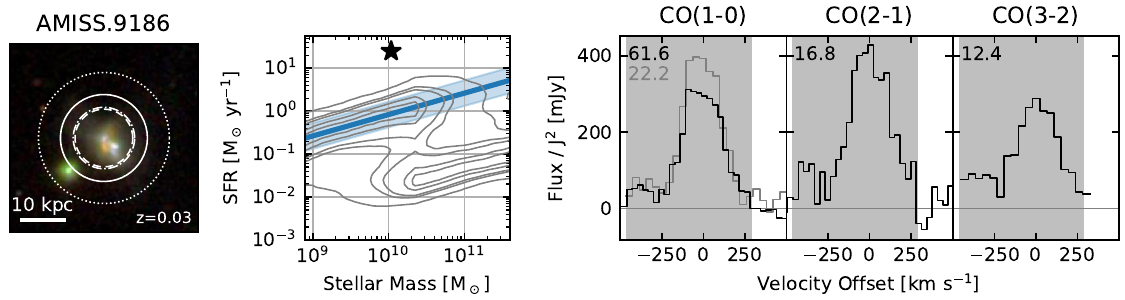}
\figsetgrpnote{Left column: SDSS cutouts of each target. Solid, dashed, and dotted lines show the 
beam sizes of the SMT for CO(2--1), the SMT for CO(3--2) (and the IRAM 30m for CO(1--0)) 
and the 12m for CO(1--0) respectively. The scale bar in the lower left shows 10 
kiloparsecs. 
Middle column: contours show the distribution of star formation rates at a given stellar 
mass, while the blue line and filled region show the main sequence of star forming 
galaxies. The stellar mass and star formation rate of the target galaxy is marked by a 
star. 
Right column: CO(1--0) spectra from AMISS (gray) and xCOLD~GASS (black), CO(2--1) spectra 
from AMISS, and CO(3--2) spectra from AMISS. Numbers in the upper right corner give the 
signal to noise ratio for each line. When a CO line is detected, the gray band indicates 
the region used to measure the line flux. The scale of the $y$-axis is such that lines 
would have the same amplitude in each transition for thermalized CO emission. The relative 
amplitudes of each spectrum give a sense of the luminosity ratios between the different 
lines.}
\figsetgrpend

\figsetgrpstart
\figsetgrpnum{15.166}
\figsetgrptitle{AMISS.9187}
\figsetplot{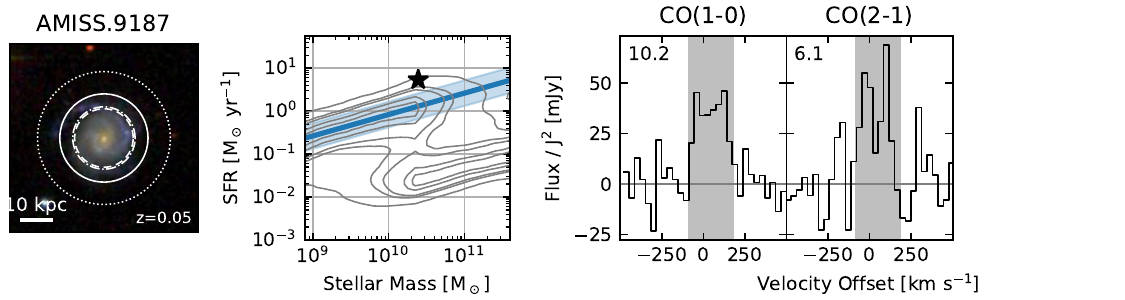}
\figsetgrpnote{Left column: SDSS cutouts of each target. Solid, dashed, and dotted lines show the 
beam sizes of the SMT for CO(2--1), the SMT for CO(3--2) (and the IRAM 30m for CO(1--0)) 
and the 12m for CO(1--0) respectively. The scale bar in the lower left shows 10 
kiloparsecs. 
Middle column: contours show the distribution of star formation rates at a given stellar 
mass, while the blue line and filled region show the main sequence of star forming 
galaxies. The stellar mass and star formation rate of the target galaxy is marked by a 
star. 
Right column: CO(1--0) spectra from AMISS (gray) and xCOLD~GASS (black), CO(2--1) spectra 
from AMISS, and CO(3--2) spectra from AMISS. Numbers in the upper right corner give the 
signal to noise ratio for each line. When a CO line is detected, the gray band indicates 
the region used to measure the line flux. The scale of the $y$-axis is such that lines 
would have the same amplitude in each transition for thermalized CO emission. The relative 
amplitudes of each spectrum give a sense of the luminosity ratios between the different 
lines.}
\figsetgrpend

\figsetgrpstart
\figsetgrpnum{15.167}
\figsetgrptitle{AMISS.9189}
\figsetplot{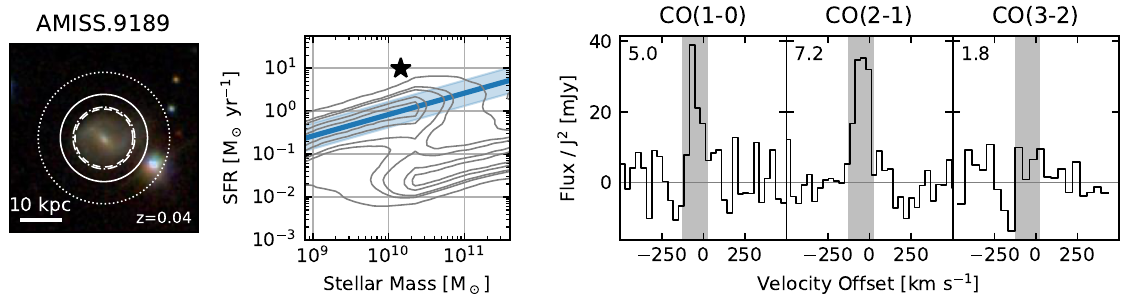}
\figsetgrpnote{Left column: SDSS cutouts of each target. Solid, dashed, and dotted lines show the 
beam sizes of the SMT for CO(2--1), the SMT for CO(3--2) (and the IRAM 30m for CO(1--0)) 
and the 12m for CO(1--0) respectively. The scale bar in the lower left shows 10 
kiloparsecs. 
Middle column: contours show the distribution of star formation rates at a given stellar 
mass, while the blue line and filled region show the main sequence of star forming 
galaxies. The stellar mass and star formation rate of the target galaxy is marked by a 
star. 
Right column: CO(1--0) spectra from AMISS (gray) and xCOLD~GASS (black), CO(2--1) spectra 
from AMISS, and CO(3--2) spectra from AMISS. Numbers in the upper right corner give the 
signal to noise ratio for each line. When a CO line is detected, the gray band indicates 
the region used to measure the line flux. The scale of the $y$-axis is such that lines 
would have the same amplitude in each transition for thermalized CO emission. The relative 
amplitudes of each spectrum give a sense of the luminosity ratios between the different 
lines.}
\figsetgrpend

\figsetgrpstart
\figsetgrpnum{15.168}
\figsetgrptitle{AMISS.9190}
\figsetplot{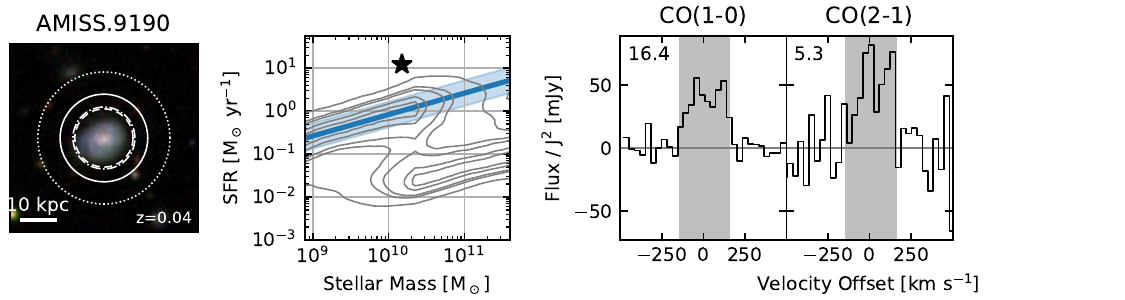}
\figsetgrpnote{Left column: SDSS cutouts of each target. Solid, dashed, and dotted lines show the 
beam sizes of the SMT for CO(2--1), the SMT for CO(3--2) (and the IRAM 30m for CO(1--0)) 
and the 12m for CO(1--0) respectively. The scale bar in the lower left shows 10 
kiloparsecs. 
Middle column: contours show the distribution of star formation rates at a given stellar 
mass, while the blue line and filled region show the main sequence of star forming 
galaxies. The stellar mass and star formation rate of the target galaxy is marked by a 
star. 
Right column: CO(1--0) spectra from AMISS (gray) and xCOLD~GASS (black), CO(2--1) spectra 
from AMISS, and CO(3--2) spectra from AMISS. Numbers in the upper right corner give the 
signal to noise ratio for each line. When a CO line is detected, the gray band indicates 
the region used to measure the line flux. The scale of the $y$-axis is such that lines 
would have the same amplitude in each transition for thermalized CO emission. The relative 
amplitudes of each spectrum give a sense of the luminosity ratios between the different 
lines.}
\figsetgrpend

\figsetgrpstart
\figsetgrpnum{15.169}
\figsetgrptitle{AMISS.9191}
\figsetplot{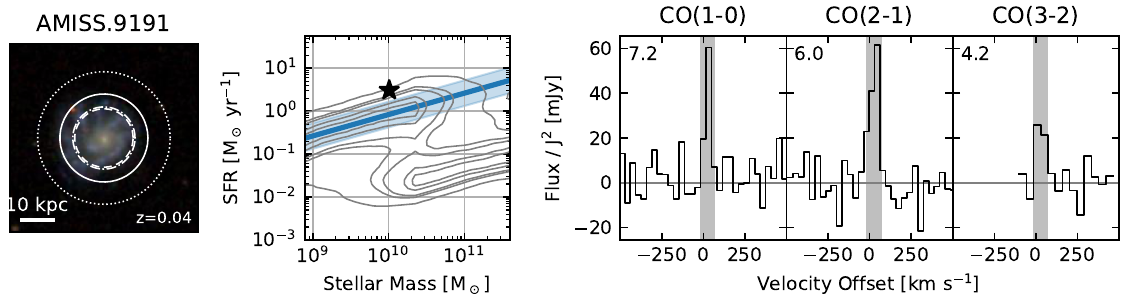}
\figsetgrpnote{Left column: SDSS cutouts of each target. Solid, dashed, and dotted lines show the 
beam sizes of the SMT for CO(2--1), the SMT for CO(3--2) (and the IRAM 30m for CO(1--0)) 
and the 12m for CO(1--0) respectively. The scale bar in the lower left shows 10 
kiloparsecs. 
Middle column: contours show the distribution of star formation rates at a given stellar 
mass, while the blue line and filled region show the main sequence of star forming 
galaxies. The stellar mass and star formation rate of the target galaxy is marked by a 
star. 
Right column: CO(1--0) spectra from AMISS (gray) and xCOLD~GASS (black), CO(2--1) spectra 
from AMISS, and CO(3--2) spectra from AMISS. Numbers in the upper right corner give the 
signal to noise ratio for each line. When a CO line is detected, the gray band indicates 
the region used to measure the line flux. The scale of the $y$-axis is such that lines 
would have the same amplitude in each transition for thermalized CO emission. The relative 
amplitudes of each spectrum give a sense of the luminosity ratios between the different 
lines.}
\figsetgrpend

\figsetgrpstart
\figsetgrpnum{15.170}
\figsetgrptitle{AMISS.9192}
\figsetplot{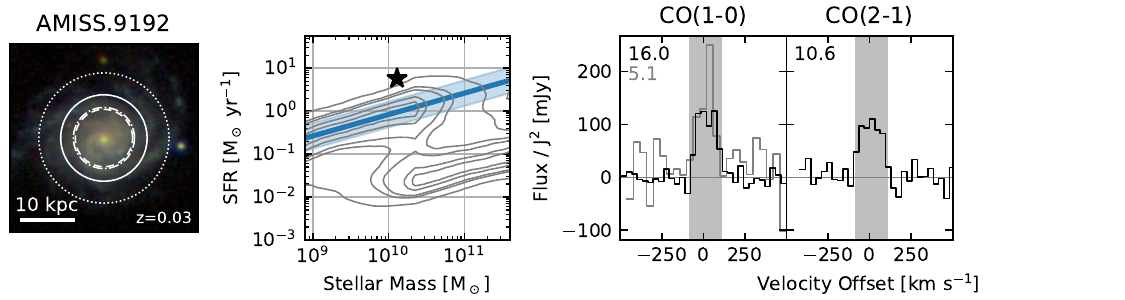}
\figsetgrpnote{Left column: SDSS cutouts of each target. Solid, dashed, and dotted lines show the 
beam sizes of the SMT for CO(2--1), the SMT for CO(3--2) (and the IRAM 30m for CO(1--0)) 
and the 12m for CO(1--0) respectively. The scale bar in the lower left shows 10 
kiloparsecs. 
Middle column: contours show the distribution of star formation rates at a given stellar 
mass, while the blue line and filled region show the main sequence of star forming 
galaxies. The stellar mass and star formation rate of the target galaxy is marked by a 
star. 
Right column: CO(1--0) spectra from AMISS (gray) and xCOLD~GASS (black), CO(2--1) spectra 
from AMISS, and CO(3--2) spectra from AMISS. Numbers in the upper right corner give the 
signal to noise ratio for each line. When a CO line is detected, the gray band indicates 
the region used to measure the line flux. The scale of the $y$-axis is such that lines 
would have the same amplitude in each transition for thermalized CO emission. The relative 
amplitudes of each spectrum give a sense of the luminosity ratios between the different 
lines.}
\figsetgrpend

\figsetgrpstart
\figsetgrpnum{15.171}
\figsetgrptitle{AMISS.9193}
\figsetplot{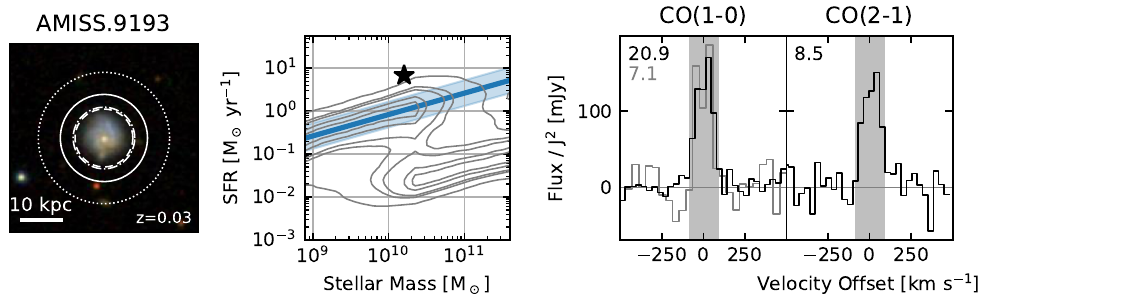}
\figsetgrpnote{Left column: SDSS cutouts of each target. Solid, dashed, and dotted lines show the 
beam sizes of the SMT for CO(2--1), the SMT for CO(3--2) (and the IRAM 30m for CO(1--0)) 
and the 12m for CO(1--0) respectively. The scale bar in the lower left shows 10 
kiloparsecs. 
Middle column: contours show the distribution of star formation rates at a given stellar 
mass, while the blue line and filled region show the main sequence of star forming 
galaxies. The stellar mass and star formation rate of the target galaxy is marked by a 
star. 
Right column: CO(1--0) spectra from AMISS (gray) and xCOLD~GASS (black), CO(2--1) spectra 
from AMISS, and CO(3--2) spectra from AMISS. Numbers in the upper right corner give the 
signal to noise ratio for each line. When a CO line is detected, the gray band indicates 
the region used to measure the line flux. The scale of the $y$-axis is such that lines 
would have the same amplitude in each transition for thermalized CO emission. The relative 
amplitudes of each spectrum give a sense of the luminosity ratios between the different 
lines.}
\figsetgrpend

\figsetgrpstart
\figsetgrpnum{15.172}
\figsetgrptitle{AMISS.9198}
\figsetplot{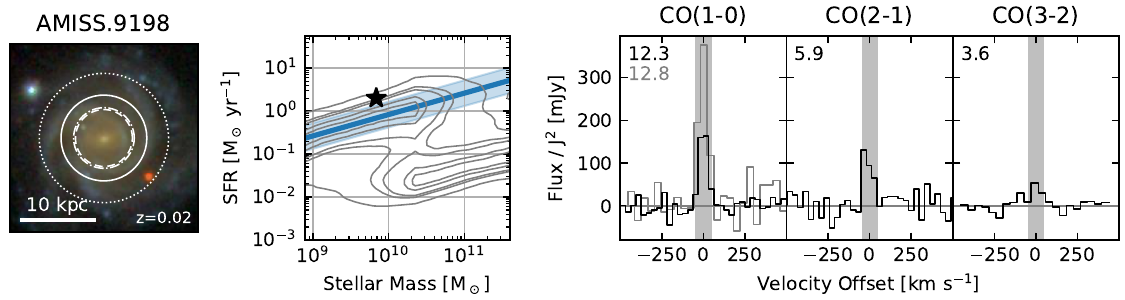}
\figsetgrpnote{Left column: SDSS cutouts of each target. Solid, dashed, and dotted lines show the 
beam sizes of the SMT for CO(2--1), the SMT for CO(3--2) (and the IRAM 30m for CO(1--0)) 
and the 12m for CO(1--0) respectively. The scale bar in the lower left shows 10 
kiloparsecs. 
Middle column: contours show the distribution of star formation rates at a given stellar 
mass, while the blue line and filled region show the main sequence of star forming 
galaxies. The stellar mass and star formation rate of the target galaxy is marked by a 
star. 
Right column: CO(1--0) spectra from AMISS (gray) and xCOLD~GASS (black), CO(2--1) spectra 
from AMISS, and CO(3--2) spectra from AMISS. Numbers in the upper right corner give the 
signal to noise ratio for each line. When a CO line is detected, the gray band indicates 
the region used to measure the line flux. The scale of the $y$-axis is such that lines 
would have the same amplitude in each transition for thermalized CO emission. The relative 
amplitudes of each spectrum give a sense of the luminosity ratios between the different 
lines.}
\figsetgrpend

\figsetgrpstart
\figsetgrpnum{15.173}
\figsetgrptitle{AMISS.9202}
\figsetplot{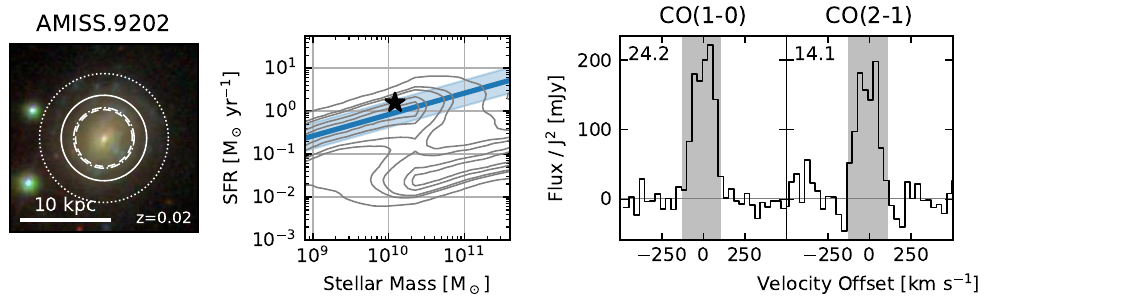}
\figsetgrpnote{Left column: SDSS cutouts of each target. Solid, dashed, and dotted lines show the 
beam sizes of the SMT for CO(2--1), the SMT for CO(3--2) (and the IRAM 30m for CO(1--0)) 
and the 12m for CO(1--0) respectively. The scale bar in the lower left shows 10 
kiloparsecs. 
Middle column: contours show the distribution of star formation rates at a given stellar 
mass, while the blue line and filled region show the main sequence of star forming 
galaxies. The stellar mass and star formation rate of the target galaxy is marked by a 
star. 
Right column: CO(1--0) spectra from AMISS (gray) and xCOLD~GASS (black), CO(2--1) spectra 
from AMISS, and CO(3--2) spectra from AMISS. Numbers in the upper right corner give the 
signal to noise ratio for each line. When a CO line is detected, the gray band indicates 
the region used to measure the line flux. The scale of the $y$-axis is such that lines 
would have the same amplitude in each transition for thermalized CO emission. The relative 
amplitudes of each spectrum give a sense of the luminosity ratios between the different 
lines.}
\figsetgrpend

\figsetgrpstart
\figsetgrpnum{15.174}
\figsetgrptitle{AMISS.9209}
\figsetplot{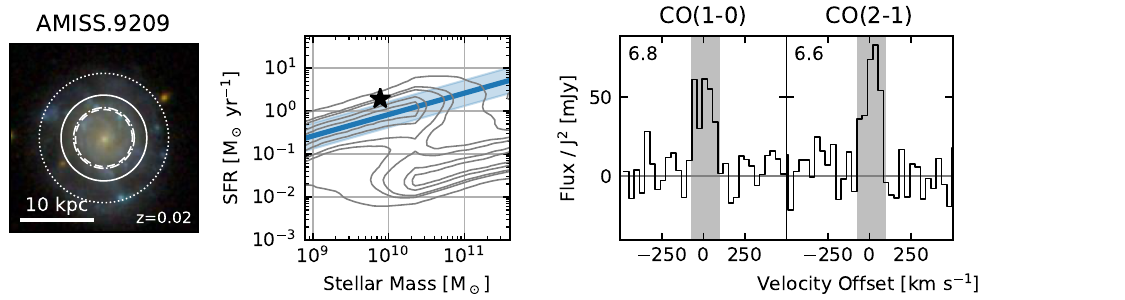}
\figsetgrpnote{Left column: SDSS cutouts of each target. Solid, dashed, and dotted lines show the 
beam sizes of the SMT for CO(2--1), the SMT for CO(3--2) (and the IRAM 30m for CO(1--0)) 
and the 12m for CO(1--0) respectively. The scale bar in the lower left shows 10 
kiloparsecs. 
Middle column: contours show the distribution of star formation rates at a given stellar 
mass, while the blue line and filled region show the main sequence of star forming 
galaxies. The stellar mass and star formation rate of the target galaxy is marked by a 
star. 
Right column: CO(1--0) spectra from AMISS (gray) and xCOLD~GASS (black), CO(2--1) spectra 
from AMISS, and CO(3--2) spectra from AMISS. Numbers in the upper right corner give the 
signal to noise ratio for each line. When a CO line is detected, the gray band indicates 
the region used to measure the line flux. The scale of the $y$-axis is such that lines 
would have the same amplitude in each transition for thermalized CO emission. The relative 
amplitudes of each spectrum give a sense of the luminosity ratios between the different 
lines.}
\figsetgrpend

\figsetgrpstart
\figsetgrpnum{15.175}
\figsetgrptitle{AMISS.9213}
\figsetplot{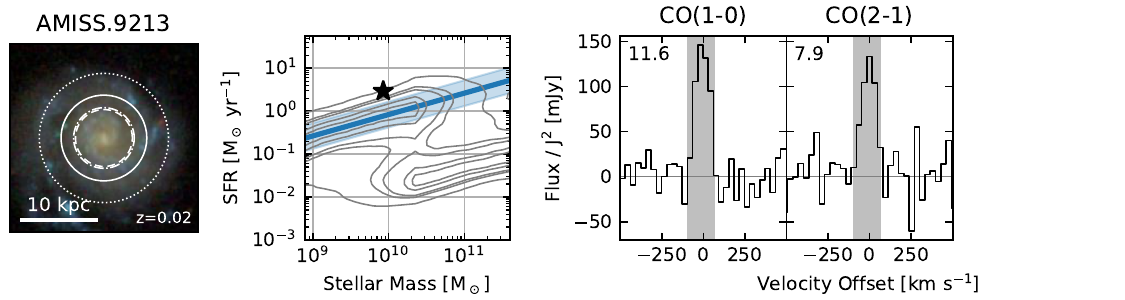}
\figsetgrpnote{Left column: SDSS cutouts of each target. Solid, dashed, and dotted lines show the 
beam sizes of the SMT for CO(2--1), the SMT for CO(3--2) (and the IRAM 30m for CO(1--0)) 
and the 12m for CO(1--0) respectively. The scale bar in the lower left shows 10 
kiloparsecs. 
Middle column: contours show the distribution of star formation rates at a given stellar 
mass, while the blue line and filled region show the main sequence of star forming 
galaxies. The stellar mass and star formation rate of the target galaxy is marked by a 
star. 
Right column: CO(1--0) spectra from AMISS (gray) and xCOLD~GASS (black), CO(2--1) spectra 
from AMISS, and CO(3--2) spectra from AMISS. Numbers in the upper right corner give the 
signal to noise ratio for each line. When a CO line is detected, the gray band indicates 
the region used to measure the line flux. The scale of the $y$-axis is such that lines 
would have the same amplitude in each transition for thermalized CO emission. The relative 
amplitudes of each spectrum give a sense of the luminosity ratios between the different 
lines.}
\figsetgrpend

\figsetgrpstart
\figsetgrpnum{15.176}
\figsetgrptitle{AMISS.9215}
\figsetplot{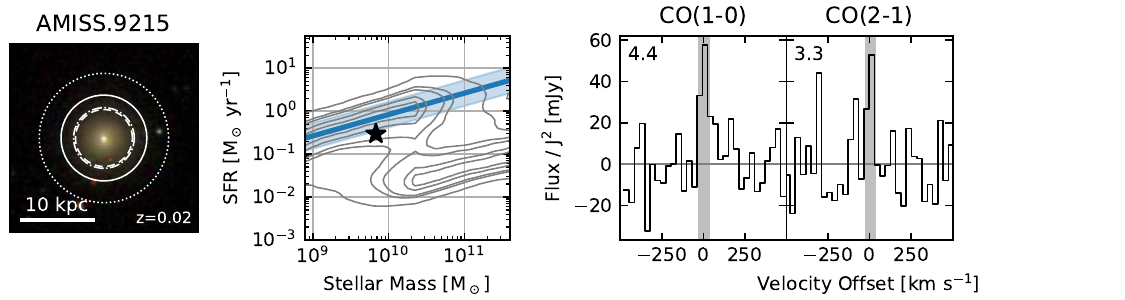}
\figsetgrpnote{Left column: SDSS cutouts of each target. Solid, dashed, and dotted lines show the 
beam sizes of the SMT for CO(2--1), the SMT for CO(3--2) (and the IRAM 30m for CO(1--0)) 
and the 12m for CO(1--0) respectively. The scale bar in the lower left shows 10 
kiloparsecs. 
Middle column: contours show the distribution of star formation rates at a given stellar 
mass, while the blue line and filled region show the main sequence of star forming 
galaxies. The stellar mass and star formation rate of the target galaxy is marked by a 
star. 
Right column: CO(1--0) spectra from AMISS (gray) and xCOLD~GASS (black), CO(2--1) spectra 
from AMISS, and CO(3--2) spectra from AMISS. Numbers in the upper right corner give the 
signal to noise ratio for each line. When a CO line is detected, the gray band indicates 
the region used to measure the line flux. The scale of the $y$-axis is such that lines 
would have the same amplitude in each transition for thermalized CO emission. The relative 
amplitudes of each spectrum give a sense of the luminosity ratios between the different 
lines.}
\figsetgrpend

\figsetgrpstart
\figsetgrpnum{15.177}
\figsetgrptitle{AMISS.9219}
\figsetplot{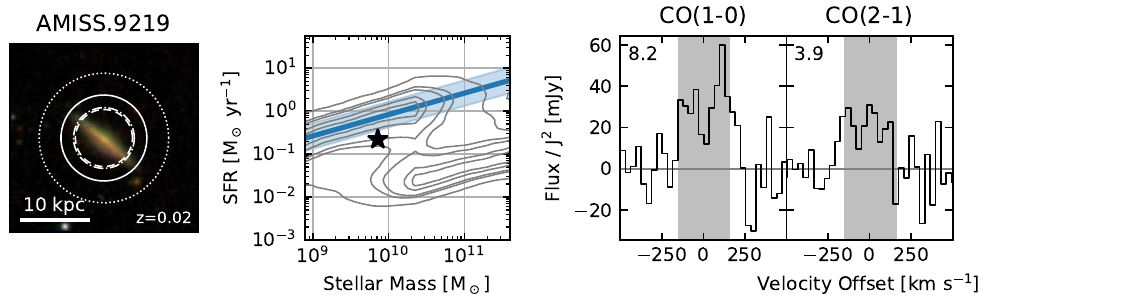}
\figsetgrpnote{Left column: SDSS cutouts of each target. Solid, dashed, and dotted lines show the 
beam sizes of the SMT for CO(2--1), the SMT for CO(3--2) (and the IRAM 30m for CO(1--0)) 
and the 12m for CO(1--0) respectively. The scale bar in the lower left shows 10 
kiloparsecs. 
Middle column: contours show the distribution of star formation rates at a given stellar 
mass, while the blue line and filled region show the main sequence of star forming 
galaxies. The stellar mass and star formation rate of the target galaxy is marked by a 
star. 
Right column: CO(1--0) spectra from AMISS (gray) and xCOLD~GASS (black), CO(2--1) spectra 
from AMISS, and CO(3--2) spectra from AMISS. Numbers in the upper right corner give the 
signal to noise ratio for each line. When a CO line is detected, the gray band indicates 
the region used to measure the line flux. The scale of the $y$-axis is such that lines 
would have the same amplitude in each transition for thermalized CO emission. The relative 
amplitudes of each spectrum give a sense of the luminosity ratios between the different 
lines.}
\figsetgrpend
\figsetend

    \begin{deluxetable*}{chc|ccc|cccc}
    
        \tablecaption{AMISS CO line catalog\label{tab:results}}
        \tablehead{
            \multicolumn{3}{c}{Source ID} & \colhead{Redshift} & \colhead{Stellar Mass} & \colhead{SFR} & \colhead{$L_{\rm CO(1-0),30m}^\prime$} & \colhead{$L_{\rm CO(1-0),12m}^\prime$} & \colhead{$L_{\rm CO(2-1)}^\prime$} & \colhead{$L_{\rm CO(3-2)}^\prime$} \\
            \colhead{AMISS} & & \colhead{Sample} & \colhead{} & \colhead{[M$_\odot$]} & \colhead{[M$_\odot$ yr$^{-1}$]} & \colhead{[K km s$^{-1}$ pc$^2$]} & \colhead{[K km s$^{-1}$ pc$^2$]} & \colhead{[K km s$^{-1}$ pc$^2$]} & \colhead{[K km s$^{-1}$ pc$^2$]} \\
            \colhead{(1)} & & \colhead{(2)} & \colhead{(3)} & \colhead{(4)} & \colhead{(5)} & \colhead{(6)} & \colhead{(7)} & \colhead{(8)} & \colhead{(9)}
        }
        \startdata
            1000 & J001934.54+161215.0 & Primary & 0.037 & $7.0\times10^{10}$ & $2.1\times10^{-1}$ & $<1.2\times10^{8}$ & & $<8.9\times10^{7}$ & \\
            1001 & J010905.96+144520.8 & Primary & 0.039 & $2.2\times10^{10}$ & $5.2\times10^{-1}$ & $<2.8\times10^{8}$ & & $(7.0\pm2.4)\times10^{7}$ & \\
            1002 & J014326.66+131913.0 & Primary & 0.028 & $3.6\times10^{10}$ & $4.6\times10^{-2}$ & $<9.6\times10^{7}$ & & $<8.8\times10^{7}$ & \\
            1003 & J020939.47+135859.4 & Primary & 0.049 & $2.1\times10^{11}$ & $1.2\times10^{0}$ & $(1.4\pm0.2)\times10^{9}$ & $(1.6\pm0.2)\times10^{9}$ & $(8.8\pm1.2)\times10^{8}$ & $<9.5\times10^{8}$ \\
            1004 & J021139.06+140830.3 & Primary & 0.027 & $2.1\times10^{10}$ & $2.1\times10^{-2}$ & $<7.3\times10^{7}$ & & $<8.5\times10^{7}$ & \\
            1005 & J021419.24+135611.2 & Primary & 0.040 & $3.3\times10^{10}$ & $4.9\times10^{-1}$ & $<2.9\times10^{8}$ & & $(1.3\pm0.3)\times10^{8}$ & \\
            1006 & J075559.95+125853.2 & Primary & 0.044 & $1.9\times10^{11}$ & $1.3\times10^{-1}$ & $<5.5\times10^{8}$ & & $<9.6\times10^{7}$ & \\
            1007 & J093236.58+095025.9 & Primary & 0.049 & $7.2\times10^{10}$ & $5.3\times10^{0}$ & $(1.3\pm0.1)\times10^{9}$ & & $(1.0\pm0.2)\times10^{9}$ & \\
            1008 & J093953.62+034850.2 & Primary & 0.029 & $2.3\times10^{10}$ & $8.8\times10^{-1}$ & $(4.0\pm0.3)\times10^{8}$ & & $(3.9\pm0.7)\times10^{8}$ & \\
            1009 & J101600.20+061505.2 & Primary & 0.046 & $2.1\times10^{11}$ & $1.6\times10^{-1}$ & $<5.5\times10^{8}$ & & $<1.1\times10^{8}$ & \\
            1010 & J101941.29+125034.7 & Primary & 0.033 & $1.8\times10^{10}$ & $1.8\times10^{0}$ & $(5.6\pm0.7)\times10^{8}$ & & $(7.0\pm1.1)\times10^{8}$ & \\
            1011 & J102316.42+115120.4 & Primary & 0.045 & $4.4\times10^{10}$ & $4.6\times10^{0}$ & $(1.4\pm0.1)\times10^{9}$ & $(1.4\pm0.3)\times10^{9}$ & $(1.2\pm0.1)\times10^{9}$ & $<8.5\times10^{8}$ \\
            1012 & J104805.79+060114.4 & Primary & 0.029 & $1.2\times10^{10}$ & $3.6\times10^{-2}$ & $<8.8\times10^{7}$ & & $<9.2\times10^{7}$ & \\
            1013 & J110037.27+102613.9 & Primary & 0.036 & $1.2\times10^{11}$ & $3.5\times10^{-1}$ & $<3.6\times10^{8}$ & & $<8.5\times10^{7}$ & \\
            1014 & J110818.34+131327.5 & Primary & 0.034 & $1.0\times10^{11}$ & $2.0\times10^{-1}$ & $<3.3\times10^{8}$ & & $<1.1\times10^{8}$ & \\
            1015 & J112017.79+041913.3 & Primary & 0.049 & $2.8\times10^{11}$ & $1.3\times10^{-1}$ & $<3.9\times10^{8}$ & & $<3.1\times10^{8}$ & \\
            1016 & J112029.23+040742.1 & Primary & 0.049 & $2.1\times10^{11}$ & $9.6\times10^{-1}$ & $(1.1\pm0.1)\times10^{9}$ & & $(5.1\pm1.4)\times10^{8}$ & \\
            1017 & J112920.69+083608.3 & Primary & 0.027 & $1.9\times10^{10}$ & $1.2\times10^{0}$ & $(4.2\pm0.5)\times10^{8}$ & & $(2.4\pm0.4)\times10^{8}$ & \\
            1018 & J122030.18+112027.3 & Primary & 0.043 & $1.2\times10^{11}$ & $8.2\times10^{-1}$ & $(7.4\pm1.0)\times10^{8}$ & & $(2.4\pm0.6)\times10^{8}$ & \\
            1019 & J123553.79+054539.8 & Primary & 0.042 & $7.8\times10^{10}$ & $3.0\times10^{-1}$ & $<2.5\times10^{8}$ & & $<2.3\times10^{8}$ & \\
            1020 & J123708.06+142426.9 & Primary & 0.031 & $1.9\times10^{10}$ & $4.1\times10^{-2}$ & $<1.1\times10^{8}$ & & $<7.7\times10^{7}$ & \\
            1021 & J124309.36+033452.2 & Primary & 0.049 & $5.8\times10^{10}$ & $7.0\times10^{0}$ & $(2.4\pm0.2)\times10^{9}$ & $(3.4\pm0.6)\times10^{9}$ & $(1.9\pm0.3)\times10^{9}$ & $(8.1\pm2.4)\times10^{8}$ \\
            1022 & J124622.67+115235.7 & Primary & 0.044 & $2.3\times10^{11}$ & $2.2\times10^{-1}$ & $<3.4\times10^{8}$ & & $<2.2\times10^{8}$ & \\
            1023 & J125055.79+031149.3 & Primary & 0.048 & $1.6\times10^{11}$ & $2.3\times10^{-1}$ & $<3.5\times10^{8}$ & & $<2.1\times10^{8}$ & \\
            1024 & J130415.04+091324.4 & Primary & 0.035 & $8.9\times10^{10}$ & $5.3\times10^{0}$ & $(2.0\pm0.1)\times10^{9}$ & $(2.6\pm0.2)\times10^{9}$ & $(1.9\pm0.3)\times10^{9}$ & $(5.5\pm1.0)\times10^{8}$ \\
        \enddata

        \tablecomments{Source IDs, galaxy properties, and CO luminosities for galaxies targeted by AMISS. A full, machine readable version of this table is available as part of the online material -- here we show only the first 25 rows and a limited subset of the columns. Columns are: (1) AMISS source ID; (2) subsample in which the target was included -- primary, star forming (SF), or filler; (3) redshift from the catalog of \citet{saintonge+17} -- CO redshifts are used when available, optical spectroscopic redshifts from SDSS are given otherwise; (4) stellar mass as reported in the xCOLD~GASS catalog; (5) star formation rate reported in the xCOLD~GASS catalog, derived following \citet{janowiecki+17}; (6) CO(1--0) luminosity from the AMISS re-processing of IRAM 30m spectra; (7) CO(1--0) luminosity from new AMISS observations with the ARO 12m; (8) CO(2--1) luminosity from AMISS observations with the SMT; (9) CO(3--2) luminosity from AMISS observations with the SMT. In columns (6)--(9) $3\sigma$ upper limits are given for undetected sources. 
        }

    \end{deluxetable*}

\subsection{CO Disk Sizes} \label{ss:results-disk}

\begin{figure*}[t]
    \centering
    \includegraphics[width=\textwidth]{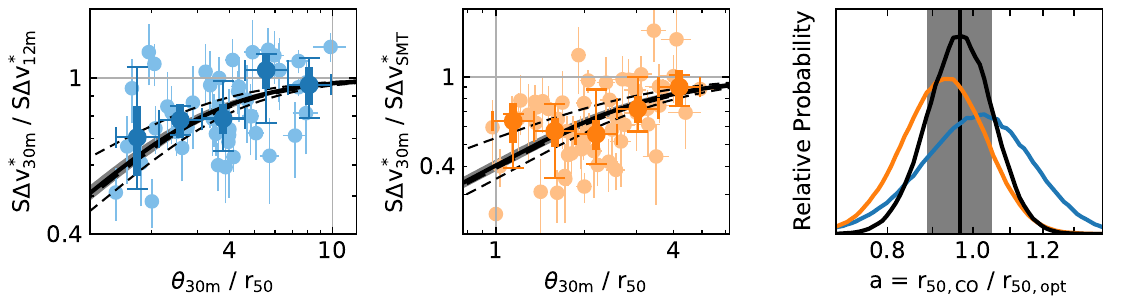}
    \caption{Left: ratios of the integrated CO(1--0) flux (with no aperture correction) from the IRAM 30m and ARO 12m telescopes as a function of the fractional size of the IRAM beam relative to the galaxy optical half-light radius, $r_{50}$. Lighter points give the values for individual galaxies. Dark points and vertical error bars show the median and 16th-84th percentile range for bins along the $x$-axis. The black dashed lines show the expected ratio for an exponential disk with a half light radius equal to the optical half light radius inclined at 10, 45, and 80 degrees. The solid line and filled gray regions show the best fitting exponential disk model with a ratio of CO to optical radii of $0.97\pm0.08$.
    Middle: same as left but for the ratio of CO(2--1) fluxes from the IRAM 30m and SMT. Right: Posterior probability distributions (PDFs) for the ratio between CO and optical half light radii fitted to our telescope-telescope luminosity ratios. Fit results are shown for the CO(1--0) data (blue), CO(2--1) data (orange), and both datasets combined (black). The black vertical line shows the median of the PDF for the fit using all data. The gray filled region shows the $1\sigma$ confidence interval and corresponds to the range of models shown by the gray filled region in the left and center panels.}
    \label{fig:disk_size}
\end{figure*}

With measurements of a single emission line by two telescopes of differing beam size it is possible to constrain the size of the emitting region \citep{elfhag+96,lavezzi+99,french+15}. In practice, deviations from the assumed disk profile, pointing errors, and the limited dynamic range in smaller--larger beam luminosity ratios make this measurement very difficult to realize and interpret for individual galaxies. However, with a large enough galaxy sample measured it may be possible to carry out a statistical study to determine the \textit{average} CO disk size and emission profile.

Ratios between two measurements of the same CO line made with different telescopes can be written as
\begin{equation}\label{eq:telratios}
    \frac{(S\Delta v)^*_1}{(S\Delta v)^*_2} = \frac{\int\int b_1(x^\prime-x,y^\prime-y) s(x,y) dxdy}{\int\int b_2(x^\prime-x,y^\prime-y) s(x,y) dxdy} \times F
\end{equation}
where $s$ is the normalized flux distribution of the source, $(S\Delta v)^*_i$ is the integrated flux seen by telescope $i$ with no aperture correction applied, $b_i$ is the normalized beam power pattern, and $F$ is a potentially unknown offset between the flux scales of the two telescopes. If the beam pattern of each telescope is known, then the ratio can be used to constrain $s$.

In the following, we use ratios of fluxes for galaxies where a given CO transition was detected by both AMISS and xCOLD~GASS. This amounts to 47 galaxies with CO(1--0) detections from both the ARO 12m and IRAM 30m, and 56 galaxies with CO(2--1) detections from the SMT and IRAM 30m. We assume the beam profiles are Gaussians of known width -- $1.06 \lambda/D$ for the 12m, $1.17\lambda/D$ for the SMT, and $1.16 \lambda/D$ for the 30m -- and that the CO emission of each galaxy can be described by an an exponential disk (Equation~\ref{eq:fluxprofile}) with a CO half light radius radius related to the optical half light radius as
\begin{equation}
    r_{50,{\rm CO}} = a r_{50,{\rm opt}} .
\end{equation}

In Figure~\ref{fig:disk_size} we plot 30m--12m and 30m--SMT flux ratios as a function of the ratio between the IRAM 30m beam size and the optical half light radius. When both the IRAM and 12m/SMT beams are small relative to the source, the flux ratio will match the ratio of the area of the 12m/SMT beam and the 30m beam, while for sources which are compact compared to both beams the flux ratio will be one. Between these limits, the ratio will change as the smaller and larger telescope beams couple differently to the source flux. 

The black dashed lines show the expected scaling for exponential disks with $a=1$ and inclinations of 10, 45, and 80 degrees. Individual measurements show a large scatter, however the median ratios in bins along the $x$-axis, fall close to the expected trends. Fitting all 103 ratios with Equation~\ref{eq:telratios} and our exponential disk model gives $a=0.97\pm0.08$.\footnote{Because we have already corrected the xCOLD~GASS flux scale to match AMISS (Section~\ref{sec:xcg}), we impose a tight log-normal prior on $F$, centered at unity and with a width of $0.02$~dex -- roughly the uncertainty in the median 30m--12m and 30m--SMT luminosity ratios used to derive the flux scale correction. We revisit this in Appendix~\ref{ap_ss:xcg_flux_disks}} The solid line and gray filled regions show the expected scaling for the fitted range of $a$ and a 45~degree inclination. 

These results indicate that, on average, CO emission arises from regions with extents comparable to the optical extent of each galaxy. This validates our use of optical sizes for determining aperture corrections, and is consistent with results based on mapped CO disks from \citet{leroy+09,bolatto+17,leroy+22}.

\subsection{Search for Extended CO Reservoirs Beyond the IRAM 30m Beam}\label{ss:results-amiss10}

\begin{deluxetable*}{lll|c|cc|cc}
    \tablecaption{Comparison of CO(1--0) luminosities for two sources overlapping between the xCOLD~GASS, MASCOT, and AMISS CO(1--0) surveys \label{tab:mascot_comp}}
    \tablehead{
        \multicolumn{3}{c}{Source ID} & \colhead{} & \multicolumn{4}{c}{CO(1--0) Luminosity [K km s$^{-1}$ pc$^2$]} \\
        \colhead{AMISS} & \colhead{xCG} & \colhead{MASCOT} & \colhead{r$_{50}$} & \colhead{xCG 30m\tablenotemark{a}} & \colhead{AMISS 30m\tablenotemark{a}} & \colhead{MASCOT 12m} & \colhead{AMISS 12m} 
        }
    
    \startdata
        9177 & 32619 & 8987-3701 & 2.8" & $(4.9\pm1.0) \times 10^8$ & $(5.6\pm0.6) \times 10^8$ & $(3.3\pm1.1) \times 10^8$ & $(6.5\pm1.2) \times 10^8$ \\
        9175 & 31775 & 9491-6101 & 4.5" & $(14.2\pm2.6) \times 10^8$ & $(15.5\pm1.2) \times 10^8$ & $(20.4\pm2.2) \times 10^8$ & $(17.4\pm3.3) \times 10^8$ \\
    \enddata
    \tablenotetext{a}{We report CO luminosities from \citet{saintonge+17} our own recalibration of the same 30m data separately}
\end{deluxetable*}

\begin{figure*}
    \centering
    \includegraphics[width=\textwidth]{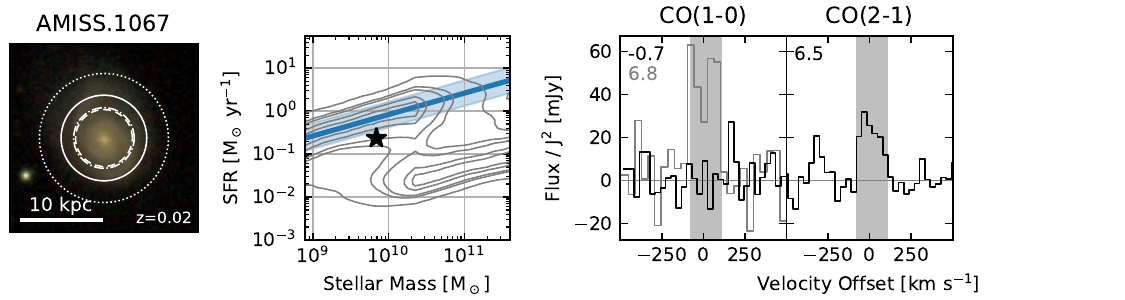}
    \caption{Similar to Figure~\ref{fig:spectra} for AMISS.1067, a source with no CO(1--0) detection from xCOLD~GASS but $6.5\sigma$ detections in both CO(1--0) and CO(2--1) from AMISS.}
    \label{fig:amiss1067}
\end{figure*}

\citet{wylezalek+22} compared IRAM 30m and ARO 12m CO(1--0) observations of two galaxies shared between the xCOLD~GASS and MASCOT surveys, and found some evidence to suggest that the 30m measurement underestimated the luminosity of the larger galaxy of the pair (xCOLD~GASS 31775). This small comparison sample and modest signal to noise did not allow definitive conclusions, but \citet{wylezalek+22} suggest that some galaxies may contain significant gas reservoirs beyond the extent of the IRAM beam, which are not accounted for by aperture corrections.

Our CO(1--0) observations allow us to explore this possibility over a much larger sample. In Section~\ref{ss:errors-othertels} we find that the beam corrected CO(1--0) luminosities determined in 30m and 12m observations generally agree within their expected errors. We find only one source for which the 12m luminosity exceeds the 30m luminosity at more than $2\sigma$ significance (comparable to the significance of the difference found in \citealt{wylezalek+22}). For our sample of 53 galaxies we would expect one or two $2\sigma$ differences due to noise alone. 

We obtained additonal ARO~12m observations of both MASCOT-xCOLD~GASS overlap sources as part of our CO(1--0) survey. Table~\ref{tab:mascot_comp} lists the CO(1--0) luminosities reported in \citet{saintonge+17} and \citet{wylezalek+22}, along with our re-calibration of the IRAM 30m luminosity and our new 12m measurement. Our new measurement for xCOLD~GASS 31775 sits between the xCOLD~GASS and MASCOT measurements, and is consistent with both within their respective uncertainties. The difference between the 30m and 12m luminosities found for this source in \citet{wylezalek+22} can likely be attributed to the 12m and 30m measurements that are respectively around $+1\sigma$ and $-1\sigma$ from the true value.

Our sample includes one galaxy detected in CO(1--0) with the ARO~12m but not the IRAM 30m -- AMISS.1067/xCOLD~GASS 113024. For this source we found a CO(1--0) luminosity of $(1.4\pm0.2)\times10^8$~K~km~s$^{-1}$~pc$^2$ with the 12m, and a CO(2--1) luminosity of $(0.7\pm0.1)\times10^8$~K~km~s$^{-1}$~pc$^2$ with the SMT. On the other hand the 30m CO(1--0) observations give a $3\sigma$ upper limit of $0.6\times10^8$~K~km~s$^{-1}$~pc$^2$ -- incompatible with the 12m result at $5\sigma$ significance. We show an image of the source and the CO spectra in Figure~\ref{fig:amiss1067}. The 12m CO(1--0) luminosity would imply a $\sim 6\times10^8$~M$_\odot$ reservoir of molecular gas lying in the outskirts of the galaxy. The optical image shows no distinctive features in this region. We believe the most plausible explanation of the 30m non-detection is an error in the observations such as a mispointing of the telescope, rather than an unusual gas distribution in the galaxy.

Among 52 galaxies detected in CO(1--0) with both the IRAM 30m and ARO 12m telescopes we find little evidence for large reservoirs of molecular gas in the outskirts of galaxies. These results, along with our findings in Section~\ref{ss:results-disk}, suggest that single pointing observations can accurately recover the total CO emission of galaxies with optical half light radii comparable or smaller than the full width at half maximum of the beam using straightforward aperture corrections. Individual sources which diverge from these expectations may exist, but are rare, and will not bias statistical studies of gas properties in large samples.

\subsection{CO Line Ratios} \label{ss:results-ratios}

\begin{figure*}
    \centering
    \includegraphics[width=\textwidth]{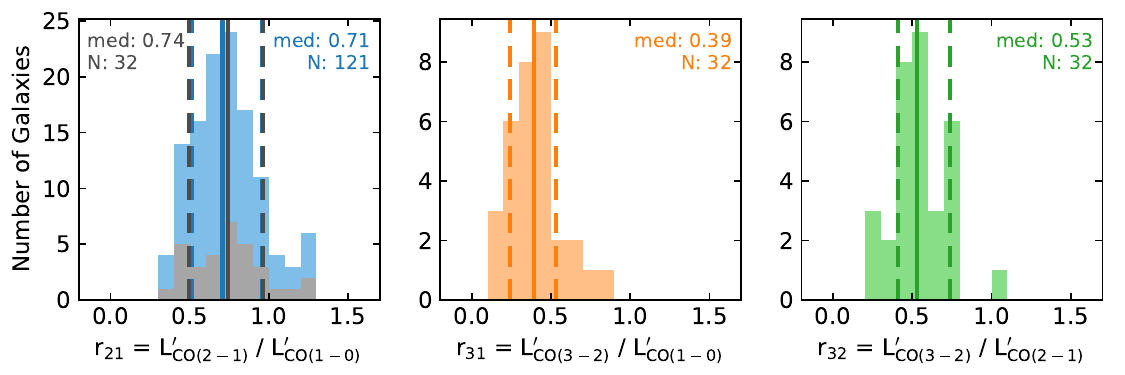}
    \caption{The distribution of $r_{21}$ (left), $r_{31}$ (center), and $r_{32}$ (right) ratios measured by AMISS. Solid vertical lines indicate the median values, and dashed lines show the 16th to 84th percentile ranges among each sample. For $r_{21}$ we show the distribution for all galaxies detected in CO(2--1) and CO(1--0) in blue, and the distribution for the galaxies also detected in CO(3--2) (i.e. those shown in the $r_{31}$ and $r_{32}$ panels) in grey.}
    \label{fig:ratios}
\end{figure*}

A key goal of AMISS is to determine the ratios between CO(1--0), CO(2--1), and CO(3--2) for a representative sample of galaxies. We define the ratio of luminosities between CO transitions with upper energy levels $j$ and $k$ as
\begin{equation}
    r_{jk} = \frac{L_{{\rm CO}(j-(j-1))}^\prime}{L_{{\rm CO}(k-(k-1))}^\prime} .
\end{equation}
Figure~\ref{fig:ratios} shows the distribution of $r_{21}$, $r_{31}$, and $r_{32}$ for galaxies detected at $3\sigma$ in both of the relevant lines. For $r_{21}$ we measure a median of $0.71$ and a 16th to 84th percentile range of $0.51$ to $0.96$. For $r_{31}$ we measure a median of $0.39$ and a range of $0.24$ to $0.53$, and for $r_{32}$ we measure a median of $0.53$ and a range of $0.41$ to $0.74$. These results suggest that the low-$J$ transitions of CO are subthermally excited in star forming galaxies in the nearby universe, and differ from the high excitation for these transitions found in infrared selected samples \citep[e.g.][]{papadopoulos+12,montoya-arroyave+23}.

Our median $r_{32}$ value is consistent with the recent study by \citet{leroy+22} which found a median $r_{32}=0.50$ using resolved maps of disk galaxies in the local universe. Our $r_{21}$ and $r_{31}$ values are higher than medians of $r_{21}=0.61$ and $r_{31}=0.31$ reported by \citet{leroy+22}, as well as $r_{21}=0.59$ from a literature survey in \citet{denbrok+21}, but lower than the $r_{31}=0.49$ reported by \citet{lamperti+20}. Our results are in excellent agreement with the $r_{21}=0.71$ found in \citet{saintonge+17} using roughly beam-matched IRAM 30m CO(1--0) and APEX CO(2--1) data for 28 xCOLD~GASS galaxies.\footnote{We have rescaled the \citet{saintonge+17} and \citet{lamperti+20} ratios to match our re-calibration of the IRAM 30m CO(1--0) data, which is used in all three of our studies.} Discrepancies between these studies may be attributable to differences in sample selection, with many recent works finding that line ratios vary systematically with galaxy properties \citep{kamenetzky+16,lamperti+20,leroy+22}, or to residual differences in calibration between the various surveys. We explore the CO line ratios in detail in the companion paper \citet{keenan+24b}.


\section{Conclusion}\label{sec:conclusion}

The Arizona Molecular ISM Survey with the SMT (AMISS), a multi-line study of sub/millimeter carbon monoxide emission in 176 $z\sim0$ galaxies. The project was designed to enable precise study of CO line excitation and molecular gas conditions through measurements of ratios between CO(1--0), CO(2--1), and CO(3--2). To accomplish these objectives, careful sample selection and data calibration are required. 
Here we have introduced the survey and described the steps taken to ensure the necessary data quality. A companion paper \citep{keenan+24b} provides an in depth study of variations in CO line ratios across the galaxy population.

AMISS represents two key advances over previous multi-line surveys targeting the lowest CO transitions:
\begin{enumerate}[noitemsep,nolistsep]
    \item The survey has been carried out with a uniform observing methodology, utilizing only one telescope per transition, and where possible measuring each line with beams of closely matched sizes. This simplifies many of the challenges related to calibration of millimeter data, and makes it possible to study luminosity ratios with limited dynamic range while avoiding significant uncertainties involved in combining inhomogeneous datasets.
    \item The AMISS sample builds upon the xCOLD~GASS sample to provide multi-line data for a representative sample of nearby galaxies spanning stellar masses from $10^9$ to $10^{11.5}$~M$_\odot$ and star formation rates from $0.001$ to $40$~M$_\odot$~yr$^{-1}$. Targets from our primary sample are selected only on the basis of galaxy mass, with no reference to properties such as IR luminosity, eliminating potential biases related to targeting intrinsically CO luminous sources.
\end{enumerate}
We make our CO line catalog publicly available as a resource for future studies. To our knowledge, this represents the largest homogeneous sample of galaxy-integrated multi-line $^{12}$CO observations for nearby galaxies.

In the paper we have provided a detailed overview of our survey design, sample selection, and data processing. In addition to our catalog, key results of this work are the following:
\begin{enumerate}[noitemsep,nolistsep]
    \item We have provided an exhaustive accounting of statistical and systematic uncertainties in our data. Aperture corrections required to account for flux falling outside the primary beam of our single-pointing observations represent the largest source of systematic uncertainty, but are less than $20\%$ for all galaxies and typically less 6\%. Relative calibration errors across our survey are at the level of $\sim 5\%$, and the uncertainty in the flux scale is also $\sim5\%$.
    \item The spatial distribution of CO emission is, on average, well described by an exponential disk of comparable size to the stellar disk of each galaxy. The ratio between CO half light radii and r-band half light radii is $0.97\pm0.08$. 
    \item For 56 galaxies we have CO(1--0) measurements from both the IRAM 30m and ARO 12m telescopes. The 12m beam covers an area $\sim 5$ times larger than that of the 30m, making it possible to search for extended gas disks or molecular gas located away from the optical galaxy. We find no evidence for such features. Instead we conclude that observations covering as little as half the optical diameter of a galaxy can, on average, recover the integrated CO luminosity of a galaxy via simple aperture corrections.
    \item We report CO line ratios for our sample of $r_{21}=0.71$, $r_{31}=0.39$, and $r_{32}=0.53$, indicating that the low-$J$ CO transitions are subthermally excited in typical star forming galaxies.
\end{enumerate}


\acknowledgments

We would like to thank the Arizona Radio Observatory operators, engineering and management staff, without whom this research would not have been possible. We would also like to thank the anonymous referee, whose feedback helped clarify and improve many aspects of this paper.

RPK was supported by the National Science Foundation through Graduate Research Fellowship grant DGE-1746060. DPM, RPK, and ECM were supported by the National Science Foundation through CAREER grant AST-1653228.

This paper makes use of data collected by the the UArizona ARO Submillimeter Telescope and the UArizona ARO 12-meter Telescope, the IRAM 30m telescope, and the Sloan Digital Sky Survey. The UArizona ARO 12m Telescope on Kitt Peak  on Mt. Graham are operated by the Arizona Radio Observatory (ARO), Steward Observatory, University of Arizona. IRAM is supported by INSU/CNRS (France), MPG (Germany) and IGN (Spain). Funding for the SDSS and SDSS-II was provided by the Alfred P. Sloan Foundation, the Participating Institutions, the National Science Foundation, the U.S. Department of Energy, the National Aeronautics and Space Administration, the Japanese Monbukagakusho, the Max Planck Society, and the Higher Education Funding Council for England. The SDSS Web Site is \url{www.sdss.org}.


\appendix

\section{The IRAM 30m Flux Scale}\label{ap:xcg_flux}

In Section~\ref{sec:xcg} we rescale the fluxes of the IRAM 30m spectra from the xCOLD~GASS to account for an apparent difference in the flux scales of the ARO telescopes and the IRAM 30m. We find three lines of evidence that these corrections are necessary: 
\begin{enumerate}[noitemsep,nolistsep]
    \item When no flux scale correction is applied to the 30m data, a comparison of CO luminosities measured at the 12m and SMT with those measured at the 30m reveals that 30m measurements are systematically low, even after accounting for differences in source-beam coupling (Section~\ref{ap_ss:xcg_flux_comparisons});
    \item $r_{21}$ line ratios determined using SMT CO(2--1) and unscaled 30m CO(1--0) data are consistently higher than values reported in the literature, with many individual ratios outside the theoretical range for emission from the cold ISM (Section~\ref{ap_ss:xcg_flux_r21});
    \item Repeating the disk size analysis in Section~\ref{ss:results-disk} but fitting for the flux scale offset between telescopes (i.e. removing the tight prior on the $F$ parameter) recovers a flux scale ratio consistent with our applied corrections (Section~\ref{ap_ss:xcg_flux_disks}).
\end{enumerate}
We discuss   these considerations in detail in the following subsections.

\begin{figure*}
    \centering
    \includegraphics[width=\textwidth]{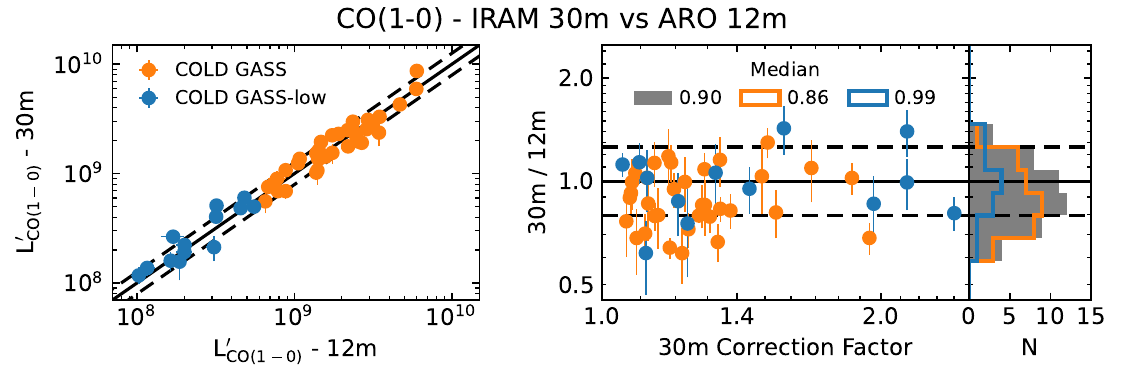}
    \includegraphics[width=\textwidth]{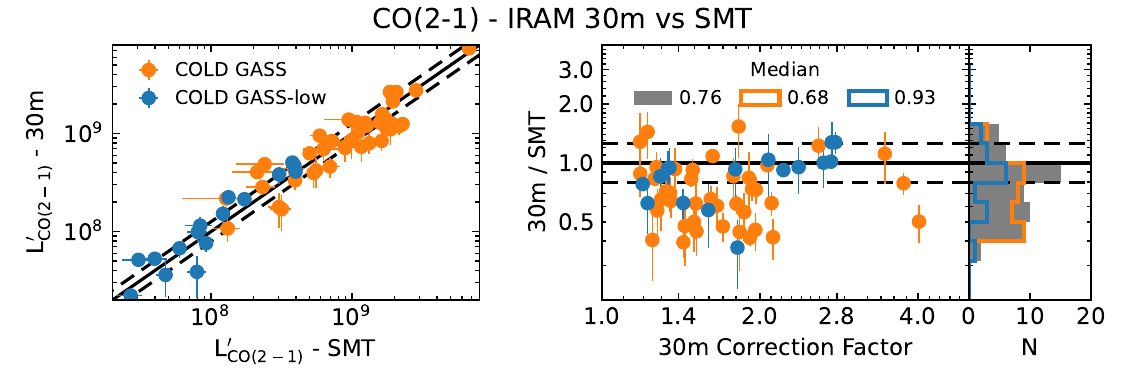}
    \caption{Comparison of the ARO 12m and IRAM 30m CO(1--0) luminosities (top) or SMT and IRAM 30m CO(2--1) luminosities (bottom) for sources observed with both telescopes. Wobbler and aperture corrections have been applied to all luminosities, but we have removed the flux scale correction applied to the 30m data in Section~\ref{sec:xcg}.
    Left: The correlation between luminosities measured with the two telescopes. The black solid line corresponds to a one-to-one ratio, while dashed lines are at $\pm0.1$~dex. The sample is divided between high mass galaxies observed at the 30m in the original COLD~GASS survey, and lower mass galaxies observed by the follow up COLD~GASS-low survey with slightly different observing parameters.
    Right: The distribution of 30m-12m (CO(1--0)) and 30m-SMT (CO(2--1)) luminosity ratios. We also plot the ratio against the aperture correction factor applied to the 30m data. We find no obvious trends between correction factor and the luminosity ratio. Error bars in both panels show statistical uncertainties of the luminosities/ratios but not systematic uncertainty.}
    \label{fig:12m30m_comp}
\end{figure*}

\subsection{Comparison of IRAM 30m and AMISS CO Luminosities}\label{ap_ss:xcg_flux_comparisons}

Figure~\ref{fig:12m30m_comp} compares CO luminosities for sources observed in CO(1--0) by both the IRAM 30m and ARO 12m telescopes (upper panels) or in CO(2--1) by both the 30m and the SMT (lower panels). In this comparison we apply the wobbler and source-beam coupling corrections discussed in Section~\ref{sec:xcg}, but do not apply any additional correction to the 30m flux scale. We compute ratios between 30m and 12m/SMT luminosities and find median (geometric mean) values of $0.90\pm0.03$ ($0.89\pm0.02$) for CO(1--0) and $0.76\pm0.04$ ($0.81\pm0.01$) for CO(2--1) indicating that the 30m luminosities are systematically lower than those of the ARO telescopes. The right panels of Figure~\ref{fig:12m30m_comp} show the distribution of ratios for each telescope pair, and plots them as a function of the aperture corrections applied to the 30m data. Because the beams for each telescope pair differ in size, a systematic error in our aperture corrections could cause the luminosity ratios to differ from 1.0. In this case, ratios would trend towards unity as the source sizes (and aperture correction) decrease. We instead find no appreciable trend, and low ratios even for galaxies with very small corrections, indicating that our aperture corrections are unlikely to be responsible for the discrepancy between the 30m data and our own. Restricting our comparison to only galaxies with correction factors less than 1.3 gives median 30m--AMISS luminosity ratios of $0.87\pm0.04$ for CO(1--0) and $0.83\pm0.11$ for CO(2--1) -- consistent with the values computed for all galaxies.

\subsection{Physical and Observational Expectations for CO Line Ratios}\label{ap_ss:xcg_flux_r21}

When no rescaling is applied to the 30m data, $r_{21}$ values computed using SMT CO(2--1) and 30m CO(1--0) luminosities have a median of 0.8 and a significant fraction of results lie above 1.0. CO line ratios may vary as a function of galaxy properties, and we find evidence of significant variations within the AMISS sample \citep{keenan+24b}. However, such systematically high ratios are difficult to reconcile with theory or other observations. For optically thick CO emission, the maximum $r_{21}$ is 1.0 when the gas is thermally excited with a relatively high temperature. \citet{leroy+22} point out that for cold gas, $r_{21}<1.0$ even in local thermal equilibrium, with $r_{21}\sim0.85$ for $T=20$~K. Hydrodynamical simulations of molecular clouds, post-processed to account for radiative transfer, suggest typical $r_{21}$ values between 0.6 and 0.8, and very rarely greater than 1.0 \citep{penaloza+17,penaloza+18,gong+20}. These expectations are born out by observations. \citet{denbrok+21} measure $r_{21}$ in a compilation single pointing observations of 81 disk galaxies, finding $r_{21}>1.0$ in only 5 cases. \citet{yajima+21}, \citet{denbrok+21}, and \citet{leroy+22} each study $r_{21}$ in resolved maps of nearby galaxies, finding typical $r_{21}$ of 0.6 to 0.7 on galaxy integrated scales, and identifying only a small fraction of individual map cells with $r_{21}>1.0$. \citet{leroy+22} compiled CO line ratios from many recent studies. Those using xCOLD~GASS CO(1--0) data lie at the upper end of the range of observations for both $r_{21}$ and $r_{31}$ (e.g., their Figure~4), consistent with an underestimation of the CO(1--0) luminoisites measured with the 30m.

\subsection{CO Disk Size Fits Including an Unknown Flux Offset}\label{ap_ss:xcg_flux_disks}

\begin{figure}
    \centering
    \includegraphics{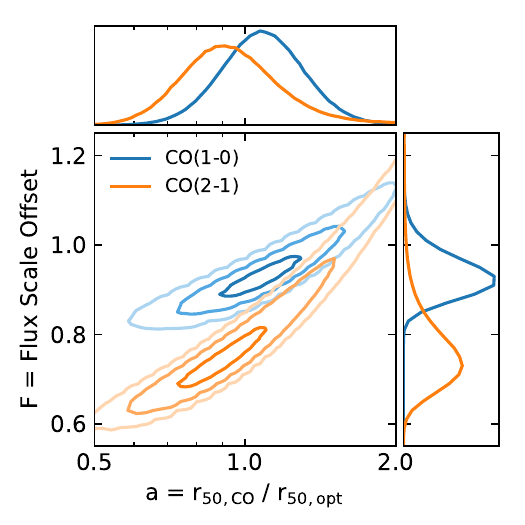}
    \caption{The posterior probability distribution for our simultaneous fit of the CO-to-optical disk size ratio ($a$) and the flux scale offset between the 30m and the  12m/SMT ($F$). Results for CO(1--0) (30m vs 12m) and CO(2--1) (30m vs SMT) are shown in blue and orange respectively. The main panel shows the two dimensional probability distributions with contours for the 1, 2, and 3$\sigma$ confidence regions. Small panels on the edges show the marginalized distribution for each parameter.}
    \label{fig:disk_and_flux}
\end{figure}

The disk size fits performed in Section~\ref{ss:results-disk} allow for an unknown multiplicative offset between flux scales of the telescopes ($F$). In Section~\ref{ss:results-disk} we placed a tight prior on $F$ based on the median luminosity ratios reported above. Here we re-run the fits, using un-scaled 30m luminosities and a uniform prior on $\log F$ in the range $F<1.5$ (this upper limit is necessary to exclude a tail of unphysically large solutions). Figure~\ref{fig:disk_and_flux} shows the posterior probability distribution for these fits. Best fitting CO--optical disk size ratios are still consistent with unity. The best fit ($1\sigma$ range) scale factors are $0.93$ ($0.89$--$0.98$) for the 30m--12m CO(1--0) flux scale ratio and $0.76$ ($0.69$--$0.86$) for the 30m--SMT CO(2--1) flux scale ratio. These values are in agreement with the median 30m--12m and 30m--SMT luminosity ratios used to derive the correction factors in Section~\ref{sec:xcg}. On the other hand, assuming that no flux scale correction (i.e. forcing $F=1$) would require CO--optical size ratios that are significantly larger than those found in resolved CO maps of nearby galaxies \citep{leroy+09}.

\subsection{Conclusion and Application of Corrections}\label{ap_ss:xcg_flux_conclusion}

Based on the above considerations, we elect to scale the 30m fluxes such that the median luminosity ratios between CO lines observed by both the 30m and the 12m or SMT are unity. This results in scale factors of 1.12 for CO(1--0) and 1.32 for CO(2--1) spectra from the 30m. We choose to scale the 30m data (rather than the 12m and SMT data) because our detailed review of ARO calibration procedures (Section~\ref{ss:errors-fluxcal}) and the good match between SMT data and samples from APEX and JCMT (Section~\ref{ss:errors-othertels}) gives us a higher degree of confidence in the ARO calibrations.

We have reviewed the calibration procedures for all three telescopes, and have been unable to conclusively identify a cause of the discrepancy in fluxes. It seems likely that differences in procedures in determining the flux scales at the facilities play a role. In particular we note that point source sensitivities for the 30m are derived from observations made during the commissioning of the EMIR receiver\footnote{https://publicwiki.iram.es/Iram30mEfficiencies, version dated 2016-11-03}, and pre-date the most recent generation of planet models used for flux calibration at the ARO. This may introduce differences of order 10\%. In our analysis of SMT beam efficiencies, we also find that the beam efficiency can change by $\sim$5\% as a result adjustments made to the telescope in the course of routine maintenance. If a similar variability exists for the 30m, then the documented point source sensitivities may have changed between the commissioning of the 30m receiver and the conclusion of xCOLD~GASS observations.

Interestingly, 30m observations for galaxies above and below $M_*\sim10^{10}$~M$_\odot$ were carried out in separate projects at different time periods and with different backends. Considering the data for the two COLD~GASS samples separately, we find that the median ratios between the 30m and 12m/SMT measurements are closer to unity for the the low mass sources observed later on (Figure~\ref{fig:12m30m_comp}), providing some support for the idea that 30m beam efficiency has varied over time.

In this work and its companion paper \citep{keenan+24b}, we primarily use the xCOLD~GASS CO(1--0) observations. The choice to re-scale the CO(1--0) data affects the normalization of trends between CO luminosities and/or line ratios and other galaxy properties, but not the slopes of these trends. This has the greatest impact on comparisons of AMISS CO line ratios with theoretical and observational results from the literature (e.g., Section~\ref{ap_ss:xcg_flux_r21}), where the resultant decrease in the average $r_{21}$ brings our data into much better agreement with other results. On the other hand, our findings in \citet{keenan+24b} of a correlation between $r_{21}$ and galaxy properties such as SFR and of different slopes for e.g., the Kennicutt-Schmidt law as measured with CO(2--1) and CO(1--0) are unaffected.

\section{Baseline Error for Platform Fits}\label{ap:blerr}

Our integrated fluxes are determined by summing the total flux within a window containing our spectral line. The zero-level of each spectrum is not perfectly known, but determined by assuming that the flux outside the integration window is zero. This zero level has its own uncertainty, which contributes to the total error in our measurements. Formally,
$$S_\nu 
    = \frac{\sum_{\rm win} S_{\nu,{\rm ch}}}{N_{\rm win}} -  \frac{\sum_{\rm bl} S_{\nu,{\rm ch}}}{N_{\rm bl}}$$
where baseline fitting and removal sets the second term equal to zero. The uncertainty is then
$$\sigma_{S_\nu} 
    = \sqrt{\sigma_{\rm win}^2 + \sigma_{\rm bl}^2}$$
with $\sigma_{\rm win} = \sigma_{\rm ch} N_{\rm win}^{-1/2}$, $\sigma_{\rm bl} = \sigma_{\rm ch} N_{\rm bl}^{-1/2}$. The resulting uncertainty in our line fluxes is 
$$\sigma_{S_\nu} 
    = \sigma_{\rm ch} \sqrt{\frac{1}{N_{\rm win}} + \frac{1}{N_{\rm bl}}} 
    = \sigma_{\rm ch} \sqrt{\frac{N_{\rm win}+N_{\rm bl}}{N_{\rm win}N_{\rm bl}}}$$
we are interested in the integrated flux, $S\Delta v = S_\nu N_{\rm win} \Delta v_{\rm ch}$, which has an uncertainty given by $\sigma_{S\Delta v} = \sigma_{S_\nu} N_{\rm win} \Delta v_{\rm ch}$ so we have
$$\sigma_{S\Delta v} 
    = \sigma_{\rm ch} \Delta v_{\rm ch} \sqrt{\frac{N_{\rm win}^2+N_{\rm win}N_{\rm bl}}{N_{\rm bl}}}$$

For $N_{\rm win}<<N_{\rm bl}$ this reduces to $\sigma_{S\Delta v} = \sigma_{\rm ch} \Delta v_{\rm ch} \sqrt{N_{\rm win}}$, which is the uncertainty often quoted for integrated fluxes. However for $N_{\rm win}\sim N_{\rm bl}$ the uncertainty on the integrated flux rises. For example, in the case that $N_{\rm win}=N_{\rm bl}$ we have $\sigma_{S\Delta v} = \sigma_{\rm ch} \Delta v_{\rm ch} \sqrt{2 N_{\rm win}}$.

Fitting a platforming baseline introduces an additional complication since the zero levels of the upper and lower half of the spectrum are now computed independently. Splitting our spectra into upper ($U$) and lower ($L$) parts, we can modify the above derivation as follows
$$S_\nu 
    = \frac{N_{\rm win,U}}{N_{\rm win}}\Big(\frac{\sum_{\rm win,U} S_{\nu,{\rm ch}}}{N_{\rm win,U}} -  \frac{\sum_{\rm bl,U} S_{\nu,{\rm ch}}}{N_{\rm bl,U}}\Big) + \frac{N_{\rm win,L}}{N_{\rm win}}\Big(\frac{\sum_{\rm win,L} S_{\nu,{\rm ch}}}{N_{\rm win,L}} -  \frac{\sum_{\rm bl,L} S_{\nu,{\rm ch}}}{N_{\rm bl,L}}\Big)$$
$$\sigma_{S_\nu}
    = \sqrt{\frac{N_{\rm win,U}^2}{N_{\rm win}^2}(\sigma_{\rm win,U}^2 + \sigma_{\rm bl,U}^2) + \frac{N_{\rm win,L}^2}{N_{\rm win}^2}(\sigma_{\rm win,L}^2 + \sigma_{\rm bl,L}^2)} 
    = \frac{\sigma_{\rm ch}}{N_{\rm win}} \sqrt{N_{\rm win} + \frac{N_{\rm win,U}^2}{N_{\rm bl,U}} + \frac{N_{\rm win,L}^2}{N_{\rm bl,L}}}$$
$$\sigma_{S\Delta v} = \sigma_{\rm ch}\Delta v_{\rm ch} \sqrt{N_{\rm win} + \frac{N_{\rm win,U}^2}{N_{\rm bl,U}} + \frac{N_{\rm win,L}^2}{N_{\rm bl,L}}}$$

In the case that $N_{\rm win,L}=N_{\rm win,U}$ and $N_{\rm bl,U}=N_{\rm bl,L}$ (i.e. the window is symmetric around the center of the bandpass), this reduces to the same result as the single baseline case.



\bibliography{refs}{}
\bibliographystyle{aasjournal}

\end{document}